\documentclass[letterpaper,11pt]{article}
\usepackage[margin=0.9in]{geometry}
\usepackage{graphicx}
\usepackage{microtype}
\usepackage{booktabs}
\usepackage{cmap}
\usepackage{multirow}
\usepackage{amssymb}
\usepackage[utf8]{inputenc}
\usepackage[english]{babel}
\usepackage{algorithm}
\usepackage{amssymb}
\usepackage[noend]{algpseudocode}
\usepackage[all]{xy}

\usepackage{pifont}
\usepackage{makecell}
\usepackage{mathtools}
\usepackage{amsthm, enumitem}
\usepackage[T1]{fontenc}
\usepackage{amsmath}
\usepackage{amsfonts}
\usepackage{bm}
\usepackage[dvipsnames]{xcolor}
\definecolor{blueviolet}{rgb}{0.2, 0.2, 0.6}
\definecolor{webgreen}{rgb}{0,.5,0}
\definecolor{webbrown}{rgb}{.6,0,0}
\usepackage{setspace}
\usepackage[pdftex,
	bookmarks=false,
	colorlinks=true,
    allcolors=blue,
	urlcolor=blue, 
	citecolor=webgreen,
	pdfstartpage=1,
	pdfstartview={FitH},  % FitBH
	bookmarksopen=false,
	hypertexnames=false
	]{hyperref}
\allowdisplaybreaks
\usepackage{tikz}
\usepackage{braket}
\usepackage[numbers,sort&compress]{natbib}
\usepackage{amsthm}
\usepackage{dsfont}
\usepackage{listings}
\usepackage[capitalize]{cleveref}
\usepackage{appendix}
\newcommand{\CNOT}{\text{CNOT}}

\newcommand{\refapp}[1]{\hyperref[#1]{\ref*{#1}}}
\newcommand{\plog}{p_\mathrm{log}}
\newcommand{\Tdec}{T_{\mathrm{dec}}}

\usepackage{authblk}

\numberwithin{equation}{section}

\numberwithin{lemc}{section}
\numberwithin{propc}{section}
\numberwithin{coroc}{section}

\newcommand{\LT}{\operatorname{LT}}

\newtheorem{theorem}{Theorem}[section]
\newtheorem{corollary}[theorem]{Corollary}

\newtheorem{definition}[theorem]{Definition}
\newtheorem{lemma}[theorem]{Lemma}
\newtheorem{proposition}[theorem]{Proposition}

\theoremstyle{definition}
\newtheorem{remark}[theorem]{Remark}
\usepackage{tikz}
\usepackage{amsmath,amssymb,amsbsy}
\usepackage{bm}
\usepackage[all]{xy}
\usepackage{xcolor}
\usetikzlibrary{arrows.meta, patterns, shadings, decorations.pathmorphing, 3d, calc}

\definecolor{moatfill}{RGB}{255, 230, 230}
\definecolor{moatedge}{RGB}{179, 0, 0}
\definecolor{kclustfill}{RGB}{255, 204, 204}
\definecolor{kclustedge}{RGB}{179, 0, 0}
\definecolor{jclustfill}{RGB}{255, 230, 178}
\definecolor{jclustedge}{RGB}{179, 115, 0}
\definecolor{jmoatfill}{RGB}{255, 245, 220}

\definecolor{msgblue}{RGB}{200, 210, 248}
\definecolor{msgblueedge}{RGB}{120, 135, 220}
\definecolor{msgbluedark}{RGB}{60, 70, 160}
\definecolor{msggreen}{RGB}{200, 228, 200}
\definecolor{msggreenedge}{RGB}{90, 155, 90}
\definecolor{msgred}{RGB}{248, 200, 200}
\definecolor{msgrededge}{RGB}{220, 120, 120}
\definecolor{msgoverlap}{RGB}{215, 195, 245}
\definecolor{msgoverlapedge}{RGB}{135, 100, 185}
\definecolor{msgteal}{RGB}{190, 228, 225}
\definecolor{msgtealedge}{RGB}{60, 145, 135}
\definecolor{msggray}{RGB}{160, 160, 160}
\definecolor{msggrayedge}{RGB}{110, 110, 110}

\definecolor{lblblue}{RGB}{45, 70, 155}
\definecolor{lblgreen}{RGB}{45, 115, 50}
\definecolor{lblpurple}{RGB}{135, 100, 185}
\definecolor{lblviolet}{RGB}{135, 100, 185}
\definecolor{lblred}{RGB}{175, 55, 55}

\definecolor{votefill}{RGB}{195, 225, 240}
\definecolor{voteedge}{RGB}{160, 160, 160}
\definecolor{lattice}{RGB}{128, 128, 128}

\definecolor{anyoncore}{RGB}{230, 230, 230}
\definecolor{anyonmid}{RGB}{220, 140, 30}
\definecolor{anyonrim}{RGB}{180, 100, 0}

\definecolor{moatfill}{RGB}{255, 230, 230}
\definecolor{moatedge}{RGB}{179, 0, 0}
\definecolor{kclustfill}{RGB}{255, 204, 204}
\definecolor{kclustedge}{RGB}{179, 0, 0}
\definecolor{jclustfill}{RGB}{255, 230, 178}
\definecolor{jclustedge}{RGB}{179, 115, 0}
\definecolor{jmoatfill}{RGB}{255, 245, 220}

\definecolor{msgblue}{RGB}{200, 210, 248}
\definecolor{msgblueedge}{RGB}{120, 135, 220}
\definecolor{msgbluedark}{RGB}{60, 70, 160}
\definecolor{msggreen}{RGB}{200, 228, 200}
\definecolor{msggreenedge}{RGB}{90, 155, 90}
\definecolor{msgred}{RGB}{248, 200, 200}
\definecolor{msgrededge}{RGB}{220, 120, 120}
\definecolor{msgoverlap}{RGB}{215, 195, 245}
\definecolor{msgoverlapedge}{RGB}{135, 100, 185}
\definecolor{msgteal}{RGB}{190, 228, 225}
\definecolor{msgtealedge}{RGB}{60, 145, 135}
\definecolor{msggray}{RGB}{160, 160, 160}
\definecolor{msggrayedge}{RGB}{110, 110, 110}

\definecolor{lblblue}{RGB}{45, 70, 155}
\definecolor{lblgreen}{RGB}{45, 115, 50}
\definecolor{lblpurple}{RGB}{135, 100, 185}
\definecolor{lblviolet}{RGB}{135, 100, 185}
\definecolor{lblred}{RGB}{175, 55, 55}

\definecolor{votefill}{RGB}{195, 225, 240}
\definecolor{voteedge}{RGB}{160, 160, 160}
\definecolor{lattice}{RGB}{128, 128, 128}

\definecolor{anyoncore}{RGB}{230, 230, 230}
\definecolor{anyonmid}{RGB}{220, 140, 30}
\definecolor{anyonrim}{RGB}{180, 100, 0}

\definecolor{stabxy}{RGB}{100, 100, 210}
\definecolor{stabxz}{RGB}{95, 165, 95}
\definecolor{stabyz}{RGB}{210, 100, 100}
\definecolor{stabcube}{RGB}{255, 230, 178}
\definecolor{stabcubeedge}{RGB}{179, 115, 0}

\tikzset{bubble/.style={fill=white, draw=black!60, rounded corners=3pt, inner sep=3pt, line width=0.6pt, text=black}}

\newcommand{\blockarrow}[5]{
  \begin{scope}[shift={(#2,#3)}, rotate=#4]
    \fill[white, draw=#1, line width=1.2pt]
      (0, 0.07) -- ({#5-0.28}, 0.07) -- ({#5-0.28}, 0.17) -- (#5, 0) -- ({#5-0.28}, -0.17) -- ({#5-0.28}, -0.07) -- (0, -0.07) -- cycle;
  \end{scope}
}

\newcommand{\anyon}{
  \shade[inner color=white!25, outer color=black!68] (0,0) circle (7pt);
  \draw[black, line width=0.6pt] (0,0) circle (7pt);
}

\newcommand{\stripedrect}[6]{
  \begin{scope}
    \clip (#3, #4) rectangle (#5, #6);
    \fill[#1] (#3, #4) rectangle (#5, #6);
    \foreach \i in {-20,-19,...,20} {
      \fill[#2] ({#3 + \i*0.2 - 0.05}, {#4 - 0.5}) -- ({#3 + \i*0.2 + 0.05}, {#4 - 0.5})
        -- ({#3 + \i*0.2 + 0.05 + 1.5}, {#6 + 0.5}) -- ({#3 + \i*0.2 - 0.05 + 1.5}, {#6 + 0.5}) -- cycle;
    }
  \end{scope}
}

\newcommand{\drawgenerator}[8]{%
\xymatrix@!0{%
& #8 \ar@{-}[ld]\ar@{.}[dd] \ar@{-}[rr] & & #7 \ar@{-}[ld]  \\%
#1 \ar@{-}[rr] \ar@{-}[dd] &  & #2 \ar@{-}[dd] &            \\%
& #6 \ar@{.}[ld] &  & #5 \ar@{-}[uu] \ar@{.}[ll]       \\%
#3 \ar@{-}[rr] &  & #4 \ar@{-}[ru]                       %
}%
}

\definecolor{anyonouter}{RGB}{100, 100, 100}
\definecolor{anyoninner}{RGB}{230, 230, 230}
\definecolor{panelbg}{RGB}{245, 247, 250}
\definecolor{heromsgblue}{RGB}{160, 175, 240}
\definecolor{heromsggreen}{RGB}{145, 195, 145}
\definecolor{heromsgred}{RGB}{235, 160, 160}

\newcommand{\heroanyon}{%
  \shade[inner color=anyoninner, outer color=anyonouter] (0,0) circle (6.3pt);%
  \draw[black, line width=0.7pt] (0,0) circle (6.3pt);%
}
\newcommand{\heroanyonat}[2]{%
  \begin{scope}[shift={(#1,#2)}]\heroanyon\end{scope}%
}
\newcommand{\msgwaves}[3]{% x, y, num_rings
  \foreach \i in {1,...,#3} {%
    \pgfmathsetmacro{\rad}{0.25 + 0.35*\i}%
    \draw[heromsgblue, line width=2.2pt, opacity=1.0, line cap=round, dashed]
      ({#1+\rad*cos(7.5)},{#2+\rad*sin(7.5)}) arc[start angle=7.5, end angle=82.5, radius=\rad cm];%
    \draw[heromsggreen, line width=2.2pt, opacity=1.0, line cap=round, dashed]
      ({#1+\rad*cos(97.5)},{#2+\rad*sin(97.5)}) arc[start angle=97.5, end angle=172.5, radius=\rad cm];%
    \draw[heromsgred, line width=2.2pt, opacity=1.0, line cap=round, dashed]
      ({#1+\rad*cos(277.5)},{#2+\rad*sin(277.5)}) arc[start angle=277.5, end angle=352.5, radius=\rad cm];%
  }%
}
\newcommand{\herodrawgrid}[2]{%
  \foreach \x in {-1,...,#1} { \draw[black, line width=0.4pt] (\x, -1) -- (\x, #2); }%
  \foreach \y in {-1,...,#2} { \draw[black, line width=0.4pt] (-1, \y) -- (#1, \y); }%
}

\definecolor{pastelorange}{RGB}{226,121,50}
\definecolor{pastelblue}{RGB}{58,138,219}
\definecolor{stringred}{RGB}{218,122,122}
\newcommand{\greyanyon}[1]{% \anyon (figure 2-3 style) with parameterized radius
  \shade[inner color=white!25, outer color=black!68] (0,0) circle (#1);%
  \draw[black, line width=0.6pt] (0,0) circle (#1);%
}
\newcommand{\greyanyonat}[3]{%
  \begin{scope}[shift={(#1,#2)}]\greyanyon{#3}\end{scope}%
}

\newcounter{cond}

\newcommand{\vertiii}[1]{{\left\vert\kern-0.25ex\left\vert\kern-0.25ex\left\vert #1 \right\vert\kern-0.25ex\right\vert\kern-0.25ex\right\vert}}

\newcommand{\uvec}[1]{\hat{\mathbf{#1}}}

\newcommand{\rom}[1]{\mathtt{\uppercase\expandafter{\romannumeral #1\relax}}}

\DeclareMathOperator{\poly}{poly}
\DeclareMathOperator{\polylog}{polylog}
\DeclareMathOperator{\polyloglog}{polyloglog}

\newcommand{\norm}[1]{\left\lVert#1\right\rVert}

\newcommand{\Z}{\mathbb Z}
\newcommand{\Q}{\mathbb Q}

\renewcommand{\l}{\left}
\renewcommand{\r}{\right}
\newcommand{\R}{\mathbb R}

\newcommand{\F}{\mathbb F}
\newcommand{\im}{\operatorname{im}}
\renewcommand{\Pr}{\mathbb P}

\newcommand{\ep}{\epsilon}

\newcommand{\vc}{\mathbf}

\newcommand{\uvc}[1]{\vc{\hat{#1}}}

\newcommand{\mcn}{\mathcal{N}}
\newcommand{\supp}{\mathrm{supp}}

\newcommand{\g}{\gamma}

\newcommand{\mcs}{\mathcal{S}}

\usepackage[font=small]{caption}

\usepackage{float}        % enables [H] float placement
\usepackage{adjustbox}    % \adjustbox{max width=...}
\usepackage{subcaption}   % subfigure environment (loaded after caption)
\usetikzlibrary{patterns.meta, bending, decorations.pathreplacing}
\definecolor{pastelbluefig}{RGB}{170,200,235}
\definecolor{pastelred}{RGB}{245,178,178}
\definecolor{injpurple}{RGB}{200,130,205}
\definecolor{redneon}{RGB}{255,150,150}
\definecolor{redneondark}{RGB}{170,20,20}
\definecolor{datagreen}{RGB}{150,200,155}
\definecolor{auxgreen}{RGB}{216,238,218}
\definecolor{datapurple}{RGB}{188,151,200}
\definecolor{datagold}{RGB}{122,158,138}
\definecolor{auxgold}{RGB}{222,233,225}
\definecolor{boxwash}{RGB}{208,220,238}
\definecolor{auxpurple}{RGB}{232,216,238}
\definecolor{compfill}{RGB}{185,199,220}
\definecolor{compline}{RGB}{84,108,147}
\definecolor{steelblue}{RGB}{49,108,175}
\definecolor{lightblue}{RGB}{120,165,215}
\definecolor{brick}{RGB}{178,53,42}
\definecolor{l0grey}{RGB}{170,170,170}
\definecolor{modblue}{RGB}{120,160,215}
\definecolor{modred}{RGB}{232,140,148}
\definecolor{goldb}{RGB}{218,160,33}
\colorlet{zfill}{pastelbluefig}
\colorlet{xfill}{pastelred}
\colorlet{xtemp}{modred}
\tikzset{
  xstab/.style={fill=pastelbluefig},
  zstab/.style={fill=pastelred},
  stabline/.style={draw=black, line width=0.6pt, line join=round},
  dataq/.style={circle, draw=black, fill=white, line width=0.5pt, inner sep=0pt, minimum size=6.5pt},
  magicq/.style={circle, draw=black, fill=injpurple, line width=0.7pt, inner sep=0pt, minimum size=6.5pt},
  stripeq/.style={circle, draw=black, line width=0.5pt, inner sep=0pt, minimum size=6.5pt, fill=white, path picture={\fill[pattern={Lines[angle=45,distance=2.5pt,line width=1.25pt]}, pattern color=black!35] (path picture bounding box.south west) rectangle (path picture bounding box.north east);}},
  diag/.style={black, line width=1.1pt},
  arealbl/.style={font=\large\bfseries, text=black, inner sep=1pt},
  dimarrow/.style={{Latex[length=4.5pt]}-{Latex[length=4.5pt]}, line width=0.6pt},
  thinguide/.style={gray, line width=0.4pt},
  rgnoutline/.style={draw=black, line width=1.1pt, rounded corners=2.5pt},
  zstripe/.style={draw=blue, line width=3.2pt, opacity=0.45, line cap=round},
  xstripe/.style={draw=red,  line width=3.2pt, opacity=0.45, line cap=round},
}
\newcommand{\autscale}{1.62}
\newcommand{\aut}[2]{%
  \begin{scope}[shift={(#1,#2)}, scale=\autscale]
    \draw[fill=compfill, draw=black, line width=0.324pt, rounded corners=0.5pt]
      (-0.075,-0.075) rectangle (0.075,0.075);
    \fill[compline] (-0.0365,-0.0365) rectangle (0.0365,0.0365);
  \end{scope}%
}
\def\Win{8}\def\figscale{0.72}\def\wirewd{0.7pt}%
\definecolor{ctrlpurple}{RGB}{135,100,185}%
\def\minicomp#1#2#3{\begin{scope}[shift={(#1,#2)}, scale=1.6]
  \draw[fill=#3!25, draw=#3!70!black, line width=0.4pt, rounded corners=0.4pt] (-0.075,-0.075) rectangle (0.075,0.075);
  \fill[#3!85!black] (-0.0365,-0.0365) rectangle (0.0365,0.0365);
\end{scope}}%

            \def\minifoldpatch{%
            \begin{scope}
            \clip (0,0) -- (4,0) -- (0,4) -- cycle;
            \fill[pastelred] (0.7373,1.1827) -- (0.1827,1.7373) arc[start angle=-55.1821, end angle=-90, radius=0.32] -- (0,0.3200) arc[start angle=90, end angle=55.1821, radius=0.32] -- (0.7373,0.8173) arc[start angle=-145.1821, end angle=-214.8179, radius=0.32] -- cycle;
            \begin{scope}\clip (0.7373,1.1827) -- (0.1827,1.7373) arc[start angle=-55.1821, end angle=-90, radius=0.32] -- (0,0.3200) arc[start angle=90, end angle=55.1821, radius=0.32] -- (0.7373,0.8173) arc[start angle=-145.1821, end angle=-214.8179, radius=0.32] -- cycle;
            \foreach \sk in {-3.93,-2.948,...,3.95}{\draw[pastelbluefig, line width=3.491pt] (\sk-9,-9) -- (\sk+9,9);}
            \end{scope}
            \fill[pastelred] (0.7373,3.1827) -- (0.1827,3.7373) arc[start angle=-55.1821, end angle=-90, radius=0.32] -- (0,2.3200) arc[start angle=90, end angle=55.1821, radius=0.32] -- (0.7373,2.8173) arc[start angle=-145.1821, end angle=-214.8179, radius=0.32] -- cycle;
            \begin{scope}\clip (0.7373,3.1827) -- (0.1827,3.7373) arc[start angle=-55.1821, end angle=-90, radius=0.32] -- (0,2.3200) arc[start angle=90, end angle=55.1821, radius=0.32] -- (0.7373,2.8173) arc[start angle=-145.1821, end angle=-214.8179, radius=0.32] -- cycle;
            \foreach \sk in {-3.93,-2.948,...,3.95}{\draw[pastelbluefig, line width=3.491pt] (\sk-9,-9) -- (\sk+9,9);}
            \end{scope}
            \fill[pastelred] (1.7373,0.1827) -- (1.1827,0.7373) arc[start angle=-55.1821, end angle=-124.8179, radius=0.32] -- (0.2627,0.1827) arc[start angle=34.8179, end angle=0, radius=0.32] -- (1.6800,0) arc[start angle=180, end angle=145.1821, radius=0.32] -- cycle;
            \begin{scope}\clip (1.7373,0.1827) -- (1.1827,0.7373) arc[start angle=-55.1821, end angle=-124.8179, radius=0.32] -- (0.2627,0.1827) arc[start angle=34.8179, end angle=0, radius=0.32] -- (1.6800,0) arc[start angle=180, end angle=145.1821, radius=0.32] -- cycle;
            \foreach \sk in {-3.93,-2.948,...,3.95}{\draw[pastelbluefig, line width=3.491pt] (\sk-9,-9) -- (\sk+9,9);}
            \end{scope}
            \fill[pastelred] (1.7373,2.1827) -- (1.1827,2.7373) arc[start angle=-55.1821, end angle=-124.8179, radius=0.32] -- (0.2627,2.1827) arc[start angle=34.8179, end angle=-34.8179, radius=0.32] -- (0.8173,1.2627) arc[start angle=124.8179, end angle=55.1821, radius=0.32] -- (1.7373,1.8173) arc[start angle=-145.1821, end angle=-214.8179, radius=0.32] -- cycle;
            \begin{scope}\clip (1.7373,2.1827) -- (1.1827,2.7373) arc[start angle=-55.1821, end angle=-124.8179, radius=0.32] -- (0.2627,2.1827) arc[start angle=34.8179, end angle=-34.8179, radius=0.32] -- (0.8173,1.2627) arc[start angle=124.8179, end angle=55.1821, radius=0.32] -- (1.7373,1.8173) arc[start angle=-145.1821, end angle=-214.8179, radius=0.32] -- cycle;
            \foreach \sk in {-3.93,-2.948,...,3.95}{\draw[pastelbluefig, line width=3.491pt] (\sk-9,-9) -- (\sk+9,9);}
            \end{scope}
            \fill[pastelred] (1.6800,4.0000) -- (0.3200,4.0000) arc[start angle=0, end angle=-34.8179, radius=0.32] -- (0.8173,3.2627) arc[start angle=124.8179, end angle=55.1821, radius=0.32] -- (1.7373,3.8173) arc[start angle=-145.1821, end angle=-180, radius=0.32] -- cycle;
            \begin{scope}\clip (1.6800,4.0000) -- (0.3200,4.0000) arc[start angle=0, end angle=-34.8179, radius=0.32] -- (0.8173,3.2627) arc[start angle=124.8179, end angle=55.1821, radius=0.32] -- (1.7373,3.8173) arc[start angle=-145.1821, end angle=-180, radius=0.32] -- cycle;
            \foreach \sk in {-3.93,-2.948,...,3.95}{\draw[pastelbluefig, line width=3.491pt] (\sk-9,-9) -- (\sk+9,9);}
            \end{scope}
            \fill[pastelred] (2.7373,1.1827) -- (2.1827,1.7373) arc[start angle=-55.1821, end angle=-124.8179, radius=0.32] -- (1.2627,1.1827) arc[start angle=34.8179, end angle=-34.8179, radius=0.32] -- (1.8173,0.2627) arc[start angle=124.8179, end angle=55.1821, radius=0.32] -- (2.7373,0.8173) arc[start angle=-145.1821, end angle=-214.8179, radius=0.32] -- cycle;
            \begin{scope}\clip (2.7373,1.1827) -- (2.1827,1.7373) arc[start angle=-55.1821, end angle=-124.8179, radius=0.32] -- (1.2627,1.1827) arc[start angle=34.8179, end angle=-34.8179, radius=0.32] -- (1.8173,0.2627) arc[start angle=124.8179, end angle=55.1821, radius=0.32] -- (2.7373,0.8173) arc[start angle=-145.1821, end angle=-214.8179, radius=0.32] -- cycle;
            \foreach \sk in {-3.93,-2.948,...,3.95}{\draw[pastelbluefig, line width=3.491pt] (\sk-9,-9) -- (\sk+9,9);}
            \end{scope}
            \fill[pastelred] (2.7373,3.1827) -- (2.1827,3.7373) arc[start angle=-55.1821, end angle=-124.8179, radius=0.32] -- (1.2627,3.1827) arc[start angle=34.8179, end angle=-34.8179, radius=0.32] -- (1.8173,2.2627) arc[start angle=124.8179, end angle=55.1821, radius=0.32] -- (2.7373,2.8173) arc[start angle=-145.1821, end angle=-214.8179, radius=0.32] -- cycle;
            \begin{scope}\clip (2.7373,3.1827) -- (2.1827,3.7373) arc[start angle=-55.1821, end angle=-124.8179, radius=0.32] -- (1.2627,3.1827) arc[start angle=34.8179, end angle=-34.8179, radius=0.32] -- (1.8173,2.2627) arc[start angle=124.8179, end angle=55.1821, radius=0.32] -- (2.7373,2.8173) arc[start angle=-145.1821, end angle=-214.8179, radius=0.32] -- cycle;
            \foreach \sk in {-3.93,-2.948,...,3.95}{\draw[pastelbluefig, line width=3.491pt] (\sk-9,-9) -- (\sk+9,9);}
            \end{scope}
            \fill[pastelred] (3.7373,0.1827) -- (3.1827,0.7373) arc[start angle=-55.1821, end angle=-124.8179, radius=0.32] -- (2.2627,0.1827) arc[start angle=34.8179, end angle=0, radius=0.32] -- (3.6800,0) arc[start angle=180, end angle=145.1821, radius=0.32] -- cycle;
            \begin{scope}\clip (3.7373,0.1827) -- (3.1827,0.7373) arc[start angle=-55.1821, end angle=-124.8179, radius=0.32] -- (2.2627,0.1827) arc[start angle=34.8179, end angle=0, radius=0.32] -- (3.6800,0) arc[start angle=180, end angle=145.1821, radius=0.32] -- cycle;
            \foreach \sk in {-3.93,-2.948,...,3.95}{\draw[pastelbluefig, line width=3.491pt] (\sk-9,-9) -- (\sk+9,9);}
            \end{scope}
            \fill[pastelred] (3.7373,2.1827) -- (3.1827,2.7373) arc[start angle=-55.1821, end angle=-124.8179, radius=0.32] -- (2.2627,2.1827) arc[start angle=34.8179, end angle=-34.8179, radius=0.32] -- (2.8173,1.2627) arc[start angle=124.8179, end angle=55.1821, radius=0.32] -- (3.7373,1.8173) arc[start angle=-145.1821, end angle=-214.8179, radius=0.32] -- cycle;
            \begin{scope}\clip (3.7373,2.1827) -- (3.1827,2.7373) arc[start angle=-55.1821, end angle=-124.8179, radius=0.32] -- (2.2627,2.1827) arc[start angle=34.8179, end angle=-34.8179, radius=0.32] -- (2.8173,1.2627) arc[start angle=124.8179, end angle=55.1821, radius=0.32] -- (3.7373,1.8173) arc[start angle=-145.1821, end angle=-214.8179, radius=0.32] -- cycle;
            \foreach \sk in {-3.93,-2.948,...,3.95}{\draw[pastelbluefig, line width=3.491pt] (\sk-9,-9) -- (\sk+9,9);}
            \end{scope}
            \fill[pastelred] (3.6800,4.0000) -- (2.3200,4.0000) arc[start angle=0, end angle=-34.8179, radius=0.32] -- (2.8173,3.2627) arc[start angle=124.8179, end angle=55.1821, radius=0.32] -- (3.7373,3.8173) arc[start angle=-145.1821, end angle=-180, radius=0.32] -- cycle;
            \begin{scope}\clip (3.6800,4.0000) -- (2.3200,4.0000) arc[start angle=0, end angle=-34.8179, radius=0.32] -- (2.8173,3.2627) arc[start angle=124.8179, end angle=55.1821, radius=0.32] -- (3.7373,3.8173) arc[start angle=-145.1821, end angle=-180, radius=0.32] -- cycle;
            \foreach \sk in {-3.93,-2.948,...,3.95}{\draw[pastelbluefig, line width=3.491pt] (\sk-9,-9) -- (\sk+9,9);}
            \end{scope}
            \fill[pastelred] (3.8173,1.7373) -- (3.2627,1.1827) arc[start angle=34.8179, end angle=-34.8179, radius=0.32] -- (3.8173,0.2627) arc[start angle=124.8179, end angle=90, radius=0.32] -- (4.0000,1.6800) arc[start angle=-90, end angle=-124.8179, radius=0.32] -- cycle;
            \begin{scope}\clip (3.8173,1.7373) -- (3.2627,1.1827) arc[start angle=34.8179, end angle=-34.8179, radius=0.32] -- (3.8173,0.2627) arc[start angle=124.8179, end angle=90, radius=0.32] -- (4.0000,1.6800) arc[start angle=-90, end angle=-124.8179, radius=0.32] -- cycle;
            \foreach \sk in {-3.93,-2.948,...,3.95}{\draw[pastelbluefig, line width=3.491pt] (\sk-9,-9) -- (\sk+9,9);}
            \end{scope}
            \fill[pastelred] (3.8173,3.7373) -- (3.2627,3.1827) arc[start angle=34.8179, end angle=-34.8179, radius=0.32] -- (3.8173,2.2627) arc[start angle=124.8179, end angle=90, radius=0.32] -- (4.0000,3.6800) arc[start angle=-90, end angle=-124.8179, radius=0.32] -- cycle;
            \begin{scope}\clip (3.8173,3.7373) -- (3.2627,3.1827) arc[start angle=34.8179, end angle=-34.8179, radius=0.32] -- (3.8173,2.2627) arc[start angle=124.8179, end angle=90, radius=0.32] -- (4.0000,3.6800) arc[start angle=-90, end angle=-124.8179, radius=0.32] -- cycle;
            \foreach \sk in {-3.93,-2.948,...,3.95}{\draw[pastelbluefig, line width=3.491pt] (\sk-9,-9) -- (\sk+9,9);}
            \end{scope}
            \draw[stabline] (0.7373,1.1827) -- (0.1827,1.7373) arc[start angle=-55.1821, end angle=-90, radius=0.32] -- (0,0.3200) arc[start angle=90, end angle=55.1821, radius=0.32] -- (0.7373,0.8173) arc[start angle=-145.1821, end angle=-214.8179, radius=0.32] -- cycle;
            \draw[stabline] (0.7373,3.1827) -- (0.1827,3.7373) arc[start angle=-55.1821, end angle=-90, radius=0.32] -- (0,2.3200) arc[start angle=90, end angle=55.1821, radius=0.32] -- (0.7373,2.8173) arc[start angle=-145.1821, end angle=-214.8179, radius=0.32] -- cycle;
            \draw[stabline] (1.7373,0.1827) -- (1.1827,0.7373) arc[start angle=-55.1821, end angle=-124.8179, radius=0.32] -- (0.2627,0.1827) arc[start angle=34.8179, end angle=0, radius=0.32] -- (1.6800,0) arc[start angle=180, end angle=145.1821, radius=0.32] -- cycle;
            \draw[stabline] (1.7373,2.1827) -- (1.1827,2.7373) arc[start angle=-55.1821, end angle=-124.8179, radius=0.32] -- (0.2627,2.1827) arc[start angle=34.8179, end angle=-34.8179, radius=0.32] -- (0.8173,1.2627) arc[start angle=124.8179, end angle=55.1821, radius=0.32] -- (1.7373,1.8173) arc[start angle=-145.1821, end angle=-214.8179, radius=0.32] -- cycle;
            \draw[stabline] (1.6800,4.0000) -- (0.3200,4.0000) arc[start angle=0, end angle=-34.8179, radius=0.32] -- (0.8173,3.2627) arc[start angle=124.8179, end angle=55.1821, radius=0.32] -- (1.7373,3.8173) arc[start angle=-145.1821, end angle=-180, radius=0.32] -- cycle;
            \draw[stabline] (2.7373,1.1827) -- (2.1827,1.7373) arc[start angle=-55.1821, end angle=-124.8179, radius=0.32] -- (1.2627,1.1827) arc[start angle=34.8179, end angle=-34.8179, radius=0.32] -- (1.8173,0.2627) arc[start angle=124.8179, end angle=55.1821, radius=0.32] -- (2.7373,0.8173) arc[start angle=-145.1821, end angle=-214.8179, radius=0.32] -- cycle;
            \draw[stabline] (2.7373,3.1827) -- (2.1827,3.7373) arc[start angle=-55.1821, end angle=-124.8179, radius=0.32] -- (1.2627,3.1827) arc[start angle=34.8179, end angle=-34.8179, radius=0.32] -- (1.8173,2.2627) arc[start angle=124.8179, end angle=55.1821, radius=0.32] -- (2.7373,2.8173) arc[start angle=-145.1821, end angle=-214.8179, radius=0.32] -- cycle;
            \draw[stabline] (3.7373,0.1827) -- (3.1827,0.7373) arc[start angle=-55.1821, end angle=-124.8179, radius=0.32] -- (2.2627,0.1827) arc[start angle=34.8179, end angle=0, radius=0.32] -- (3.6800,0) arc[start angle=180, end angle=145.1821, radius=0.32] -- cycle;
            \draw[stabline] (3.7373,2.1827) -- (3.1827,2.7373) arc[start angle=-55.1821, end angle=-124.8179, radius=0.32] -- (2.2627,2.1827) arc[start angle=34.8179, end angle=-34.8179, radius=0.32] -- (2.8173,1.2627) arc[start angle=124.8179, end angle=55.1821, radius=0.32] -- (3.7373,1.8173) arc[start angle=-145.1821, end angle=-214.8179, radius=0.32] -- cycle;
            \draw[stabline] (3.6800,4.0000) -- (2.3200,4.0000) arc[start angle=0, end angle=-34.8179, radius=0.32] -- (2.8173,3.2627) arc[start angle=124.8179, end angle=55.1821, radius=0.32] -- (3.7373,3.8173) arc[start angle=-145.1821, end angle=-180, radius=0.32] -- cycle;
            \draw[stabline] (3.8173,1.7373) -- (3.2627,1.1827) arc[start angle=34.8179, end angle=-34.8179, radius=0.32] -- (3.8173,0.2627) arc[start angle=124.8179, end angle=90, radius=0.32] -- (4.0000,1.6800) arc[start angle=-90, end angle=-124.8179, radius=0.32] -- cycle;
            \draw[stabline] (3.8173,3.7373) -- (3.2627,3.1827) arc[start angle=34.8179, end angle=-34.8179, radius=0.32] -- (3.8173,2.2627) arc[start angle=124.8179, end angle=90, radius=0.32] -- (4.0000,3.6800) arc[start angle=-90, end angle=-124.8179, radius=0.32] -- cycle;
            \end{scope}
            \node[stripeq] at (0,0) {};
            \node[stripeq] at (0,2) {};
            \node[stripeq] at (1,1) {};
            \node[stripeq] at (2,0) {};
            \node[dataq] at (0,4) {};
            \node[dataq] at (1,3) {};
            \node[dataq] at (2,2) {};
            \node[dataq] at (3,1) {};
            \node[dataq] at (4,0) {};
            }%

\usetikzlibrary{external}
\ifnum\pdfshellescape=1
\fi

\newcommand{\blfootnote}[1]{%
  \begingroup
  \renewcommand\thefootnote{}\footnote{#1}%
  \addtocounter{footnote}{-1}%
  \endgroup
}

\begin{document}

    \title{Local decoders for fault-tolerant quantum computation and translation-invariant stabilizer codes}
    
    \author[1]{Nathaniel Selub}
    \author[2]{Aditya Bhardwaj}
    \author[1]{Ethan Lake}
    \affil[1]{\normalsize{\textit{Department of Physics, University of California, Berkeley, Berkeley, CA 94720, USA}}}
    \affil[2]{\normalsize{\textit{Department of Computing and Mathematical Sciences and Institute for Quantum Information and Matter, California Institute of Technology, Pasadena, CA 91125, USA}}}

    \date{}

    \setlength{\affilsep}{5pt}
    \maketitle
    \vspace{-57pt}
    \renewcommand{\abstractname}{}
    \begin{abstract}
        \normalsize
        We construct the first fully spatially local fault-tolerant quantum computer based on topological codes in fewer than four spatial dimensions. Our construction is a two-dimensional architecture that uses only geometrically local quantum and classical operations, bounded-speed classical communication and computation, and a constant density of quantum and classical resources. The core component is a new time-translation-invariant cellular-automaton decoder for the surface code. This decoder preserves logical information for a time stretched-exponential in the code distance and operates continuously during state injection, stabilizer-state preparation, lattice surgery, and transversal readout. We also prove that every translation-invariant topological Pauli stabilizer code is locally decodable under phenomenological noise.
    \end{abstract}

    \blfootnote{Visualizations of the decoders studied in this work are available at {\hypersetup{urlcolor=red}{\url{https://local-decoders.github.io}}}.}
    
    \tableofcontents

    \section{Introduction}\label{sec:intro}
        Quantum computation requires quantum error correction to remain useful in the presence of noise. This motivates the study of scalable fault-tolerant quantum computation in three spatial dimensions and below. Here, we adopt a rather stringent definition of scalability, calling a scheme scalable if and only if it
        \begin{itemize}
            \item has a constant threshold noise rate $p_c$, below which the logical error rate vanishes in the infinite-system-size limit $L \to \infty$;\footnote{Throughout, we consider only families of codes whose distance diverges as $L\to\infty$.}
            \item uses only geometrically local interactions;
            \item employs only bounded-speed classical communication and computation;
            \item and requires a density of quantum and classical resources that remains \emph{finite} as $L\rightarrow \infty$.
        \end{itemize}
        In practice, not all of these requirements will be of equal experimental relevance (see, e.g., Refs.~\cite{reilly2019challenges, bluvstein2024logical, battistel2023real, google_qec}). Regardless, we focus on schemes satisfying this definition because the conditions it requires are, at least in principle, the most favorable for large-scale quantum computation.
        
        Many leading approaches to building quantum computers are based on two-dimensional architectures that store logical information in topological codes. The simplicity, geometric locality, and relatively high thresholds of these codes place them among the most attractive candidates for near-term implementation.

        It is by now well understood how to use topological codes to perform fault-tolerant quantum computation in two and three spatial dimensions, provided one relaxes some of the scalability constraints above. The most widely studied approaches outsource decoding to a centralized classical computer, whose computation speed and/or resource density must diverge as $L \to \infty$ in order for fault-tolerant computation to be performed~\cite{Dennis_2002, Fowler_2012, PhysRevLett.98.190504, Horsman_2012, RevModPhys.87.307, bombin2015single, Litinski_2019, skoric2023parallel, tan2023scalable, bombin2023modular}.
        
        Local decoders have the potential to avoid this centralized bottleneck, but existing constructions can only decode quantum memories and have yet to address the challenge of decoding during quantum computation~\cite{Harrington2004, paletta2026localdecodertoriccode, winter2026highperformancecellularautomatondecoders, dauphinais2017fault, lake2025fastofflinedecodinglocal, lake2025localactiveerrorcorrection, balasubramanian2025localautomaton2dtoric, schotte2022faulttoleranterrorcorrectionuniversal, Vasmer_2021, Kubica_2019, san2023cellular, lang2018strictly, breuckmann2016local, hastings2013decoding, Chirame_2025, Herold_2015, Herold_2017, guedes2024quantum}. Separately, a less frequently discussed scalability concern is whether a constant density of bounded-bandwidth, bounded-propagation-speed wires suffices to route the measurement data and per-qubit control pulses required for fault tolerance based on low-dimensional topological codes. Thus, it is an open question whether the classical infrastructure necessary for fault-tolerant quantum computation with these codes can \emph{itself} be implemented in a strictly scalable manner.

\begin{figure}[t]
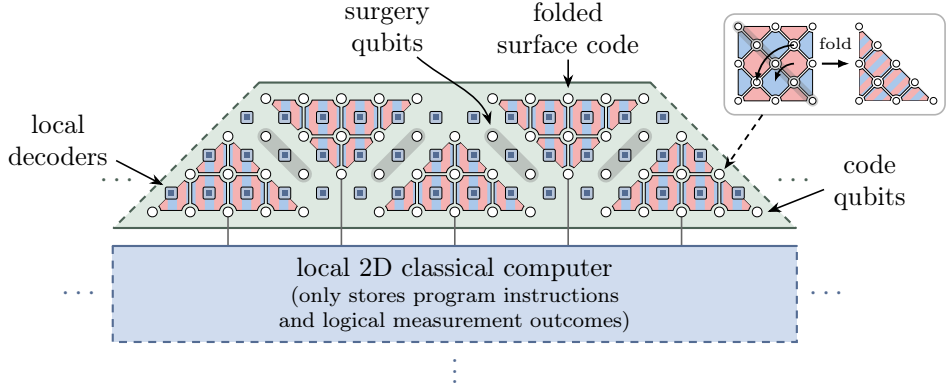
\centering
            \adjustbox{max width=\textwidth}{%
            \tikzsetnextfilename{architecture}
%
            }
            \caption{A two-dimensional fault-tolerant quantum computer that uses only geometrically local quantum and classical operations, bounded-speed classical communication and computation, and a constant density of quantum and classical resources. Logical operations are performed on a line of surface-code patches decoded by a network of constant-size, noiseless classical processors (local decoders). Each patch is folded along one of its diagonals: stripes indicate spatially overlapping $X$- and $Z$-type stabilizers; qubits overlap in pairs, except along the fold. Logical $H$ gates are implemented fold-transversally, and logical $\CNOT$ gates between neighboring patches are implemented via lattice surgery. State injection, stabilizer-state preparation, and transversal readout are used to complete this gate set to a universal set of fault-tolerant primitives. The local decoders run continuously during all of these primitives. The patches lie along the boundary of a noiseless, local, two-dimensional classical computer whose only functions are to control which primitive each patch executes at each step of the quantum program and to receive decoded logical measurement outcomes. This classical computer is never used for decoding.}
            \label{fig:architecture}
        \end{figure}

        In this paper, we resolve this question by constructing the first scalable fault-tolerant quantum computer in two dimensions based on topological codes (Fig.~\ref{fig:architecture}). We first design a cellular-automaton (CA) decoder for the surface code that can be implemented using a \emph{constant} number of noiseless classical bits per site. It operates by inducing simple dynamics: within each cluster of defects created by the noise, defects move toward a designated corner of the cluster until they meet and annihilate (Fig.~\ref{fig:hero}; see Section~\ref{subsec:intro:prior-work} for a comparison to earlier constructions). We then choose a set of fault-tolerant primitives that are both sufficient for universal quantum computation and capable of running concurrently with our decoder. Finally, we combine these primitives with our decoder to obtain a two-dimensional quantum computer that uses strictly local quantum and classical operations and only a constant density of quantum and classical resources.
        
        This result has three immediate implications. First, to the best of our knowledge, all prior local decoders for two-dimensional topological codes require per-site classical resources that grow at least polylogarithmically with system size. Our constant-resource-density (hereafter constant-density) CA decoder is the first to break this barrier.
        
        Second, all prior fault-tolerant quantum computers built from topological codes in fewer than four spatial dimensions rely on non-local classical computation and communication. Our construction removes these non-local ingredients, yielding the first local fault-tolerant quantum-computing architecture based on topological codes in three spatial dimensions and below.

        Third, it is currently believed that superconducting-qubit and quantum-dot platforms will need to devote considerable wiring to moving syndrome data off chip for decoding. Our architecture changes this picture: the simplicity of our decoder allows error correction to be performed, in principle, entirely on chip at cryogenic temperatures, so that each logical qubit requires only a single constant-bandwidth input/output (IO) wire.
        
        Taken together, these results lay a rigorous foundation for scalable quantum computation based on low-dimensional topological codes and open the door to quantum computers with substantially reduced engineering overhead.

        Next, we turn to a separate question: which quantum memories are locally decodable? This question is closely tied to fundamental problems in condensed-matter physics, including the classification of non-equilibrium quantum phases of matter~\cite{cubitt2015stability,sang2024mixed,rakovszky2024defining}. For example, coupling a quantum memory to a local decoder provides a versatile method for realizing stable\footnote{Here, \emph{stable} is taken to mean stable against noise in the quantum components of the system, without requiring stability against noise in the classical automaton performing the decoding.} non-equilibrium phases of matter. Moreover, expanding the class of locally decodable quantum memories may open new paths to scalable fault-tolerant quantum computation. However, a general answer is currently unknown, and even in Euclidean space, the only quantum memories known to be locally decodable are those based on concatenated codes or conventional topological order.

        In this work, we prove that every translation-invariant topological Pauli stabilizer code (hereafter translation-invariant stabilizer code) is locally decodable under phenomenological noise.\footnote{This result also applies to translation-invariant classical stabilizer codes whose classical distance grows as $\Omega(L^{\alpha})$ for some $\alpha > 0$.} The implications of this result are particularly interesting for codes associated with fracton phases of matter~\cite{Vijay_2016, haah2013latticequantumcodesexotic}. Since individual fractons cannot move without creating additional defects, it is a priori unclear whether such codes admit local decoders. Our result shows that fracton models can be locally decoded despite the kinetic constraints of their excitations, yielding the first local decoders for such codes.
        
        Interestingly, the decoders furnished by our proof all operate according to the same principle that underlies our surface-code decoder, that is, by driving the defects in each error cluster toward a designated corner of the cluster until they annihilate.\footnote{This principle was first employed in Ref.~\cite{Bravyi_2013} as a syndrome neutrality-testing subroutine for a global decoder.} Our work thus considerably expands the class of quantum memories known to be locally decodable and provides a unifying principle for constructing local decoders.
    
    \subsection{Summary and overview}\label{subsec:intro:summary}

        \subsubsection{Results}\label{subsubsec:intro:results}

            In this paper, we construct local decoders for the one-dimensional repetition code and the two-dimensional toric and surface codes. We also construct a local, two-dimensional fault-tolerant quantum computer based on folded surface codes~\cite{Kubica_2015, Moussa_2016}. Finally, we prove that every translation-invariant stabilizer code is locally decodable. Throughout, we assume that all classical operations are noiseless.

            First, we summarize our results in the code-capacity setting. Here, our decoders use only a constant number of classical bits per site and can be implemented either synchronously in discrete time via a CA update rule or asynchronously in continuous time via a geometrically local Lindbladian. 

            In this setting, we establish a non-zero threshold $p_*$ against $p$-bounded Pauli noise.\footnote{Noise is said to be $p$-bounded if for every set $S$ of qubits, the probability that all qubits in $S$ suffer an error is at most $p^{|S|}$.} Specifically, we show that when $p < p_*$, the logical error rate is $O((p/p_*)^{L^\alpha})$ for some $\alpha>0$ and the average decoding time is $O(\log^\eta{L})$ for some $\eta > 0$, where $L$ is the linear system size and the average is taken with respect to the noise distribution. We give rigorous proofs for the repetition, toric, and surface codes under synchronous, discrete-time dynamics; the same rigorous guarantees hold under asynchronous, continuous-time dynamics via the ``marching soldiers'' scheme of Berman and Simon~\cite{berman1988investigations, lake2025fastofflinedecodinglocal}. We prove the same guarantees for general translation-invariant stabilizer codes.\footnote{The constant-density code-capacity decoders furnished by our general construction break translation invariance; translation invariance can be restored at the cost of $\poly(\log L)$ classical bits per site.} We also corroborate our results for the repetition, toric, and surface codes numerically.

            \begin{figure}[!t]
                \centering
                \resizebox{\textwidth}{!}{%
                \tikzsetnextfilename{hero}
                \begin{tikzpicture}
                
                \def\gmax{5}
                \def\pad{0}
                
                % ---- Frame 1:  errors create defects ----
                \begin{scope}[shift={(0, 0)}]
                \fill[panelbg] (-1,-1) rectangle ({\gmax},{\gmax});
                \begin{scope}
                    \clip (-1,-1) rectangle ({\gmax},{\gmax});
                    \herodrawgrid{\gmax}{\gmax}
                
                    \heroanyonat{0.5}{3.5}  \heroanyonat{3.5}{2.5}
                    \heroanyonat{2.5}{0.5}  \heroanyonat{0.5}{0.5}
                
                    \draw[black, dashed, line width=1.4pt, rounded corners=4pt]
                    (0.13,0.13) rectangle (3.87,3.87);
                \end{scope}
                \draw[black, line width=0.8pt] (-1,-1) rectangle ({\gmax},{\gmax});
                \node[bubble, font=\small\bfseries, anchor=south, align=center] at (2,{\gmax+0.25})
                    {errors create a neutral\\configuration of defects};
                \end{scope}
                
                % Arrow 1 → 2
                \draw[-{Stealth[length=8pt,width=7pt]}, line width=2.5pt, black]
                (5.2, 2) -- (6.2, 2);
                
                % ---- Frame 2:  messages propagate ----
                \begin{scope}[shift={(7.4, 0)}]
                \fill[panelbg] (-1,-1) rectangle ({\gmax},{\gmax});
                \herodrawgrid{\gmax}{\gmax}
                
                \msgwaves{0.5}{3.5}{3}
                \msgwaves{3.5}{2.5}{3}
                \msgwaves{2.5}{0.5}{3}
                \msgwaves{0.5}{0.5}{3}
                
                \heroanyonat{0.5}{3.5}  \heroanyonat{3.5}{2.5}
                \heroanyonat{2.5}{0.5}  \heroanyonat{0.5}{0.5}
                
                \draw[black, line width=0.8pt] (-1,-1) rectangle ({\gmax},{\gmax});
                \node[bubble, font=\small\bfseries, anchor=south, align=center] at (2,{\gmax+0.25})
                    {messages grow into\\three different quadrants};
                \end{scope}
                
                % Arrow 2 → 3
                \draw[-{Stealth[length=8pt,width=7pt]}, line width=2.5pt, black]
                (12.6, 2) -- (13.6, 2);
                
                % ---- Frame 3:  defects move ----
                \begin{scope}[shift={(14.8, 0)}]
                \fill[panelbg] (-1,-1) rectangle ({\gmax},{\gmax});
                \begin{scope}
                    \clip (-1,-1) rectangle ({\gmax},{\gmax});
                    \herodrawgrid{\gmax}{\gmax}
                
                    \heroanyonat{0.5}{2.5}  \heroanyonat{2.5}{2.5}
                    \heroanyonat{1.5}{0.5}  \heroanyonat{0.5}{0.5}
                
                    \draw[black, dashed, line width=1.4pt, rounded corners=4pt]
                    (0.13,0.13) rectangle (2.87,2.87);
                
                    \draw[-{Triangle[length=6pt,width=5pt]}, line width=1.5pt, black]
                    (0.5,2.12) -- (0.5,{2.12-0.625});
                
                    \draw[-{Triangle[length=6pt,width=5pt]}, line width=1.5pt, black]
                    (1.21,0.5) -- (0.81,0.5);
                
                    \draw[-{Triangle[length=6pt,width=5pt]}, line width=1.5pt, black]
                    (2.15,2.15) -- ({2.15-0.625*0.707},{2.15-0.625*0.707});
                \end{scope}
                \draw[black, line width=0.8pt] (-1,-1) rectangle ({\gmax},{\gmax});
                \node[bubble, font=\small\bfseries, anchor=south, align=center]
                    at (2,{\gmax+0.25})
                    {defects move toward\\the bottom-left corner};
                \end{scope}
                
                \end{tikzpicture}
                }
                \caption{Toric-code decoding mechanism in the code-capacity setting. \textbf{Left:}~A single cluster of errors creates a charge-neutral configuration of defects, enclosed in a bounding box. More generally, the global error configuration always decomposes uniquely into well-separated clusters (see Fig.~\ref{fig:clustering}). The decoder never explicitly identifies clusters or their associated bounding boxes; it merely exploits the fact that these structures always exist. \textbf{Center:}~Each defect sources three types of messages that grow into three different quadrants. \textbf{Right:}~Messages (suppressed here for clarity) are used to monotonically move defects toward the bottom-left corner of the initial bounding box: a defect moves left whenever it sees a blue or red message immediately to its left and otherwise moves down whenever it sees a blue or green message immediately below it. The movement rules are such that defects never leave their initial bounding box. Furthermore, once all defects have come into causal contact, the dynamics force every defect that is not simultaneously leftmost and bottommost to move down or to the left at a constant, non-zero average speed. In the worst case, no defects annihilate en route, and all of them meet at the bottom-left corner of the initial bounding box, where they are guaranteed to annihilate by charge neutrality. Defects in all clusters undergo these dynamics in parallel. Similar mechanisms underlie our other decoders.}
                \label{fig:hero}
            \end{figure}
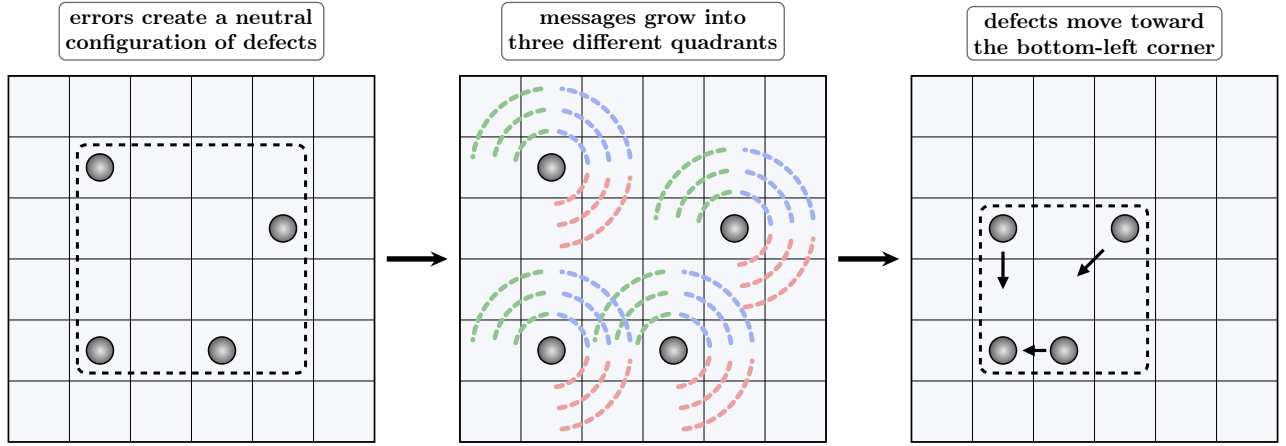

            Next, we summarize our results in the phenomenological-noise setting. For the repetition, toric, and surface codes, we construct translation-invariant streaming decoders that use $\poly(\log{\log{L}})$ classical bits per site; for general translation-invariant stabilizer codes, we construct analogous decoders that use $\poly(\log{L})$ bits per site. By streaming decoders, we mean decoders that begin processing syndromes as soon as they arrive, rather than waiting to act on a complete or windowed syndrome history. For the toric and surface codes, we also construct hierarchical streaming decoders that use only a \emph{constant} number of classical bits per site, at the cost of breaking translation invariance.\footnote{The one-dimensional analogs of these constructions are not constant density because there is not enough room in one dimension to route the necessary wiring. Separately, it may be possible to adapt the constant-density toric-code construction to general translation-invariant stabilizer codes; the main complication is that generic codes, unlike the toric code, are not invariant under coarse-graining.} All of these decoders operate in synchronous discrete time.\footnote{We believe that these decoders can be extended to operate asynchronously in continuous time using the techniques of Ref.~\cite{lake2025localactiveerrorcorrection}.}

            In this setting, we establish a non-zero threshold $p_*'$ against $p$-bounded Pauli and stabilizer-measurement noise. Specifically, we show that when $p < p_*'$, the probability that a logical error occurs after $T$ syndrome-extraction rounds is at most $T \cdot O\l((p/p_*')^{L^{\alpha'}}\r)$ for some $\alpha' > 0$. We give rigorous proofs for all of the constructions described above. We also numerically estimate the threshold of the translation-invariant streaming decoder for the surface code under i.i.d.\ bit-flip and stabilizer-measurement noise and find it to be $p_c \approx 1.1\%$.

            We now turn to our results on fault-tolerant quantum computation. We construct a local, two-dimensional quantum computer that realizes the Clifford$+T$ universal gate set using lattice surgery, magic-state distillation~\cite{knill2004faulttolerantpostselectedquantumcomputation, Bravyi_2005}, and $Y$-state distillation~\cite{Raussendorf_2007}, and we prove a threshold theorem for our architecture under phenomenological noise.\footnote{Our proof readily generalizes to circuit-level noise by applying standard results.} The computer operates in synchronous discrete time and consists of a line\footnote{Our construction can easily be modified to have two-dimensional grid-like connectivity between logical qubits.} of folded surface-code patches.\footnote{The only reason we fold our surface codes is to have access to Hadamard gates that are compatible with lattice surgery and do not require patch rotation. We do not want patch rotation because locally decoding it, while presumably possible, would not be particularly simple. The folded geometry allows us to perform fold-transversal Hadamards, which are particularly simple to locally decode, while keeping the two boundary types accessible for lattice surgery after each gate, without patch rotation. Separately, we note that there is no obstruction to using our framework to design local decoders for, and then computing with, rotated surface codes; we work with unrotated surface codes throughout because they happen to be simpler to locally decode.}

            Each patch is decoded using the constant-density streaming decoder introduced above, augmented with subroutines that allow it to operate continuously during state injection, stabilizer-state preparation, lattice surgery, and transversal readout. We prove that the same decoder performance guarantees established in the memory setting also hold during each of these procedures. Furthermore, we prove that our injection protocol prepares an encoded state with logical error rate $O(p)$ from an unencoded state prepared with physical error rate $p$. Our threshold theorem then follows from these two guarantees together with known results.

            We also prove that our computer requires only a constant density of bounded-bandwidth, bounded-propagation-speed classical wires. Moreover, we show that each logical qubit can be controlled and read out using a single constant-bandwidth IO wire. The wire is used only to send program instructions (e.g., ``perform a logical Hadamard'') and return logical measurement outcomes during program execution. This is possible because decoding is performed entirely within each patch, so no syndrome data ever needs to be transmitted over the wire.

            Our quantum computer still requires a separate classical computer, but only for reasons that are unavoidable. A quantum computer built from surface codes of linear size $L$ can reliably execute quantum programs of length $\exp(\Omega(L^{\alpha'}))$. In the worst case, such programs require a comparable number of classical bits to specify their instructions and store their logical measurement outcomes; neither fits within a constant-density $L \times L$ patch and thus must be stored elsewhere. Because this classical computer is not used for decoding, however, it is spared the pressures of the backlog problem~\cite{RevModPhys.87.307} and is subject to no latency requirements beyond delivering program instructions synchronously.\footnote{Even if we wanted to use the separate classical computer for decoding, the most obvious route to doing so---shipping all syndromes to it for decoding---would not be scalable. Each patch generates $\Theta(L^2)$ bits of syndrome data per round, but in two dimensions, a constant density of bounded-bandwidth wires can carry only $O(L)$ bits per round across the patch boundary. Therefore, this route would require wires whose bandwidth diverges as $L \to \infty$.} Consequently, it trivially admits a local, two-dimensional, constant-density implementation using only bounded-speed classical communication and computation.

        \subsubsection{Construction}

            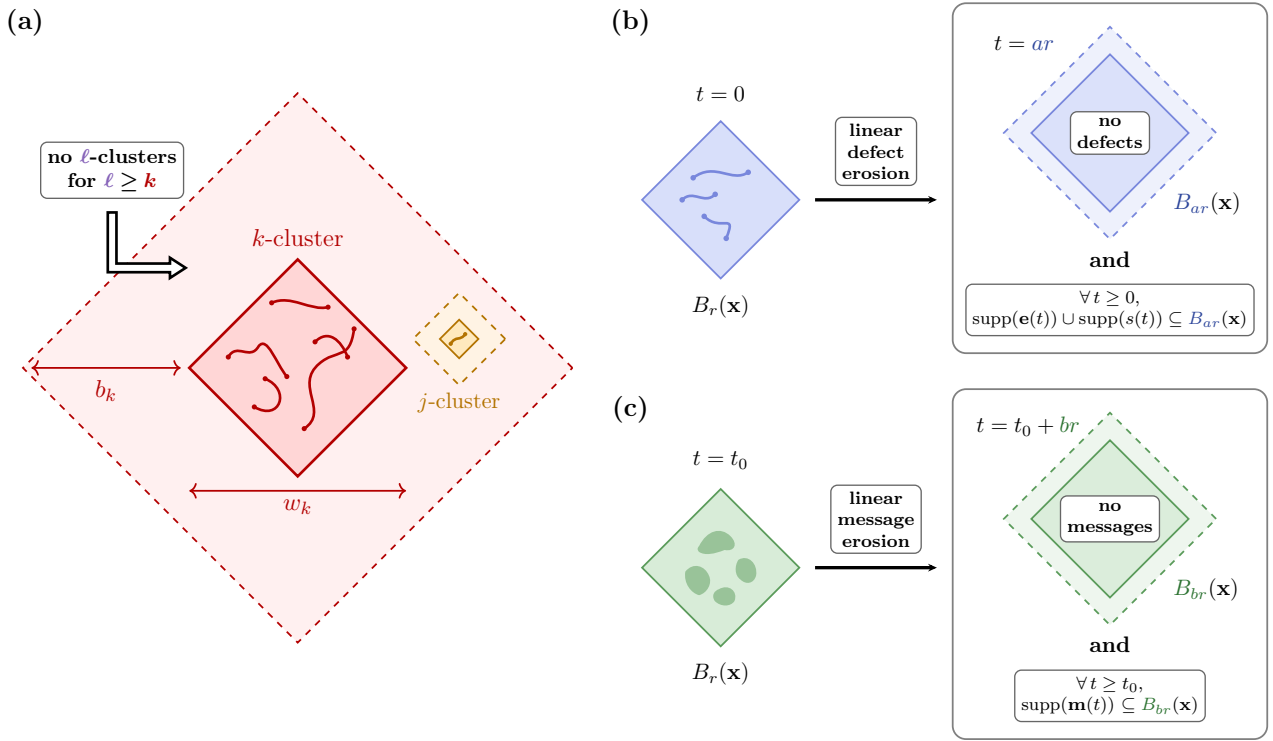
\begin{figure}[!t]
                \centering
                \resizebox{\textwidth}{!}{%
                \tikzsetnextfilename{clustering}
                \begin{tikzpicture}
                
                % ============================================================
                % PANEL (a): Clustering
                % ============================================================
                \begin{scope}[shift={(-3.5,-0.15)}, scale=0.38]
                
                % === K-CLUSTER ===
                \def\kcx{0}
                \def\kcy{0}
                \def\kr{4.5}
                \def\moatw{6.9}
                \pgfmathsetmacro{\kmr}{\kr+\moatw}
                
                % === J-CLUSTER ===
                \def\jcx{6.7}
                \def\jcy{1.2}
                \def\jr{0.8}
                \def\jmoat{1.116}
                \pgfmathsetmacro{\jmr}{\jr+\jmoat}
                
                % ---- Moat region ----
                \fill[moatfill, opacity=0.6]
                    (\kcx+\kmr, \kcy) -- (\kcx, \kcy+\kmr) -- (\kcx-\kmr, \kcy) -- (\kcx, \kcy-\kmr) -- cycle;
                
                % ---- k-cluster diamond ----
                \fill[kclustfill, opacity=0.7]
                    (\kcx+\kr, \kcy) -- (\kcx, \kcy+\kr) -- (\kcx-\kr, \kcy) -- (\kcx, \kcy-\kr) -- cycle;
                
                % ---- j-cluster moat diamond ----
                \fill[jmoatfill, opacity=0.6]
                    (\jcx+\jmr, \jcy) -- (\jcx, \jcy+\jmr) -- (\jcx-\jmr, \jcy) -- (\jcx, \jcy-\jmr) -- cycle;
                
                % ---- j-cluster diamond ----
                \fill[jclustfill, opacity=0.75]
                    (\jcx+\jr, \jcy) -- (\jcx, \jcy+\jr) -- (\jcx-\jr, \jcy) -- (\jcx, \jcy-\jr) -- cycle;
                
                % ---- Anyon string 1 ----
                \draw[kclustedge, very thick] (-1.08, 2.7) .. controls (-0.27, 3.06) and (0.54, 2.34) .. (1.26, 2.52);
                \fill[kclustedge] (-1.08, 2.7) circle (3pt);
                \fill[kclustedge] (1.26, 2.52) circle (3pt);
                
                % ---- Anyon string 2 ----
                \draw[kclustedge, very thick] (-2.88, 0.45) .. controls (-2.52, 1.08) and (-1.98, 0.27) .. (-1.62, 0.9) .. controls (-1.26, 1.35) and (-0.9, 0.18) .. (-0.45, -0.36);
                \fill[kclustedge] (-2.88, 0.45) circle (3pt);
                \fill[kclustedge] (-0.45, -0.36) circle (3pt);
                
                % ---- Anyon string 3 ----
                \draw[kclustedge, very thick]
                    (0.27, -2.52) .. controls (1.35, -1.8) and (-0.18, -0.72) ..
                    (0.9, 0.18) .. controls (1.62, 0.81) and (2.16, 0.27) ..
                    (2.34, 1.62);
                \fill[kclustedge] (0.27, -2.52) circle (3pt);
                \fill[kclustedge] (2.34, 1.62) circle (3pt);
                
                % ---- Anyon string 4 ----
                \draw[kclustedge, very thick] (-1.8, -1.62) .. controls (-0.72, -2.25) and (-0.27, -0.9) .. (-1.35, -0.45);
                \fill[kclustedge] (-1.8, -1.62) circle (3pt);
                \fill[kclustedge] (-1.35, -0.45) circle (3pt);
                
                % ---- Anyon string 5 ----
                \draw[kclustedge, very thick] (0.72, 1.08) .. controls (1.26, 1.62) and (1.8, 1.17) .. (2.07, 0.45);
                \fill[kclustedge] (0.72, 1.08) circle (3pt);
                \fill[kclustedge] (2.07, 0.45) circle (3pt);
                
                % ---- Anyon string in j-cluster ----
                \draw[jclustedge, very thick] (6.35, 1.0) .. controls (6.55, 1.35) and (6.75, 1.1) .. (6.95, 1.4);
                \fill[jclustedge] (6.35, 1.0) circle (2.5pt);
                \fill[jclustedge] (6.95, 1.4) circle (2.5pt);
                
                % ---- Borders ----
                \draw[kclustedge, very thick]
                    (\kcx+\kr, \kcy) -- (\kcx, \kcy+\kr) -- (\kcx-\kr, \kcy) -- (\kcx, \kcy-\kr) -- cycle;
                \draw[moatedge, dashed, thick]
                    (\kcx+\kmr, \kcy) -- (\kcx, \kcy+\kmr) -- (\kcx-\kmr, \kcy) -- (\kcx, \kcy-\kmr) -- cycle;
                \draw[jclustedge, thick]
                    (\jcx+\jr, \jcy) -- (\jcx, \jcy+\jr) -- (\jcx-\jr, \jcy) -- (\jcx, \jcy-\jr) -- cycle;
                \draw[jclustedge, dashed, thick]
                    (\jcx+\jmr, \jcy) -- (\jcx, \jcy+\jmr) -- (\jcx-\jmr, \jcy) -- (\jcx, \jcy-\jmr) -- cycle;
                
                % ---- Labels ----
                \node[kclustedge, above] at (0, \kr+0.2) {$k$-cluster};
                \node[jclustedge, font=\small, below] at (\jcx, \jcy-\jmr+0.05) {$j$-cluster};
                
                % ---- Width annotations ----
                \draw[<->, thick, kclustedge] (-\kr, {-\kr-0.6}) -- (\kr, {-\kr-0.6});
                \node[below, kclustedge] at (0, {-\kr-0.6}) {$w_k$};
                
                \draw[<->, thick, moatedge] ({-\kr-0.35}, 0) -- ({-\kmr+0.35}, 0);
                \node[below, moatedge] at ({(-\kr-\kmr)/2}, -0.15) {$b_k$};
                
                % Moat label
                \node[bubble, font=\footnotesize\bfseries, align=center]
                    at (-7.7, 8.2) {\textbf{no} $\boldsymbol{\textcolor{lblviolet}{\ell}}$\textbf{-clusters}\\\textbf{for} $\boldsymbol{\textcolor{lblviolet}{\ell} \geq \textcolor{kclustedge}{k}}$};
                \fill[white, draw=black, line width=1.2pt]
                (-7.52, 6.45) -- (-7.52, 4.33) -- (-5.30, 4.33) -- (-5.30, 4.57)
                -- (-4.65, 4.15) -- (-5.30, 3.73) -- (-5.30, 3.97)
                -- (-7.88, 3.97) -- (-7.88, 6.45) -- cycle;
                
                \end{scope}
                
                % Panel (a) label
                \node[font=\large\bfseries] at (-7.8, 5.3) {(a)};
                
                % ============================================================
                % PANEL (b): Linear Defect Erosion
                % ============================================================
                \begin{scope}[shift={(2.55,2.5)}, scale=0.62]
                
                \begin{scope}[shift={(1.0, 0)}]
                \def\rsmall{2.0}
                \fill[msgblue, opacity=0.7]
                    (\rsmall, 0) -- (0, \rsmall) -- (-\rsmall, 0) -- (0, -\rsmall) -- cycle;
                \draw[msgblueedge, thick]
                    (\rsmall, 0) -- (0, \rsmall) -- (-\rsmall, 0) -- (0, -\rsmall) -- cycle;
                
                \draw[msgblueedge, very thick] (-0.7, 0.56) .. controls (-0.14, 0.98) and (0.28, 0.42) .. (0.7, 0.7);
                \fill[msgblueedge] (-0.7, 0.56) circle (2pt);
                \fill[msgblueedge] (0.7, 0.7) circle (2pt);
                
                \draw[msgblueedge, very thick] (-0.42, -0.42) .. controls (0.0, -0.84) and (0.42, -0.28) .. (0.14, -0.98);
                \fill[msgblueedge] (-0.42, -0.42) circle (2pt);
                \fill[msgblueedge] (0.14, -0.98) circle (2pt);
                
                \draw[msgblueedge, very thick] (-0.98, 0.0) .. controls (-0.7, 0.28) and (-0.42, -0.14) .. (-0.14, 0.14);
                \fill[msgblueedge] (-0.98, 0.0) circle (2pt);
                \fill[msgblueedge] (-0.14, 0.14) circle (2pt);
                
                \node[font=\small\bfseries, below] at (0, -\rsmall-0.2) {$B_r(\mathbf{x})$};
                \node[font=\small, anchor=south] at (0, \rsmall+0.3) {$t = 0$};
                \end{scope}
                
                \draw[-{Stealth[length=4pt,width=3.5pt]}, line width=1.5pt] (3.4, 0) -- (6.5, 0);
                \node[bubble, font=\scriptsize\bfseries, align=center, above] at (4.95, 0.35) {linear\\defect\\erosion};
                
                \begin{scope}[shift={(10.9, -1.0)}]
                \begin{scope}[shift={(0, 2.7)}]
                \def\rlarge{2.7}
                \fill[msgblue, opacity=0.3]
                    (\rlarge, 0) -- (0, \rlarge) -- (-\rlarge, 0) -- (0, -\rlarge) -- cycle;
                \draw[msgblueedge, thick, dashed]
                    (\rlarge, 0) -- (0, \rlarge) -- (-\rlarge, 0) -- (0, -\rlarge) -- cycle;
                
                \def\rsmall{2.0}
                \fill[msgblue, opacity=0.5]
                    (\rsmall, 0) -- (0, \rsmall) -- (-\rsmall, 0) -- (0, -\rsmall) -- cycle;
                \draw[msgblueedge, thick]
                    (\rsmall, 0) -- (0, \rsmall) -- (-\rsmall, 0) -- (0, -\rsmall) -- cycle;
                
                \node[msgblueedge!50, font=\tiny] at (-0.3, 0.4) {$\cdot$};
                \node[msgblueedge!50, font=\tiny] at (0.4, -0.2) {$\cdot$};
                \node[msgblueedge!50, font=\tiny] at (0.1, -0.5) {$\cdot$};
                \node[msgblueedge!50, font=\tiny] at (-0.5, -0.1) {$\cdot$};
                \node[msgblueedge!50, font=\tiny] at (0.6, 0.5) {$\cdot$};
                
                \node[bubble, font=\scriptsize\bfseries, align=center] at (0, 0) {no\\defects};
                \node[font=\small, anchor=south east] at (-1.2, \rlarge-0.8) {$t = \textcolor{lblblue}{ar}$};
                \node[font=\small\bfseries, anchor=north west] at (1.4, -\rlarge+1.4) {$\textcolor{lblblue}{B_{ar}}(\mathbf{x})$};
                \end{scope}
                
                \node[font=\small\bfseries] at (0, -0.5) {and};
                \node[bubble, font=\scriptsize, align=center] at (0, -1.8) {$\forall\, t \geq 0$,\\$\mathrm{supp}(\mathbf{e}(t)) \cup \mathrm{supp}(s(t)) \subseteq \textcolor{lblblue}{B_{ar}}(\mathbf{x})$};
                \draw[black!50, thick, rounded corners=6pt] (-4.0, -2.9) rectangle (4.0, 6.0);
                \end{scope}
                
                \end{scope}
                
                % Panel (b) label
                \node[font=\large\bfseries] at (1.75, 5.3) {(b)};
                
                % ============================================================
                % PANEL (c): Linear Message Erosion
                % ============================================================
                \begin{scope}[shift={(2.55,-3.3)}, scale=0.62]
                
                \begin{scope}[shift={(1.0, 0)}]
                \def\rsmall{2.0}
                \fill[msggreen, opacity=0.7]
                    (\rsmall, 0) -- (0, \rsmall) -- (-\rsmall, 0) -- (0, -\rsmall) -- cycle;
                \draw[msggreenedge, thick]
                    (\rsmall, 0) -- (0, \rsmall) -- (-\rsmall, 0) -- (0, -\rsmall) -- cycle;
                
                \fill[msggreenedge, opacity=0.5]
                (-0.55, 0.55) .. controls (-0.35, 0.95) and (0.0, 1.05) .. (0.25, 0.8) .. controls (0.45, 0.6) and (0.35, 0.4) .. (0.1, 0.45) .. controls (-0.15, 0.35) and (-0.4, 0.3) .. (-0.55, 0.4) .. controls (-0.6, 0.45) and (-0.6, 0.5) .. cycle;
                \fill[msggreenedge, opacity=0.5]
                (0.45, 0.2) .. controls (0.7, 0.35) and (0.95, 0.1) .. (0.9, -0.2) .. controls (0.85, -0.4) and (0.65, -0.45) .. (0.5, -0.3) .. controls (0.35, -0.2) and (0.35, 0.05) .. cycle;
                \fill[msggreenedge, opacity=0.5]
                (-0.9, -0.1) .. controls (-0.7, 0.15) and (-0.4, 0.05) .. (-0.3, -0.2) .. controls (-0.2, -0.4) and (-0.35, -0.65) .. (-0.55, -0.7) .. controls (-0.75, -0.65) and (-0.95, -0.4) .. cycle;
                \fill[msggreenedge, opacity=0.5]
                (-0.05, -0.55) .. controls (0.2, -0.4) and (0.45, -0.6) .. (0.35, -0.85) .. controls (0.25, -1.0) and (0.0, -1.0) .. (-0.15, -0.85) .. controls (-0.25, -0.7) and (-0.15, -0.58) .. cycle;
                
                \node[font=\small\bfseries, below] at (0, -\rsmall-0.2) {$B_r(\mathbf{x})$};
                \node[font=\small, anchor=south] at (0, \rsmall+0.3) {$t = t_0$};
                \end{scope}
                
                \draw[-{Stealth[length=4pt,width=3.5pt]}, line width=1.5pt] (3.4, 0) -- (6.5, 0);
                \node[bubble, font=\scriptsize\bfseries, align=center, above] at (4.95, 0.35) {linear\\message\\erosion};
                
                \begin{scope}[shift={(10.9, -1.46)}]
                \begin{scope}[shift={(0, 2.7)}]
                \def\rlarge{2.7}
                \fill[msggreen, opacity=0.3]
                    (\rlarge, 0) -- (0, \rlarge) -- (-\rlarge, 0) -- (0, -\rlarge) -- cycle;
                \draw[msggreenedge, thick, dashed]
                    (\rlarge, 0) -- (0, \rlarge) -- (-\rlarge, 0) -- (0, -\rlarge) -- cycle;
                
                \def\rsmall{2.0}
                \fill[msggreen, opacity=0.5]
                    (\rsmall, 0) -- (0, \rsmall) -- (-\rsmall, 0) -- (0, -\rsmall) -- cycle;
                \draw[msggreenedge, thick]
                    (\rsmall, 0) -- (0, \rsmall) -- (-\rsmall, 0) -- (0, -\rsmall) -- cycle;
                
                \node[msggreenedge!50, font=\tiny] at (-0.3, 0.4) {$\cdot$};
                \node[msggreenedge!50, font=\tiny] at (0.4, -0.2) {$\cdot$};
                \node[msggreenedge!50, font=\tiny] at (0.1, -0.5) {$\cdot$};
                \node[msggreenedge!50, font=\tiny] at (-0.5, -0.1) {$\cdot$};
                \node[msggreenedge!50, font=\tiny] at (0.6, 0.5) {$\cdot$};
                
                \node[bubble, font=\scriptsize\bfseries, align=center] at (0, 0) {no\\messages};
                \node[font=\small, anchor=south east] at (-0.55, \rlarge-0.8) {$t = t_0 + \textcolor{lblgreen}{br}$};
                \node[font=\small\bfseries, anchor=north west] at (1.4, -\rlarge+1.4) {$\textcolor{lblgreen}{B_{br}}(\mathbf{x})$};
                \end{scope}
                
                \node[font=\small\bfseries] at (0, -0.5) {and};
                \node[bubble, font=\scriptsize, align=center] at (0, -1.8) {$\forall\, t \geq t_0$,\\$\mathrm{supp}(\mathbf{m}(t))$ $\subseteq \textcolor{lblgreen}{B_{br}}(\mathbf{x})$};
                \draw[black!50, thick, rounded corners=6pt] (-4.0, -2.9) rectangle (4.0, 6.0);
                \end{scope}
                
                \end{scope}
                
                % Panel (c) label
                \node[font=\large\bfseries] at (1.75, -0.8) {(c)};
                
                \end{tikzpicture}
                }
                \caption{Clustering and linear erosion. \textbf{(a)}~Clustering structure. Any error configuration decomposes uniquely into well-separated clusters. Each $k$-cluster (darker red region) of diameter $w_k$ is surrounded by a buffer (lighter red region) of width $b_k = \Omega(w_k)$ that is free of $\ell$-clusters for all $\ell \geq k$. A smaller $j$-cluster and its buffer may nevertheless lie within this buffer (as shown) or within the $k$-cluster itself (not shown). Strings represent errors, and their endpoints represent anyons; in general, errors need not be string-like and defects need not be anyonic. \textbf{(b)}~Linear defect erosion. There exists a constant $a \geq 1$ such that if the initial error configuration is contained in an $\ell_1$-ball $B_r(\vc{x})$, then (1) all defects are eliminated by time $ar$, and (2) for all $t \geq 0$, both the correction applied so far, $\vc{e}(t)$, and the surviving defects, $s(t)$, are supported within the enlarged ball $B_{ar}(\vc{x})$. \textbf{(c)}~Linear message erosion. There exists a constant $b \geq 1$ such that if all defects have been eliminated by time $t_0$ and all non-trivial messages, $\vc{m}(t_0)$, are contained within $B_r(\vc{x})$, then (1) all messages become trivial by time $t_0 + br$, and (2) for all $t \geq t_0$, their support remains within $B_{br}(\vc{x})$. Any decoder that satisfies both of these erosion properties decodes each cluster in any error configuration independently and can fail only when decoding clusters of diameter at least of order the linear system size, which are extremely rare below threshold.}
                \label{fig:clustering}
            \end{figure}
            
            We begin by describing the two key properties of our cellular-automaton dynamics and then explain how these properties imply a threshold in the code-capacity setting. These properties are illustrated in panels (b) and (c) of Fig.~\ref{fig:clustering}. In what follows, all balls are defined with respect to the $\ell_1$ norm. For simplicity, we state these properties for an infinite system; in a finite system, they hold whenever the balls involved have diameter at most a constant fraction of the linear system size.
            
            First, we design the decoders so that if the initial set of errors is contained within a ball of diameter $W$, then (1)~all defects---that is, flipped parity checks---are eliminated in time $O(W)$, and (2)~for all $t \geq 0$, all defects and correction flips remain within an enlarged ball of diameter $O(W)$. We call this property \emph{linear defect erosion}.

            Next, in our construction, each lattice site carries a set of auxiliary bits called \emph{messages}. A message at a given site and time is said to be \emph{trivial} if it is zero. We design the CA dynamics so that, in the absence of defects, non-trivial messages satisfy the following guarantee: if all non-trivial messages are contained within a ball of diameter $W$ at time $t_0$, then (1)~all messages become trivial by time $t_0 + O(W)$, and (2)~for all $t \geq t_0$, their support remains within an enlarged ball of diameter $O(W)$. We call this property \emph{linear message erosion}.

            Together, linear defect erosion and linear message erosion imply a property we call \emph{linear cluster erosion}: an error configuration consisting of a single cluster contained within a ball of diameter $W$ is fully corrected in time $O(W)$; any non-trivial messages created during this process are erased on the same timescale; and for all $t \geq 0$, the support of all corrections, defects, and non-trivial messages remains within an enlarged ball of diameter $O(W)$.
            
            Linear cluster erosion is sufficient to prove a threshold under code-capacity noise via a standard multiscale clustering argument originally due to G\'acs~\cite{gacs2001reliable}. Specifically, the error configuration can always be uniquely decomposed into a hierarchy of clusters across increasing length scales, where each cluster is surrounded by a buffer whose width is proportional to the cluster's diameter, and no clusters of comparable or larger size lie within this buffer. This decomposition is shown in panel (a) of Fig.~\ref{fig:clustering}. Moreover, causality and linear cluster erosion imply that a cluster of diameter $W$ can influence the decoder's dynamics only within an $O(W)$-neighborhood in spacetime. The buffer-width proportionality constant used to define clusters can then be chosen to be large enough to ensure that all clusters are eliminated before they can interact with one another. Therefore, the global correction equals the modulo-two sum of the corrections obtained by decoding each cluster independently. Consequently, for a logical error to occur, the noise must produce a cluster whose diameter is of order $L$, and below threshold, the probability that such a cluster occurs decays rapidly with $L$.

            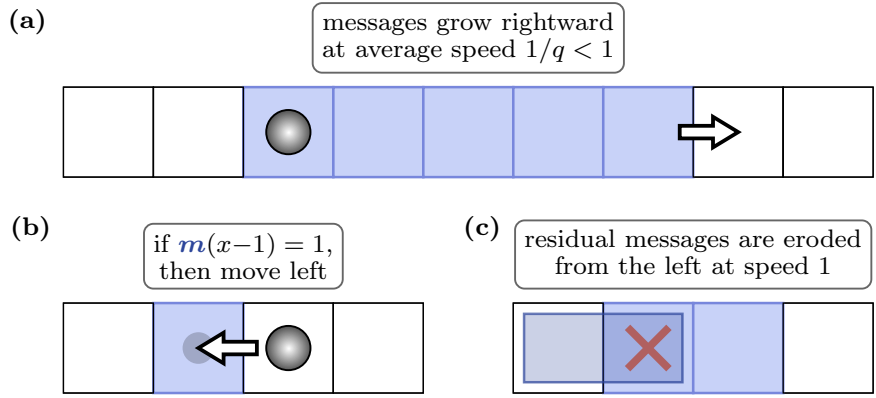
\begin{figure}[!t]
                \centering
                \resizebox{0.7\textwidth}{!}{%
                \tikzsetnextfilename{rep-rule}
                \begin{tikzpicture}[scale=1]
                
                \def\s{1}
                
                % ===== Panel (a): Message Growth in 1D =====
                \begin{scope}[shift={(-3.65, 0.6)}]
                
                \node[font=\scriptsize\bfseries, anchor=north east] at ({-3*\s - \s/2 - 0.05}, {\s/2 + 1.0}) {(a)};
                
                \foreach \x in {-3,-2,-1,0,1,2,3,4,5} {
                  \draw[black, thin] (\x*\s - \s/2, -\s/2) rectangle (\x*\s + \s/2, \s/2);
                }
                
                \foreach \x in {-1,0,1,2,3} {
                  \fill[msgblue, draw=msgblueedge, thick] (\x*\s - \s/2, -\s/2) rectangle (\x*\s + \s/2, \s/2);
                }
                
                \foreach \x in {-3,-2,4,5} {
                  \draw[black, thin] (\x*\s - \s/2, -\s/2) rectangle (\x*\s + \s/2, \s/2);
                }
                
                \begin{scope}[shift={({-1*\s}, 0)}]
                  \anyon
                \end{scope}
                
                \blockarrow{black}{3.35*\s}{0}{0}{0.65}
                
                \node[bubble, font=\scriptsize\rmfamily, anchor=south, align=center] at ({1*\s}, {\s/2 + 0.15}) {messages grow rightward\\[-1pt]at average speed $1/q < 1$};
                
                \end{scope}
                
                % ===== Panel (b): Defect Movement in 1D =====
                \begin{scope}[shift={(-5.65, -1.8)}]
                
                \node[font=\scriptsize\bfseries, anchor=north east] at ({-1*\s - \s/2}, {\s/2 + 1.1}) {(b)};
                
                \foreach \x in {-1,0,1,2} {
                  \draw[black, thin] (\x*\s - \s/2, -\s/2) rectangle (\x*\s + \s/2, \s/2);
                }
                
                \fill[msgblue, draw=msgblueedge, thick] ({0*\s - \s/2}, -\s/2) rectangle ({0*\s + \s/2}, \s/2);
                
                \foreach \x in {-1,1,2} {
                  \draw[black, thin] (\x*\s - \s/2, -\s/2) rectangle (\x*\s + \s/2, \s/2);
                }
                
                \fill[black, opacity=0.2] ({0*\s}, 0) circle (5pt);
                
                \blockarrow{black}{0.65*\s}{0}{180}{0.65}
                
                \begin{scope}[shift={({1*\s}, 0)}]
                  \anyon
                \end{scope}
                
                \node[bubble, font=\scriptsize\rmfamily, anchor=south, align=center] at ({0.5*\s}, {\s/2 + 0.15}) {if \textcolor{lblblue}{$\bm{m}$}$(x{-}1) = 1$,\\[-1pt]then move left};
                
                \end{scope}
                
                % ===== Panel (c): Residual Message Erosion in 1D =====
                \begin{scope}[shift={(-0.65, -1.8)}]
                
                \node[font=\scriptsize\bfseries, anchor=north east] at ({-1*\s - \s/2}, {\s/2 + 1.1}) {(c)};
                
                \foreach \x in {-1,0,1,2} {
                  \draw[black, thin] (\x*\s - \s/2, -\s/2) rectangle (\x*\s + \s/2, \s/2);
                }
                
                \foreach \x in {0,1} {
                  \fill[msgblue, draw=msgblueedge, thick] (\x*\s - \s/2, -\s/2) rectangle (\x*\s + \s/2, \s/2);
                }
                
                \foreach \x in {-1,2} {
                  \draw[black, thin] (\x*\s - \s/2, -\s/2) rectangle (\x*\s + \s/2, \s/2);
                }
                
                \def\g{0.12}
                \fill[lblblue, opacity=0.25]
                  ({-1*\s - \s/2 + \g}, {\s/2 - \g}) -- ({0*\s + \s/2 - \g}, {\s/2 - \g})
                  -- ({0*\s + \s/2 - \g}, {-\s/2 + \g}) -- ({-1*\s - \s/2 + \g}, {-\s/2 + \g}) -- cycle;
                \draw[lblblue, thick, opacity=0.7]
                  ({-1*\s - \s/2 + \g}, {\s/2 - \g}) -- ({0*\s + \s/2 - \g}, {\s/2 - \g})
                  -- ({0*\s + \s/2 - \g}, {-\s/2 + \g}) -- ({-1*\s - \s/2 + \g}, {-\s/2 + \g}) -- cycle;
                
                \fill[msgrededge!80!black]
                  ({0*\s - 0.26}, -0.2) -- ({0*\s - 0.06}, 0) -- ({0*\s - 0.26}, 0.2) -- ({0*\s - 0.2}, 0.26) -- ({0*\s}, 0.06) -- ({0*\s + 0.2}, 0.26) -- ({0*\s + 0.26}, 0.2) -- ({0*\s + 0.06}, 0) -- ({0*\s + 0.26}, -0.2) -- ({0*\s + 0.2}, -0.26) -- ({0*\s}, -0.06) -- ({0*\s - 0.2}, -0.26) -- cycle;
                
                \node[bubble, font=\scriptsize\rmfamily, anchor=south, align=center] at ({0.5*\s}, {\s/2 + 0.15}) {residual messages are eroded\\[-1pt]from the left at speed $1$};
                
                \end{scope}
                
                \end{tikzpicture}
                }
                \caption{Repetition-code decoder rule in the code-capacity setting. The toric-code decoder (Fig.~\ref{fig:hero}) is a two-dimensional generalization of this one-dimensional decoder. \textbf{(a)}~Message growth. A defect sources messages that grow rightward at average speed $1/q$, where $q \geq 2$ is the period of a constant-size clock stored at each site. \textbf{(b)}~Defect movement. A defect at site $x$ moves left whenever it sees a message at site $x-1$. For simplicity, only the message to the left of the defect is shown. \textbf{(c)}~Message erasure. A message at a defect-free site $x$ is erased whenever there is no message at site $x-1$ and no defect at site $x+1$. Therefore, once all defects produced by a cluster have been eliminated, the left boundary of the cluster's residual message support advances rightward at speed at least $1$. Since the right boundary grows only at average speed $1/q < 1$, the left boundary eventually catches up to it, erasing all residual messages along the way.}
                \label{fig:rep_ca_decoder_rule}
            \end{figure}

\begin{figure}
                \centering
                \includegraphics[width=0.9\linewidth]{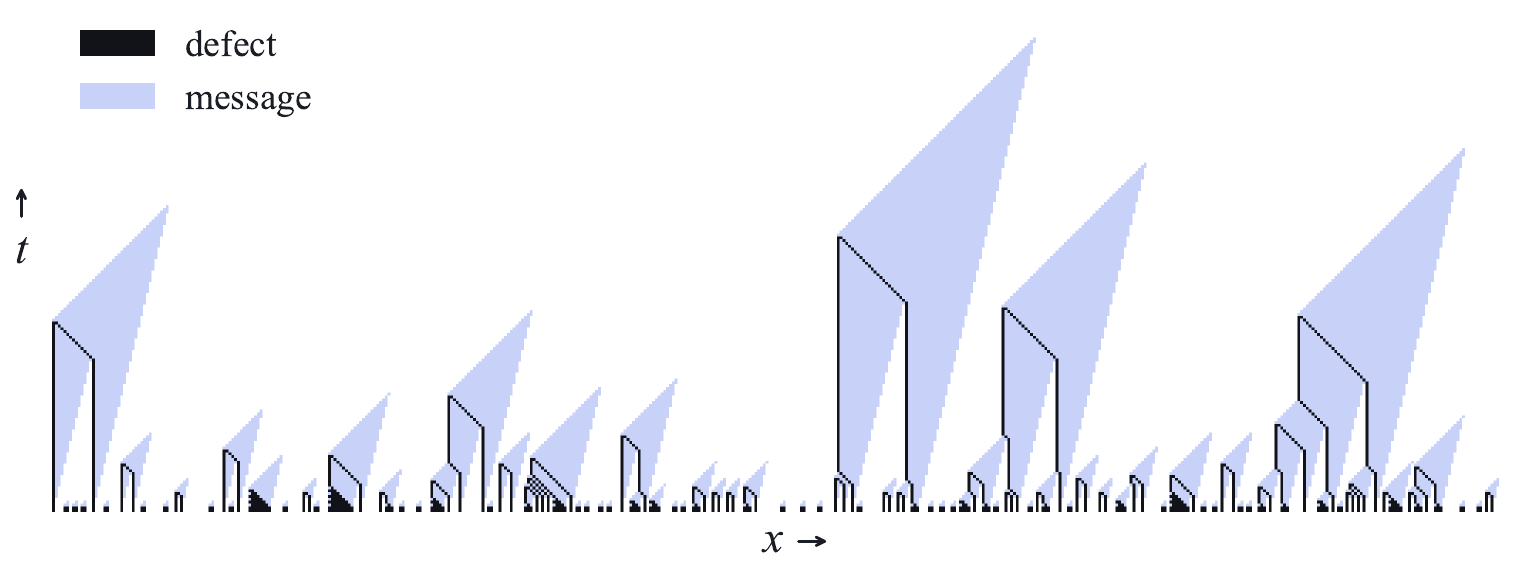}
                \caption{Spacetime history of the repetition-code decoder dynamics in the code-capacity setting. The bottom row shows the initial defects created by i.i.d.\ bit-flip errors. Defects source messages that grow rightward; within each cluster, these messages cause defects to move monotonically leftward but never beyond the cluster's initial leftmost defect. As a result, the bounding interval of the cluster's defects steadily shrinks; in the worst case, it collapses to a single point, and by charge neutrality, this suffices to annihilate every defect in the cluster. Each cluster is eliminated within a time proportional to its initial diameter, so only large clusters can have large temporal extent. Once all of a cluster's defects have been annihilated, all of its messages are guaranteed to be erased faster than they can grow, creating triangle-like regions in spacetime. When a site contains both a defect and a message, only the defect is displayed.}
                \label{fig:rep_st_diag}
            \end{figure}

            Next, we describe our general strategy for achieving linear defect erosion and linear message erosion. The simplest realization of this strategy is the repetition-code decoder, whose rule and dynamics are illustrated in Figs.~\ref{fig:rep_ca_decoder_rule} and~\ref{fig:rep_st_diag}, respectively. The toric-code decoder of Fig.~\ref{fig:hero} realizes the same strategy in two dimensions. Our strategy for achieving linear defect erosion in the presence of periodic boundary conditions\footnote{Linear defect erosion can be achieved in systems with open boundary conditions using a simple modification of this strategy.} is to construct dynamics that are guaranteed to monotonically move defects toward a designated corner of their original bounding box at a constant, non-zero average speed. Concretely, when defects move, they can only take steps along $-\uvc{x}$ in one dimension and along $-\uvc{x}$ or $-\uvc{y}$ in two dimensions. Moreover, messages always attempt to steer defects toward a coordinate-extremal defect, meaning a defect that is extremal in at least one coordinate: the leftmost defect in one dimension and the leftmost or bottommost defect in two dimensions. The dynamics are also such that defects cannot leave their original bounding box. Together, these properties cause the defects' bounding box to shrink at a linear rate. In the worst case, the bounding box shrinks to a single point by time $O(W)$; since each error cluster creates a charge-neutral set of defects, this ensures that all defects are annihilated by that time.

            This strategy reduces constructing a CA decoder that satisfies linear defect erosion to designing dynamics that cause the defects' bounding box to shrink at a linear rate. For the repetition and toric codes, we construct such dynamics by hand. For general translation-invariant stabilizer codes, we prove that suitably defined generalizations of such dynamics always exist using Haah's polynomial formalism~\cite{Haah_2013}. In this formalism, the set of realizable defect configurations corresponds to a submodule of a free module over a ring of (Laurent) polynomials in $D$ variables. We first reduce the existence of such dynamics to the existence of a sufficiently fast algorithm for polynomial division on these submodules, and then prove that such an algorithm exists using Gr\"obner-basis techniques.

            To obtain linear message erosion, we use one strategy for the repetition and toric codes, and another strategy for general translation-invariant stabilizer codes. For the repetition and toric codes, we define local message erasure rules. These rules allow a message to be erased only after the defect that created it has been annihilated. The rules are dimension-dependent: in one dimension, the rule is a simple shearing mechanism, and in two dimensions, the rules use rotated variants of Toom's rule. In each dimension, the message erosion speed is set to be strictly greater than the message growth speed. This ensures that, in the absence of defects, any non-trivial messages initially supported in a ball of diameter $W$ are erased in time $O(W)$. This strategy, however, has two limitations: it does not erase messages aggressively enough for use in our phenomenological-noise decoders, and even in the code-capacity setting, rigorously proving that it works in three dimensions and beyond appears unwieldy.

            For general translation-invariant stabilizer codes, our strategy is to intermittently erase all messages in the system, with increasingly large time intervals between erasure rounds. This strategy provably yields linear message erosion in any dimension and, as we describe below, extends to our phenomenological-noise decoders. A priori, implementing it requires timers with a growing number of classical bits per site, but this overhead can be made constant at the cost of breaking translation invariance.

            We now describe how to extend these ideas to phenomenological noise. The simplest (non-local) option is to perform sliding-window decoding on the spacetime history of detector events---sites where two consecutive (potentially faulty) stabilizer measurements disagree---using higher-dimensional versions of the decoders described above.\footnote{In practice, this option may be quite useful: our decoders offer a very low-overhead way of performing window decoding extremely quickly, at the cost of reduced (but respectable) performance relative to non-local decoders. This makes them natural candidates for use as predecoders or, when augmented with a measure of decoder confidence, as the first tier of a multi-decoder pipeline.} A second option is to use these higher-dimensional decoders to construct streaming window decoders; this is the approach taken in Ref.~\cite{lake2025localactiveerrorcorrection}.

            \begin{figure}
                \centering
                \tikzsetnextfilename{online-decoder}
                \begin{tikzpicture}[scale=1.0, every node/.style={font=\small}]
                    \def\W{6}%
                    \def\sliceH{0.62}%
                    \def\gap{0.08}%
                    \def\bwOffset{0.55}%
                    % quantum system
                    \fill[pastelblue!18] (0,-0.75) rectangle (\W,-0.15);
                    \draw[black, line width=0.8pt] (0,-0.75) rectangle (\W,-0.15);
                    \node[pastelblue] at (\W/2,-0.45) {quantum system};
                    % z slices
                    \draw[black, line width=0.6pt] (0,0) rectangle (\W,\sliceH);
                    \draw[black, line width=0.6pt] (0,\sliceH+\gap) rectangle (\W,2*\sliceH+\gap);
                    \draw[black, line width=0.6pt] (0,2*\sliceH+2*\gap) rectangle (\W,3*\sliceH+2*\gap);
                    % final slice
                    \draw[black, line width=0.6pt] (0,3*\sliceH+3*\gap+\bwOffset) rectangle (\W,4*\sliceH+3*\gap+\bwOffset);
                    % Z axis on the left (tick marks + labels)
                    \def\axisx{-0.3}%
                    \draw[line width=0.8pt] (\axisx,\sliceH/2) -- (\axisx,2*\sliceH+2*\gap+\sliceH/2+0.3);
                    \draw[line width=0.8pt] (\axisx,3*\sliceH+3*\gap+\bwOffset+\sliceH/2-0.3) -- (\axisx,3*\sliceH+3*\gap+\bwOffset+\sliceH/2);
                    \draw[line width=0.6pt] (\axisx-0.08,\sliceH/2) -- (\axisx+0.08,\sliceH/2);
                    \node[font=\small, anchor=east] at (\axisx-0.12,\sliceH/2) {$0$};
                    \draw[line width=0.6pt] (\axisx-0.08,\sliceH+\gap+\sliceH/2) -- (\axisx+0.08,\sliceH+\gap+\sliceH/2);
                    \node[font=\small, anchor=east] at (\axisx-0.12,\sliceH+\gap+\sliceH/2) {$1$};
                    \draw[line width=0.6pt] (\axisx-0.08,2*\sliceH+2*\gap+\sliceH/2) -- (\axisx+0.08,2*\sliceH+2*\gap+\sliceH/2);
                    \node[font=\small, anchor=east] at (\axisx-0.12,2*\sliceH+2*\gap+\sliceH/2) {$2$};
                    \node[font=\large] at (\axisx,3*\sliceH+2.5*\gap+\bwOffset/2+0.08) {$\vdots$};
                    \node[font=\Large] at (\W/2,3*\sliceH+2.5*\gap+\bwOffset/2+0.08) {$\vdots$};
                    \draw[line width=0.6pt] (\axisx-0.08,3*\sliceH+3*\gap+\bwOffset+\sliceH/2) -- (\axisx+0.08,3*\sliceH+3*\gap+\bwOffset+\sliceH/2);
                    \node[font=\small, anchor=east] at (\axisx-0.12,3*\sliceH+3*\gap+\bwOffset+\sliceH/2) {$K-1$};
                    \node[name=bufferlbl, font=\scriptsize, align=center, anchor=east] at (\axisx-0.55,2*\sliceH+2*\gap+\sliceH/2) {stack of code-capacity\\decoders\\(slices)};
                    \def\sw{0.081}%
                    \def\hw{0.171}%
                    % defect pairs across slices
                    \def\r{3pt}%
                    \draw[stringred, line width=1.0pt] (1.4,\sliceH/2) -- (1.4,-0.15);
                    \greyanyonat{1.4}{\sliceH/2}{\r}
                    \draw[stringred, line width=1.0pt] (4.9,\sliceH/2) -- (5.25,\sliceH/2);
                    \greyanyonat{4.9}{\sliceH/2}{\r}
                    \greyanyonat{5.25}{\sliceH/2}{\r}
                    \coordinate (arrbot2) at (2.1,\sliceH+\gap+\sliceH/2+0.2);
                    \filldraw[stringred!45, draw=black!50, line width=0.5pt,
                              line join=round]
                      ([xshift=-\sw cm]arrbot2) --
                      ([xshift=\sw cm]arrbot2) --
                      ([xshift=\sw cm, yshift=0.2475cm]arrbot2) --
                      ([xshift=\hw cm, yshift=0.2475cm]arrbot2) --
                      ([yshift=0.45cm]arrbot2) --
                      ([xshift=-\hw cm, yshift=0.2475cm]arrbot2) --
                      ([xshift=-\sw cm, yshift=0.2475cm]arrbot2) -- cycle;
                    \draw[stringred, line width=1.0pt] (2.6,\sliceH/2) -- (2.95,\sliceH/2);
                    \greyanyonat{2.6}{\sliceH/2}{\r}
                    \greyanyonat{2.95}{\sliceH/2}{\r}
                    \draw[stringred, line width=1.0pt] (1.0,\sliceH+\gap+\sliceH/2) -- (2.1,\sliceH+\gap+\sliceH/2);
                    \greyanyonat{1.0}{\sliceH+\gap+\sliceH/2}{\r}
                    \greyanyonat{2.1}{\sliceH+\gap+\sliceH/2}{\r}
                    \draw[stringred, line width=1.0pt] (4.25,2*\sliceH+2*\gap+\sliceH/2) -- (5.4,2*\sliceH+2*\gap+\sliceH/2) -- (5.4,\sliceH+\gap+\sliceH/2);
                    \greyanyonat{4.25}{2*\sliceH+2*\gap+\sliceH/2}{\r}
                    \greyanyonat{5.4}{\sliceH+\gap+\sliceH/2}{\r}
                    \draw[stringred, line width=1.0pt] (2.73,3*\sliceH+3*\gap+\bwOffset+\sliceH/2) -- (5.2,3*\sliceH+3*\gap+\bwOffset+\sliceH/2);
                    \greyanyonat{2.73}{3*\sliceH+3*\gap+\bwOffset+\sliceH/2}{\r}
                    \greyanyonat{5.2}{3*\sliceH+3*\gap+\bwOffset+\sliceH/2}{\r}
                    \def\bx{\W+0.4}%
                    \def\bh{0.32}%
                    {
                      \node[font=\scriptsize, pastelorange, align=center] at (\bx+1.2,3*\sliceH+2.5*\gap+\bwOffset/2) {maximum time spent\\in slice $k < K-1$};
                      \fill[pastelorange!65] (\bx,\sliceH/2-\bh/2) rectangle (\bx+0.3,\sliceH/2+\bh/2);
                      \node[anchor=west, font=\scriptsize] at (\bx+0.4,\sliceH/2) {$t_0 \sim 1$};
                      \fill[pastelorange!65] (\bx,\sliceH+\gap+\sliceH/2-\bh/2) rectangle (\bx+0.82,\sliceH+\gap+\sliceH/2+\bh/2);
                      \node[anchor=west, font=\scriptsize] at (\bx+0.92,\sliceH+\gap+\sliceH/2) {$t_1 \sim e$};
                      \fill[pastelorange!65] (\bx,2*\sliceH+2*\gap+\sliceH/2-\bh/2) rectangle (\bx+2.22,2*\sliceH+2*\gap+\sliceH/2+\bh/2);
                      \node[anchor=west, font=\scriptsize] at (\bx+2.32,2*\sliceH+2*\gap+\sliceH/2) {$t_2 \sim e^2$};
                    }
                    % new-defect tag: leader from below the quantum bar to the freshly inserted defect
                    \draw[-{Triangle[length=1.6mm,width=1.4mm]}, black!50, line width=0.7pt]
                      ([rotate around={-30:(1.4,\sliceH/2)}]1.28,-1.16) --
                      ([rotate around={-30:(1.4,\sliceH/2)}]1.33,\sliceH/2-0.13);
                    \node[font=\scriptsize, black!60, anchor=east, align=center] at (0.5,-1.2) {new defects are inserted\\when detector events occur};
                    % callout above the final slice: the long surviving chain evidences the sorting claim
                    \node[font=\scriptsize, black!60, anchor=south, align=center] at (3.96,3.45) {only defects from large, rare clusters\\can reach the final slice};
                    \draw[-{Triangle[length=1.6mm,width=1.4mm]}, black!50, line width=0.7pt] (3.96,3.42) -- (3.96,3.05);
                  \end{tikzpicture}
                  \caption{Translation-invariant streaming decoder for the repetition code under phenomenological noise. The decoder consists of $K$ coupled code-capacity decoders---called \emph{slices}---stacked along an auxiliary dimension. A new defect is inserted in slice $k=0$ whenever two consecutive (potentially faulty) stabilizer measurements at the same site disagree. Defects correspond to the ends of error chains in spacetime (abstractly represented by red strings). Each slice runs a slightly modified version of the code-capacity repetition-code decoder dynamics. Additionally, if a defect survives in slice $k < K-1$ for $t_k\sim e^k$ time steps, it is promoted to the same spatial location in slice $k+1$. As an example, the red arrow depicts where the defect below it would go if it were promoted. Apart from defect promotion, slices evolve independently of one another. The dynamics are such that only defects from large, rare error clusters can survive long enough to reach the final slice. When $K = \Theta(\log L)$, where $L$ is the linear system size, this slice-and-promotion construction ensures that spacetime versions of linear defect erosion and linear message erosion (Fig.~\ref{fig:clustering}) hold, so that each cluster in any spacetime error configuration is decoded independently and logical errors can arise only from clusters of diameter $\Omega(L)$. Similar constructions underlie our other streaming decoders.}
                  \label{fig:online-decoder}
            \end{figure}
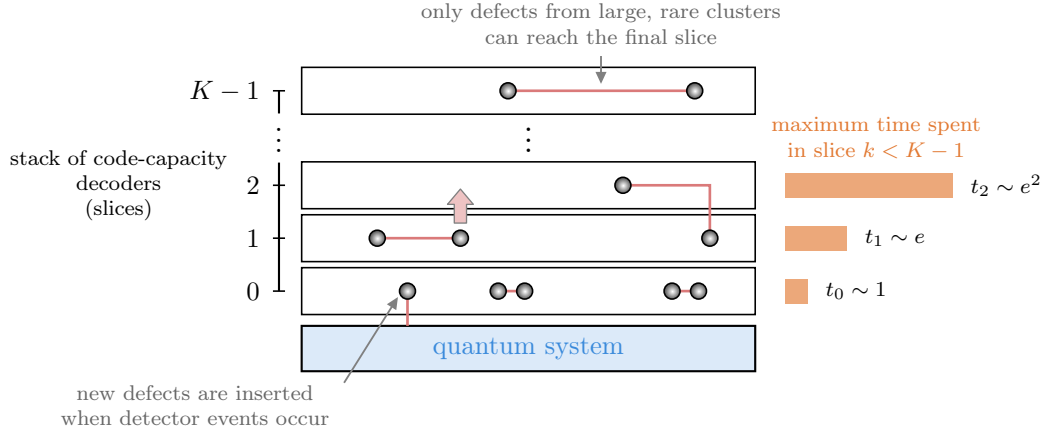
 
            Here, we instead pursue a third option and build decoders that can be viewed as streaming window decoders whose time dimension has been drastically compressed. For each of the codes above, we construct a translation-invariant streaming decoder by stacking $K$ coupled copies of its code-capacity decoder---called \emph{slices}---along an auxiliary dimension (Fig.~\ref{fig:online-decoder}). At each time step, newly observed detector events are inserted as defects in the bottom slice, and each slice runs a slightly modified version of the code-capacity decoder dynamics. In addition, each defect carries a local timer that follows it as it moves: if a defect survives in slice $k < K-1$ for $t_k \sim e^k$ time steps, it is promoted to the same spatial location in slice $k+1$ and its timer is reset (defects in the final slice carry no timer). The timers are chosen so that every cluster of diameter $O(W)$ in the spacetime error configuration is eliminated before its defects can be promoted beyond slice $k \sim \log W$. Apart from defect promotion, slices evolve independently of one another.

            The slices have the effect of sorting clusters by scale. Specifically, the smallest clusters are eliminated in the bottom slice before their defects' timers can expire, and only defects from larger, rarer clusters can survive long enough to begin ascending the stack. When $K = \Theta(\log L)$, this sorting ensures that spacetime versions of linear defect erosion and linear message erosion hold, so that all clusters are decoded independently and logical errors can arise only from clusters of diameter $\Omega(L)$. We use a spacetime version of the multiscale clustering argument to formalize this picture and prove that $\Theta(\log L)$ slices---and thus $\poly(\log L)$ classical bits per site---suffice to achieve a memory lifetime that scales as a stretched exponential in $L$.

            We then show that this already-low overhead can be reduced even further. When $K = o(\log L)$, clusters whose defects live long enough to reach the final slice are no longer guaranteed to be decoded independently of one another. Nevertheless, by combining our construction with renormalization-style arguments from Ref.~\cite{lake2025localactiveerrorcorrection}, we prove that $K = \Theta(\log\log L)$ slices---and thus $\poly(\log\log L)$ classical bits per site---suffice to control the effects of these interactions and preserve the decoder's stretched-exponential memory lifetime scaling.

            Both constructions extend to every translation-invariant stabilizer code. We prove rigorous guarantees for the $\poly(\log L)$ construction in every case. For the $\poly(\log\log L)$ construction, we give rigorous proofs only for the repetition, toric, and surface codes, but physical arguments suggest that this construction should work far more broadly.
            
            The thresholds of our decoders generically decrease with the dimension $D$ of the decoding problem, and numerics indicate that our decoders are likely impractical for $D > 3$.\footnote{This phenomenon can be understood as follows. The probability of a cluster capable of causing a logical error is essentially independent of $D$ and scales as $e^{-\Omega(L^{\alpha})}$~\cite{lake2025fastofflinedecodinglocal}, while the entropy of such clusters grows with $D$.}

            To obtain a two-dimensional, constant-density decoder for the toric and surface codes, we replace the $K$ full-resolution slices with a hierarchy of progressively coarser slices, in which a slice-$k$ site represents a block of linear size $\sim e^k$. The coarse-graining keeps the total number of sites in the hierarchy proportional to the number of physical sites and allows every local timer to take values in a constant range, at the cost of breaking spatial translation invariance. We then show that this construction can be implemented in two dimensions using a constant density of bounded-bandwidth, bounded-propagation-speed wires.

            Finally, we use the constant-density surface-code decoder to construct a local, two-dimensional fault-tolerant quantum computer. We do so by designing simple, local subroutines that allow the decoder to operate in the time-dependent, non-uniform geometry that arises during state injection, stabilizer-state preparation, and lattice surgery. These procedures also create stabilizers whose initial measurement outcomes are both noisy and intrinsically random; the same subroutines enable the decoder to locally determine a globally consistent interpretation of these outcomes on the fly. We prove that the dynamics of these subroutines satisfy linear cluster erosion, so that the stretched-exponential lifetime guarantees from the memory setting also hold during each of these procedures.

            We then prove that our injection protocol prepares an encoded state with logical error rate $O(p)$. This guarantee follows from linear cluster erosion via a simple energy--entropy argument: a cluster of diameter $W$ can cause a logical error during injection only if it occurs within spacetime distance $O(W)$ of the injection point, so at most $\poly(W)$ clusters contribute at each scale, while below threshold each such cluster occurs with probability at most $(p/p_*')^{\Omega(W^{\alpha'})}$. The logical error probability is therefore at most $\sum_{W \geq 1} \poly(W)\,(p/p_*')^{\Omega(W^{\alpha'})}$, which, for sufficiently small $p < p_*'$, is dominated by its first term and is thus $O(p)$.\footnote{The same argument implies that the $O(p)$ logical error rate guarantee extends to a large class of injection protocols for topological codes on Euclidean lattices whenever decoding is performed by a global decoder satisfying an analog of linear defect erosion. The Bravyi--Haah renormalization-group decoder is one such decoder: it runs in polynomial time and has a provable threshold for any code family defined on hypergraphs with polynomial volume growth~\cite{Bravyi_2013}. By contrast, existing proofs of $O(p)$ logical error rate in this setting rely on minimum-weight decoding, which is generically inefficient.}

            \begin{figure}[t]
                \centering
                % ---------------------------------------------------------------
                \begin{subfigure}[t]{0.48\textwidth}
                    \centering
                    \tikzsetnextfilename{layout-intra}
                    \begin{tikzpicture}[scale=\figscale]
                    % --- streets, clipped to the window ---
                    \begin{scope}
                        \clip (0,0) rectangle (\Win,\Win);
                        \foreach \p in {0.5,1.5,...,7.5}{
                        \draw[l0grey, line width=\wirewd] (\p,0)--(\p,\Win);
                        \draw[l0grey, line width=\wirewd] (0,\p)--(\Win,\p);
                        }
                        \foreach \p in {1,3,5,7}{
                        \draw[steelblue, line width=\wirewd] (\p,0)--(\p,\Win);
                        \draw[steelblue, line width=\wirewd] (0,\p)--(\Win,\p);
                        }
                        \foreach \p in {2,6}{
                        \draw[brick, line width=\wirewd] (\p,0)--(\p,\Win);
                        \draw[brick, line width=\wirewd] (0,\p)--(\Win,\p);
                        }
                    \end{scope}
                    % --- sites at street intersections ---
                    \foreach \x in {0.5,1.5,...,7.5}{
                        \foreach \y in {0.5,1.5,...,7.5}{
                        \minicomp{\x}{\y}{l0grey}}}
                    \foreach \x in {1,3,5,7}{
                        \foreach \y in {1,3,5,7}{
                        \minicomp{\x}{\y}{steelblue}}}
                    \foreach \x in {2,6}{
                        \foreach \y in {2,6}{
                        \minicomp{\x}{\y}{brick}}}
                    \draw[black!55, line width=0.6pt] (0,0) rectangle (\Win,\Win);
                    \node[font=\scriptsize] at (\Win/2,\Win+0.45) {local decoders + intra-slice wiring};
                    % --- legend (self-centering text node) ---
                    \node[font=\scriptsize] at (\Win/2,-0.75) {\textcolor{l0grey}{\rule[0.55ex]{12pt}{0.7pt}}\,\,slice $0$\hspace{1.729em}\textcolor{steelblue}{\rule[0.55ex]{12pt}{0.7pt}}\,\,slice $1$\hspace{1.729em}\textcolor{brick}{\rule[0.55ex]{12pt}{0.7pt}}\,\,slice $2$};
                    \end{tikzpicture}
                \end{subfigure}
                \hfill
                % ---------------------------------------------------------------
                \begin{subfigure}[t]{0.48\textwidth}
                    \centering
                    \tikzsetnextfilename{layout-inter}
                    \begin{tikzpicture}[scale=\figscale]
                    \draw[black!55, line width=0.6pt] (0,0) rectangle (\Win,\Win);
                    \node[font=\scriptsize] at (\Win/2,\Win+0.45) {local decoders + inter-slice and I/O wiring};
                    % --- level 0 -> 1 promotion wires (one funnel per level-1 site) ---
                    \foreach \Rx in {0,1,2,3}{\foreach \Ry in {0,1,2,3}{
                        \pgfmathsetmacro\px{2*\Rx+1}
                        \pgfmathsetmacro\py{2*\Ry+1}
                        \draw[lightblue, line width=\wirewd] (\px,\py-0.5) -- (\px,\py+0.5);
                        \foreach \ax in {0,1}{\foreach \ay in {0,1}{
                        \pgfmathsetmacro\cx{2*\Rx+\ax+0.5}
                        \pgfmathsetmacro\cy{2*\Ry+\ay+0.5}
                        \draw[lightblue, line width=\wirewd]
                            (\cx,\cy) -- (\px,\cy);
                        }}
                    }}
                    % --- level 1 -> 2 promotion wires (one funnel per level-2 site) ---
                    \foreach \Rx in {0,1}{\foreach \Ry in {0,1}{
                        \pgfmathsetmacro\ppx{4*\Rx+2}
                        \pgfmathsetmacro\ppy{4*\Ry+2}
                        \draw[brick, line width=\wirewd] (\ppx,\ppy-1) -- (\ppx,\ppy+1);
                        \foreach \ax in {0,1}{\foreach \ay in {0,1}{
                        \pgfmathsetmacro\cx{4*\Rx+2*\ax+1}
                        \pgfmathsetmacro\cy{4*\Ry+2*\ay+1}
                        \draw[brick, line width=\wirewd]
                            (\cx,\cy) -- (\ppx,\cy);
                        }}
                    }}
                    % --- wires from the four level-2 sites into the central control site ---
                    \draw[ctrlpurple, line width=\wirewd] (4,2) -- (4,6);
                    \foreach \cx/\cy in {2/2, 2/6, 6/2, 6/6}{
                        \draw[ctrlpurple, line width=\wirewd]
                        (\cx,\cy) -- (4,\cy);
                    }
                    % --- I/O port at the center ---
                    \filldraw[fill=white, draw=ctrlpurple, line width=0.9pt] (\Win/2,\Win/2) circle (0.11);
                    % --- sites on top ---
                    \foreach \x in {0.5,1.5,...,7.5}{
                        \foreach \y in {0.5,1.5,...,7.5}{
                        \minicomp{\x}{\y}{l0grey}}}
                    \foreach \x in {1,3,5,7}{
                        \foreach \y in {1,3,5,7}{
                        \minicomp{\x}{\y}{steelblue}}}
                    \foreach \x in {2,6}{
                        \foreach \y in {2,6}{
                        \minicomp{\x}{\y}{brick}}}
                    % --- legend (self-centering text node) ---
                    \node[font=\scriptsize] at (\Win/2,-0.75) {\textcolor{lightblue}{\rule[0.55ex]{12pt}{0.7pt}}\,\,slice $0\!\to\!1$\hspace{1.729em}\textcolor{brick}{\rule[0.55ex]{12pt}{0.7pt}}\,\,slice $1\!\to\!2$\hspace{1.729em}\textcolor{ctrlpurple}{\rule[0.55ex]{9pt}{0.7pt}\,$\circ$}\,\,I/O};
                    \end{tikzpicture}
                \end{subfigure}
                % ---------------------------------------------------------------
                \caption{Layout of the local decoders within a surface-code patch. (A folded version of this layout is drawn schematically in Fig.~\ref{fig:architecture}.) The classical processors implement a hierarchical, surface-code version of the streaming decoder of Fig.~\ref{fig:online-decoder}. Sites in higher slices are sparser and run increasingly coarse-grained versions of the dynamics of Fig.~\ref{fig:hero}. This sparsification is why the decoder admits a constant-density two-dimensional embedding even though the number of slices diverges with the code distance. Both panels show the same grid of processors and differ only in which wiring is drawn. \textbf{Left:}~Intra-slice wiring, which implements each slice's internal dynamics. \textbf{Right:}~Inter-slice wiring, which carries promoted defects from one slice to the next, together with wires that connect the final-slice sites to a single constant-bandwidth input/output port at the center of the patch. The patch is controlled and read out entirely through this port. Program instructions of constant size (e.g., ``perform a logical Hadamard'') fan out from the port along the tree-like inter-slice wiring to individual sites. Logical measurement outcomes---parities of many local measurement outcomes---are aggregated along the same wiring and read out at the port.}
                \label{fig:toric-coarse-grained-layout}
            \end{figure}
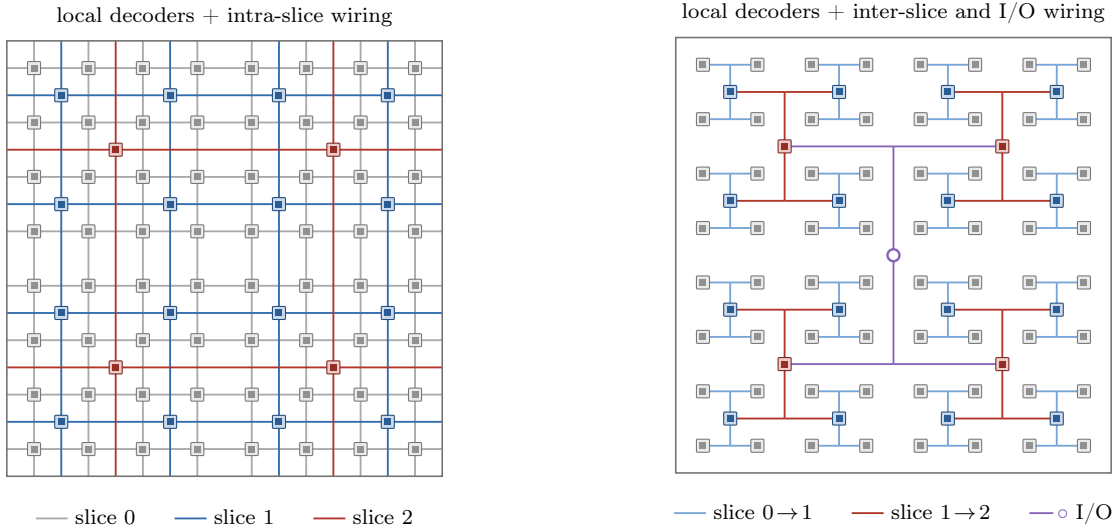

            The decoder's hierarchical layout (Fig.~\ref{fig:toric-coarse-grained-layout}) also serves as a control and readout system: each patch's single constant-bandwidth IO wire attaches to the final slice of the hierarchy at a dedicated port. Program instructions fan out from there, traversing the tree-like inter-slice wiring down to individual sites, and logical measurement outcomes---parities of many local measurement outcomes---are aggregated upward through the same wiring and read out at the port.

    \subsection{Comparison with prior work}\label{subsec:intro:prior-work}

        \begin{table}[!t]
            \centering
            \newcommand{\yes}{{\bfseries\color{ForestGreen}\scalebox{1.1}{$\checkmark$}}}
            \newcommand{\no}{{\bfseries\color{red}\scalebox{1.1}{\ding{55}}}}
            \newcommand{\maybe}{{\bfseries\scalebox{1.1}{?}}}
            {
            \setlength{\tabcolsep}{3pt}
            \small
            \begin{tabular}{l c c c c c}
                \toprule
                \bf{decoder} & \bf{codes} & \bf{\makecell{classical\\overhead}} &
                \bf{\makecell{robust to\\noisy\\class.\@ comp.\@}} & \bf{\makecell{local\\FTQC}} &
                \bf{\makecell{threshold\\value}} \\
                \midrule
                Lake~\cite{lake2025fastofflinedecodinglocal} & \makecell{mobile\\point-like defects} &
                $\poly(\log L)$ & \no & \no & \makecell{1.5\%} \\
                \midrule
                \multirow{2}{*}[-1.5ex]{\makecell[l]{Balasubramanian\\et al.~\cite{balasubramanian2025localautomaton2dtoric}}}
                & 2D toric & $\Theta(\log L)$ & \yes & \no & $\lesssim 0.01\%$ \\
                \cmidrule(lr){2-6}
                & \makecell{3D stack of\\2D toric} & $O(1)$ & \yes & \no & $\lesssim 0.01\%$ \\
                \midrule
                \makecell[l]{D\"unnweber\\et al.~\cite{dunnweber2026quantummemoryautonomouscomputation}} &
                2D concatenated & $O(1)$ & \yes & \yes & \maybe \\
                \midrule
                \textbf{this work} & \makecell{translation-invariant\\stabilizer codes} & $O(1)^*$ & \no & \yes$^*$ & \makecell{1.1\%}\\
                \bottomrule
            \end{tabular}}
            \caption{Comparison of local decoders. \emph{Codes:} code families or defect types each decoder handles. \emph{Classical overhead:} amount of classical resources per data qubit. \emph{Robust to noisy classical computation (class.\ comp.):} whether the decoder still has a threshold under noisy classical computation. \emph{Local fault-tolerant quantum computation (FTQC):} whether the code--decoder architecture supports local fault-tolerant quantum computation, rather than quantum memory alone. \emph{Threshold value:} reported threshold noise strength of quantum memories under phenomenological noise and synchronous updates. For D\"unnweber et al.'s construction, the threshold is unknown, but its value is expected to be impractically small. The starred classical-overhead entry indicates that, while our toric- and surface-code decoders have $O(1)$ classical overhead, our construction for completely general translation-invariant stabilizer codes requires $\poly(\log{L})$ overhead. The starred local-FTQC entry applies to the surface code only. The threshold value quoted for this work is for surface-code memory experiments using our $\poly(\log L)$-overhead surface-code decoder. \emph{Mobile point-like defects} refers to defects that are point-like and freely mobile, such as those in the repetition and toric codes and, more generally, Abelian anyons. The qualifier \emph{mobile} is used to exclude fractons, which are point-like defects that are subject to kinetic constraints and thus not freely mobile.}
            \label{tab:decoder-comparison}
        \end{table}

        In this subsection, we compare our constructions with prior work (see Table~\ref{tab:decoder-comparison}).\footnote{We do not consider local decoders for extended objects (e.g., membranes), which can be handled using variants of Toom's rule~\cite{Kubica_2019}.} Ultimately, the only other scalable constructions in low dimensions that support local fault-tolerant quantum computation are those based on either concatenated quantum circuits or G\'acs' hierarchical self-simulation techniques~\cite{gacs2001reliable}.

        We first discuss space- and time-translation-invariant, non-hierarchical decoders for codes with mobile point-like defects (e.g., the repetition and toric codes). Each of these decoders assumes noiseless classical computation.

        \textbf{GKL automaton.} Our repetition-code automata are similar to the GKL automaton and its variants~\cite{gacs1978one}. These constant-density rules decode the repetition code under code-capacity noise by eroding domains of errors. Although they can be modified to avoid measuring individual bits, this modification does not readily generalize to the quantum setting. Separately, their erosion mechanism is specific to one dimension. Our repetition-code automata, by contrast, use only conventional stabilizer measurements and a dimension-independent erosion principle. This allows them to generalize directly to quantum error-correcting codes in higher dimensions.

        \textbf{Lake's construction.} Lake designed a CA decoder for codes with mobile point-like defects and proved that it has a threshold under phenomenological noise. It implements a local approximation to the rule ``move each defect toward its nearest neighbor in spacetime'' by having defects broadcast messages about their locations~\cite{lake2025fastofflinedecodinglocal,lake2025localactiveerrorcorrection}. For maximal error suppression, it requires $\Theta(\log L)$ bits per qubit to store messages and $\poly(\log L)$ bits per qubit to store a spacetime window of height $\poly(\log L)$.\footnote{In the appendix of Ref.~\cite{lake2025fastofflinedecodinglocal}, Lake also constructed a constant-density code-capacity decoder that was numerically found, but not proven, to have a threshold.}

        \textbf{Paletta et al.'s construction.} Paletta et al.\ constructed a local decoder for the repetition code~\cite{paletta2025highperformancelocaldecodersdefect}, which Paletta later extended to the toric code~\cite{paletta2026localdecodertoriccode}. These decoders are similar to the GKL automaton but require $\Theta(\log L)$ bits per qubit. Both were simulated under phenomenological noise; the repetition-code decoder was additionally proven to have a code-capacity threshold.

        \textbf{Winter et al.'s construction.} Winter et al.\ constructed local, constant-density, GKL-like decoders for the repetition and toric codes~\cite{winter2026highperformancecellularautomatondecoders}. Both were numerically found to have code-capacity thresholds and were also simulated under phenomenological noise.

        \textbf{Field-based decoders.} Field-based decoders operate by moving defects along the gradient of a potential governed by Poisson's equation. For maximal error suppression, they require $\Theta(L)$ bits per qubit and a classical communication speed that diverges as $\Theta(L^2)$.

        The GKL automaton and its variants, as well as field-based decoders, provably lack a threshold under phenomenological noise~\cite{park1997ergodicity,lake2025localactiveerrorcorrection}. The constructions of Refs.~\cite{paletta2026localdecodertoriccode,winter2026highperformancecellularautomatondecoders} also appear to lack a threshold: for the toric code, their error suppression under phenomenological noise seems to improve with system size only up to a finite, noise-dependent scale, beyond which it plateaus. This is consistent with arguments from Ref.~\cite{lake2025localactiveerrorcorrection}, which suggest that these decoders should not have a threshold under phenomenological noise because the decoding dynamics are screened by new defects that are continually created by the environment.\footnote{Ref.~\cite{winter2026highperformancecellularautomatondecoders} also reports simulations of its decoders under classical computation noise. To be clear, the arguments of Ref.~\cite{lake2025localactiveerrorcorrection} suggest that, even with reliable classical computation, these decoders should not have a threshold under phenomenological noise.} By the same reasoning, we expect our code-capacity decoders to also lack a threshold under phenomenological noise.

        Our decoders improve upon these works in several ways. First, in the code-capacity setting, we prove that our constant-density toric-code decoder has a threshold, whereas Refs.~\cite{lake2025fastofflinedecodinglocal, winter2026highperformancecellularautomatondecoders} provide only numerical evidence for a threshold in the same resource regime. Second, our translation-invariant streaming decoders for phenomenological noise, which have density $\poly(\log\log L)$, have lower overhead than the translation-invariant construction of Ref.~\cite{lake2025localactiveerrorcorrection}, which has density $\poly(\log L)$, while satisfying the same rigorous memory-lifetime bounds and achieving comparable numerical performance. Third, our hierarchical constructions have rigorous thresholds under phenomenological noise while using only a constant density of resources; none of these prior works exhibits threshold behavior under phenomenological noise in this resource regime, even numerically. Fourth, unlike the decoders of Refs.~\cite{paletta2026localdecodertoriccode, winter2026highperformancecellularautomatondecoders}, our decoders handle open boundary conditions, which are essential for lattice-surgery-based computation. Fifth, the decoders in all of these prior works apply only to codes with mobile point-like defects, whereas our decoding framework extends to all translation-invariant stabilizer codes.

        We emphasize that none of these decoders directly support fault-tolerant quantum {\it computation}, and extending any of them to do so (especially with low overhead) requires proof techniques substantially beyond those used in the works above.

        Next, we discuss hierarchical decoders.

        \textbf{Broom algorithm.} Bravyi and Haah introduced a renormalization-group decoder for general topological codes on Euclidean lattices~\cite{Bravyi_2013}. For the toric code and Haah's code, this decoder uses a subroutine called the broom algorithm, which operates by sweeping defects in each error cluster toward a designated corner of the cluster. Haah subsequently proved that this sweeping strategy successfully decodes all translation-invariant local stabilizer codes on Euclidean lattices~\cite{Haah_2013}. 

        We make three modifications to the broom algorithm that enable it to locally decode all translation-invariant stabilizer codes. First, the original algorithm relies on global information and non-local communication. We design a local version of the algorithm that instead gathers the necessary information on the fly using message passing. Second, the original algorithm erases clusters in time linear in their volume, which makes it too slow to be used for local decoding. We parallelize the algorithm over all defects, which allows it to erase clusters in time linear in their diameter, making the algorithm fast enough to be used for local decoding. Finally, applying the original algorithm to phenomenological noise requires window decoding, and thus $\Omega(L)$ classical bits per site. Our construction instead extends the modified broom algorithm to the phenomenological-noise setting via a streaming decoder that uses only $\poly(\log{L})$ bits per site.

        \textbf{Tsirelson's automaton and its quantum analog.} Tsirelson~\cite{10.1007/BFb0070081} introduced a hierarchical local decoder for the repetition code and proved that it has a threshold. Balasubramanian et al.~\cite{balasubramanian2025localautomaton2dtoric} constructed an analog for the toric code and proved the corresponding result. Both constructions break space- and time-translation invariance, use strictly local interactions, and do not require reliable classical computation. In each construction, the per-site classical resources necessary to store the local update rules grow as $\Theta(\log L)$.\footnote{Assuming access to noiseless local classical computation does not immediately allow one to remove this overhead. In particular, even if one could simulate noiseless local classical computation with constant spacetime overhead using only noisy local operations, it is not clear how to exploit such a primitive to make the per-site classical resources constant.} This diverging resource overhead can be removed but only by increasing the spatial dimension by one and decoding a $(D+1)$-dimensional stack of $D$-dimensional codes. Our decoders, by contrast, achieve constant density in only two spatial dimensions---rather than the three required by the higher-dimensional variant of Ref.~\cite{balasubramanian2025localautomaton2dtoric}---and support fault-tolerant quantum computation in addition to quantum memory, albeit under the assumption of noiseless classical computation.
        
        \textbf{Harrington's decoder.} Harrington's renormalization-group-based decoder for the toric code~\cite{Harrington2004} can be viewed as a precursor to our hierarchical streaming decoders. It uses bounded-speed classical communication, requires $\poly(\log L)$ classical bits per qubit, breaks space- and time-translation invariance, and assumes noiseless classical computation. As noted in Ref.~\cite{balasubramanian2025localautomaton2dtoric}, the original argument may not fully rigorously establish fault tolerance in the presence of transient faults, although fault tolerance was later proven for the same decoder under the additional assumption of instantaneous classical communication~\cite{dauphinais2017fault}. Our hierarchical streaming constructions improve upon Harrington's construction in several ways. First, we rigorously prove that our constructions have a threshold without assuming instantaneous classical communication. Second, we show that our toric- and surface-code decoders admit constant-density two-dimensional embeddings. Third, our decoders accommodate open boundary conditions (for the repetition and toric codes), and our translation-invariant streaming decoders extend to all translation-invariant stabilizer codes. Fourth, our constructions apply not only to quantum memory but also to fault-tolerant quantum computation.

        Finally, we discuss prior constructions that support scalable fault-tolerant computation.

        \textbf{Concatenated circuits.} Scalable fault-tolerant quantum computation in three spatial dimensions and below can be achieved using circuit concatenation. The threshold theorems of Aharonov and Ben-Or~\cite{aharonov1996faulttolerantquantumcomputation} and Knill, Laflamme, and Zurek~\cite{knill1998resilient} were originally proven using concatenated codes, and the construction of Aharonov and Ben-Or becomes fully local---without assuming reliable classical computation---once the classical processing responsible for error correction is treated as part of the circuit being concatenated~\cite{gottesman2024surviving}. Forthcoming work along these lines constructs a two-dimensional universal fault-tolerant quantum computer that requires only a constant number of control lines~\cite{pattison_fault_tolerant}. These approaches therefore achieve locality by absorbing all decoding into the fault-tolerant circuit itself, whereas ours delegates decoding to a noiseless classical cellular automaton running alongside the code.

        \textbf{G\'acs' self-simulating constructions.} In a series of works, G\'acs has developed an incredibly sophisticated technique for performing scalable, noise-robust universal classical computation using a hierarchy of self-simulating Turing machines that successively perform different levels of error-corrected information processing in classical concatenated codes~\cite{gacs1989self,gacs2001reliable}. In two or more dimensions, errors are corrected using Toom's rule; in one dimension, the mechanism is more obscure.
        
        \textbf{D\"unnweber et al.'s quantum extension.} D\"unnweber et al.\ recently showed that G\'acs' two-dimensional construction can be generalized to perform scalable fault-tolerant \emph{quantum} computation~\cite{dunnweber2026quantummemoryautonomouscomputation}. Apart from ours and those based on circuit concatenation, this is the only construction capable of performing scalable fault-tolerant quantum computation in three spatial dimensions and below. Moreover, unlike our approach, it does not require reliable classical computation. The main drawback of this approach is that---at least in its present form---its complicated nature likely renders it extremely impractical (even numerically estimating the value of the threshold appears to be a rather daunting task). Additionally, the fault-tolerant preparation of the scheme's requisite initial state using the same fixed interactions that subsequently perform the computation remains an open problem.
        
    \subsection{Outlook}\label{subsec:intro:outlook}
    
        Our work advances two intertwined programs: the characterization of noise-robust quantum phases of matter and the development of scalable fault-tolerant quantum computers. Our general result establishes that far more quantum phases of matter than previously thought can be stabilized by local error-correcting dynamics, while our architecture shows that one of the leading approaches to fault-tolerant quantum computation---that based on topological codes and lattice surgery---can be implemented in a provably scalable, fully local manner.

        We conclude by surveying some open questions and offering perspective on our results.

        \textbf{Optimizing practical performance.} Our constructions are primarily intended as proofs of principle, and there is substantial room for practical improvement, e.g., by using the neural-cellular-automaton techniques developed in Ref~\cite{Pajouheshgar_2026}. Our architecture also employs the most basic forms of lattice surgery and state distillation; incorporating more recent advancements could substantially reduce the spacetime overhead of these primitives in our scheme.

        \textbf{Toward an absolutely stable architecture.} Our architecture assumes access to both a reliable global clock and noiseless classical computation. Removing these assumptions would yield an \emph{absolutely stable} fault-tolerant quantum computer based on topological codes.

        There are at least two reasons to construct such a computer. First, from the perspective of non-equilibrium quantum phases of matter, the decoding mechanism underlying this architecture could lead to a variety of quantum phases quite different from the one realized by D\"unnweber et al.'s absolutely stable quantum computer based on concatenated codes~\cite{dunnweber2026quantummemoryautonomouscomputation}, especially if this mechanism proves to be as broadly applicable to different types of topological order as the decoding principle presented in this work.\footnote{Achieving such broad applicability would also require constant-density decoders for general translation-invariant stabilizer codes. This problem is interesting in its own right. In particular, it would be valuable to understand whether such constant-density decoders are possible in general or whether there is some fundamental barrier for codes that are not invariant under coarse-graining, e.g., fracton codes. The entanglement renormalization group~\cite{Vidal_2007, Haah_2014} may prove useful for addressing this question.}

        Second, from a practical perspective, removing the need for noiseless classical computation would enable the error-correcting dynamics to be implemented in a fully coherent manner (i.e., in a ``measurement-free'' way). Such an implementation could increase the speed of the quantum computer by eliminating the latency associated with performing measurement and classical processing before applying feedback. Because such an implementation would forgo the use of reliable classical hardware, it would likely achieve weaker error suppression than our current protocol---a poor trade at present. As quantum hardware improves, however, speed will become increasingly important relative to error suppression, so this trade may eventually become worthwhile. D\"unnweber et al.'s construction already achieves such a coherent implementation, but qualitative features suggest that its threshold is impractically small; a coherent implementation based on our architecture could have a much higher threshold. Indeed, even topological-code decoders that likely lack a threshold under classical computation noise already perform quite well at practically relevant system sizes, so similar decoders that do have a threshold could plausibly perform at least as well.

        We now assess the feasibility of removing these assumptions. Of the two, access to a reliable global clock may be the easier one to remove. Specifically, we believe that our streaming decoders can be desynchronized using techniques similar to those of Ref.~\cite{lake2025localactiveerrorcorrection}, and these same techniques could also serve as a starting point for making the entire quantum computer robust to desynchronization.

        Removing the assumption of noiseless classical computation is likely more challenging. One possible route is to implement our decoding principle using G\'acs' self-simulation techniques; another is to employ von Neumann-style classical fault tolerance~\cite{von1956probabilistic}. Alternatively, one could start from the hierarchical construction of Balasubramanian et al.~\cite{balasubramanian2025localautomaton2dtoric}, which establishes a threshold for the toric code without assuming access to reliable classical computation, and attempt to use our ideas to modify their construction to perform universal quantum computation. In particular, one could modify their construction to operate on surface codes and use our techniques to prove an $O(p)$ logical error rate guarantee for state injection. Clifford operations could then be implemented using transversal CNOT and fold-transversal Hadamard gates on a three-dimensional stack of folded surface codes, with non-Clifford gates supplied through state injection and distillation. A central unresolved issue in this approach, however, is how to implement the acceptance, rejection, and routing decisions required by distillation when the classical control responsible for these decisions is itself noisy. 

        \textbf{Local just-in-time decoding for non-Abelian codes.} Our architecture implements non-Clifford gates (and $S$-gates) using state distillation. Recent work has opened an alternative route to two-dimensional universal fault-tolerant quantum computation, in which non-Clifford gates are performed by interfacing the surface code with non-Abelian codes~\cite{Davydova_2026, delafuente2026highthresholddecodingnonpaulicodes}. Because these schemes avoid state distillation, they have the potential to substantially reduce the spacetime overhead of fault-tolerant quantum computation. For example, the unitary $T$ gate of Ref.~\cite{Davydova_2026} could be used to remove $Y$-state and magic state distillation from our architecture entirely, immediately making its spacetime overhead competitive with that of architectures based on topological codes and state-of-the-art distillation protocols.

        These schemes, however, require \emph{just-in-time} decoders, which must commit to corrections on the fly before the full syndrome history is available. Therefore, realizing such reductions in our architecture would require \emph{local} just-in-time decoders. It is likely that our framework can be used to construct such decoders, since it natively handles the tasks at the core of just-in-time decoding: our streaming decoders already commit to corrections on the fly and locally resolve intrinsically random measurement outcomes into globally consistent interpretations. One difference is that corrections could no longer be tracked in software and would instead have to be applied physically. Another likely difference is that one would need to run separate just-in-time decoders for charges and fluxes and interface them so that they operate in concert, whereas in our case the decoders for charges and fluxes are completely decoupled.

        Separately, it would be interesting to explore whether our framework extends to fault-tolerant topological quantum computing schemes based on non-Abelian anyons, such as the ungauging approach of Ref.~\cite{lyons2026quantumcomputinganyonsfault}.

    \subsection{Structure of paper}\label{subsubsec:intro:structure}

            The structure of the remainder of this paper is as follows. In Section~\ref{sec:preliminaries}, we present definitions and notation for the cellular automata and codes we study, and we introduce the multiscale cluster decomposition used throughout our analysis. In Section~\ref{sec:threshold}, we formally define the linear erosion properties described in the overview and show how these properties imply both a non-zero threshold and polylogarithmic average decoding time for any code-capacity decoder that satisfies them. In Section~\ref{sec:rep-code}, we construct a constant-density code-capacity decoder for the repetition code---the simplest realization of our decoding strategy and the template for all subsequent constructions---and prove that it satisfies the desired linear erosion properties; we also numerically study its threshold and behavior under desynchronization. In Section~\ref{sec:toric-code}, we carry out an analogous construction and analysis for the toric and surface codes.

            In Section~\ref{sec:online-decoding}, we construct translation-invariant streaming decoders for the repetition, toric, and surface codes under phenomenological noise. We first prove that these decoders achieve stretched-exponential memory lifetimes while using only $\poly(\log{L})$ classical bits per site, and then refine the analysis to show that $\poly(\log\log{L})$ bits suffice; we also show how to use these decoders to perform stabilizer-state preparation and, for the surface code, state injection. We conclude with numerical simulations of surface-code memory experiments.

            In Section~\ref{sec:constant-density}, we construct streaming decoders for the toric and surface codes that use only a constant density of classical bits and bounded-bandwidth, bounded-propagation-speed wires. These decoders are two-dimensional but not translation-invariant. We prove that they achieve stretched-exponential memory lifetimes and describe how to adapt the preparation and injection procedures of Section~\ref{sec:online-decoding} to the constant-density setting. We also show how these decoders can be controlled and read out through a single constant-bandwidth wire.
 
            In Section~\ref{sec:decoding-during-computation}, we use our constant-density surface-code streaming decoder to construct a local, two-dimensional fault-tolerant quantum computer, and we prove a threshold theorem for this architecture. In particular, we show how to perform local decoding during lattice surgery and Hadamard gates and how to control these operations through each patch's single constant-bandwidth wire.

            Finally, in Section~\ref{sec:general}, we use Haah's polynomial formalism~\cite{Haah_2013, Haah_2017} to generalize our constructions to arbitrary translation-invariant stabilizer codes. We prove that every such code admits a translation-invariant code-capacity CA decoder with a non-zero threshold and polylogarithmic average decoding time, as well as a translation-invariant streaming decoder with a non-zero threshold under phenomenological noise. Both decoders use $\poly(\log{L})$ classical bits per site; for the code-capacity decoder, this overhead can be reduced to a constant, at the cost of breaking translation invariance.
            
\section{Preliminaries}\label{sec:preliminaries}

    In this section, we introduce the definitions needed to describe discrete-time cellular-automaton decoders and the probabilistic tools necessary to analyze their performance. Readers who are familiar with this material are encouraged to skip ahead to the next section.
    
    \subsection{Cellular automata}\label{subsec:preliminaries:cellular-automata}
        For the majority of this work, we consider only translation-invariant geometrically local CSS codes on $D$-dimensional Euclidean lattices. Furthermore, we restrict to decoders that decode $X$- and $Z$-type errors independently. The lone exception is Section~\ref{sec:general}, which treats general (non-CSS) Pauli stabilizer codes. When a duality relates the two error types---or, as in the repetition code, when only one type of error is correctable---we treat only the decoding of $X$-type errors using $Z$-type stabilizers. For simplicity, we state the definitions in this section for the code-capacity setting; generalizations to phenomenological noise and to fault-tolerant quantum computation will be introduced as needed throughout the paper.

        Let $\Lambda = \Z_L^D$ be a finite $D$-dimensional lattice, where each site corresponds to a unit cell. All lattices are assumed to have periodic boundary conditions unless stated otherwise. For the codes we study, each site $\vc{x}\in\Lambda$ is associated with a unique parity check $s(\vc x)$ and a finite set of qubits indexed by a set $J$. This gives the identifications $\Lambda_s \coloneqq \Lambda$ and $\Lambda_e \coloneqq \Lambda \times J$, where $\Lambda_s$ is the set of checks and $\Lambda_e$ is the set of qubits. We denote the qubit at site $\vc{x}$ with index $j\in J$ as $(\vc{x},j)$, and throughout this section, we assume $J = \{1,\dots,n_e\}$ for some integer $n_e\ge 1$. In general, we do not distinguish between checks and sites.
        
        To illustrate our notation, we briefly consider the one-dimensional repetition code. In this case, we have $J = \{1\}$, so $\Lambda_s=\Lambda_e=\Z_L$. The qubit at site $i$ is denoted by $(i,1)\in\Lambda_e$, and the check $Z_i Z_{i+1}$ is denoted by $s(i)\in\Lambda_s$, where $i$ and $i+1$ are understood modulo $L$.

        Throughout, all distances are measured using the $\ell_1$-distance $d$ on $\Lambda$. The closed ball of radius $r$ centered at site $\vc x$ is defined as
        \[
            B_r(\vc x)\coloneqq \{\vc y\in\Lambda : d(\vc x,\vc y)\le r\,\},
        \]
        and all balls are assumed to be closed unless stated otherwise. The distance between a check and a qubit, or any two checks or qubits, is defined as the distance between their respective sites. By geometric locality of the code, each check involves only qubits located at sites within a finite distance $d_0 > 0$ of the check.

        In this work, we consider only synchronous, discrete-time CA decoders. In the code-capacity setting, any such decoder can be converted into an asynchronous, continuous-time decoder without sacrificing performance (either in the logical error rate or in the scaling of the expected decoding time) using the so-called ``marching soldiers'' scheme~\cite{berman1988investigations, cook2008self}. For an explanation of this conversion procedure, see Ref.~\cite{lake2025fastofflinedecodinglocal}. In the phenomenological noise setting, converting a synchronous decoder into an asynchronous one is not as straightforward; in our case, it may still be possible using the techniques of Ref.~\cite{lake2025localactiveerrorcorrection}, but we do not pursue this possibility here.

        To synchronize the dynamics across different spatial locations, we assume the existence of a reliable discrete global clock with non-negative integer times $t \in \Z_{\geq 0}$, so our spacetime is $\Lambda \times \Z_{\geq 0}$. Often, we use non-integer times for notational convenience, but these should be interpreted as floored unless stated otherwise.
        
        Let $\mcs$ be a finite set of states. A \emph{trajectory} is a function
        \[
            \zeta: \Lambda \times \Z_{\geq 0} \to \mcs,
        \]
        and a \emph{configuration} is a function
        \[
            \xi : \Lambda \to \mcs.
        \]
        A \emph{cellular automaton} is a local update rule
        \[
            \mathcal{A} : \mcs^{\mcn} \to \mcs,
        \]
        where the \emph{neighborhood} $\mcn$ is a fixed finite set of displacement vectors
        \[
            \mcn \subset \mathbb{Z}^D.
        \]
        The \emph{interaction range} $R$ of the automaton is defined as
        \[
            R \coloneqq \max_{\vc n\in\mcn} \norm{\vc n}_1,
        \]
        where $\norm{\vc n}_1$ is the $\ell_1$-norm.
        
        Unless stated otherwise, we assume that all sites are updated synchronously according to a translation-invariant cellular automaton $\mathcal{A}$. In particular, given a configuration $\xi$, the updated configuration $\xi'$ is defined for all sites $\vc x \in \Lambda$ by
        \[
            \xi'(\vc x)
            =
            \mathcal{A}\left(\xi(\vc x + \vc n)_{\vc n \in \mcn}\right),
        \]
        where the addition is understood modulo $L$ under periodic boundary conditions. Therefore, given an initial configuration $\zeta(\vc x, 0)$, the cellular automaton induces a trajectory $\zeta$ via
        \[
            \zeta(\vc x, t+1)
            =
            \mathcal{A}\left(\zeta(\vc{x}+\vc{n}, t)_{\vc{n}\in\mcn}\right).
        \]

        We parameterize the state at each site $\vc x$ as
        \[
            \xi(\vc{x})=\bigl(e_1(\vc{x}),\ldots,e_{n_e}(\vc{x}), s(\vc{x}), m_1(\vc{x}),\ldots,m_{n_m}(\vc{x}), c(\vc{x})\bigr)
        \]
        for some integer $n_m \geq 1$. Each entry of $\xi(\vc x)$ is called a \emph{channel}. There are four types of channels: 
        \begin{enumerate} 
            \item correction channels $e_j(\vc x) \in \{0,1\}$, which equal $1$ iff qubit $(\vc{x},j)$ has been flipped an odd number of times by the decoding dynamics;
            \item a syndrome channel $s(\vc x) \in \{0,1\}$, which stores the value of the syndrome at location $\vc x$ and equals $1$ iff the parity check $s(\vc x)$ is violated;
            \item message channels $m_j(\vc x) \in \{0,1\}$, which store auxiliary classical information used by the cellular automata; and
            \item a clock channel $c(\vc x) \in \{0, \ldots, q-1\}$ for some integer $q \ge 2$, which increments by $1$ modulo $q$ at each time step.
        \end{enumerate} 
        When $n_e=1$, we write $e(\vc{x})$ instead of $e_1(\vc{x})$, and when $n_m=1$, we write $m(\vc{x})$ instead of $m_1(\vc{x})$.
        
        A check $s(\vc x)\in\Lambda_s$ with $s(\vc x)=1$ is called a \emph{defect}, and the contents of message channels are called messages. A message or syndrome at a given site at time $t$ is said to be trivial if it is zero. When the syndrome configuration is identically zero, we say the syndrome configuration is trivial, and when all messages are trivial, we say the message configuration is trivial. We will abuse notation by saying that errors, corrections, and defects are supported on balls in $\Lambda$, even though qubits live in $\Lambda_e$ and checks live in $\Lambda_s$; formally, it is the corresponding sites in $\Lambda$ that lie in those balls.

        We assume a fixed \emph{default initial configuration} in which all message, correction, and clock channels are zero at every site. Thus, at time $t=0$, the only variable data are the initial error configuration and the corresponding initial syndrome configuration. Because of this, the clock channels at all sites also remain synchronized; we write $c(t)$ for their common value.
        
        Next, we assume there exists a finite distance $d_1 > 0$ such that $e_j(\vc x,t)$ may change at time $t+1$ only if there exists a defect $s(\vc y, t) = 1$ with $d(\vc x, \vc y) \le d_1$; we call this the \emph{defect-gated flip property}. Intuitively, this means that a qubit can be flipped by the decoder only if there is an active defect nearby.
        
        We assume that the cellular-automaton dynamics are constrained to produce syndromes equal to the syndrome of the modulo-two sum of the initial error configuration and current correction configuration. Together, the defect-gated flip property and this syndrome-consistency assumption imply that if the syndrome configuration is identically zero at time $t$, then for all times $t' \geq t$, the syndromes remain zero and all correction bits remain unchanged; we refer to this as the \emph{no-spontaneous-defects property}. We further assume that the automaton reads correction bits only when computing syndrome information; all other dependence on the correction configuration is mediated by the syndrome channels. Finally, we assume that trivial messages remain trivial at any site whose neighborhood has trivial syndromes and messages. 
 
        A cellular automaton that is defined for a translation-invariant geometrically local CSS code and satisfies all of the assumptions stated in this subsection is called a \emph{cellular-automaton decoder}.

        Although we only consider geometrically local CSS codes, our framework can be used to perform code-capacity decoding for any translation-invariant non-Abelian anyon model. The main differences are that defects must be moved using ribbon operators and that anyons must be remeasured at each time step due to non-determinism in fusion outcomes. For more details, see Ref.~\cite{lake2025fastofflinedecodinglocal}.

    \subsection{Clustering}\label{sec:clustering}

        Here, we describe the sparsity theorem---originally due to G\'acs (see, e.g., Ref.~\cite{gacs1989self}), and employed in a diverse range of work in quantum error correction---which decomposes an error configuration into a hierarchy of well-separated clusters of increasing size, with noise events rapidly becoming rare at higher levels. This decomposition provides the basic organizing principle for our decoding strategy. For simplicity, we phrase the definitions in terms of qubit errors in space, but the framework applies unchanged to the $(D+1)$-dimensional spacetime error configurations that arise under phenomenological noise or during fault-tolerant quantum computation.
        
        The results in this subsection are stated without proof and are not new. Our exposition closely follows that of Appendix A of Ref.~\cite{lake2025fastofflinedecodinglocal}.

        Going forward, we will often identify elements of $\Lambda_e$ with error locations and refer to these elements as \emph{errors}. For a subset $S \subseteq \Lambda_e$, we define its diameter by
        \[
            \operatorname{diam}(S) \coloneqq \max_{u,v\in S} d(u,v).
        \]

        \begin{definition}[$p$-bounded noise]
            A \emph{noise realization} $N$ is a random subset of $\Lambda_e$. We say that the noise is \emph{$p$-bounded} if, for all finite subsets $A \subset \Lambda_e$, the marginal probability for errors to occur on all locations in $A$ is exponentially small in the size of $A$: 
            \begin{equation}
                \label{pbound} \Pr[A \subseteq N] \leq p^{|A|}.
            \end{equation}
        \end{definition}
        In what follows, we will always assume the existence of a finite $0 < p < 1$ for which the above holds. A slight extension of this definition broadens our discussion to encompass a very general range of error models, including ones that are non-Markovian, continuous-time, and coherent (see, e.g., Ref.~\cite{gottesman2024surviving}). For simplicity of presentation, however, we focus on the standard (incoherent) case throughout.

        \begin{definition}[clustering]
            Let $N$ be a noise realization. A $(W,B)$\emph{-cluster} $C_{(W,B)}$ is a subset of $N$ such that
            \begin{enumerate}
                \item $\operatorname{diam}(C_{(W,B)}) \leq W$, and
                \item $C_{(W,B)}$ is separated from all other errors in $N$ by a buffer distance of size greater than $B$:
                \[
                    d(N \setminus C_{(W,B)}, C_{(W,B)}) > B,
                \]
                where we have defined the distance between two sets $A_1,A_2$ as
                \[
                    d(A_1,A_2) = \min_{x\in A_1,y\in A_2} d(x,y).
                \]
            \end{enumerate}
            An error $x\in N$ is called $(W,B)$\emph{-clustered} if it is a member of a $(W,B)$-cluster, and a subset of $\Lambda_e$ is said to be $(W,B)$-clustered if all errors contained in it are $(W,B)$-clustered.
        \end{definition}
        
        We now fix a triple of positive constants $(\beta, \gamma, n)$, with $n$ a positive integer, satisfying
        \begin{equation}
            \label{paramineqs} \frac{2\gamma}{1 - 1/n} < \beta < \gamma n
        \end{equation}
        and define the variables
        \[
            w_k = w_0 n^k, \qquad b_k = b_0 n^k \qquad \text{for }k \geq 1,
        \]
        where $w_0 = 2\beta$ and $b_0 = \gamma n - \beta$. Note that if $(\beta, \gamma, n)$ satisfy the inequalities \eqref{paramineqs}, then $(\beta, \gamma, n')$ satisfy these inequalities for any integer $n' \geq n$. Furthermore, we have that
        \begin{equation}
            \label{bkwk}
            \frac{b_k}{w_k} = \frac{(\gamma/\beta)n-1}{2},
        \end{equation}
        which grows linearly with $n$. Therefore, any triple $(\beta, \gamma, n)$ that satisfies the inequalities \eqref{paramineqs} gives rise to an entire family of triples $(\beta,\gamma,n')$ that satisfy \eqref{paramineqs} and for which $b_k/w_k$ can be made arbitrarily large by choosing $n'$ sufficiently large. We will not use this property in this section, but we will make heavy use of it in nearly all later sections.
        
        These parameters control the scales involved in different levels of the hierarchical structure we use to coarse-grain the noise, in the manner made precise in the following definitions:
        \begin{definition}[level-$k$ noise and error sets]
            Let $N_0 = N$. The \emph{$(k+1)$th level noise set $N_{k+1}$} is defined as the set obtained by deleting from $N_k$ all $(w_k, b_k)$-clustered errors.

            The subset of $N$ that is deleted when passing from level $k$ to level $k+1$ defines the \emph{level-$k$ clustered error set}, which we write as $E_k$:
            \[
                E_k = N_{k} \setminus N_{k+1}.
            \]
        \end{definition}

        A $(w_k, b_k)$-cluster $C_{(w_k,b_k)}$ will be called a $k$\emph{-cluster} and will often be denoted by $C_k$. 

        \begin{proposition}
            If $b_k \geq w_k$, then the clusters removed when passing from level $k$ to level $k+1$ are disjoint:
            \[
                E_k = \bigsqcup_i C_k^{(i)},
            \]
            where $\bigsqcup$ denotes a disjoint union. Since $N = \bigsqcup_k E_k$, this gives a decomposition of $N$ into disjoint clusters as
            \[
                N = \bigsqcup_{k,i} C_k^{(i)}.
            \]
        \end{proposition}
        We will always assume $b_k \geq w_k$ for all $k$ in order to make use of the above proposition, which we will need throughout this work; however, this assumption is not necessary for the sparsity theorem to hold.

        \begin{definition}
            The \emph{level-$k$ error rate} $p_k$ is the largest probability with which a particular error belongs to $N_k$:
            \[
                p_k = \max_{x \in \Lambda_e} \Pr[x\in N_k].
            \]
        \end{definition}

        If $p$ is sufficiently small, then the quantities $p_k$ decay very rapidly with $k$.
        \begin{theorem}[sparsity bound]\label{thm:sparse}
            Suppose $(\beta, \gamma, n)$ satisfy the inequalities \eqref{paramineqs}. Then, for any $p$-bounded noise model on $\Lambda_e = \Z^D \times J$, the level-$k$ error rate satisfies
            \[
                p_k \leq C\left(\frac{p}{p_*}\right)^{2^k},
            \]
            where $p_*$ and $C$ are positive, $k$-independent constants that depend only on $D$, $\beta$, $\gamma$, $n$, and the number $|J|$ of qubits per site.
        \end{theorem}

        \begin{remark}
            The sparsity theorem as stated above is for $\Lambda_e = \Z^D \times J$, i.e., for an infinite system. The theorem also holds, with the same constants, for all finite systems $\Lambda_e = \Z_L^D \times J$.
        \end{remark}

        We now briefly sketch why the framework introduced here is useful. Let $L$ be the linear system size. First, note that the sparsity theorem implies that, below threshold, the probability that a $k$-cluster with $w_k = \Omega(L)$ occurs decays rapidly with $L$. Next, let $E$ be the event that no such $k$-cluster occurs. In subsequent sections, we will show that, conditioned on $E$, our decoders always (1) annihilate defects in any cluster with other defects from the same cluster, and (2) allow clusters to expand by at most a constant factor before being annihilated. Since $k$-clusters have diameter at most $w_k$, the resulting correction will have support of diameter $O(w_k)$ and thus cannot cause a logical error unless $w_k$ is of order $L$, which we have conditioned on not happening. Therefore, the probability that a logical error occurs is at most the probability that $E$ does not occur, which, by the sparsity theorem, decays very rapidly with $L$.

\section{Code-capacity thresholds and decoding times}\label{sec:threshold}
    
    In this section, we formally define several linear erosion properties and prove that any CA decoder that satisfies them has a non-zero threshold $p_c \geq p_*$ and polylogarithmic (in $L$) average decoding time for $p < p_*$.\footnote{Straightforward extensions of arguments from Ref.~\cite{lake2025fastofflinedecodinglocal} imply an even stronger result, namely, that the decoding time of individual runs is polylogarithmic with probability at least $1 - 1/\poly(L)$.} Throughout, we work in the code-capacity setting and consider a fixed CA decoder together with a fixed code. The average decoding time result applies only to the code-capacity setting and follows from Ref.~\cite{lake2025fastofflinedecodinglocal}. The threshold result is stated for the code-capacity setting, but similar results for phenomenological noise and fault-tolerant quantum computation follow by the same proof technique.
    
    \subsection{Linear erosion properties}\label{subsec:threshold:linear-erosion}

        We begin by defining non-trivial messages, corrections, and syndromes.

        \begin{definition}[non-trivial messages, corrections, and syndromes]
            Let $\vc m(\vc x, t)$, $\vc e(\vc x, t)$, and $s(\vc x, t)$ denote the message, correction, and syndrome configurations at site $\vc x$ and time $t$. We say these are \emph{non-trivial} at site $\vc x$ if they are non-zero. Their supports at time $t$ are
            \begin{align*}
                \supp(\vc m(t)) &\coloneqq \{\vc x \in \Lambda : \vc m(\vc x, t) \neq \bf{0}\},\\
                \supp(\vc e(t)) &\coloneqq \{\vc x \in \Lambda : \vc e(\vc x, t) \neq \bf{0}\},\\
                \supp(s(t)) &\coloneqq \{\vc x \in \Lambda : s(\vc x, t) \neq 0\}.
            \end{align*}
        \end{definition}

        Next, we formally bound the speed at which non-trivial messages, corrections, and syndromes can spread. Recall that our setting has three underlying locality properties: (1) geometric locality of the code, i.e., each check only involves qubits within a distance $d_0$ of the check; (2) a finite interaction range $R$ of the CA decoder, which implies the update at each site $\vc x$ can only depend on channels at sites within a distance $R$ of $\vc x$; and (3) the defect-gated flip property, i.e., the correction $e_j(\vc x, t)$ at a qubit $(\vc x, j)$ can only change at time $t+1$ if there is a defect within a distance $d_1$ of $\vc x$ at time $t$.

        The following lightcone bound holds as an immediate consequence of the above three locality properties. It states that the support of non-trivial messages, corrections, and syndromes can only grow by at most $\max(R,d_0 + d_1)$ per time step.

        \begin{lemma}[lightcone bound]\label{lem:lightcone}
            If the initial error configuration is supported in $B_r(\vc x)$ for some site $\vc x \in \Lambda$ and radius $r \geq 0$, then for all $t \geq 0$,
            \[
                \supp(\vc m(t)), \supp(\vc e(t)), \supp(s(t)) \subseteq B_{r + d_0 + vt}(\vc x),
            \]
            where $v = \max(R, d_0 + d_1) > 0$.
        \end{lemma}

        Consequently, for every point $(\vc x, t)$ in spacetime, the set of all spacetime points that could have affected the state $\zeta(\vc x, t)$ is well-defined. We call this set of points the past lightcone of $(\vc x, t)$.

        We now define linear defect erosion, linear message erosion, and linear cluster erosion.

        \begin{definition}[linear defect erosion]
            A CA decoder is said to satisfy linear defect erosion if there exists a constant $a \geq 1$ such that, for every radius $r \geq 1$ and site $\vc x \in \Lambda$, the following holds whenever the initial error configuration is supported in $B_r(\vc x)$:
            \begin{enumerate}
                \item All defects are eliminated by some time $t \leq ar$.
                \item For all $t \geq 0$, the correction and syndrome configurations are supported in $B_{ar}(\vc x)$.
            \end{enumerate}
        \end{definition}

        \begin{definition}[linear message erosion]
            A CA decoder is said to satisfy linear message erosion if there exists a constant $b \geq 1$ such that, for every time $t_0$, radius $r \geq 1$, and site $\vc x \in \Lambda$, the following holds whenever the syndrome configuration is trivial for all times $t \geq t_0$ and $\supp(\vc m(t_0)) \subseteq B_r(\vc x)$:
            \begin{enumerate}
                \item $\supp(\vc m(t)) = \varnothing$ for all $t \geq t_0 + br$.
                \item $\supp(\vc m(t)) \subseteq B_{br}(\vc x)$ for all $t \geq t_0$.
            \end{enumerate}
        \end{definition}

        \begin{definition}[linear cluster erosion]\label{def:linear_cluster_erosion}
            A CA decoder is said to satisfy linear cluster erosion if there exists a constant $\lambda \geq 1$ such that the following holds for every radius $r \geq 1$, site $\vc x \in \Lambda$, and initial error configuration supported in $B_r(\vc x)$:
            \begin{enumerate}
                \item The entire spacetime region where the syndrome and message configurations may be non-trivial is contained in $B_{\lambda r}(\vc x) \times [0, \lambda r]$.
                \item The correction configuration is supported in $B_{\lambda r}(\vc x)$ for all $t \geq 0$ and remains constant for all $t \geq \lambda r$.
            \end{enumerate}
        \end{definition}

        \begin{lemma}
            If a CA decoder satisfies linear defect erosion and linear message erosion, then it also satisfies linear cluster erosion.
        \end{lemma}

        \begin{proof}
            Suppose the initial error configuration is supported in $B_r(\vc x)$ with no other errors present. By linear defect erosion, all defects are eliminated by time $ar$ and for all $t \geq 0$, the correction and syndrome configurations are supported in $B_{ar}(\vc x)$. By the defect-gated flip property, the correction configuration remains constant for all $t \geq ar$, and by the no-spontaneous-defects property, the syndrome configuration remains trivial for all $t \geq ar$. Next, by the lightcone bound, $\supp(\vc m(t)) \subseteq B_{r + d_0 + vt}(\vc x)$ for all $t \geq 0$, so in particular $\supp(\vc m(ar)) \subseteq B_{(1+av)r + d_0}(\vc x)$. Finally, linear message erosion implies $\supp(\vc m(t)) \subseteq B_{b((1+av)r + d_0)}(\vc x)$ for all $t \geq ar$ and $\supp(\vc m(t)) = \varnothing$ for all $t \geq ar + b((1+av)r + d_0)$. Let $\lambda = a + b((1+av) + d_0)$. Then, the entire spacetime region where the syndrome and message configurations may be non-trivial is contained in $B_{\lambda r}(\vc x) \times [0, \lambda r]$. Furthermore, the correction configuration is supported in $B_{\lambda r}(\vc x)$ for all $t \geq 0$ and remains constant for all $t \geq \lambda r$.
        \end{proof}

        \begin{definition}[net correction]
            Suppose a CA decoder satisfies linear cluster erosion. For any trajectory $\zeta$ with a finite initial error configuration, define the decoder's \emph{net correction} $\Delta\zeta \in \F_2^{\Lambda_e}$ by
            \[
                \Delta\zeta \coloneqq \lim_{t \to \infty}\vc e(t).
            \]
        \end{definition}
        Note that, by linear cluster erosion, the decoder eventually stops flipping bits, so the net correction is well-defined.

        \begin{definition}[cluster decoupling]\label{def:cluster_decoupling}
            A CA decoder is said to satisfy cluster decoupling if there exists a constant $\mu \geq 1$ such that the following holds for every radius $r \geq 1$, site $\vc x \in \Lambda$, and initial error configuration with finitely many errors and no errors in the annulus $B_{\mu r}(\vc x) \setminus B_r(\vc x)$: the net correction $\Delta\zeta$ decomposes as
            \[
                \Delta\zeta = \Delta\zeta^{\mathrm{in}} + \Delta\zeta^{\mathrm{out}},
            \]
            where $\zeta$, $\zeta^{\mathrm{in}}$, and $\zeta^{\mathrm{out}}$ are the trajectories whose initial error configurations are the full initial error configuration, its restriction to $B_r(\vc x)$, and its restriction to $\Lambda \setminus B_{\mu r}(\vc x)$, respectively.
        \end{definition}

        \begin{lemma}
            If a CA decoder satisfies linear cluster erosion, then it also satisfies cluster decoupling.
        \end{lemma}

        \begin{proof}
            Suppose the initial error configuration has no errors in the annulus $B_{\mu r}(\vc x) \setminus B_r(\vc x)$. By the lightcone bound, the non-trivial messages, corrections, and syndromes due to the inner errors are contained in $B_{r + vt + d_0}(\vc x)$ at time $t$, while those due to the outer errors remain outside $B_{\mu r - d_0 - vt}(\vc x)$. We call $B_{r + vt + d_0}(\vc x)$ the inner region at time $t$ and $\Lambda \setminus B_{\mu r - d_0 - vt}(\vc x)$ the outer region at time $t$. Consequently, by the lightcone bound and causality, whenever $r + 2vt + 2d_0 < \mu r$, no site in the inner region at time $t$ is affected by the outer errors, and no site in the outer region is affected by the inner errors. Choose $\mu > 1 + 2v\lambda + 2v + 2d_0$. Then this condition holds for all times $t \leq \lambda r$, so up to time $\lambda r$, sites in the inner region evolve exactly as in $\zeta^{\mathrm{in}}$ and sites in the outer region evolve exactly as in $\zeta^{\mathrm{out}}$. By linear cluster erosion, the inner region is fully resolved before the two regions interact, which implies the net correction decomposes as a sum of inner and outer corrections in the desired manner.
        \end{proof}

    \subsection{Bounds on logical error rates and decoding times}\label{subsec:threshold:logical_error_rates_and_decoding_times}

        In this subsection, we assume that the CA decoder satisfies all the properties defined in the previous subsection for some set of constants $a$, $b$, $\lambda$, and $\mu$. We also fix a triple of constants $(\beta, \gamma, n)$ such that (i) the inequalities \eqref{paramineqs} hold; (ii) $b_k \geq (1+\mu)w_k$ for all $k$; and (iii) $w_0 \geq 1$. We also assume a finite system with linear size $L$.

        Strictly speaking, in a finite system, the erosion properties will not hold for all initial error configurations. However, the configurations on which they can fail to hold are precisely those containing clusters large enough to potentially cause a logical error, which are the very configurations whose probability we bound in our threshold proof. A more proper treatment would condition on the event that no such cluster occurs, then bound the probability of this event using the sparsity theorem. Since this conditioning is straightforward but tedious, we simplify the presentation by assuming that the erosion properties hold for all initial error configurations.

        \begin{theorem}[hierarchical cluster decoupling]\label{thm:hierarchical_decoupling}
            Let $N$ be a noise realization, and let
            \[
                N=\bigsqcup_{k,i} C_k^{(i)}
            \]
            be the decomposition of $N$ into disjoint $k$-clusters $C_k^{(i)}$. Next, let $\zeta$ denote the trajectory with initial error configuration $N$ and let $\zeta_k^{(i)}$ denote the trajectory with initial error configuration $C_k^{(i)}$. Then, the net correction $\Delta \zeta$ satisfies
            \[
                \Delta\zeta \;=\; \sum_{k,i}\Delta\zeta_k^{(i)}.
            \]
        \end{theorem}

        \begin{proof}
            For each $k \geq 0$, recall that $N_k$ denotes the set of errors remaining after removing all $j$-clusters for $j < k$, and that $N_k = E_k \sqcup N_{k+1}$, where $E_k = \bigsqcup_i C_k^{(i)}$ is the disjoint union of $k$-clusters. We claim that for every $k \geq 0$,
            \begin{equation}\label{eq:delta_peeling}
                \Delta\zeta_k = \sum_i \Delta\zeta_k^{(i)} + \Delta\zeta_{k+1},
            \end{equation}
            where $\zeta_k$ denotes the trajectory with initial error configuration $N_k$. Suppose \eqref{eq:delta_peeling} holds for all $k \geq 0$. Since $N$ is finite, there exists some smallest $k_{\max}$ such that $N_{k_{\max} + 1} = \varnothing$. Iterating \eqref{eq:delta_peeling} from $k = 0$ to $k = k_{\max}$ and using $\zeta_0 = \zeta$ yields the desired decomposition.

            It remains to prove \eqref{eq:delta_peeling}. Fix $k \geq 0$. Each $k$-cluster $C_k^{(i)}$ has diameter at most $w_k$ and is therefore contained in the ball $B_{w_k}\l(\vc x_k^{(i)}\r)$ for some site $\vc x_k^{(i)} \in C_k^{(i)}$. Moreover, $C_k^{(i)}$ is separated from all other errors in $N_k$ by a distance greater than $b_k \geq (1 + \mu) w_k$, so the annulus $B_{\mu w_k}\l(\vc x_k^{(i)}\r) \setminus B_{w_k}\l(\vc x_k^{(i)}\r)$ is free of $N_k$ errors. Applying cluster decoupling to each $k$-cluster in $N_k$ then gives
            \[
                \Delta\zeta_k = \sum_i\Delta\zeta_k^{(i)} + \Delta \zeta_{k+1},
            \]
            which is precisely \eqref{eq:delta_peeling}.
        \end{proof}

        The following corollary of Theorem~\ref{thm:hierarchical_decoupling} states that if each $k$-cluster in $N$ is small enough that a ball of radius $\lambda w_k$ cannot support a logical operator, then decoding cannot produce a logical error.
        \begin{definition}[logical diameter and cutoff level]
            Let $d_X$ denote the minimum diameter of any non-trivial $X$-type logical operator, and let $k_L$ be the largest non-negative integer such that $2\lambda w_k < d_X$ for all $k < k_L$.
        \end{definition}
        Note that for all $k < k_L$, the ball $B_{\lambda w_k}(\vc x)$ cannot support a non-trivial $X$-type logical operator.

        \begin{corollary}\label{cor:no_large_cluster_correctness}
            If $N_{k_L} = \varnothing$, i.e., every cluster in the decomposition $N = \bigsqcup_{k,i} C_k^{(i)}$ has level $k < k_L$, then the decoder succeeds: the residual error after decoding is an $X$-type stabilizer.
        \end{corollary}
        \begin{proof}
            By Theorem~\ref{thm:hierarchical_decoupling}, $\Delta\zeta = \sum_{k,i} \Delta\zeta_k^{(i)}$. Each $\Delta\zeta_k^{(i)} + C_k^{(i)}$ is supported in $B_{\lambda w_k}\l(\vc x_k^{(i)}\r)$ by linear cluster erosion, and since $k < k_L$, this ball cannot support a non-trivial $X$-type logical operator. Hence, $\Delta\zeta + N$ is an $X$-type stabilizer.
        \end{proof}

        Let $\mathsf{Fail}$ denote the event that decoding produces a non-trivial $X$-type logical error.

        \begin{theorem}[threshold scaling]
            Assume $d_X = \Omega(L)$, so that $k_L = \Omega(\log L)$. Then, there exist constants $\alpha > 0$ and $p_* > 0$ such that for $p < p_*$,
            \[
                \Pr[\mathsf{Fail}] \leq \l(\frac{p}{p_*}\r)^{\Omega(L^\alpha)}.
            \]
        \end{theorem}

        \begin{proof}
            Let $p_*$ be the threshold lower bound from the sparsity theorem (Theorem~\ref{thm:sparse}). If $N_{k_L} = \varnothing$, then all clusters have level $k < k_L$, so decoding succeeds by Corollary~\ref{cor:no_large_cluster_correctness}. Therefore, $\mathsf{Fail} \subseteq \{N_{k_L} \neq \varnothing\}$. By a union bound and the definition of $p_{k_L}$,
            \[
                \Pr[N_{k_L} \neq \varnothing] \leq \sum_{x \in \Lambda_e} \Pr[x \in N_{k_L}] \leq |\Lambda_e| \, p_{k_L}.
            \]
            By the sparsity theorem, $p_{k_L} \leq C(p/p_*)^{2^{k_L}}$. Since $k_L \geq c \log L$ for some $c > 0$, we have $2^{k_L} \geq L^{c \log 2}$. The polynomial prefactor $|\Lambda_e| = O(L^D)$ can then be absorbed into the exponent, which gives
            \[
                \Pr[\mathsf{Fail}] \leq \l(\frac{p}{p_*}\r)^{\Omega(L^\alpha)}
            \]
            for $\alpha = c \log 2$.
        \end{proof}

        \begin{theorem}[average decoding time]\label{thm:avg_time}
            Let $N$ be a noise realization with largest cluster level
            \[
                k_{\max} \coloneqq \max\{k : \exists\, i \text{ with } C_k^{(i)} \neq \varnothing\}.
            \]
            Assume that $k_L = \Omega(\log L)$ and that the decoder terminates in time $O(w_{k_{\max}})$ when $k_{\max} < k_L$ and in $\poly(L)$ time when $k_{\max} \geq k_L$. Then, there exists $\eta > 0$ such that when $p < p_*$, the expected decoding time is $O(\log^\eta L)$, where the expectation is over noise realizations drawn from the $p$-bounded error model.
        \end{theorem}

        We do not prove this here; the proof technique is the same as in Ref.~\cite{lake2025fastofflinedecodinglocal}. The weaker $\poly(L)$ bound when $k_{\max} \geq k_L$ suffices because the events where $k_{\max} \geq k_L$ are stretched-exponentially (in $L$) rare and contribute negligibly to the expectation, so a decoder that deals with them in a finite-time but otherwise arbitrary way is guaranteed not to pay significantly for this arbitrariness in its average decoding time.\footnote{This relaxation is useful because there may exist rare initial error configurations on which the simplest decoder does not terminate on its own, formally resulting in infinite expected decoding time. In fact, this will be the case for all the decoders we construct for systems with periodic boundary conditions. One formal remedy to these rare events is to add a timer that initiates random moves if decoding does not terminate by time $O(L)$, resulting in diffusion-limited decoding that terminates in $\poly(L)$ time with high probability. The rare realizations on which even diffusion fails to terminate in polynomial time contribute negligibly to the expected decoding time. When we say that the decoders in this paper have polylogarithmic average decoding time, we mean that, with this modification in place, the expected decoding time is $O(\log^\eta L)$ for some $\eta > 0$. In practice, however, it is preferable not to add a timer at all: non-termination serves as a strong signal that the computation should be discarded (or, at the very least, that a more accurate global decoder should be used).}

        Note that polylogarithmic average decoding time is guaranteed only for $p < p_*$. Since $p_*$ is a lower bound on the true threshold $p_c$, the decoder may in principle exhibit parametrically larger average decoding time in the regime $p_* \leq p < p_c$.

\section{Repetition code under code-capacity noise} \label{sec:rep-code}

    In this section, we construct a code-capacity CA decoder for the repetition code and prove that it has a non-zero threshold $p_*$ and polylogarithmic average decoding time for $p < p_*$. Our proof strategy is to establish linear defect erosion and linear message erosion; the desired conclusions then follow from Section~\ref{sec:threshold}. We begin by studying the repetition code because, although it is a classical error-correcting code, its decoder uses nearly every decoding mechanism in this paper and does so in the simplest possible manner. The toric- and surface-code decoders of Section~\ref{sec:toric-code} are direct two-dimensional generalizations of the construction given here.

    We work at infinite system size for simplicity. The generalization to finite system size is straightforward and omitted: one establishes both erosion properties for all clusters small enough that periodic boundary conditions can be neglected; since the largest such clusters have diameter $W = \Omega(L)$, this suffices to imply the desired results.

    We now give an informal overview of the decoder. To achieve linear defect erosion, we move all defects in a cluster monotonically toward the cluster's leftmost defect. This is achieved as follows. A single message channel carries messages that travel rightward at average speed $1/q$ for some fixed integer $q \geq 2$. A defect moves left whenever it sees a message immediately to its left; we will design the dynamics so that no defect ever moves past the original leftmost defect in the cluster. Messages flood a cluster of width $W$ in time $O(W)$, after which every defect except the current leftmost one is guaranteed to move left at every time step. This ensures that after the $O(W)$ flooding time, all defects annihilate within $O(W)$ additional steps---though defects may be eliminated sooner by merging with each other during their leftward motion. Once all defects have been annihilated, residual messages occupy an enlarged $O(W)$-diameter defect-free region, since the message fronts expand rightward beyond the original cluster boundary.

    To achieve linear message erosion, we design a local erasure rule, which, in the absence of defects, is as follows: a message ``dies'' (is set to zero) if no non-trivial message lies immediately to its left. A clock channel is used to ensure that messages grow more slowly than they are erased, so the erasure rule catches up to rightward-growing message fronts. In particular, the erasure rule eliminates messages in a defect-free region in time linear in the messages' extent, so linear message erosion is satisfied. This mechanism is similar in spirit to that of the implicit message erosion present in the GKL CA~\cite{gacs1978one}.

    \subsection{Definitions}\label{subsec:rep-code:definitions}

        We consider the one-dimensional repetition code on $\Lambda = \Z$ with $J = \{1\}$, so $\Lambda_s = \Lambda_e = \Z$ and the parity check at site $x$ is $s(x) = Z_x Z_{x+1}$. The state of the automaton at each site consists of four channels:
        \[
            \xi(x) = \bigl(e(x),\, s(x),\, m(x),\, c(x)\bigr),
        \]
        where $e(x) \in \{0,1\}$ is the correction channel, $s(x) \in \{0,1\}$ is the syndrome channel, $m(x) \in \{0,1\}$ is a single message channel, and $c(x) \in \{0, \ldots, q-1\}$ is the clock channel for some fixed integer $q \geq 2$.

        The local update rule is specified in Algorithm~\ref{alg:rep-code}. We recommend reading the following description of the algorithm before inspecting Algorithm~\ref{alg:rep-code} directly. We make the same recommendation for all subsequent decoders: the informal descriptions preceding each algorithm are designed to be read first and provide the intuition needed to parse the pseudocode. The clock increments as $c(x, t+1) = (c(x,t) + 1) \bmod q$. The correction channel updates as
        \[
            e(x, t+1) = e(x, t) \oplus \bigl[s(x, t) = 1 \;\text{and}\; m(x-1, t) = 1\bigr],
        \]
        where $\oplus$ denotes addition modulo $2$. Since $e(x)$ can only flip when there is a defect at site $x$, and flipping $e(x)$ toggles the syndromes at sites $x$ and $x-1$, the only effect of a flip is to move a defect one step to the left. The syndrome at time $t+1$ is then obtained by updating the syndrome at time $t$ according to the changes in the correction configuration. The message channel updates as $m(x, t+1) = 1$ if and only if at least one of the following conditions holds:
        \begin{enumerate}[label=(\roman*)]
            \item \emph{Defect persistence:} $s(x, t) = 1$;
            \item \emph{Defect arrival:} $s(x+1, t) = 1$ and $m(x, t) = 1$;
            \item \emph{Message growth:} $s(x, t) = 0$, $m(x, t) = 0$, $c(x, t) = 0$, and $m(x-1, t) = 1$;
            \item \emph{Message persistence:} $s(x, t) = 0$, $m(x, t) = 1$, and $m(x-1, t) = 1$.
        \end{enumerate}
        Otherwise, $m(x, t+1) = 0$. 
        
        Condition (i) ensures that every defect maintains a message at its own site. Condition (ii) preserves the message at the new defect location after a leftward move: a defect at $x+1$ moves left precisely when it sees a message at $x$, and this message is retained to prevent premature erasure and the ensuing delay in repopulating the message. Condition (iii) allows messages to spread rightward, but only when the clock reads zero. This causes messages to grow at average speed $1/q$. Condition (iv) maintains an existing message provided a message to its left still exists; when $m(x-1, t) = 0$ and neither condition (i) nor (ii) applies, the message at $x$ is erased. This implements the erasure rule described in the overview: in the absence of defects, a message dies if no message lies immediately to its left.

        The update rule in Algorithm~\ref{alg:rep-code} is written as two loops for notational convenience: one over all sites, and one over $S_{\mathrm{inc}}$, the set of sites that receive an incoming defect during the time step. The second loop sets the message at each site in $S_{\mathrm{inc}}$ to $1$, implementing condition (ii); this can equivalently be implemented locally within the first loop, making the entire rule a single synchronous pass over all sites. We adopt the two-loop convention for all of our decoders because it is convenient in higher dimensions, where writing the incoming syndrome update as a local rule is straightforward but tedious.

        Note that the local update rule satisfies our formal definition of a CA decoder. Each of the decoders we define in later sections for other codes is likewise a formal CA decoder; the reader can verify this in each case.

        \begin{algorithm}[H]
            \caption{Repetition code update from time $t$ to time $t+1$.}
            \label{alg:rep-code}
            \begin{algorithmic}[1]
                \State $S_{\mathrm{inc}} \gets \emptyset$
                \ForAll{$x \in \Lambda$}
                    \State $m(x,\,t+1) \gets 0$; \;\; $e(x,\,t+1) \gets e(x,\,t)$
                    \Statex
                    \If{$s(x,t) = 1$} \Comment{\textbf{Part 1: Defect dynamics}}
                        \State $m(x,\,t+1) \gets 1$
                        \If{$m(x-1,t) = 1$}
                            \State $e(x,\,t+1) \gets e(x,t) \oplus 1$; \;\; $S_{\mathrm{inc}} \gets S_{\mathrm{inc}} \cup \{x-1\}$ \Comment{Move left}
                        \EndIf
                    \Statex
                    \Else \Comment{\textbf{Part 2: Message dynamics}}
                        \If{$m(x,t) = 0$ \textbf{and} $c(x,t) = 0$ \textbf{and} $m(x-1,t) = 1$}
                            \State $m(x,\,t+1) \gets 1$ \Comment{Growth}
                        \ElsIf{$m(x,t) = 1$}
                            \State $m(x,\,t+1) \gets m(x-1,\,t)$ \Comment{Persistence or erasure}
                        \EndIf
                    \EndIf
                    \Statex
                    \State $c(x,\,t+1) \gets (c(x,t) + 1) \bmod q$
                \EndFor
                \Statex
                \ForAll{$x' \in S_{\mathrm{inc}}$} \Comment{\textbf{Part 3: Defect arrival}}
                    \State $m(x',\,t+1) \gets 1$
                \EndFor
            \end{algorithmic}
        \end{algorithm}
 
    \subsection{Linear erosion}\label{subsec:rep-code:linear-erosion}

        We now prove linear defect erosion and linear message erosion. We begin with the former. First, we show that the number of defects is non-increasing. Next, we fix a convention for tracking defect worldlines and show that the number of messages to the right of a defect grows linearly in time. These facts imply that, by time $O(W)$, messages have flooded the entire cluster, and thus every non-leftmost defect in a cluster of width $W$ moves at every time step thereafter.
        
        The following proposition is immediate from the construction of the decoder.
        \begin{proposition}[defect monotonicity]\label{prop:rep-code:defect-monotone}
            The number of defects is non-increasing in time.
        \end{proposition}

        Next, we fix a convention that uniquely determines all defect worldlines. At $t = 0$, all defects are well-defined. We then impose the following association rules for all later times $t \geq 1$:
        \begin{itemize}
            \item \textbf{Move into an empty site.} If a defect moves left into a site that is unoccupied at time $t$, or whose occupying defect simultaneously moves left, its worldline moves to the new position.
            \item \textbf{Stationary defect.} If a defect remains at its site and no defect enters during the time step, its worldline continues at that position.
            \item \textbf{Annihilation.} If a defect moves left into a site occupied by a stationary defect, the two defects annihilate: both worldlines terminate at time $t$, and neither defect is considered alive at time $t+1$.
        \end{itemize}
        With these rules, each defect has a well-defined worldline. Note that since the number of defects is non-increasing, no new defect worldlines are created by the dynamics.
        
        For the remainder of this section, we assume that the initial error configuration is confined to an interval of width $W$. We will prove that, in this case, all defects are annihilated within time $O(W)$.
        
        Fix a single defect worldline $\{X(t)\}$ with each $X(t) \in \Z$. We will assume that the defect is created at $t = 0$ and is alive only at times $t = 0, 1, \ldots, T$, where $T$ is finite; it will be clear a posteriori that assuming $T$ is finite entails no loss of generality.
        
        By construction of the decoder, every defect moves monotonically to the left. In particular, the position of every defect is non-increasing in time.
        \begin{proposition}[directed movement property]\label{prop:rep-code:directed-movement}
            For every $t \in \{0, 1, \ldots, T-1\}$,
            \[
                X(t+1) \in \{X(t),\, X(t) - 1\}.
            \]
        \end{proposition}

        Next, we show that for all times $t \geq 1$ at which the defect is alive, the message at its current site is $1$ at both time $t$ and time $t+1$.
        \begin{lemma}[trivial message persistence]\label{lem:rep-code:trivial-persistence}
            For every $t \in \{1, \ldots, T\}$,
            \[
                m(X(t),\, t) = m(X(t),\, t+1) = 1.
            \]
        \end{lemma}
        \begin{proof}
            Fix $t \in \{1, \ldots, T\}$. By definition, $s(X(t-1), t-1) = 1$. If $X(t) = X(t-1)$, then the defect persists at the same site, so $m(X(t), t) = 1$ by condition (i). If $X(t) = X(t-1) - 1$, then the defect has moved left, so $m(X(t), t) = 1$ by the incoming syndrome update. Finally, since $s(X(t), t) = 1$, condition (i) gives $m(X(t), t+1) = 1$.
        \end{proof}

        Next, we show that once a site has been visited by the defect, its message remains non-trivial for at least as long as the defect is alive. Intuitively, this holds because as long as the defect is alive, every previously visited site either still hosts the defect or has a message to its left, either of which prevents erasure.

        \begin{lemma}[message persistence]\label{lem:rep-code:message-persistence}
            For every $t \in \{1, \ldots, T\}$ and every $t' \in [t, T]$,
            \[
                m(X(t),\, t') = 1.
            \]
        \end{lemma}
        \begin{proof}
            The proof is by backward induction on $t$ and forward induction on $t'$.
        
            \textbf{Base case.} For $t = T$, Lemma~\ref{lem:rep-code:trivial-persistence} gives $m(X(T), T) = 1$, and no forward induction is needed.
        
            \textbf{Inductive step.} Suppose the result holds for all $\tau \in [t+1, T]$; we prove it for $t$ by forward induction on $t' \in [t, T]$. The forward base case $t' = t$ follows from Lemma~\ref{lem:rep-code:trivial-persistence}. For the forward inductive step, assume $m(X(t), \tau') = 1$ for all $\tau' \in [t, t'-1]$, where $t' \in [t+1, T]$; we wish to show $m(X(t), t') = 1$.
        
            If $X(t'-1) = X(t)$, then the defect is at site $X(t)$ at time $t'-1$, so Lemma~\ref{lem:rep-code:trivial-persistence} immediately gives $m(X(t), t') = 1$.
        
            Otherwise, $X(t'-1) \neq X(t)$, so the defect has moved left of $X(t)$ by time $t'-1$. By the directed movement property, there exists some time $t'' \in (t, t')$ at which $X(t'') = X(t) - 1$. By the backward inductive hypothesis (applied at time $t''$), we have $m(X(t) - 1, \tau) = 1$ for all $\tau \in [t'', T]$; in particular, $m(X(t) - 1, t'-1) = 1$. By the forward inductive hypothesis, $m(X(t), t'-1) = 1$. Since both $m(X(t), t'-1) = 1$ and $m(X(t) - 1, t'-1) = 1$, condition (iv) ensures $m(X(t), t') = 1$.
        \end{proof}

        We now show that a persistent message at a site causes messages to spread to the right at average speed $1/q$.
        \begin{lemma}[message growth]\label{lem:rep-code:message-growth}
            Fix a site $x \in \Z$ and times $t' \geq t$. Define the number of growth opportunities between $t$ and $t'$ by
            \[
                N(t, t') \coloneqq \bigl|\{\tau \in [t, t'-1] : c(\tau) = 0\}\bigr|.
            \]
            Then $N(t, t') \geq \frac{t' - t}{q} - 1$. Furthermore, if $m(x, \tau) = 1$ for all $\tau \in [t, t']$, then for every $\tau \in [t, t']$ and every integer $i \in [0, N(t, \tau)]$,
            \[
                m(x + i,\, \tau) = 1.
            \]
        \end{lemma}

        \begin{proof}
            \textbf{Growth rate bound.} Write $\Delta = t' - t$. Since $c(\tau)$ cycles with period $q$ and equals $0$ exactly once per period, the number of times $c(\tau) = 0$ in any interval of length $\Delta$ is at least $\lfloor \Delta / q \rfloor \geq \Delta/q - 1$.
        
            \textbf{Message growth.} We prove the result by induction on $\tau \in [t, t']$. For $\tau = t$, we have $N(t, t) = 0$, so the only claim is $m(x, t) = 1$, which holds by assumption. Suppose the result holds for $\tau - 1$; we show it for $\tau$. Fix $i \in [0, N(t, \tau)]$. If $i = 0$, then $m(x, \tau) = 1$ by assumption. If $i \geq 1$ and $N(t, \tau) = N(t, \tau-1)$, then $i \leq N(t, \tau-1)$, so $m(x+i, \tau-1) = 1$ by the inductive hypothesis. Since $m(x+i-1, \tau-1) = 1$ as well, condition (iv) (or condition (i) if a defect occupies the site) gives $m(x+i, \tau) = 1$. If $i \geq 1$ and $N(t, \tau) = N(t, \tau-1) + 1$, then either $i \leq N(t, \tau-1)$, in which case the same argument applies, or $i = N(t, \tau)$. In the latter case, $m(x+i-1, \tau-1) = 1$ by the inductive hypothesis and $c(\tau-1) = 0$, so one of conditions (i), (iii), or (iv) guarantees that $m(x+i, \tau) = 1$.
        \end{proof}

        \begin{theorem}[linear defect erosion]\label{thm:rep-code:linear-defect-erosion}
            There exists a constant $a \geq 1$ such that the following holds. Suppose the initial error configuration is supported in a ball of radius $r \geq 1$ centered at $x$ and the initial message configuration is trivial. Then:
            \begin{enumerate}
                \item All defects are eliminated by time $t \leq ar$.
                \item For all $t \geq 0$, the correction and syndrome configurations are supported in $B_{ar}(x)$.
            \end{enumerate}
        \end{theorem}
        \begin{proof}
            By Proposition~\ref{prop:rep-code:defect-monotone}, no new defects are created by the dynamics, so it suffices to consider the at most $O(r)$ defects present at $t = 0$ and their worldlines. By Lemmas~\ref{lem:rep-code:trivial-persistence},~\ref{lem:rep-code:message-persistence}, and~\ref{lem:rep-code:message-growth}, there exists a constant $c_1 > 0$ such that by time $c_1 r$, if a defect is still alive, all sites within distance $2r$ to its right carry non-trivial messages.
            
            After time $c_1 r$, every surviving defect either sees a message to its left and moves left, or is the current leftmost defect and has no message to its left.\footnote{A defect that is currently the leftmost may still move left if there exist residual messages to its left from a previously annihilated defect.} Each defect can move left at most $O(r)$ times before reaching the left endpoint of the original interval, since no defect can move right by the directed movement property, and no defect can cross the left endpoint by construction of the dynamics. Separately, at each time step in which more than one defect survives, all but the leftmost are guaranteed to move left, so there can be at most $O(r)$ time steps during which there exists a stationary defect. Combining these two bounds, all defects are annihilated within $O(r)$ additional time steps, which establishes (1). By the directed movement property, the lightcone bound, and the defect-gated flip property, all correction flips and syndromes remain within $B_{O(r)}(x)$ throughout, which establishes (2). Therefore, both (1) and (2) hold for $a$ sufficiently large.
        \end{proof}

        \begin{theorem}[linear message erosion]\label{thm:rep-code:linear-message-erosion}
            There exists a constant $b \geq 1$ such that the following holds. Suppose the syndrome configuration is trivial for all times $t \geq t_0$ and $\supp(m(t_0))$ is contained in a ball of radius $r \geq 1$ centered at $x_0$. Then:
            \begin{enumerate}
                \item $\supp(m(t)) = \varnothing$ for all $t \geq t_0 + br$.
                \item $\supp(m(t)) \subseteq B_{br}(x_0)$ for all $t \geq t_0$.
            \end{enumerate}
        \end{theorem}
        \begin{proof}
            Since there are no defects, conditions (i) and (ii) never apply. Messages can only persist via condition (iv), which requires $m(x-1, t) = 1$, and can only spread via condition (iii), which requires $m(x-1, t) = 1$, $m(x, t) = 0$, and $c(x, t) = 0$. In particular, the leftmost message at any given time has no message to its left, so it is erased at the next time step. Therefore, the left boundary of the message support advances to the right at speed at least $1$. Meanwhile, condition (iii) allows the right boundary to advance at average speed at most $1/q$. Since $q \geq 2$, the left boundary catches up to the right boundary in $O(r)$ time, and the right boundary expands by at most $O(r)$ throughout. Both (1) and (2) follow by choosing $b$ sufficiently large.
        \end{proof}

    \subsection{Open boundary conditions}\label{subsec:rep-code:open-boundary-conditions}

        We now describe how to modify our construction to decode the one-dimensional repetition code with open boundary conditions.\footnote{The approach we take to decoding under open boundary conditions is different from that of Ref.~\cite{lake2025fastofflinedecodinglocal}. There, the dynamics are modified so that the condensing boundaries themselves source messages that attract defects to them. The same modification would enable our construction to decode the repetition code with open boundary conditions (and the surface code) in the code-capacity setting.

        Under phenomenological noise, however, this approach is problematic for our construction. In Ref.~\cite{lake2025fastofflinedecodinglocal}, messages are never erased, but the original dynamics are such that messages from distant boundaries are automatically ignored whenever messages from nearby defects are present. Consequently, even when boundary-sourced messages flood the entire system, defects in clusters far from the boundary still pair up among themselves rather than flowing all the way to the boundary, preserving (a suitably defined generalization of) linear cluster erosion.

        In our construction, by contrast, there is no native mechanism for prioritizing messages from nearby defects over those from distant boundaries, so without further modification, messages continuously sourced by the boundaries would flood the entire system and cause all defects to move in one direction. The modified dynamics of this section provide an alternative mechanism for decoding under open boundary conditions that avoids this problem.}
    
        We begin by describing the modified dynamics. First, we partition the system into a left half $\Lambda_L$, consisting of all sites with $x$-coordinate at most $\lfloor L/2 \rfloor$, and a right half $\Lambda_R$, consisting of all sites with $x$-coordinate greater than $\lfloor L/2 \rfloor$.
 
        The decoder runs the usual CA dynamics\footnote{For message erasure dynamics, neighboring sites that lie outside the system are treated as having trivial messages.} at every time step, but at times that are multiples of a fixed integer $q_s$, it also performs a \emph{splitting step}. Specifically, whenever $t \in q_s \Z$, we perform the following procedure.\footnote{This can be implemented locally by having each site carry an additional clock channel that counts modulo $q_s$.} First, let $s(x, t)$, $m(x, t)$, and $e(x, t)$ denote the syndrome, message, and correction configurations before the splitting step. We then recalculate these configurations based on the splitting step, producing updated configurations $s'(x, t)$, $m'(x, t)$, and $e'(x, t)$. Finally, we compute $s(x, t+1)$, $m(x, t+1)$, and $e(x, t+1)$ by applying the CA update rule to the updated configurations.

        During the splitting step, all syndromes and messages on the left half are translated one site to the left, and all syndromes and messages on the right half are translated one site to the right, with the following exceptions:
        \begin{itemize}
            \item \emph{Center sites.} At sites with $x$-coordinate $\lfloor L/2\rfloor$ and $x$-coordinate $\lfloor L/2\rfloor + 1$, the syndrome is translated outward as usual, but the messages are \emph{copied} rather than translated---that is, after the step, both the original site and its outward neighbor carry a copy of the message.
            \item \emph{Boundary sites.} At sites with $x$-coordinate $1$ or $L$, messages and defects translated out of the system are annihilated.
        \end{itemize}

        We now prove that linear defect erosion and linear message erosion hold under the modified dynamics for all clusters of width $W \leq cL$ for some constant $c > 0$, provided $q_s$ is chosen appropriately. Since the clustering hierarchy and sparsity theorem are unaffected by the presence of boundaries, this suffices to establish a threshold and polylogarithmic average decoding time by the results of Section~\ref{sec:threshold}.
         
        Suppose all the initial errors are confined to an interval of width $W$. We claim that there exists a sufficiently small constant $c > 0$ such that, whenever $W \leq cL$, the cluster falls into exactly one of the following three cases. The existence of such a $c$ will be clear a posteriori: each case specifies conditions on the cluster's position relative to the center and the boundary, and for $c$ sufficiently small, these conditions are mutually exclusive and exhaustive.

        \paragraph{Clusters far from the center and the boundary.} If the cluster and all of its messages are entirely contained in $\Lambda_L$ or entirely in $\Lambda_R$ throughout the decoding process, then the dynamics are identical to the periodic case up to spatial translation, so the analysis of Section~\ref{subsec:rep-code:linear-erosion} applies without modification.

        \paragraph{Clusters near the center.} Suppose the cluster's dynamics are modified due to being near the center of the system. In this case, the splitting step pushes defects on opposite sides of the center apart at an effective speed of at most $2/q_s$. The message replication mechanism at the center ensures that messages are not prematurely erased by the splitting: messages that cross the center are copied rather than torn apart, so the message persistence properties (Lemmas~\ref{lem:rep-code:trivial-persistence}--\ref{lem:rep-code:message-persistence}) continue to hold. Furthermore, messages still grow at average speed $1/q$ by the usual growth mechanism (Lemma~\ref{lem:rep-code:message-growth}), and are occasionally duplicated by the replication step. Therefore, provided $q_s$ is chosen large enough that the message growth speed $1/q$ exceeds the splitting speed $2/q_s$ (e.g., $q_s > 2q$), messages flood the entire cluster in $O(W)$ time. After this time, all but the current leftmost defect are guaranteed to move to the left at every non-splitting step. The leftward motion of defects must overcome the splitting drift; choosing $q_s$ large enough (e.g., $q_s> 2$) ensures that defects still annihilate in $O(W)$ time. Once all defects have been annihilated, the erasure front advances at speed $1$ during regular steps, while messages grow at speed $1/q$ and the splitting step expands the message support by at most two sites every $q_s$ steps; hence the erasure front catches up to the growth front provided $1/q + 2/q_s < 1$, which holds for $q, q_s$ sufficiently large. Therefore, linear message erosion holds.

        \paragraph{Clusters near the boundary.} Finally, suppose a cluster of width $W$ is at a distance $\ell = O(W)$ from the boundary and is sufficiently far from the center that center effects can be neglected. (If $\ell \gg W$, the cluster resolves via its own internal dynamics in $O(W)$ time before being affected by the boundary, reducing to the first case.) The splitting step translates the cluster toward the boundary at speed $1/q_s$, so in the worst case, all defects in the cluster reach the boundary and annihilate after $O(\ell q_s) = O(W)$ steps. After this time, all residual messages are contained in an interval of size $O(W)$, which is contained in a single half of the system by assumption, and are thus cleaned up by linear message erosion as in the first case.

        Combining the three cases and choosing $q$ and $q_s$ such that $q_s > 2q$ and $1/q + 2/q_s < 1$, we see a posteriori that there exists a constant $c > 0$ such that both erosion properties hold for all clusters of width $W \leq cL$. Therefore, our decoder has a non-zero threshold and polylogarithmic average decoding time by the results of Section~\ref{sec:threshold}.
    
    \subsection{Numerics}\label{subsec:rep-code:numerics}

        \begin{figure}
            \centering
            \includegraphics[width=\linewidth]{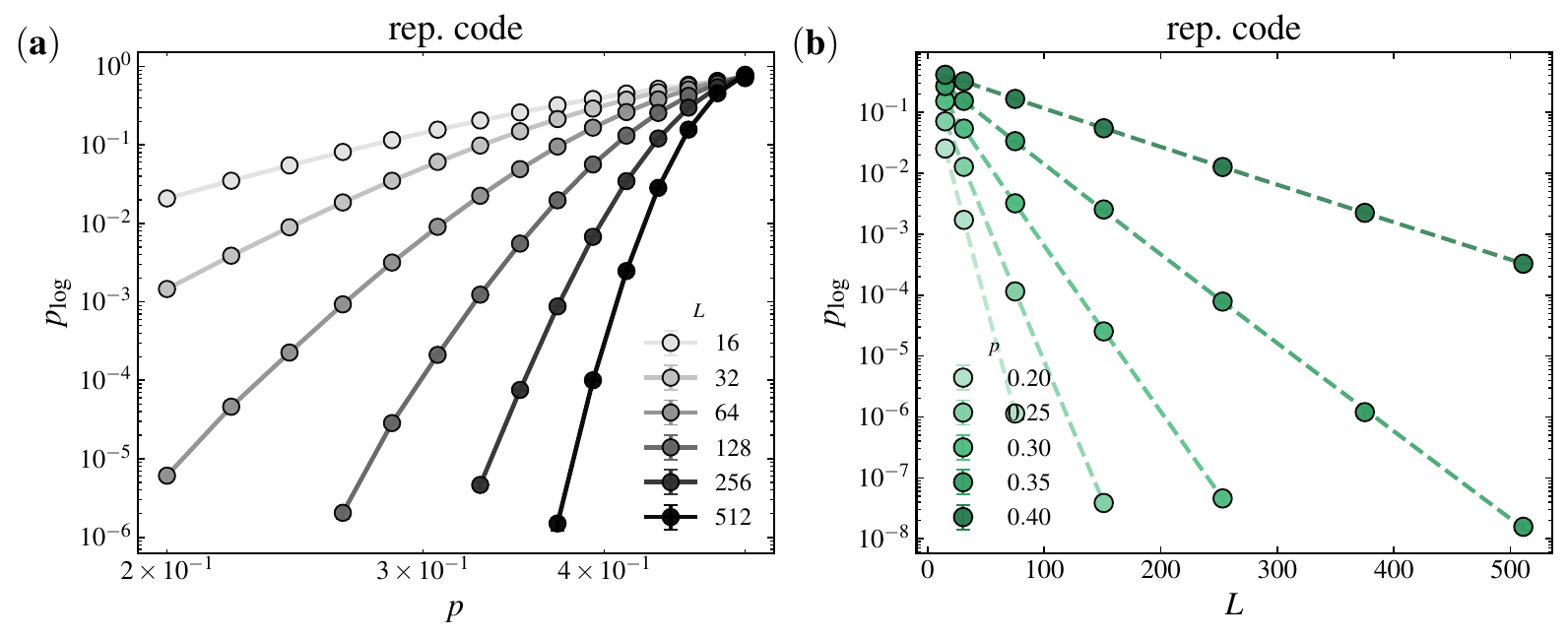}
            \caption{Performance of the code-capacity repetition-code decoder. \textbf{(a)} Logical error rate $\plog$ as a function of the physical error rate $p$ under i.i.d.\ bit-flip noise. The data are consistent with a threshold of $p_c = 1/2$. \textbf{(b)} Sub-threshold scaling of the logical error rate $\plog$ with $L$. Dashed lines are fits to the form $\plog = C(p/p_c)^{\gamma L^{\alpha}}$ with a single exponent $\alpha$ shared across all $p$. The fitted $\alpha = 0.97 \pm 0.01$ is consistent with exponential decay of $\plog$ in $L$, and the fitted $\gamma$ increases from $0.08$ to $0.2$ as $p$ decreases from $0.4$ to $0.2$.}
            \label{fig:repcodethreshold}
        \end{figure}

        \begin{figure}
            \centering
            \includegraphics[width=0.5\linewidth]{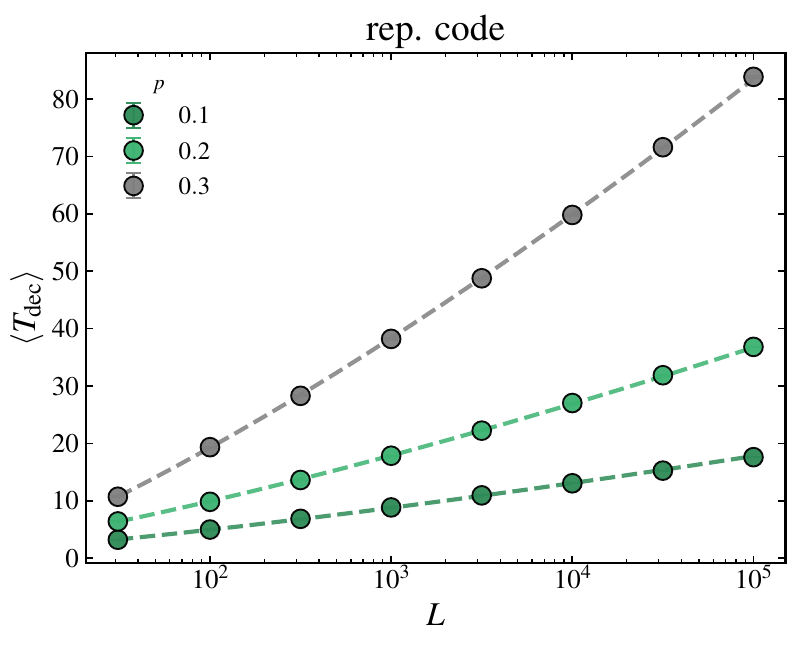}
            \caption{Average decoding time versus $L$ for the repetition-code decoder. Dashed lines are fits to the form $\langle \Tdec \rangle = A + B \log^\eta{L}$, where $A$ and $B$ are $p$-dependent constants. We find $\eta \approx 1.4$ at the simulated values of $p$.}
            \label{fig:rep_avg_steps}
        \end{figure}
        
        \begin{figure}
            \centering
            \includegraphics[width=\linewidth]{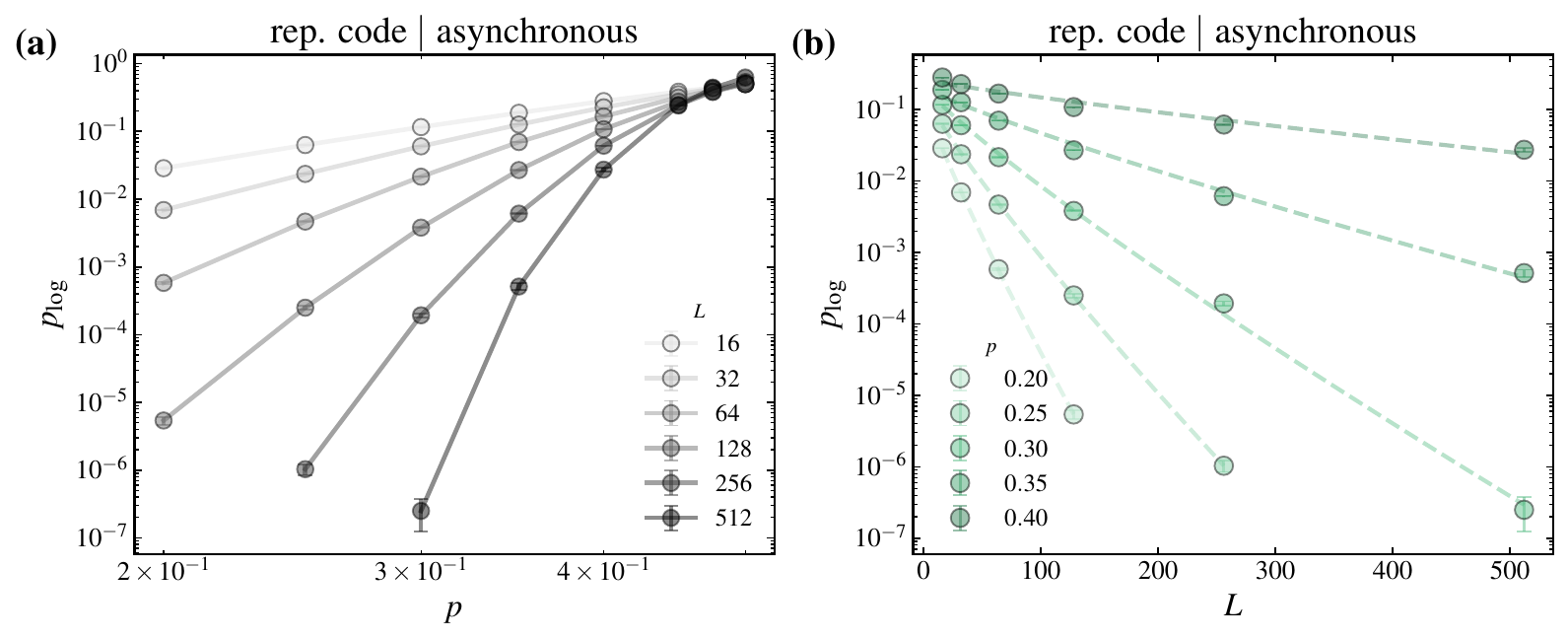}
            \caption{Performance of the code-capacity repetition-code decoder under completely asynchronous, uncoordinated updates. \textbf{(a)} Logical error rate $\plog$ as a function of the physical error rate $p$ under i.i.d.\ bit-flip noise. The data are consistent with a threshold of $p_c = 1/2$. \textbf{(b)} Sub-threshold scaling of the logical error rate $\plog$ with $L$. Dashed lines are fits to the form $\plog = C(p/p_c)^{\gamma L^{\alpha}}$ with a single exponent $\alpha$ shared across all $p$. The fitted $\alpha = 0.87 \pm 0.03$ is consistent with stretched exponential decay of $\plog$ in $L$, and the fitted $\gamma$ increases from $0.04$ to $0.16$ as $p$ decreases from $0.4$ to $0.2$.}
            \label{fig:repcode_uncoord}
        \end{figure}

        We study the performance of our decoder under i.i.d.\ bit-flip noise using Monte Carlo simulations. Figs.~\ref{fig:repcodethreshold} and~\ref{fig:rep_avg_steps} show the synchronous results: we observe a threshold consistent with $p_c = 1/2$ and find that the average decoding time grows polylogarithmically with $L$ at the simulated values of $p$.\footnote{Runs that have not terminated within an $O(L)$ window are counted as failures in $\plog$ and excluded from $\langle \Tdec \rangle$; at the sizes and error rates of Fig.~\ref{fig:rep_avg_steps}, such runs constituted a negligible fraction of the total number of simulations.} The code used to generate the figures in this section and the sections to follow is available in the linked repository~\cite{localdecodersrepo}.

        Like Lake's decoder, ours achieves the same threshold as the optimal decoder given by minimum-weight matching. A priori, it is surprising that a local decoder can have the same threshold as the optimal global decoder. However, the following simple heuristic in the spirit of Ref.~\cite{lake2025fastofflinedecodinglocal} suggests that local decoders for the one-dimensional repetition code are capable of achieving $p_c = 1/2$: a defect pair has length $\ell$ with probability $(1-p)p^{\ell - 1}$, giving an average pair length of $\langle \ell_{\mathrm{pair}} \rangle = 1/(1-p)$, while the average separation between pairs is $\langle s_{\mathrm{pair}} \rangle = 1/p$. Since $\langle \ell_{\mathrm{pair}} \rangle \leq \langle s_{\mathrm{pair}} \rangle$ as long as $p \leq 1/2$, this provides heuristic evidence that a threshold of $p_c = 1/2$ is possible for local decoders that move defects toward (left-)nearest neighbors.

        One difference between our decoder and that of Ref.~\cite{lake2025fastofflinedecodinglocal} is that our decoder's average decoding time appears to grow as $\log^\eta L$ for $\eta > 1$, whereas the average decoding time of the decoder of Ref.~\cite{lake2025fastofflinedecodinglocal} appears to grow as $\log L$. One explanation for the value of $\eta$ is as follows. Linear cluster erosion implies that the decoding time of a noise realization is at most a constant times the diameter of its largest cluster. Separately, the sparsity theorem bounds the probability of a $k$-cluster by $C(p/p_*)^{2^k}$, and in fact, the numerics of Ref.~\cite{lake2025fastofflinedecodinglocal} suggest that this bound is sharp. Therefore, the largest typical cluster has diameter $\Theta(\log^{\log_2 n} L)$, so $\langle \Tdec \rangle = O(\log^{\log_2 n} L)$ for every value of the sparsity-theorem parameter $n$ for which linear cluster erosion holds for our decoder. This decoding-time bound is sharp if the largest typical separation between defects in such a cluster is of order its diameter $w_k$, so that decoding of the cluster takes time of order $w_k$. We therefore expect $\eta \approx \log_2 n$ for the value of $n$ at which both linear cluster erosion and this condition hold. Since this value is set by the decoder's cluster erosion and expansion constants and can differ between decoders with the same threshold, we do not expect the decoding-time exponents of our decoder and that of Ref.~\cite{lake2025fastofflinedecodinglocal} to coincide. Moreover, the clustering argument for our decoder is more stringent than that of Ref.~\cite{lake2025fastofflinedecodinglocal}: our proof of linear cluster erosion requires the spacetime activity regions of distinct clusters to be disjoint, so that clusters never interact, whereas the argument of Ref.~\cite{lake2025fastofflinedecodinglocal} tolerates some interaction between neighboring clusters. It is therefore natural to expect that the value of $n$ for which the above argument holds should be larger for our decoder, and hence that our decoder's average-decoding-time exponent should be larger than that of the decoder of Ref.~\cite{lake2025fastofflinedecodinglocal}.

        We also study the performance of our decoder under completely asynchronous, uncoordinated updates, in which a single randomly chosen site updates at each time step. This setting is beyond the scope of the rigorous desynchronization techniques described in Ref.~\cite{lake2025fastofflinedecodinglocal}. The results, shown in Fig.~\ref{fig:repcode_uncoord}, are also consistent with a threshold of $p_c = 1/2$.

        We conclude with a brief remark on a tradeoff in our construction. In Algorithm~\ref{alg:rep-code}, the separation between message growth and message erasure speeds is enforced via a clock register: the erasure rule runs at every time step while the growth rule runs only when the clock is zero. An alternative is to eliminate the clock and instead increase the spatial communication radius of the update rule, allowing each site to consult its two (or more) nearest left neighbors. This lets the erasure front advance faster than the growth front without any temporal gating. The two approaches offer different tradeoffs: the clock-based scheme keeps the interaction range minimal at the cost of extra per-site classical overhead, while the extended-range scheme removes the clock at the cost of a larger update neighborhood.

\section{Toric and surface codes under code-capacity noise} \label{sec:toric-code}

    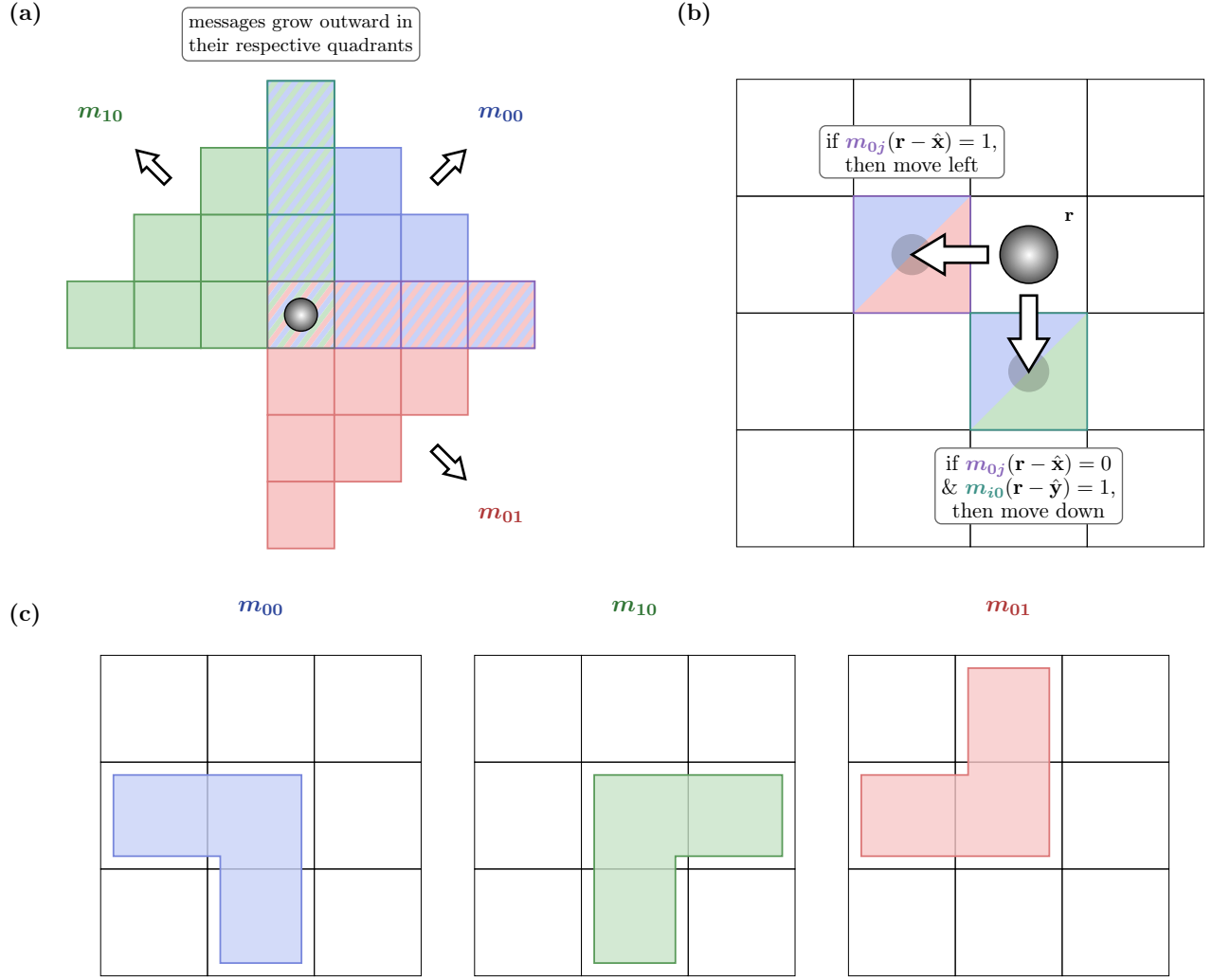
\begin{figure}[p]
        \centering
        \resizebox{\textwidth}{!}{%
        \tikzsetnextfilename{toric-rule}
        \begin{tikzpicture}[scale=1.1]
        
        \tikzset{bubble/.style={fill=white, draw=black!60, rounded corners=3pt, inner sep=3pt, line width=0.6pt, text=black}}
        
        % ===== (a) Message growth =====
        \begin{scope}[shift={(0,0)}]
        
        \def\s{1}
        
        % m_{00} grows into (+x, +y): blue
        \foreach \x/\y in {0/1, 0/2, 0/3, 1/0, 1/1, 1/2, 2/0, 2/1, 3/0} {
            \fill[msgblue, draw=msgblueedge, thick] (\x*\s - \s/2, \y*\s - \s/2) rectangle (\x*\s + \s/2, \y*\s + \s/2);
        }
        
        % m_{10} grows into (-x, +y): GREEN
        \foreach \x/\y in {0/1, 0/2, 0/3, -1/0, -1/1, -1/2, -2/0, -2/1, -3/0} {
            \fill[msggreen, draw=msggreenedge, thick] (\x*\s - \s/2, \y*\s - \s/2) rectangle (\x*\s + \s/2, \y*\s + \s/2);
        }
        
        % m_{01} grows into (+x, -y): RED
        \foreach \x/\y in {0/-1, 0/-2, 0/-3, 1/0, 1/-1, 1/-2, 2/0, 2/-1, 3/0} {
            \fill[msgred, draw=msgrededge, thick] (\x*\s - \s/2, \y*\s - \s/2) rectangle (\x*\s + \s/2, \y*\s + \s/2);
        }
        
        % Overlap on +y axis
        \foreach \y in {1, 2, 3} {
            \stripedrect{msgblue}{msggreen}{-0.5}{\y - 0.5}{0.5}{\y + 0.5}
            \draw[msgtealedge, thick] (0 - \s/2, \y*\s - \s/2) rectangle (0 + \s/2, \y*\s + \s/2);
        }
        
        % Overlap on +x axis
        \foreach \x in {1, 2, 3} {
            \stripedrect{msgblue}{msgred}{\x - 0.5}{-0.5}{\x + 0.5}{0.5}
            \draw[msgoverlapedge, thick] (\x*\s - \s/2, 0 - \s/2) rectangle (\x*\s + \s/2, 0 + \s/2);
        }
        
        % Origin cell: triple stripes
        \begin{scope}
        \clip (-\s/2, -\s/2) rectangle (\s/2, \s/2);
        \fill[white] (-\s/2, -\s/2) rectangle (\s/2, \s/2);
        \foreach \i in {-20,-19,...,20} {
            \fill[msgblue] ({-0.5 + \i*0.3 - 0.05}, -1) -- ({-0.5 + \i*0.3 + 0.05}, -1)
            -- ({-0.5 + \i*0.3 + 0.05 + 1.5}, 1) -- ({-0.5 + \i*0.3 - 0.05 + 1.5}, 1) -- cycle;
            \fill[msggreen] ({-0.5 + \i*0.3 + 0.05}, -1) -- ({-0.5 + \i*0.3 + 0.15}, -1)
            -- ({-0.5 + \i*0.3 + 0.15 + 1.5}, 1) -- ({-0.5 + \i*0.3 + 0.05 + 1.5}, 1) -- cycle;
            \fill[msgred] ({-0.5 + \i*0.3 + 0.15}, -1) -- ({-0.5 + \i*0.3 + 0.25}, -1)
            -- ({-0.5 + \i*0.3 + 0.25 + 1.5}, 1) -- ({-0.5 + \i*0.3 + 0.15 + 1.5}, 1) -- cycle;
        }
        \end{scope}
        \draw[msggrayedge, thick] (-\s/2, -\s/2) rectangle (\s/2, \s/2);
        
        % Anyon orb
        \anyon
        
        % Block arrows
        \blockarrow{black}{2.0}{2.0}{45}{0.65}
        \node[lblblue, font=\bfseries] at (3.0, 3.0) {$\bm{m_{00}}$};
        
        \blockarrow{black}{-2.0}{2.0}{135}{0.65}
        \node[lblgreen, font=\bfseries] at (-3.0, 3.0) {$\bm{m_{10}}$};
        
        \blockarrow{black}{2.0}{-2.0}{-45}{0.65}
        \node[lblred, font=\bfseries] at (3.0, -3.0) {$\bm{m_{01}}$};
        
        \node[font=\bfseries, anchor=north west] at (-4.5, 4.8) {(a)};
        
        \node[bubble, font=\footnotesize\rmfamily, align=center] at (0, 4.2) {messages grow outward in\\their respective quadrants};
        
        \end{scope}
        
        % ===== (b) Defect movement =====
        \begin{scope}[shift={(10.9, 0.9)}, scale=1.75]
        
        \def\s{1}
        \foreach \x in {-2,-1,0,1} {
        \foreach \y in {-2,-1,0,1} {
            \draw[black, thin] (\x*\s - \s/2, \y*\s - \s/2) rectangle (\x*\s + \s/2, \y*\s + \s/2);
        }
        }
        \begin{scope}
        \clip (-1*\s - \s/2, -\s/2) rectangle (-1*\s + \s/2, \s/2);
        \fill[msgblue] (-1*\s - \s/2, -\s/2) rectangle (-1*\s + \s/2, \s/2);
        \fill[msgred] (-1.5, -0.5) -- (-0.5, 0.5) -- (-0.5, -0.5) -- cycle;
        \end{scope}
        \draw[msgoverlapedge, thick] (-1*\s - \s/2, -\s/2) rectangle (-1*\s + \s/2, \s/2);
        \begin{scope}
        \clip (-\s/2, -1*\s - \s/2) rectangle (\s/2, -1*\s + \s/2);
        \fill[msgblue] (-\s/2, -1*\s - \s/2) rectangle (\s/2, -1*\s + \s/2);
        \fill[msggreen] (-0.5, -1.5) -- (0.5, -0.5) -- (0.5, -1.5) -- cycle;
        \end{scope}
        \draw[msgtealedge, thick] (-\s/2, -1*\s - \s/2) rectangle (\s/2, -1*\s + \s/2);
        \fill[black, opacity=0.2] (-1, 0) circle (5pt);
        \fill[black, opacity=0.2] (0, -1) circle (5pt);
        \blockarrow{black}{-0.35}{0}{180}{0.65}
        \blockarrow{black}{0}{-0.35}{-90}{0.65}
        \shade[inner color=white!25, outer color=black!68] (0,0) circle (7pt);
        \draw[black, line width=0.7pt] (0,0) circle (7pt);
        \node[font=\footnotesize, above right, xshift=13pt, yshift=12pt] at (0,0) {$\mathbf{r}$};
        \node[bubble, font=\small\rmfamily, above, align=center] at (-1, 0.65) {if \textcolor{msgoverlapedge}{$\bm{m_{0j}}$}$(\mathbf{r} - \hat{\mathbf{x}}) = 1$,\\[-1pt]then move left};
        \node[bubble, font=\small\rmfamily, below, align=center] at (0, -1.65) {if \textcolor{msgoverlapedge}{$\bm{m_{0j}}$}$(\mathbf{r} - \hat{\mathbf{x}}) = 0$\\[-1pt]\& \textcolor{msgtealedge}{$\bm{m_{i0}}$}$(\mathbf{r} - \hat{\mathbf{y}}) = 1$,\\[-1pt]then move down};
        
        \end{scope}
        
        \node[font=\bfseries, anchor=north west] at (5.5, 4.8) {(b)};
        
        % ===== (c) Toom votes =====
        \begin{scope}[shift={(-0.6, -7.5)}, scale=1.6]
        
        \def\s{1}
        \def\off{3.5}
        \def\g{0.12}
        
        \begin{scope}[shift={(0,0)}]
        \foreach \x in {-1,0,1} {
            \foreach \y in {-1,0,1} {
            \draw[black, thin] (\x*\s - \s/2, \y*\s - \s/2) rectangle (\x*\s + \s/2, \y*\s + \s/2);
            }
        }
        \fill[msgblue, opacity=0.8]
            (-1.5+\g, 0.5-\g) -- (0.5-\g, 0.5-\g) -- (0.5-\g, -1.5+\g) -- (-0.5+\g, -1.5+\g) -- (-0.5+\g, -0.5+\g) -- (-1.5+\g, -0.5+\g) -- cycle;
        \draw[msgblueedge, thick]
            (-1.5+\g, 0.5-\g) -- (0.5-\g, 0.5-\g) -- (0.5-\g, -1.5+\g) -- (-0.5+\g, -1.5+\g) -- (-0.5+\g, -0.5+\g) -- (-1.5+\g, -0.5+\g) -- cycle;
        \node[lblblue, font=\bfseries, above] at (0, 1.8) {$\bm{m_{00}}$};
        \end{scope}
        
        \begin{scope}[shift={(\off,0)}]
        \foreach \x in {-1,0,1} {
            \foreach \y in {-1,0,1} {
            \draw[black, thin] (\x*\s - \s/2, \y*\s - \s/2) rectangle (\x*\s + \s/2, \y*\s + \s/2);
            }
        }
        \fill[msggreen, opacity=0.8]
            (-0.5+\g, 0.5-\g) -- (1.5-\g, 0.5-\g) -- (1.5-\g, -0.5+\g) -- (0.5-\g, -0.5+\g) -- (0.5-\g, -1.5+\g) -- (-0.5+\g, -1.5+\g) -- cycle;
        \draw[msggreenedge, thick]
            (-0.5+\g, 0.5-\g) -- (1.5-\g, 0.5-\g) -- (1.5-\g, -0.5+\g) -- (0.5-\g, -0.5+\g) -- (0.5-\g, -1.5+\g) -- (-0.5+\g, -1.5+\g) -- cycle;
        \node[lblgreen, font=\bfseries, above] at (0, 1.8) {$\bm{m_{10}}$};
        \end{scope}
        
        \begin{scope}[shift={(2*\off,0)}]
        \foreach \x in {-1,0,1} {
            \foreach \y in {-1,0,1} {
            \draw[black, thin] (\x*\s - \s/2, \y*\s - \s/2) rectangle (\x*\s + \s/2, \y*\s + \s/2);
            }
        }
        \fill[msgred, opacity=0.8]
            (0.5-\g, 1.5-\g) -- (0.5-\g, -0.5+\g) -- (-1.5+\g, -0.5+\g) -- (-1.5+\g, 0.5-\g) -- (-0.5+\g, 0.5-\g) -- (-0.5+\g, 1.5-\g) -- cycle;
        \draw[msgrededge, thick]
            (0.5-\g, 1.5-\g) -- (0.5-\g, -0.5+\g) -- (-1.5+\g, -0.5+\g) -- (-1.5+\g, 0.5-\g) -- (-0.5+\g, 0.5-\g) -- (-0.5+\g, 1.5-\g) -- cycle;
        \node[lblred, font=\bfseries, above] at (0, 1.8) {$\bm{m_{01}}$};
        \end{scope}
        
        \end{scope}
        
        \node[font=\bfseries, anchor=north west] at (-4.5, -4.2) {(c)};
        
        \end{tikzpicture}
        }
        \caption{Toric-code decoder rule in the code-capacity setting. \textbf{(a)}~Message growth. A defect seeds three message fronts $m_{00}$, $m_{10}$, and $m_{01}$ that grow outward in the $(+x,+y)$, $(-x,+y)$, and $(+x,-y)$ quadrants. Overlapping regions on the axes carry multiple message flavors. \textbf{(b)}~Defect movement. A defect at site $\mathbf{r}$ moves left whenever it sees an $m_{0j}$ message at $\mathbf{r} - \hat{\mathbf{x}}$ for some $j$; otherwise, it moves down if it sees an $m_{i0}$ message at $\mathbf{r} - \hat{\mathbf{y}}$ for some $i$. \textbf{(c)}~Toom voting neighborhoods. Each message channel is erased using a rotated variant of Toom's rule, oriented so that erasure proceeds from the direction opposite to growth. The shaded L-shaped region for each $m_{ij}$ indicates the voting neighborhood used when deciding whether the message at the central site should persist.}
    \end{figure}

    In this section, we construct a code-capacity decoder for the toric code and prove that it has a non-zero threshold $p_*$ and polylogarithmic average decoding time for $p < p_*$. The proof strategy is the same as in one dimension: establish linear defect erosion and linear message erosion, then apply the results of Section~\ref{sec:threshold}. We again work at infinite system size for simplicity and omit the straightforward generalization to finite systems.

    We now give an informal overview of our toric code decoder, which can be viewed as a two-dimensional generalization of our repetition code decoder. To achieve linear defect erosion, we monotonically move all defects in a cluster downward and leftward, toward the cluster's leftmost and bottommost defects. This is achieved using three message channels $m_{ij}$ for $(i,j) \in \mathcal{M}_2 \coloneqq \{0,1\}^2 \setminus \{(1,1)\}$, where $m_{ij}$ propagates in the $((-1)^i x, (-1)^j y)$ direction at an average speed of $1/q$ for some fixed integer $q \geq 2$. The excluded channel $m_{11}$, which would propagate into the $(-x,-y)$ quadrant, is not included because it is unnecessary for our construction, similar to how left-propagating messages are unnecessary for the one-dimensional construction. A defect moves left whenever it sees an $m_{0j}$ message (for any $j$) immediately to its left; otherwise, it moves down if it sees an $m_{i0}$ message (for any $i$) immediately below it; otherwise, it remains stationary. 

    By construction, no defect ever moves left of the original leftmost defect or below the original bottommost defect in the cluster. By time $O(W)$, messages have spread sufficiently throughout a cluster of diameter $W$ so that all but at most one of the surviving defects are guaranteed to move at non-zero average speed. This ensures that all defects annihilate within $O(W)$ additional steps, although defects may be eliminated sooner by merging before reaching the leftmost or bottommost defect. Once all defects have been annihilated, residual messages occupy an enlarged $O(W)$-diameter defect-free region, since the message fronts expand beyond the original cluster boundary.

    To achieve linear message erosion, we design a local erasure rule. In the absence of defects, the rule operates roughly as follows. Each of the three message channels $m_{00}$, $m_{10}$, and $m_{01}$ is erased by a rotated variant of Toom's rule, oriented so that erasure does not impede growth. The clock channel ensures that messages grow more slowly than they are erased, so the erasure rule catches up to expanding message fronts. In particular, the erasure rule eliminates messages in a defect-free region in time linear in the messages' extent, so linear message erosion is satisfied. The erasure rules are further designed so that erasure cannot begin before the defect that sourced the message has been annihilated; this is achieved by coupling the different messages together to prevent premature erasure, in a manner made precise below.

    For the purposes of our proofs, it is convenient to have messages grow more often than defect movement can occur, so the decoder that we define and formally analyze below is slightly different from the one described above. Numerically, we find that the movement constraint is unnecessary. Moreover, removing it and increasing $q$ each independently raise the threshold and improve the decay of the logical error rate, with the largest gains obtained by doing both.

    \subsection{Definitions}\label{subsec:toric-code:definitions}

        We consider the two-dimensional toric code on $\Lambda = \Z^2$. The state at each site consists of seven channels:
        \[
            \xi(\vc{r}) = \bigl(e_x(\vc{r}),\, e_y(\vc{r}),\, s(\vc{r}),\, m_{00}(\vc{r}),\, m_{10}(\vc{r}),\, m_{01}(\vc{r}),\, c(\vc{r})\bigr),
        \]
        where $e_x(\vc{r}), e_y(\vc{r}) \in \{0,1\}$ are correction channels for the horizontal and vertical qubits in the negative $x$ and $y$ directions relative to the parity check, $s(\vc{r}) \in \{0,1\}$ is the syndrome channel, $m_{ij}(\vc{r}) \in \{0,1\}$ for $(i,j) \in \mathcal{M}_2$ are three message channels, and $c(\vc{r}) \in \{0, \ldots, q-1\}$ is the clock channel for some fixed integer $q \geq 2$, which we take to be $q = 3$ in this subsection and the next.

        The three message channels correspond to three quadrants: $m_{ij}$ propagates in the $((-1)^i x, (-1)^j y)$ direction. Explicitly, $m_{00}$ propagates in the $(+x,+y)$ direction, $m_{01}$ in the $(+x,-y)$ direction, and $m_{10}$ in the $(-x,+y)$ direction.

        The local update rule is specified in Algorithm~\ref{alg:toric-code}. The clock increments as $c(\vc{r}, t+1) = (c(\vc{r},t) + 1) \bmod q$. The correction channels update as follows. Let $\ell(\vc{r},t) \coloneqq [s(\vc{r}, t) = 1 \;\text{and}\; \exists\, j \text{ s.t.\ } m_{0j}(\vc{r}-\uvec x, t) = 1 \;\text{and}\; c(\vc{r},t) = 0]$ denote the indicator for a leftward move. Then
        \begin{align*}
            e_x(\vc{r}, t+1) &= e_x(\vc{r}, t) \oplus \ell(\vc{r},t),\\
            e_y(\vc{r}, t+1) &= e_y(\vc{r}, t) \oplus \bigl[s(\vc{r}, t) = 1 \;\text{and}\; \exists\, i \text{ s.t.\ } m_{i0}(\vc{r}-\uvec y, t) = 1 \;\text{and}\; \lnot \ell(\vc{r},t) \;\text{and}\; c(\vc{r},t) = 0\bigr].
        \end{align*}
        Since corrections can only be applied at defect sites, and flipping $e_x(\vc{r})$ moves the defect from $\vc{r}$ to $\vc{r} - \uvec{x}$ while flipping $e_y(\vc{r})$ moves it from $\vc{r}$ to $\vc{r} - \uvec{y}$, the only effect of a flip is to move a defect one step to the left or down. Leftward moves take priority over downward moves.

        The message channels update as follows. Here, $\Call{ToomVote}{i,j,\vc{r},t}$ returns the majority of $m_{ij}(\vc{r},t)$, $m_{ij}(\vc{r}+(-1)^{i+1}\uvec x,t)$, and $m_{ij}(\vc{r}+(-1)^{j+1}\uvec y,t)$. For each $(i,j) \in \mathcal{M}_2$, we set $m_{ij}(\vc{r}, t+1) = 1$ if and only if at least one of the following conditions holds:
        \begin{enumerate}[label=(\roman*)]
            \item \emph{Defect persistence:} $s(\vc{r}, t) = 1$.
            \item \emph{Defect arrival:} site $\vc{r}$ is set to receive at least one incoming defect at time $t+1$ (i.e., $\vc{r} \in S_{\mathrm{inc}}$).
            \item \emph{Message growth:} $s(\vc{r}, t) = 0$, $m_{ij}(\vc{r}, t) = 0$, $c(\vc{r}, t) \in \{0,1\}$, and $m_{ij}(\vc{r} + (-1)^{i+1} \uvec{x}, t) = 1$ or $m_{ij}(\vc{r} + (-1)^{j+1} \uvec{y}, t) = 1$.
            \item \emph{Message persistence (majority vote):} $s(\vc{r}, t) = 0$, $m_{ij}(\vc{r}, t) = 1$, and $\Call{ToomVote}{i,j,\vc{r},t} = 1$.
            \item \emph{Message persistence (coupling):} $s(\vc{r}, t) = 0$, $m_{ij}(\vc{r}, t) = 1$, $(i,j) \neq (0,0)$, $m_{00}(\vc{r}, t) = 1$, and $\Call{ToomVote}{0,0,\vc{r},t} = 1$.
        \end{enumerate}
        Otherwise, $m_{ij}(\vc{r}, t+1) = 0$. The syndrome at time $t+1$ is then obtained by updating the syndrome at time $t$ according to the changes in the correction configuration.

        Condition (i) ensures that every defect maintains all three messages at its own site. Condition (ii) preserves messages at the new defect location after a move, preventing premature erasure and the ensuing delay in repopulating the message. Condition (iii) allows each message $m_{ij}$ to grow in its designated quadrant, but only when $c(\vc{r},t) \in \{0,1\}$, so that messages propagate at average speed $2/3$. Condition (iv) implements Toom's majority rule for erasure: each message channel has its own rotated variant, oriented so that erasure proceeds from the direction opposite to growth. Condition (v) couples the erasure of $m_{01}$ and $m_{10}$ to $m_{00}$: even if the Toom vote for $m_{01}$ or $m_{10}$ would erase the message, it is preserved if $m_{00}$ at the same site survives its own Toom vote. This coupling prevents premature erasure of $m_{01}$ and $m_{10}$ while the defect that sourced them is still alive. The reason is that defects always move left or down, so their movement intrinsically preserves $m_{00}$ but not $m_{01}$ or $m_{10}$; the coupling to $m_{00}$ prevents the latter two from being prematurely erased before the defect has been annihilated. For a general clock period $q$, the rule is defined so that messages can grow whenever $c(\vc{r},t) \in \{0,\ldots,q-2\}$.

        The update rule in Algorithm~\ref{alg:toric-code} is written as two loops, as in the one-dimensional case. The second loop overwrites all messages at each site in $S_{\mathrm{inc}}$ to $1$, implementing condition (ii); as before, this can equivalently be implemented locally within the first loop.

        \begin{algorithm}[H]
            \caption{Toric code update from time $t$ to $t+1$.}
            \label{alg:toric-code}
            \begin{algorithmic}[1]
                \Function{SetAll}{$\vc{r}$}
                    \State $m_{ij}(\vc{r},t+1) \gets 1 \;\; \forall\, (i,j) \in \mathcal{M}_2$
                \EndFunction
                \Function{ToomVote}{$i,j,\vc{r},t$}
                    \State \Return $\Call{Maj}{m_{ij}(\vc{r},t),\; m_{ij}(\vc{r}+(-1)^{i+1}\uvec{x},t),\; m_{ij}(\vc{r}+(-1)^{j+1}\uvec{y},t)}$
                \EndFunction
                \Statex
                \State $S_{\mathrm{inc}} \gets \emptyset$
                \ForAll{$\vc{r} \in \Lambda$}
                    \State $m_{ij}(\vc{r},t+1) \gets m_{ij}(\vc{r},t) \;\; \forall\, (i,j) \in \mathcal{M}_2$; \;\; $e_x(\vc{r},t+1) \gets e_x(\vc{r},t)$; \;\; $e_y(\vc{r},t+1) \gets e_y(\vc{r},t)$
                    \Statex
                    \If{$s(\vc{r},t) = 1$} \Comment{\textbf{Part 1: Defect dynamics}}
                        \State $\Call{SetAll}{\vc{r}}$
                        \If{$c(\vc{r},t) = 0$}
                            \If{$\exists\, j$ s.t.\ $m_{0j}(\vc{r}-\uvec{x},t) = 1$}
                                \State $e_x(\vc{r},t+1) \gets e_x(\vc{r},t) \oplus 1$; \;\; $S_{\mathrm{inc}} \gets S_{\mathrm{inc}} \cup \{\vc{r} - \uvec{x}\}$ \Comment{Move left}
                            \ElsIf{$\exists\, i$ s.t.\ $m_{i0}(\vc{r}-\uvec{y},t) = 1$}
                                \State $e_y(\vc{r},t+1) \gets e_y(\vc{r},t) \oplus 1$; \;\; $S_{\mathrm{inc}} \gets S_{\mathrm{inc}} \cup \{\vc{r} - \uvec{y}\}$ \Comment{Move down}
                            \EndIf
                        \EndIf
                    \Statex
                    \Else \Comment{\textbf{Part 2: Message dynamics}}
                        \ForAll{$(i,j) \in \mathcal{M}_2$} \Comment{Growth}
                            \If{$m_{ij}(\vc{r},t) = 0$}
                                \State $m_{ij}(\vc{r},t+1) \gets \bigl(m_{ij}(\vc{r}+(-1)^{i+1}\uvec{x},t) \lor m_{ij}(\vc{r}+(-1)^{j+1}\uvec{y},t)\bigr) \land \bigl(c(\vc{r},t) \in \{0,1\}\bigr)$
                            \EndIf
                        \EndFor
                        \Statex
                        \State $v_{00} \gets 0$ \Comment{Erasure}
                        \If{$m_{00}(\vc{r},t) = 1$}
                            \State $v_{00} \gets \Call{ToomVote}{0,0,\vc{r},t}$
                            \State $m_{00}(\vc{r},t+1) \gets v_{00}$
                        \EndIf
                        \ForAll{$(i,j) \in \{(0,1),\,(1,0)\}$}
                            \If{$m_{ij}(\vc{r},t) = 1$}
                                \State $v_{ij} \gets \Call{ToomVote}{i,j,\vc{r},t}$
                                \State $m_{ij}(\vc{r},t+1) \gets v_{ij} \lor \bigl(m_{00}(\vc{r},t) \land v_{00}\bigr)$
                            \EndIf
                        \EndFor
                    \EndIf
                    \Statex
                    \State $c(\vc{r},t+1) \gets (c(\vc{r},t) + 1) \bmod q$
                \EndFor
                \Statex
                \ForAll{$\vc{r}' \in S_{\mathrm{inc}}$} \Comment{\textbf{Part 3: Defect arrival}}
                    \State $\Call{SetAll}{\vc{r}'}$
                \EndFor
            \end{algorithmic}
        \end{algorithm}

    \subsection{Linear erosion}

        We now prove linear defect erosion and linear message erosion, beginning with the former. The proof strategy for linear defect erosion is a direct generalization of the one-dimensional case. First, we show that the number of defects is non-increasing. Next, we fix a convention for tracking defect worldlines and show that messages grow linearly in time in the relevant regions around each defect. These facts imply that, by time $O(W)$, messages have spread enough throughout a cluster of diameter $W$ so that every defect that is not simultaneously the leftmost and bottommost moves at non-zero average speed for all time steps thereafter.

        As in the one-dimensional case, the following proposition is immediate from the construction of the decoder.
        \begin{proposition}[defect monotonicity]\label{prop:toric-code:defect-monotone}
            The number of defects is non-increasing in time.
        \end{proposition}
        
        Next, we fix a convention that uniquely determines all defect worldlines. At $t = 0$, each defect site contains a single well-defined defect. We then impose the following association rules for all later times $t \geq 1$:
        \begin{itemize}
            \item \textbf{Move into an empty or vacated site.} If a defect is the only defect to move into a site that is unoccupied at time $t$, or whose occupying defect simultaneously moves left or down, its worldline continues at the new position.
            \item \textbf{Stationary defect.} If a defect remains at its site and either no defect enters or two defects enter during the time step, its worldline continues at that position.
            \item \textbf{Annihilation.} If two defects move into the same site, they annihilate. Similarly, if only one defect moves into a site occupied by a stationary defect, they annihilate. In each case, both worldlines terminate at the previous time step, and neither defect is considered alive at the time of annihilation.
        \end{itemize}
        With these rules, each defect has a well-defined worldline. Since the number of defects is non-increasing, no new defect worldlines are created by the dynamics.
        
        In what follows, we assume that the initial error configuration is confined to a cluster of diameter $W$. Fix a single defect worldline $\{\vc{X}(t)\}$ with each $\vc{X}(t) = (X_1(t), X_2(t)) \in \Z^2$. We will assume that the defect is created at $t = 0$ and is alive only at times $t = 0, 1, \ldots, T$, where $T$ is finite; it will be clear a posteriori that assuming $T$ is finite entails no loss of generality.
        
        By construction, every defect moves monotonically to the left and downward. In particular, both coordinates of every defect are non-increasing in time.
        \begin{proposition}[directed movement property]\label{prop:toric-code:directed-movement}
            For every $t \in \{0, 1, \ldots, T-1\}$,
            \[
                \vc{X}(t+1) \in \bigl\{\vc{X}(t),\; \vc{X}(t) - \uvec{x},\; \vc{X}(t) - \uvec{y}\bigr\}.
            \]
        \end{proposition}

        By the same argument as in Lemma~\ref{lem:rep-code:trivial-persistence}, for every $t \in \{1, \ldots, T\}$ all messages at the defect's current site are non-trivial at both time $t$ and time $t+1$.
        \begin{lemma}[trivial message persistence]\label{lem:toric-code:trivial-persistence}
            For every $t \in \{1, \ldots, T\}$ and every $(i,j) \in \mathcal{M}_2$,
            \[
                m_{ij}(\vc{X}(t),\, t) = m_{ij}(\vc{X}(t),\, t+1) = 1.
            \]
        \end{lemma}

        Next, we show that once a site has been visited by the defect, all of its messages remain non-trivial for at least as long as the defect is alive.
        \begin{lemma}[message persistence]\label{lem:toric-code:message-persistence}
            For every $t \in \{1, \ldots, T\}$, every $t' \in [t, T]$, and every $(i,j) \in \mathcal{M}_2$,
            \[
                m_{ij}(\vc{X}(t),\, t') = 1.
            \]
        \end{lemma}

        \begin{proof}
            The proof is by backward induction on $t \in \{1, \ldots, T\}$ and forward induction on $t' \in [t, T]$.
        
            \textbf{Base case.} For $t = T$, Lemma~\ref{lem:toric-code:trivial-persistence} gives $m_{ij}(\vc{X}(T), T) = 1$ for all $(i,j) \in \mathcal{M}_2$, and no forward induction is needed.
        
            \textbf{Inductive step.} Suppose the result holds for all $\tau \in [t+1, T]$; we prove it for $t$ by forward induction on $t' \in [t, T]$. The forward base case $t' = t$ follows from Lemma~\ref{lem:toric-code:trivial-persistence}. For the forward inductive step, assume $m_{ij}(\vc{X}(t), \tau') = 1$ for all $\tau' \in [t, t'-1]$ and all $(i,j) \in \mathcal{M}_2$, where $t' \in [t+1, T]$; we wish to show $m_{ij}(\vc{X}(t), t') = 1$ for all $(i,j) \in \mathcal{M}_2$.
        
            If $\vc{X}(t'-1) = \vc{X}(t)$, then the defect is at site $\vc{X}(t)$ at time $t'-1$, so Lemma~\ref{lem:toric-code:trivial-persistence} immediately gives $m_{ij}(\vc{X}(t), t') = 1$ for all $(i,j) \in \mathcal{M}_2$.
        
            Otherwise, $\vc{X}(t'-1) \neq \vc{X}(t)$, so the defect has moved away from $\vc{X}(t)$ by time $t'-1$. By the directed movement property, there exists $t'' \in (t, t')$ such that $\vc{X}(t'')$ is either $\vc{X}(t) - \uvec{x}$ or $\vc{X}(t) - \uvec{y}$. By the backward inductive hypothesis, $m_{ij}(\vc{X}(t''), \tau) = 1$ for all $\tau \in [t'', T]$ and all $(i,j) \in \mathcal{M}_2$. In particular, $m_{00}(\vc{X}(t''), t'-1) = 1$. By the forward inductive hypothesis, $m_{ij}(\vc{X}(t), t'-1) = 1$ for all $(i,j) \in \mathcal{M}_2$. We now verify that each message persists:
            \begin{itemize}
                \item For $m_{00}$: since $m_{00}(\vc{X}(t), t'-1) = 1$ and $m_{00}(\vc{X}(t''), t'-1) = 1$, where $\vc{X}(t'')$ is either to the left or below $\vc{X}(t)$, the Toom vote satisfies $\Call{ToomVote}{0,0,\vc{X}(t),t'-1} = 1$, so $m_{00}(\vc{X}(t), t') = 1$ is guaranteed by condition (iv).
                \item For $(i,j) \in \{(0,1), (1,0)\}$: since $m_{00}(\vc{X}(t), t'-1) = 1$ and $\Call{ToomVote}{0,0,\vc{X}(t),t'-1} = 1$, condition (v) guarantees $m_{ij}(\vc{X}(t), t') = 1$.
            \end{itemize}
            This completes the forward induction, and thus the backward induction.
        \end{proof}

        We now show that a persistent message at a site causes that message to spread into its designated quadrant at average speed $2/3$.

        \begin{lemma}[message growth]\label{lem:toric-code:message-growth}
            Fix a site $\vc{x} \in \Z^2$ and times $t' \geq t$. Define the number of growth opportunities between $t$ and $t'$ by
            \[
                N(t, t') \coloneqq \bigl|\{\tau \in [t, t'-1] : c(\tau) \in \{0,1\}\}\bigr|.
            \]
            Then $N(t, t') \geq \frac{2(t' - t)}{3} - 1$. Furthermore, fix $(i,j) \in \mathcal{M}_2$. If $m_{ij}(\vc{x}, \tau) = 1$ for all $\tau \in [t, t']$, then for every $\tau \in [t, t']$ and every $(a, b) \in \Z_{\geq 0}^2$ with $a + b \leq N(t, \tau)$,
            \[
                m_{ij}\bigl(\vc{x} + (-1)^i a\, \uvec{x} + (-1)^j b\, \uvec{y},\, \tau\bigr) = 1.
            \]
        \end{lemma}
        \begin{proof}
            \textbf{Growth rate bound.} Write $\Delta = t' - t$. Since $c(\tau)$ cycles with period $3$ and takes the value $2$ (the unique non-growth value) exactly once per period, the number of non-growth times in any interval of length $\Delta$ is at most $\lceil \Delta/3 \rceil$. Therefore,
            \[
                N(t, t') = \Delta - \bigl|\{\tau \in [t, t'-1] : c(\tau) = 2\}\bigr| \geq \Delta - \left\lceil \frac{\Delta}{3} \right\rceil \geq \frac{2\Delta}{3} - 1.
            \]
            \textbf{Cone growth.} We prove the result for $m_{00}$; the cases $m_{01}$ and $m_{10}$ follow by the same argument after reflecting coordinates to match their growth neighborhoods. For $m_{00}$, the claim is that $m_{00}(\vc{x} + (a, b), \tau) = 1$ for all $(a, b) \in \Z_{\geq 0}^2$ with $a + b \leq N(t, \tau)$.
            We proceed by induction on $\tau \in [t, t']$. For $\tau = t$, we have $N(t, t) = 0$, so the only claim is $m_{00}(\vc{x}, t) = 1$, which holds by assumption. Suppose the result holds for $\tau - 1$; we show it for $\tau$. Fix $(a, b) \in \Z_{\geq 0}^2$ with $a + b \leq N(t, \tau)$.
            If $a + b = 0$, then $m_{00}(\vc{x}, \tau) = 1$ by assumption. If $a + b \geq 1$ and $N(t, \tau) = N(t, \tau - 1)$, then $a + b \leq N(t, \tau - 1)$, so $m_{00}(\vc{x} + (a,b), \tau - 1) = 1$ by the inductive hypothesis. Since at least one of $m_{00}(\vc{x} + (a-1, b), \tau - 1)$ or $m_{00}(\vc{x} + (a, b-1), \tau - 1)$ also equals $1$ by the inductive hypothesis, the Toom vote guarantees $m_{00}(\vc{x} + (a,b), \tau) = 1$ by condition (iv).
            If $a + b \geq 1$ and $N(t, \tau) = N(t, \tau - 1) + 1$, then either $a + b \leq N(t, \tau - 1)$, in which case the same argument applies, or $a + b = N(t, \tau)$. In the latter case, at least one of $m_{00}(\vc{x} + (a-1, b), \tau - 1)$ or $m_{00}(\vc{x} + (a, b-1), \tau - 1)$ equals $1$ by the inductive hypothesis, and $c(\tau - 1) \in \{0,1\}$, so condition (iii) guarantees $m_{00}(\vc{x} + (a,b), \tau) = 1$.
        \end{proof}

        We now show that the message support around each defect grows linearly in time. Define the northeast half-ball by
        \[
            B_\rho^{\nearrow}(\vc{x}) \coloneqq B_\rho(\vc{x}) \cap \bigl\{\vc{x}' \in \Z^2 : (x_2' - x_2) \geq -(x_1' - x_1)\bigr\},
        \]
        and decompose it into the sectors:
        \begin{align*}
            B_{\rho,10}^{\nearrow}(\vc{x}) &\coloneqq B_\rho^{\nearrow}(\vc{x}) \cap \{\vc{x}' : x_1' \leq x_1\}, \\
            B_{\rho,01}^{\nearrow}(\vc{x}) &\coloneqq B_\rho^{\nearrow}(\vc{x}) \cap \{\vc{x}' : x_2' \leq x_2\},
        \end{align*}
        where the subscripts indicate the message flavor that will cover each sector: $m_{10}$ (which grows in the $(-x,+y)$ direction) covers the upper-left sector, and $m_{01}$ (which grows in the $(+x,-y)$ direction) covers the lower-right sector. In the remaining upper-right region, we guarantee only that at least one of the three messages is non-trivial at each site.

        The following lemma shows that a filled half-ball is preserved whenever the defect takes a step. Its hypotheses are stated in terms of the message configuration alone, but the proof uses two facts about how that configuration arose. First, it uses the fact that the update from $t-1$ to $t$ is a growth-opportunity step, because the defect moved. Second, it uses the fact that every site of the shifted half-ball that already carries a message retains it, because its oriented Toom neighborhood lies inside the filled half-ball. We only ever apply the lemma to the half-ball of messages sourced by the defect itself along its own worldline, as in Lemma~\ref{lem:halfball-filling}, where both of these conditions hold.

        \begin{lemma}[half-ball shift]\label{lem:halfball-shift}
            Suppose that at time $t-1$, the half-ball $B_\rho^{\nearrow}(\vc{X}_{t-1})$ is filled in the sense that
            \begin{enumerate}
                \item $B_\rho^{\nearrow}(\vc{X}_{t-1}) \subseteq \supp(\vc{m}(t-1))$,
                \item $B_{\rho,10}^{\nearrow}(\vc{X}_{t-1}) \subseteq \{\vc{x} : m_{10}(\vc{x}, t-1) = 1\}$,
                \item and $B_{\rho,01}^{\nearrow}(\vc{X}_{t-1}) \subseteq \{\vc{x} : m_{01}(\vc{x}, t-1) = 1\}$.
            \end{enumerate}
            If $\vc{X}_t \in \{\vc{X}_{t-1} - \uvec{x},\, \vc{X}_{t-1} - \uvec{y}\}$, then the same three inclusions hold at time $t$ with $\vc{X}_t$ in place of $\vc{X}_{t-1}$.
        \end{lemma}

        \begin{proof}
            We treat the case $\vc{X}_t = \vc{X}_{t-1} - \uvec{x}$; the case $\vc{X}_t = \vc{X}_{t-1} - \uvec{y}$ follows by symmetry.
            
            First, observe that since $\vc X_t \neq \vc X_{t-1}$, it follows that the update from $t-1$ to $t$ is a growth-opportunity step (i.e., $c(t-1) \in \{0,1\}$). Next, observe that every point $\vc{z} \in B_\rho^{\nearrow}(\vc{X}_t) \setminus B_\rho^{\nearrow}(\vc{X}_{t-1})$ has a neighbor in the growth direction of whichever message flavor covers its sector, and that neighbor lies in the corresponding filled region of $B_\rho^{\nearrow}(\vc{X}_{t-1})$. Specifically:

            \textbf{$m_{10}$ sector.} Let $\vc{z} \in B_{\rho,10}^{\nearrow}(\vc{X}_t) \setminus B_{\rho,10}^{\nearrow}(\vc{X}_{t-1})$. Then $\vc{w} \coloneqq \vc{z} + \uvec{x}$ satisfies $\vc{w} \in B_{\rho,10}^{\nearrow}(\vc{X}_{t-1})$: the distance to $\vc{X}_{t-1}$ drops by one, the northeast half-space inequality is preserved, and $w_1 = z_1 + 1 \leq (X_t)_1 + 1 = (X_{t-1})_1$. By hypothesis, $m_{10}(\vc{w}, t-1) = 1$. Since $\vc{w} = \vc{z} + \uvec{x} = \vc{z} - (-1)^1 \uvec{x}$ is a growth-direction neighbor of $\vc{z}$ for $m_{10}$, and the update is a growth-opportunity step, the growth rule guarantees $m_{10}(\vc{z}, t) = 1$.

            \textbf{$m_{01}$ sector.} Let $\vc{z} \in B_{\rho,01}^{\nearrow}(\vc{X}_t) \setminus B_{\rho,01}^{\nearrow}(\vc{X}_{t-1})$. Since $\vc{X}_t$ and $\vc{X}_{t-1}$ share the same $x_2$-coordinate, such $\vc{z}$ must lie on the new diagonal boundary. If $z_1 \geq (X_{t-1})_1$, then $\vc{w} \coloneqq \vc{z} + \uvec{y}$ lies in $B_{\rho,01}^{\nearrow}(\vc{X}_{t-1})$, and the $m_{01}$ growth rule (which grows in the $-y$ direction) gives $m_{01}(\vc{z}, t) = 1$. If $z_1 < (X_{t-1})_1$, then the diagonal constraint forces $\vc{z} = \vc{X}_t$, and $m_{01}(\vc{X}_t, t) = 1$ by the incoming defect rule (condition (ii)).

            \textbf{Upper-right sector.} Points in $B_\rho^{\nearrow}(\vc{X}_t)$ with $z_1 > (X_t)_1$ and $z_2 > (X_t)_2$ already lie in $B_\rho^{\nearrow}(\vc{X}_{t-1})$: the distance to $\vc{X}_{t-1}$ is one less than the distance to $\vc{X}_t$, and the northeast diagonal constraint for $\vc{X}_{t-1}$ is satisfied by inspection. These points are therefore covered by hypothesis and message persistence. Combining with the $m_{10}$ and $m_{01}$ sectors, it follows that $B_\rho^{\nearrow}(\vc{X}_t) \subseteq \supp(\vc{m}(t))$.
        \end{proof}

        \begin{lemma}[half-ball filling]\label{lem:halfball-filling}
            Let $\{\vc{X}(\tau)\}_{\tau=0}^{T}$ be a defect worldline. Then, for every $t \in [1, T]$ and every $\tau \in [t, T]$,
            \begin{enumerate}
                \item $B_{\rho_\tau}^{\nearrow}(\vc{X}(t)) \subseteq \supp(\vc{m}(\tau))$,
                \item $B_{\rho_\tau,10}^{\nearrow}(\vc{X}(t)) \subseteq \supp(m_{10}(\tau))$,
                \item $B_{\rho_\tau,01}^{\nearrow}(\vc{X}(t)) \subseteq \supp(m_{01}(\tau))$,
            \end{enumerate}
            where $\rho_\tau = \lfloor (\tau + 1)/3 \rfloor$.
        \end{lemma}

        \begin{proof}
            The proof proceeds by outer induction on $t \in [1, T]$ and inner induction on $\tau \in [t, T]$.

            \textbf{Outer base case ($t = 1$).} By message persistence (Lemma~\ref{lem:toric-code:message-persistence}), all three messages satisfy $m_{ij}(\vc{X}(1), \tau) = 1$ for all $\tau \in [1, T]$. Applying message growth (Lemma~\ref{lem:toric-code:message-growth}) with source site $\vc{x} = \vc{X}(1)$ starting at time $t = 1$, each message cone has extent at least $N(1, \tau)$ by time $\tau$. Write $\Delta = \tau - 1$. For $\Delta = 0$, we have $\rho_\tau = 0$ and the claim is trivial. For $\Delta = 1$, we have $\rho_\tau = 1$ and $N(1, \tau) \geq 1$ by inspection. For $\Delta \geq 2$, we use $N(1, \tau) \geq \Delta - \lceil \Delta/3 \rceil$, and one verifies that $\Delta - \lceil \Delta/3 \rceil \geq \lfloor (\Delta + 2)/3 \rfloor = \rho_\tau$ for all $\Delta \geq 2$. Therefore, the three cones cover $B_{\rho_\tau}^{\nearrow}(\vc{X}(1))$: the $m_{00}$ cone covers the upper-right sector, the $m_{10}$ cone covers the upper-left sector, and the $m_{01}$ cone covers the lower-right sector. This establishes all three inclusions for $t = 1$ and all $\tau \in [1, T]$.

            \textbf{Outer inductive step.} Assume the result holds for all times at most $t - 1$; we prove it for $t$. If $\vc{X}(t) = \vc{X}(t-1)$, the result follows immediately from the hypothesis at $t - 1$, so assume $\vc{X}(t) \neq \vc{X}(t-1)$. Since movement only occurs when $c(t-1) = 0$, we have $t - 1 \equiv 0 \pmod{3}$, so the update from $t-1$ to $t$ is a growth-opportunity step. Moreover, $\rho_t = \rho_{t-1}$.

            \emph{Inner base case ($\tau = t$).} The outer inductive hypothesis gives the three inclusions at time $t - 1$ for the ball centered at $\vc{X}(t-1)$ with radius $\rho_{t-1} = \rho_t$. Since the update is a growth-opportunity step, Lemma~\ref{lem:halfball-shift} shifts the filled half-ball from $\vc{X}(t-1)$ to $\vc{X}(t)$, establishing the three inclusions at time $t$ with radius $\rho_t$.

            \emph{Inner inductive step ($\tau - 1 \to \tau$).} Suppose the three inclusions hold at time $\tau - 1$ with radius $\rho_{\tau-1}$. If $\rho_\tau = \rho_{\tau-1}$, the inclusions persist by the Toom majority rule (condition (iv)) and the coupling condition (v), together with message persistence. If $\rho_\tau = \rho_{\tau-1} + 1$, then the update from $\tau - 1$ to $\tau$ is a growth-opportunity step. One checks (straightforwardly but tediously) that every point in the new outer layer $B_{\rho_\tau}^{\nearrow}(\vc{X}(t)) \setminus B_{\rho_{\tau-1}}^{\nearrow}(\vc{X}(t))$ either already lies in $B_{\rho_\tau}^{\nearrow}(\vc{X}(t-1))$ (and is thus covered by the outer inductive hypothesis), or has a neighbor in the growth direction of its sector's message flavor that lies in $B_{\rho_{\tau-1}}^{\nearrow}(\vc{X}(t))$, so the growth rule ensures it is filled. This establishes the three inclusions at time~$\tau$.
        \end{proof}

        \begin{theorem}[linear defect erosion]\label{thm:toric-code:linear-defect-erosion}
            There exists a constant $a \geq 1$ such that the following holds. Suppose the initial error configuration is supported in $B_r(\vc{x})$ for some $r \geq 1$ and $\vc{x} \in \Z^2$. Then:
            \begin{enumerate}
                \item All defects are eliminated by time $t \leq ar$.
                \item For all $t \geq 0$, the correction and syndrome configurations are supported in $B_{ar}(\vc{x})$.
            \end{enumerate}
        \end{theorem}

        \begin{proof}
            By Proposition~\ref{prop:toric-code:defect-monotone}, no new defects are created, so it suffices to track the finitely many defects present at $t = 0$ and their worldlines. Define the potential $u(\vc{x}) \coloneqq x_1 + x_2$. Since no defect can move past the leftmost or bottommost initial defect, every surviving worldline $\vc{X}_t$ satisfies $u(\vc{X}_t) \geq x_{\min} + y_{\min}$ for all $t$, where $x_{\min}$ and $y_{\min}$ are the minimum $x$- and $y$-coordinates over the initial defect set. Moreover, by the directed movement property, every defect move decreases $u$ by exactly $1$, so $u(\vc{X}_t) \leq u(\vc{X}_{t-1})$ for all $t \geq 1$ and every worldline $\vc{X}_t$. Since the initial error configuration is contained in a ball of radius $r$, every initial defect satisfies $u(\vc{x}) \leq x_{\min} + y_{\min} + C_0 r$ for some constant $C_0 > 0$. 
            
            By Lemma~\ref{lem:halfball-filling}, each surviving defect has a filled northeast half-ball whose radius grows linearly in time. Therefore, there exists a constant $C_1 > 0$ such that by time $C_1 r$, every surviving defect's half-ball has radius $\Omega(r)$ and for all $t \geq C_1 r$, every surviving defect is either contained in some other defect's half-ball or is the sole $u$-minimizer, where, by $u$-minimizer, we mean a defect that achieves the minimum value of $u$ among all surviving defects. In general, there can be multiple $u$-minimizers at any given time.

            We now show that for all $t \geq C_1 r$, any defect contained in another's half-ball is guaranteed to see messages within $O(1)$ time steps that force it to move. Suppose first that the defect is not a $u$-minimizer, so it lies in one of the three filled sectors of the other defect's half-ball.

            In the upper-left sector, the defect sees an $m_{10}$ message below it and moves down at the next movement step. By symmetry, in the lower-right sector, the defect sees an $m_{01}$ message to its left and moves left.

            The upper-right sector is the only non-trivial case. The defect is guaranteed to see a message both to its left and below, so the only configuration that could block a move is an $m_{10}$ to its left together with an $m_{01}$ below it. But this can only happen if there is also an $m_{00}$ message at the defect's own site, sourced by the other defect whose half-ball contains it. Since the defect lies strictly inside the upper-right sector, that source defect must have taken at least one more step down or left after emitting the $m_{00}$, which guarantees that there exists an $m_{00}$ message either to the left of the defect or below it, so the defect moves at the next movement step.

            This handles the case where the defect is not a $u$-minimizer. Suppose instead the defect is a $u$-minimizer contained in another's half-ball. Without loss of generality, assume it lies in the upper-left sector, so an $m_{10}$ message sits at its lower-right neighbor. Within $O(1)$ steps, this message grows into the site directly below the defect, forcing it to move.

            Therefore, for all $t \geq C_1 r$, any defect contained in another's half-ball is guaranteed to see messages within $O(1)$ time steps that force it to move.

            Consequently, for all $t \geq C_1 r$, every non-$u$-minimizer defect moves at least once every $O(1)$ time steps, with each move decreasing $u$ by $1$. Since $u$ can decrease at most $O(r)$ times, any defect can be a non-$u$-minimizer for at most $O(r)$ time steps. Similarly, every $u$-minimizer forces all other surviving defects to move at least once every $O(1)$ time steps, so a defect can be a $u$-minimizer for at most $O(r)$ time steps. After $O(r)$ additional time steps, every surviving defect therefore has to be a lone $u$-minimizer, but this is impossible by charge neutrality, which requires an even number of surviving defects. Therefore, it follows that all defects are eliminated by time $O(r)$, establishing~(1). Furthermore, by combining the above argument with the directed movement property, the lightcone bound, and the defect-gated flip property, it follows that there exists a constant $C_2 > 0$ such that all correction flips and syndromes remain within $B_{C_2 r}(\vc{x})$ for all $t \geq 0$, establishing~(2). Choosing $a$ sufficiently large so that $a \geq C_1 + O(1)$ and $a \geq C_2$ gives the desired result.
        \end{proof}

        \begin{lemma}[linear message erosion]\label{lem:toric-code:linear-message-erosion}
            There exists a constant $b \geq 1$ such that the following holds. Suppose the syndrome configuration is trivial for all times $t \geq t_0$ and $\supp(\vc{m}(t_0)) \subseteq B_r(\vc{x})$ for some $r \geq 1$ and $\vc x \in \Z^2$. Then:
            \begin{enumerate}
                \item $\supp(\vc{m}(t)) = \varnothing$ for all $t \geq t_0 + br$.
                \item $\supp(\vc{m}(t)) \subseteq B_{br}(\vc{x})$ for all $t \geq t_0$.
            \end{enumerate}
        \end{lemma}

        \begin{proof}
            We first prove erasure of $m_{00}$, then $m_{01}$ and $m_{10}$.
        
            \textbf{Erasure of $m_{00}$.} Since the syndrome configuration is trivial by assumption, conditions (i) and (ii) never apply, so $m_{00}$ evolves autonomously via clock-gated growth (condition (iii), active when $c(t) \in \{0,1\}$) and Toom majority erasure (condition (iv), active at every step) with oriented neighborhood $\{\vc{r},\, \vc{r} - \uvec{x},\, \vc{r} - \uvec{y}\}$.
        
            Since growth depends only on nearest neighbors, the support can expand by at most distance $1$ per growth step, which gives the loose bound
            \[
                \supp(m_{00}(t)) \subseteq B_{r + (t - t_0)}(\vc{x}) \qquad (t \geq t_0).
            \]
            Moreover, in any block of $3$ consecutive time steps, growth can advance the boundary by at most $2$ layers, while Toom erasure acts on all $3$ steps. Consequently, there exists some $c_1 \geq 1$ such that $m_{00}$ vanishes by some time $t_1 \leq t_0 + c_1 r$, with the support of $m_{00}$ confined to $B_{c_1 r}(\vc{x})$ throughout.
        
            \textbf{Erasure of $m_{01}$ and $m_{10}$.} For $t \geq t_1$, we have that $m_{00}$ is trivial, so the coupling effect in condition (v) becomes inactive. Each of $m_{01}$ and $m_{10}$ then evolves by the same clock-gated Toom mechanism as $m_{00}$ (up to a coordinate reflection), and the same argument gives erasure by time $t_1 + c_2 r$ with support confined to $B_{c_2 r}(\vc{x})$ throughout, for some constant $c_2 \geq 1$. Setting $b = c_1 + c_2$ gives the desired result.
        \end{proof}
    
    \subsection{Open boundary conditions}\label{subsec:toric-code:open-boundary-conditions}

        The open-boundary modification is the same as in Section~\ref{subsec:rep-code:open-boundary-conditions}. In particular, suppose that the defects under consideration condense at the $-x$ and $+x$ boundaries of the surface-code patch. We split the lattice by a vertical cut through the middle (using the same degeneracy-breaking convention for even-length systems as in the repetition-code case) and every $q_s$ steps, shift all defects and messages in the left half one site in the $-x$ direction and all defects and messages in the right half one site in the $+x$ direction. At the central cut, defects are shifted outward but messages are copied rather than translated; at the condensing boundaries, defects and messages translated out of the system are annihilated. Sites outside the system are otherwise treated as carrying trivial messages. We implement the splitting step at times $t \in q_s \Z$ in the same manner as in the repetition code. 

        The defect and message erosion proofs are essentially the same as in the open-boundary-condition repetition-code case; however, the bookkeeping in the toric code case is somewhat more involved. In the proof outlines that follow, defects are only allowed to move when $t = 0 \mod q$. Furthermore, messages are allowed to grow at all time steps except when $t = q-1 \mod q$.
        
        First, we can divide all clusters of diameter at most $cL$, for some sufficiently small constant $c > 0$, into three mutually exclusive categories: those that are purely translated, those that interact with a condensing boundary, and those whose dynamics are affected by interacting with the center line. The clusters that are purely translated or interact with a condensing boundary can be shown to satisfy linear cluster erosion by essentially the same arguments as in the repetition-code case.

        The only category we will discuss in more detail is the clusters that interact with the center, as these are slightly more complicated to formally reason about. Observe that the message erosion mechanism catches up with the growth front provided $(q-1)/q + 2/q_s < 1$, so when $q_s$ satisfies this condition, linear message erosion holds.
        
        We now outline how to rigorously establish linear defect erosion. Formally, to make the following statements well-defined, we would have to specify a defect worldline assignment convention that accounts for splitting steps, but we will not go through these (straightforward but tedious) steps; here, one can just imagine choosing some reasonable manner of doing this and then applying the following arguments to defect worldlines under this convention. First, one can check that because messages are copied, message cones produced by a defect still cannot be erased until at least the defect that sourced the message has been annihilated. Consequently, all that remains is to analyze the message growth and defect movement dynamics.
        
        Next, note that if the half-ball filling property holds at time $t$ for the messages $m_{ij}(x, t)$ surrounding an active defect in the configuration $s(x, t)$, then after the splitting step, the defect is either still alive or annihilated. If it is annihilated, then the half-ball filling property vacuously holds since the defect's worldline has been terminated. If it is alive, then the half-ball filling property still holds for the defect's new position in the post-split configuration $s'(x, t)$ and $m'_{ij}(x, t)$, except potentially along the new diagonal for at most $O(1)$ time steps because the splitting step can temporarily distort the northeast half-ball before message growth restores it. In particular, it is clear that for $q$ sufficiently large, the half-ball filling radius is guaranteed to grow at some constant non-zero average speed $c' > 0$. Furthermore, defects can move apart from each other at average speed at most $2/q_s$, so for $q$ and $q_s$ sufficiently large, the half-ball filling grows fast enough that a version of the argument in Theorem~\ref{thm:toric-code:linear-defect-erosion} goes through. Specifically, such a modified argument shows that after $O(W)$ time, all defects contained in another's half-ball will be guaranteed to see messages within $O(1)$ time steps that force them to move. This will induce some average speed $\bar v$ at which the maximal distance between defects will be reduced, and as long as $2/q_s < \bar v$, this average speed will overcome any distance-increasing effect from the splitting step.

    \subsection{Numerics}\label{subsec:numerics:toric-code}

        \begin{figure}[!t]
            \centering
            \includegraphics[width=\linewidth]{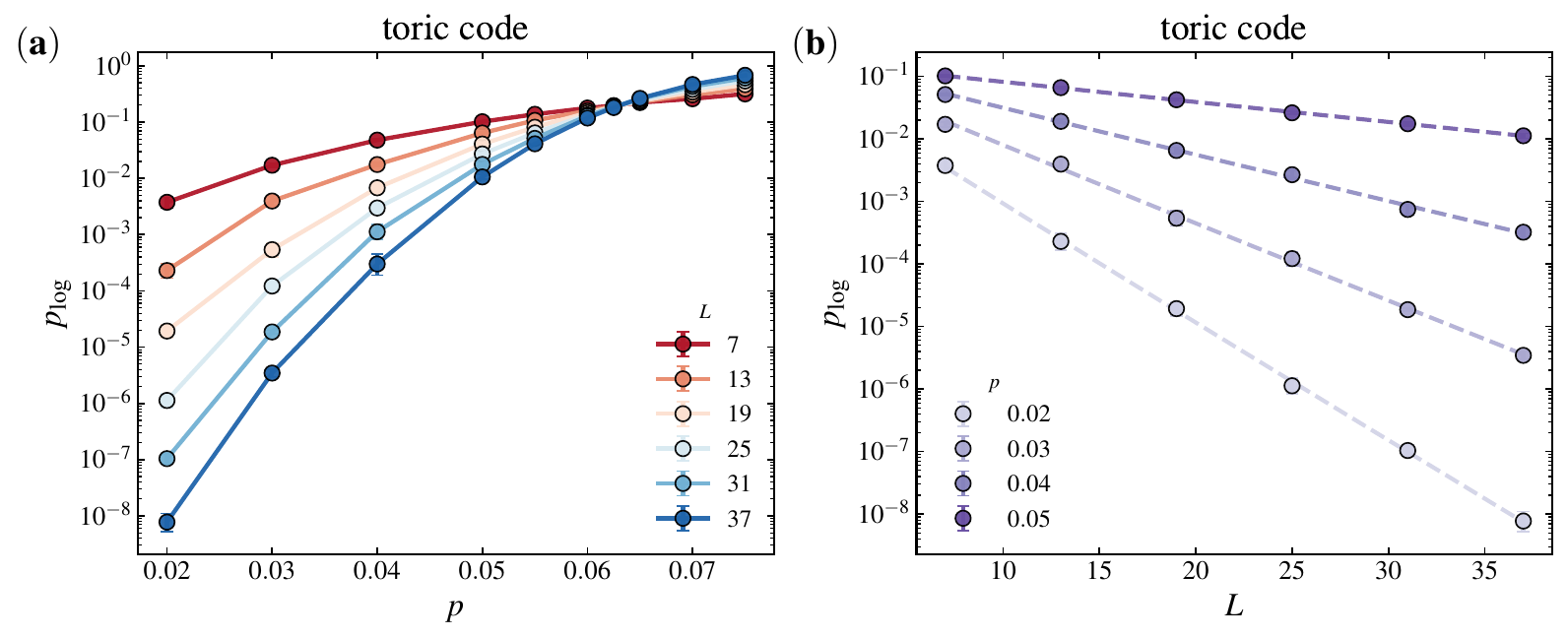}
                \caption{Performance of the code-capacity toric-code decoder, using a clock period of $q=6$ and allowing defect movement at every timestep. \textbf{(a)} Logical error rate $\plog$ as a function of the physical error rate $p$ under i.i.d.\ bit-flip noise. The data are consistent with a threshold of $p_c \approx 6.3\%$. \textbf{(b)} Sub-threshold scaling of the logical error rate $\plog$ with code distance $L$. Dashed lines are fits to the form $\plog = C(p/p_c)^{\gamma L^{\alpha}}$ with a single exponent $\alpha$ shared across all $p$. The fitted $\alpha = 0.98 \pm 0.06$ is consistent with exponential decay of $\plog$ in $L$, and the fitted $\gamma$ increases from $0.35$ to $0.42$ as $p$ decreases from $0.05$ to $0.02$.}
            \label{fig:tc_threshold}
        \end{figure}
         
        \begin{figure}[t]
            \centering
            \includegraphics[width=\linewidth]{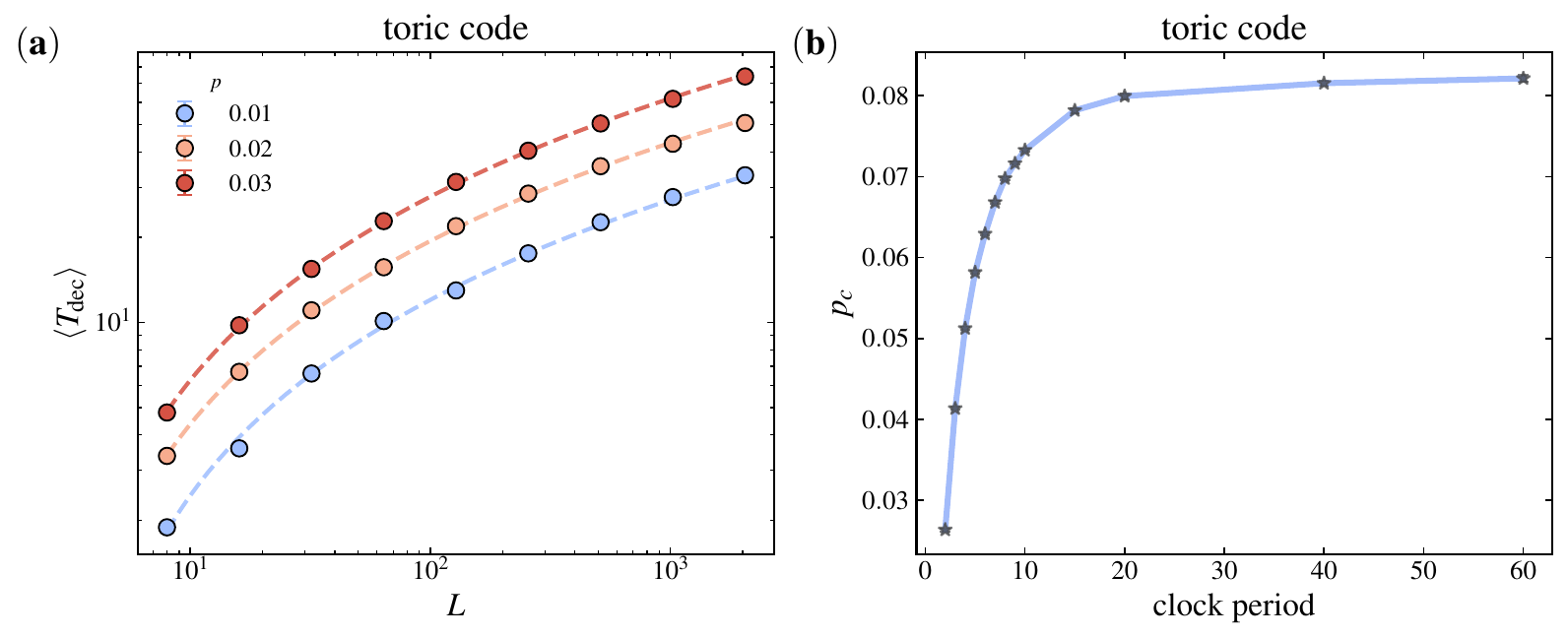}
            \caption{Average decoding time and threshold dependence on the clock period for the toric-code decoder. \textbf{(a)}~Scaling of the average decoding time (number of CA steps) as a function of system size $L$ when using a clock period of $q=6$. Dashed lines are fits to the form $\langle \Tdec \rangle = A + B \log^\eta{L}$, where $A$ and $B$ are $p$-dependent constants. We find $\eta \approx 1.9$. \textbf{(b)}~Threshold as a function of the clock period. The threshold increases monotonically with $q$ and saturates, reaching $8.2\%$ at $q = 60$.}
            \label{fig:tc_tdec_and_clock}
        \end{figure}
        
        \begin{figure}[!t]
            \centering
            \includegraphics[width=\linewidth]{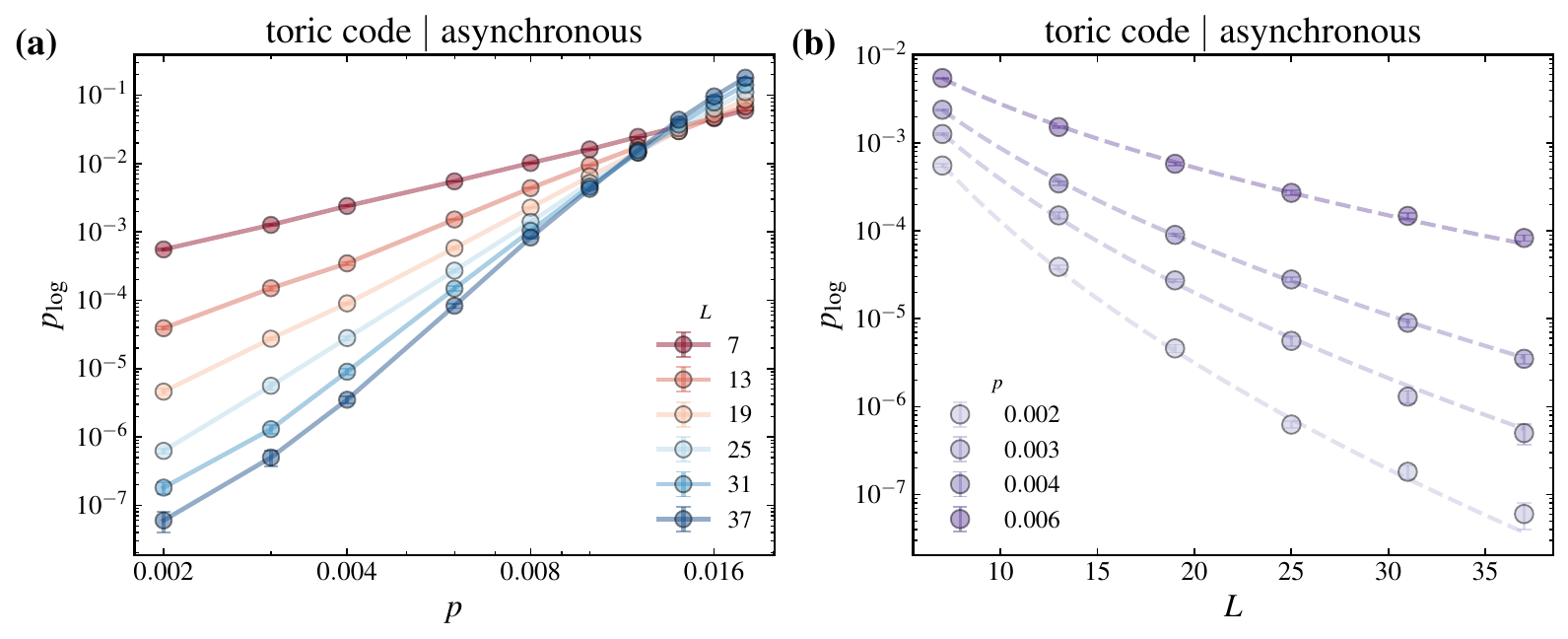}
            \caption{Performance of the code-capacity toric-code decoder under completely asynchronous, uncoordinated updates using a clock period of $q = 6$. \textbf{(a)} Logical error rate $\plog$ as a function of the physical error rate $p$ under i.i.d.\ bit-flip noise. The data are consistent with a threshold of $p_c \approx 1.1\%$. \textbf{(b)} Sub-threshold scaling of the logical error rate $\plog$ with $L$. Dashed lines are fits to the form $\plog = C(p/p_c)^{\gamma L^{\alpha}}$ with a single exponent $\alpha$ shared across all $p$. The fitted $\alpha = 0.46 \pm 0.02$ indicates stretched exponential decay of $\plog$ in $L$, with $\gamma \approx 2$ at the simulated values of $p$.}
            \label{fig:tc_uncoord}
        \end{figure}

        We now study the performance of our decoder under i.i.d.\ bit-flip noise using Monte Carlo simulations. All results are reported for a clock period of $q = 6$ (except in panel (b) of Fig.~\ref{fig:tc_tdec_and_clock}, which sweeps $q$), with defect movement allowed at every time step and message growth allowed only when $c = 0$. Fig.~\ref{fig:tc_threshold} shows the logical error rate as a function of the physical error rate and code distance under synchronous updates; the data are consistent with a threshold of $p_c \approx 6.3\%$. In Fig.~\ref{fig:tc_tdec_and_clock}, also under synchronous updates, we show that the average decoding time scales as $\langle \Tdec \rangle = O(\log^\eta L)$ for $\eta \approx 1.9$; in addition, we investigate how the threshold scales with the clock period $q$ and find that the data are consistent with an asymptotic threshold of $p_c \gtrsim 8.2\%$ as $q \to \infty$. For comparison, the code-capacity decoder of Ref.~\cite{lake2025fastofflinedecodinglocal} achieves a threshold of $7.3\%$ using $O(\log L)$ classical resources per site, and the decoder of Ref.~\cite{winter2026highperformancecellularautomatondecoders} achieves $7.5\%$ using a constant amount of classical resources per site. The threshold of our decoder surpasses the thresholds of both of these other decoders for sufficiently large clock periods, while using only a constant amount of classical memory per site.

        We also study the performance of our decoder under completely asynchronous, uncoordinated updates, in which a single randomly chosen site updates at each time step. The results, shown in Fig.~\ref{fig:tc_uncoord}, are consistent with a threshold of $p_c \approx 1.1\%$. Since each site stores and updates its clock register independently, we conclude that our decoder is robust to desynchronization of the message growth, message erasure, and defect movement phases between different sites.
        
\section{Translation-invariant streaming decoders} \label{sec:online-decoding}

    In this section, we construct translation-invariant streaming decoders for the repetition, toric, and surface codes under phenomenological noise. By phenomenological noise, we mean $p$-bounded Pauli noise on the data qubits and $p$-bounded bit-flip noise on the stabilizer-measurement outcomes.

    We build each streaming decoder by stacking $K$ coupled copies of the corresponding code-capacity decoder, called \emph{slices}, along an auxiliary dimension. These slices are indexed by $k = 0, \ldots, K-1$. At each time step, newly observed detector events are inserted as defects in the bottom slice $k = 0$, and defects that survive long enough are promoted upward in the stack.

    Promotion is controlled by timers. Each defect in slice $k < K-1$ carries a timer that follows it along its worldline; when the timer expires, the defect is promoted to the same spatial location in slice $k+1$ and its timer is reset (defects in the final slice carry no timer). The timer lengths grow exponentially with $k$:
    \[
        t_k = t_0 n^k,
    \]
    where $n$ is the parameter from the sparsity theorem and $t_0$ is a constant positive integer. The timers are chosen so that every $k$-cluster is erased before its defects can be promoted beyond slice $k$. Consequently, $K = \Theta(\log L)$ slices suffice to decode every cluster too small to support a logical operator. Because clusters large enough to support a logical operator occur with inverse stretched-exponential probability, this yields a decoder that achieves a stretched-exponential memory lifetime while using only $\poly(\log L)$ classical bits per site.\footnote{Here, we are using the fact that a timer of range $t_k$ can be represented using $O(\log t_k)$ bits and that the timer-conditioned operations we perform can be implemented using a number of bits that is polynomial in the number of timer bits.}

    The same construction can be combined with a renormalization-style argument from Ref.~\cite{lake2025localactiveerrorcorrection} to obtain a decoder with even lower resource overhead and provably the same memory-lifetime scaling. Let us refer to slice $K-1$ as the \emph{back wall}. Since lower slices erase smaller clusters before their defects can reach it, the back wall effectively experiences a sparser, renormalized noise model. In Ref.~\cite{lake2025localactiveerrorcorrection}, it was shown that to maintain a stretched-exponential memory lifetime, it suffices to ``renormalize'' the noise model by decoding all clusters of diameter at most $\polylog(L)$ before their defects can reach the back wall. Such clusters have level $O(\log\log L)$, so in the present construction, they can be decoded using only
    \[
        K = \Theta(\log\log L)
    \]
    slices. Therefore, the memory-lifetime scaling can be preserved using only $\poly(\log\log L)$ classical bits per site.

    Let us briefly compare our construction with that of Ref.~\cite{lake2025localactiveerrorcorrection}, which requires $\poly(\log L)$ classical bits per site to achieve the same memory-lifetime scaling. There, defects also ascend an analog of the slice stack, and the dynamics satisfy an analog of linear cluster erosion. However, in that work, defects ascend at unit speed. Consequently, since a cluster of diameter $W$ can survive for a time of order $W$, its defects can climb $\Theta(W)$ slices before being erased. This implies that guarding the back wall against all clusters of diameter at most $\polylog(L)$ requires $\polylog(L)$ slices. In our construction, by contrast, the timers slow the ascent of defects exponentially: a defect must survive for time $t_k = t_0 n^k$ to leave slice $k$. Therefore, since our construction satisfies linear cluster erosion, a cluster of diameter $W$ can climb only $O(\log W)$ slices before being erased, so only $\Theta(\log\log L)$ slices are needed to guard the back wall against all clusters of diameter at most $\polylog(L)$.

    In Section~\ref{sec:constant-density}, we will reduce this $\poly(\log\log L)$ overhead to a constant for the toric and surface codes, at the cost of breaking translation invariance. We begin by constructing the translation-invariant, non-constant-overhead decoders of the current section because, after doing so, this reduction will be simple to understand.\footnote{Technically, the $\Theta(\log\log L)$ construction is not needed to understand this reduction; the $\Theta(\log L)$ one alone suffices. We nevertheless present the $\Theta(\log\log L)$ construction because it is of independent interest: it improves upon the overhead of all prior phenomenological-noise decoders for two-dimensional translation-invariant stabilizer codes, which require $\poly(\log L)$ classical bits per site, while simultaneously preserving translation invariance.}

    The $\poly(\log L)$-overhead constructions for the repetition and toric codes extend directly to general translation-invariant stabilizer codes. Indeed, the constructions in this section can be viewed as instantiations of a general scheme for promoting local code-capacity decoders to local streaming decoders; this is discussed further in Section~\ref{sec:general}.

    This section is organized as follows. First, we define the decoder for the repetition code and give the analogous construction for the toric code. Then, we describe how to modify these constructions for systems with open boundary conditions and how to perform decoding during transversal readout. Next, we prove memory-lifetime bounds for both the $K=\Theta(\log L)$ and $K=\Theta(\log\log L)$ constructions. We also describe how to perform local decoding during stabilizer-state preparation (hereafter state preparation) and state injection. We conclude with numerical simulations of the $\Theta(\log L)$-slice surface-code decoder.

    \subsection{Repetition code}\label{subsec:online-preliminaries}

        We consider the repetition code on the lattice $\Lambda = \Z$, where we are again working at infinite system size for simplicity. Under phenomenological noise, the decoding problem naturally lives on the two-dimensional spacetime lattice
        \[
            \widetilde{\Lambda} \coloneqq \Lambda \times \Z_{\geq 0}
        \]
        which we equip with two types of edges. The horizontal edges are pairs $\bigl((i-1,t),(i,t)\bigr)$ for $i\in\Lambda$ and $t\in\Z_{\geq 0}$, and the vertical edges are pairs $\bigl((i,t),(i,t+1)\bigr)$. We associate each site with the edge above it and the edge to its left. 

        The spacetime lattice encodes faults as follows:
        \begin{enumerate}
            \item Each horizontal edge $((i - 1, t), (i, t))$ records whether a bit-flip error occurred at the unique data qubit associated with spatial site $i$ in between times $t-1$ and $t$.
            \item Each vertical edge $((i, t), (i, t+1))$ records whether the measurement of the stabilizer $Z_i Z_{i+1}$ was faulty at time $t$.
            \item Each spacetime site $(i, t)\in\widetilde{\Lambda}$ stores the measured syndrome difference
            \[
                \phi(i,t) \coloneqq \tilde s_t(i) \oplus\tilde s_{t-1}(i)\in\{0,1\},
            \]
            where $\tilde s_{t}(i)$ denotes the (potentially faulty) measurement outcome of the stabilizer $Z_i Z_{i+1}$ at time $t$. Here, we assume $\tilde s_{-1}(i) = 0$ for all $i$.\footnote{If the value of $\tilde{s}_{-1}(i)$ is unknown, then the situation is somewhat more complicated. This issue is important for state preparation, state injection, and lattice surgery and is discussed in more detail below.}
        \end{enumerate}

        We now describe the decoder in more detail. At each time step, newly observed syndrome differences are inserted as defects in the bottom slice. A defect in slice $k < K-1$ is promoted only after it has remained in that slice for
        \[
            t_k = t_0 n^k
        \]
        time steps; upon promotion, it is removed from slice $k$ and reinserted at the same spatial location in slice $k+1$ with its timer reset. Defects on the back wall cannot be promoted further. We also assign timers to messages and erase messages once their timers expire in order to prevent messages generated by larger clusters from interfering with the decoding of smaller clusters.

        Before its timer expires, a defect evolves according to the base CA rule within its current slice: depending on the message configuration in that slice, it either stays still or moves one step to the left. Each leftward move flips the corresponding correction bit in that slice. The correction value at a data qubit is the modulo-two sum of the correction bits at that qubit over all slices.

        At each site $x \in \Lambda$ and in each slice $k<K-1$, the defect-sector variables are
        \[
            \xi_k(x) = \bigl(e_k(x),\, s_k(x),\, \tau_k(x)\bigr),
        \]
        where $e_k(x) \in \{0,1\}$ is the correction bit, $s_k(x) \in \{0,1\}$ is the defect channel, and $\tau_k(x)$ is the defect timer. If $s_k(x)=0$, we set $\tau_k(x)=0$. For $k=K-1$, there is no defect timer, and the variables are simply
        \[
            \xi_{K-1}(x) = \bigl(e_{K-1}(x),\, s_{K-1}(x)\bigr).
        \]
        We first describe the defect-sector update; Algorithm~\ref{alg:layered-defect-update} then collects these rules in a single pseudocode summary. The first set of defect-sector variables $s_k(x, 0)$, $e_k(x, 0)$, and $\tau_k(x, 0)$ are computed by applying this update to the initial conditions $s_k(x, -1) = e_k(x, -1) = 0$ for all $x$ and $k$.

        To simplify the assignment of timers to defects, we define the update in terms of defect candidates. The only effect the environment has on the decoder is to create new defect candidates in the bottom slice. Let
        \[
            \phi(x,t) \in \{0,1\}
        \]
        denote the measured syndrome difference $\tilde s_t(x) \oplus \tilde s_{t-1}(x)$ at the spacetime point $(x, t)$. When $\phi(x,t+1)=1$, a new defect candidate is inserted at $(k, x)$, i.e., at location $x$ in slice $k = 0$, during the update from time $t$ to $t+1$. The notion of a defect candidate will be made more precise below.
        
        We now define the update from time $t$ to time $t+1$. For each existing defect $s_k(x,t)=1$ with $k<K-1$, define its incremented timer
        \[
            \tau_k^+(x,t) \coloneqq \tau_k(x,t)+1.
        \]
        If $\tau_k^+(x,t)=t_k$, then the defect is promoted vertically: it is removed from slice $k$ and contributes a defect candidate at $(k+1,x)$ at time $t+1$. The promoted defect has timer $0$ in the new slice, unless $k+1=K-1$, in which case no timer is needed. A defect whose timer expires is not allowed to move horizontally during the same step.
        
        If the timer does not expire, then the defect evolves according to the base decoder in slice $k$. Let
        \[
            \lambda_k(x,t) \in \{0,1\}
        \]
        be the indicator that the base repetition-code decoder moves the defect at $x$ one step to the left, as determined by the current message configuration in slice $k$. If $\lambda_k(x,t)=1$, then the correction bit is flipped,
        \[
            e_k(x,t+1) = e_k(x,t) \oplus 1,
        \]
        and the defect contributes a candidate at $(k,x-1)$ with timer $\tau_k^+(x,t)$. If $\lambda_k(x,t)=0$, then the defect contributes a stationary candidate at $(k,x)$ with timer $\tau_k^+(x,t)$. For defects on the back wall, there is no timer check. They simply evolve according to the base decoder in slice $K-1$.

        The timer assignment convention for defects is defined as follows. If an even number of defect candidates meet at a site, they pair-annihilate and no defect remains. If an odd number of candidates meet at a site, a defect survives and on timed slices, the timer is taken to be the minimum timer among the defect candidates.

        For the pseudocode in Algorithm~\ref{alg:layered-defect-update}, let $\mathcal{R}_k(x,t+1)$ denote the multiset of defect candidates arriving at $(k,x)$ during the update from time $t$ to time $t+1$. Each candidate is written as a pair $(a,u)$, where $a\in\{\mathrm{up},\mathrm{move},\mathrm{stay},\mathrm{env}\}$ is the action taken by the defect and $u$ is its proposed defect timer. The label $a$ is used only for bookkeeping and is ignored in the actual computation.
 
        \begin{algorithm}[H]
            \caption{Defect-sector update for the translation-invariant repetition-code streaming decoder at time $t$.}
            \label{alg:layered-defect-update}
            \begin{algorithmic}[1]
                \State $\mathcal{R}_k(x,t+1) \gets \emptyset$ for all $k,x$ \Comment{Candidates arriving at $(k,x)$}
                \Statex \Comment{Each record $(a,u)$ stores the action $a$ taken by the defect and its proposed timer $u$}
                \State $e_k(x,t+1) \gets e_k(x,t)$ for all $k,x$ \Comment{Copy corrections before applying new moves}
                \Statex
        
                \ForAll{$k=0,\ldots,K-1$} \Comment{Part 1: Update defects already present at time $t$}
                    \ForAll{$x \in \Lambda$ with $s_k(x,t)=1$}
                        \If{$k<K-1$}
                            \State $\tau^+ \gets \tau_k(x,t)+1$ \Comment{Increment the timer before choosing an action}
                            \If{$\tau^+=t_k$}
                                \State Add candidate $(\mathrm{up},0)$ to $\mathcal{R}_{k+1}(x,t+1)$
                                \Comment{Vertical promotion}
                            \ElsIf{$\lambda_k(x,t)=1$}
                                \State $e_k(x,t+1) \gets e_k(x,t+1) \oplus 1$ \Comment{Record the crossed horizontal edge}
                                \State Add candidate $(\mathrm{move},\tau^+)$ to $\mathcal{R}_k(x-1,t+1)$
                                \Comment{Horizontal move}
                            \Else
                                \State Add candidate $(\mathrm{stay},\tau^+)$ to $\mathcal{R}_k(x,t+1)$
                                \Comment{Stationary defect}
                            \EndIf
                        \Else
                            \If{$\lambda_{K-1}(x,t)=1$}
                                \State $e_{K-1}(x,t+1) \gets e_{K-1}(x,t+1) \oplus 1$ \Comment{Record the crossed horizontal edge}
                                \State Add candidate $(\mathrm{move},\varnothing)$ to $\mathcal{R}_{K-1}(x-1,t+1)$
                                \Comment{Horizontal move in final slice}
                            \Else
                                \State Add candidate $(\mathrm{stay},\varnothing)$ to $\mathcal{R}_{K-1}(x,t+1)$
                                \Comment{Stationary defect in final slice}
                            \EndIf
                        \EndIf
                    \EndFor
                \EndFor
        
                \Statex
                \ForAll{$x \in \Lambda$ with $\phi(x,t + 1)=1$} \Comment{Part 2: Insert newly observed detector events}
                    \State Add candidate $(\mathrm{env},0)$ to $\mathcal{R}_0(x,t+1)$
                    \Comment{New environmental defect}
                \EndFor
        
                \Statex
                \ForAll{$k=0,\ldots,K-1$} \Comment{Part 3: Resolve all candidate collisions}
                    \ForAll{$x \in \Lambda$}
                        \If{$|\mathcal{R}_k(x,t+1)|$ is even} \Comment{Even parity: candidates pair-annihilate}
                            \State $s_k(x,t+1) \gets 0$
                            \If{$k<K-1$}
                                \State $\tau_k(x,t+1) \gets 0$
                            \EndIf
                        \Else \Comment{Odd parity: one defect survives}
                            \State $s_k(x,t+1) \gets 1$
                            \If{$k<K-1$}
                                \State $\tau_k(x,t+1) \gets \min\{u : (a,u)\in\mathcal{R}_k(x,t+1)\}$ \Comment{Assign the smallest candidate timer}
                            \EndIf
                        \EndIf
                    \EndFor
                \EndFor
            \end{algorithmic}
        \end{algorithm}

        We next specify the message-sector update. In each timed slice $k<K-1$, the message-sector variables at site $x$ are
        \[
            \xi_k'(x)=\bigl(m_k(x),c_k(x),\theta_k(x)\bigr),
        \]
        where $m_k(x)\in\{0,1\}$ is the message channel, $c_k(x)\in\{0,\ldots,q-1\}$ is the clock channel, and $\theta_k(x)$ is the message timer. We set $\theta_k(x)=0$ when $m_k(x)=0$, while if $m_k(x)=1$ then $\theta_k(x)\in\{0,\ldots,t_k-1\}$. In the final slice, messages are untimed, so
        \[
            \xi_{K-1}'(x)=\bigl(m_{K-1}(x),c_{K-1}(x)\bigr).
        \]

        The message update is performed after the defect-sector update, so the variables $s_k(x,t+1)$, and also $\tau_k(x,t+1)$ for $k<K-1$, have already been determined. We define the update in terms of message candidates. If no message candidate is produced at a site, then no message is written there. If multiple message candidates are produced, then a message is written, and in a timed slice, its timer is the minimum timer among all candidates.

        For $k<K-1$, let $\mathcal{B}_k(x,t + 1)$ be the set of message-candidate timers at $(k,x)$ during the update from time $t$ to time $t+1$. The following candidates are generated.

        First, any defect present after the defect-sector update produces a message candidate:
        \[
            s_k(x,t+1)=1
            \quad\Longrightarrow\quad
            \tau_k(x,t+1)\in \mathcal{B}_k(x,t+1).
        \]

        Second, an old defect at $(k,x)$ produces a message candidate at its previous location unless it is promoted out of slice $k$:
        \[
            s_k(x,t)=1
            \quad\text{and}\quad
            \tau_k(x,t)+1<t_k
            \quad\Longrightarrow\quad
            \tau_k(x,t)+1\in \mathcal{B}_k(x,t+1).
        \]

        Third, the base repetition-code message rule produces candidates corresponding to persistence and growth. Define
        \[
            \pi_k(x,t)
            \coloneqq
            \bigl[
                m_k(x,t)=1
                \;\text{and}\;
                m_k(x-1,t)=1
                \;\text{and}\;
                \theta_k(x,t)<t_k-1
            \bigr],
        \]
        and
        \[
            \gamma_k(x,t)
            \coloneqq
            \bigl[
                m_k(x-1,t)=1
                \;\text{and}\;
                c_k(x,t)=0
                \;\text{and}\;
                \theta_k(x-1,t)<t_k-1
            \bigr].
        \]
        If $\pi_k(x,t)=1$, then persistence produces the candidate timer
        \[
            \theta_k(x,t)+1\in \mathcal{B}_k(x,t+1).
        \]
        If $\gamma_k(x,t)=1$, then growth from the left produces the candidate timer
        \[
            \theta_k(x-1,t)+1\in \mathcal{B}_k(x,t+1).
        \]

        After all message candidates have been generated, the message at $(k,x)$ is updated according to
        \begin{align*}
            \mathcal{B}_k(x,t+1)=\emptyset
            &\quad\Longrightarrow\quad
            m_k(x,t+1)=0,\qquad
            \theta_k(x,t+1)=0,\\
            \mathcal{B}_k(x,t+1)\neq\emptyset
            &\quad\Longrightarrow\quad
            m_k(x,t+1)=1,\qquad
            \theta_k(x,t+1)=\min \mathcal{B}_k(x,t+1).
        \end{align*}
        On the back wall, the same rule applies with all timer fields removed.
        
        Finally, in every slice,
        \[
            c_k(x,t+1)=(c_k(x,t)+1)\bmod q,
        \]
        where $c_k(x, 0) = 0$ for all $x$ and $k$.
        The message-sector update rules are summarized in pseudocode in Algorithm~\ref{alg:layered-message-update}.

        \begin{algorithm}[H]
            \caption{Message-sector update for the translation-invariant repetition-code streaming decoder at time $t$.}
            \label{alg:layered-message-update}
            \begin{algorithmic}[1]
                \ForAll{$k=0,\ldots,K-1$}
                    \ForAll{$x\in\Lambda$}

                        \If{$k<K-1$}
                            \State $\mathcal{B}_k(x,t+1)\gets
                            \{\tau_k(x,t+1): s_k(x,t+1)=1\}$
                            \Statex \hspace{1.95cm}
                            $\cup\;\{\tau_k(x,t)+1: s_k(x,t)=1 \;\text{and}\; \tau_k(x,t)+1<t_k\}$
                            \Statex \hspace{1.95cm}
                            $\cup\;\{\theta_k(x,t)+1: m_k(x,t)=1 \;\text{and}\; m_k(x-1,t)=1 \;\text{and}\; \theta_k(x,t)<t_k-1\}$
                            \Statex \hspace{1.95cm}
                            $\cup\;\{\theta_k(x-1,t)+1: m_k(x-1,t)=1 \;\text{and}\; c_k(x,t)=0 \;\text{and}\; \theta_k(x-1,t)<t_k-1\}$

                            \If{$\mathcal{B}_k(x,t+1)=\emptyset$}
                                \State $m_k(x,t+1)\gets 0$, \quad $\theta_k(x,t+1)\gets 0$
                            \Else
                                \State $m_k(x,t+1)\gets 1$, \quad $\theta_k(x,t+1)\gets \min\mathcal{B}_k(x,t+1)$
                            \EndIf

                        \Else
                            \State $m_k(x,t+1)\gets
                            [s_k(x,t+1)=1]
                            \;\vee\;
                            [s_k(x,t)=1]$
                            \Statex \hspace{2.95cm}
                            $\vee\;
                            [m_k(x,t)=1 \;\text{and}\; m_k(x-1,t)=1]$
                            \Statex \hspace{2.95cm}
                            $\vee\;
                            [m_k(x-1,t)=1 \;\text{and}\; c_k(x,t)=0]$
                        \EndIf

                        \State $c_k(x,t+1)\gets (c_k(x,t)+1)\bmod q$
                    \EndFor
                \EndFor
            \end{algorithmic}
        \end{algorithm}

    \subsection{Toric code}\label{subsec:online-toric-code}

        We now give the analogous construction for the toric code. The only substantive change is that each slice now runs the code-capacity toric-code decoder.

        We consider the toric code on the lattice $\Lambda=\Z^2$. The decoding problem lives on the three-dimensional spacetime lattice
        \[
            \widetilde{\Lambda}\coloneqq \Lambda \times \Z_{\geq 0}.
        \]
        We equip this lattice with three types of edges. The $x$-edges are pairs $\bigl((\vc{r} - \uvec{x},t),(\vc{r},t)\bigr)$, the $y$-edges are pairs $\bigl((\vc{r} - \uvec{y},t),(\vc{r},t)\bigr)$, and the vertical edges are pairs $\bigl((\vc{r},t),(\vc{r},t+1)\bigr)$. We associate each spacetime site $(\vc r, t)$ with the $x$-edge to its left, the $y$-edge below it, and the vertical edge above it.

        The lattice encodes faults as follows:
        \begin{enumerate}
            \item Each $x$-edge $((\vc{r}-\uvec{x},t),(\vc{r},t))$ records whether a bit-flip error occurred on the qubit associated with the $x$-edge at $\vc{r}$ between times $t-1$ and $t$.
            \item Each $y$-edge $((\vc{r}-\uvec{y},t),(\vc{r},t))$ records whether a bit-flip error occurred on the qubit associated with the $y$-edge at $\vc{r}$ between times $t-1$ and $t$.
            \item Each vertical edge $((\vc{r},t),(\vc{r},t+1))$ records whether the measurement of the stabilizer at $\vc{r}$ was faulty at time $t$; here, the stabilizer at $\vc{r}$ is a toric code stabilizer supported on the four qubits that correspond to the edges incident to $\vc{r}$.
            \item Each point $(\vc{r},t)\in\widetilde{\Lambda}$ stores the syndrome difference
            \[
                \phi(\vc{r},t)
                \coloneqq
                \tilde s_t(\vc{r})\oplus \tilde s_{t-1}(\vc{r})\in\{0,1\},
            \]
            where $\tilde s_t(\vc{r})$ is the potentially faulty measurement outcome of the stabilizer at $\vc{r}$ at time $t$, and we set $\tilde s_{-1}(\vc{r})=0$ for all $\vc{r}$.
        \end{enumerate}

        The decoder stores $K$ two-dimensional slices indexed by $k=0,\ldots,K-1$. A detector event $\phi(\vc{r},t)=1$ inserts a new defect candidate at position $\vc{r}$ in the bottom slice. In every timed slice $k<K-1$, a defect is promoted after
        \[
            t_k=t_0 n^k
        \]
        time steps. Defects in the final slice are not promoted further. As in the repetition-code construction, messages in the timed slices also carry timers.

        Throughout this section (excluding the numerical subsection), we use the convention that messages grow when the local clock satisfies $c\in\{0,1\}$; defects are allowed to move only when $c=0$; and leftward defect motion takes priority over downward motion.

        In each timed slice $k<K-1$, the defect-sector variables at site $\vc{r}$ are
        \[
            \xi_k(\vc{r})
            =
            \bigl(e_{x,k}(\vc{r}),\,e_{y,k}(\vc{r}),\,s_k(\vc{r}),\,\tau_k(\vc{r})\bigr),
        \]
        where $e_{x,k}(\vc{r})$ and $e_{y,k}(\vc{r})$ are the correction bits for the horizontal and vertical qubits in the negative $x$ and $y$ directions relative to $\vc{r}$, $s_k(\vc{r})$ is the defect channel, and $\tau_k(\vc{r})$ is the defect timer, respectively. If $s_k(\vc{r})=0$, we set $\tau_k(\vc{r})=0$. In the final slice, there are no defect timers, so
        \[
            \xi_{K-1}(\vc{r})
            =
            \bigl(e_{x, K-1}(\vc{r}),\,e_{y, K-1}(\vc{r}),\,s_{K-1}(\vc{r})\bigr).
        \]
        The first set of defect-sector variables $s_k(\vc r, 0)$, $e_{x, k}(\vc r, 0), e_{y, k}(\vc r, 0)$, and $\tau_k(\vc r, 0)$ are computed by applying Algorithm~\ref{alg:layered-toric-defect-update} to the initial conditions $s_k(\vc r, -1) = e_{x, k}(\vc r, -1) = e_{y, k}(\vc r, -1) = 0$ for all $\vc{r}$ and $k$. The initial condition for the clock is $c(\vc{r}, 0) = 0$ for all $\vc{r}$.

        We now describe the defect-sector update. For an existing defect $s_k(\vc{r},t)=1$ in a timed slice $k<K-1$, define
        \[
            \tau_k^+(\vc{r},t)\coloneqq \tau_k(\vc{r},t)+1.
        \]
        If $\tau_k^+(\vc{r},t)=t_k$, then the defect is promoted vertically and contributes a candidate at $(k+1,\vc{r})$ at time $t+1$ with timer $0$ in the new slice. A promoted defect does not move spatially during the same update.

        If the timer does not expire, the defect evolves according to the base toric-code rule in its current slice. Define
        \[
            \lambda^x_k(\vc{r},t)
            \coloneqq
            \bigl[
                c_k(\vc{r},t)=0
                \;\text{and}\;
                \exists\, j\in\{0,1\}
                \text{ s.t. }
                m_{0j,k}(\vc{r}-\uvec{x},t)=1
            \bigr],
        \]
        and
        \[
            \lambda^y_k(\vc{r},t)
            \coloneqq
            \bigl[
                c_k(\vc{r},t)=0
                \;\text{and}\;
                \exists\, i\in\{0,1\}
                \text{ s.t. }
                m_{i0,k}(\vc{r}-\uvec{y},t)=1
            \bigr]
            \;\text{and}\;
            \lnot \lambda^x_k(\vc{r},t).
        \]
        If $\lambda^x_k(\vc{r},t)=1$, the bit $e_{x,k}(\vc{r})$ is flipped and the defect contributes a candidate at $(k,\vc{r}-\uvec{x})$ with timer $\tau_k^+(\vc{r},t)$. If $\lambda^y_k(\vc{r},t)=1$, the bit $e_{y,k}(\vc{r})$ is flipped and the defect contributes a candidate at $(k,\vc{r}-\uvec{y})$ with timer $\tau_k^+(\vc{r},t)$. Otherwise, it contributes a stationary candidate at $(k,\vc{r})$ with timer $\tau_k^+(\vc{r},t)$. Defects in the final slice follow the same spatial rule but without any timer check.

        The timer assignment convention for defects is the same as in the repetition-code construction. In Algorithm~\ref{alg:layered-toric-defect-update}, $\mathcal{R}_k(\vc{r},t+1)$ denotes the multiset of defect candidates arriving at $(k,\vc{r})$. Each candidate is a pair $(a,u)$, where $a\in\{\mathrm{up},\mathrm{left},\mathrm{down},\mathrm{stay},\mathrm{env}\}$ is the action taken by the defect and $u$ is its proposed timer. Together, these rules give the update below.

        \begin{algorithm}[H]
            \caption{Defect-sector update for the translation-invariant toric-code streaming decoder at time $t$.}
            \label{alg:layered-toric-defect-update}
            \begin{algorithmic}[1]
                \State $\mathcal{R}_k(\vc{r},t+1) \gets \emptyset$ for all $k,\vc{r}$
                \Comment{Candidates arriving at $(k,\vc{r})$}
                \State $e_{x,k}(\vc{r},t+1) \gets e_{x,k}(\vc{r},t)$ for all $k,\vc{r}$
                \State $e_{y,k}(\vc{r},t+1) \gets e_{y,k}(\vc{r},t)$ for all $k,\vc{r}$
                \Statex
        
                \ForAll{$k=0,\ldots,K-1$}
                    \ForAll{$\vc{r}\in\Lambda$ with $s_k(\vc{r},t)=1$}
                        \If{$k<K-1$}
                            \State $\tau^+\gets \tau_k(\vc{r},t)+1$
                            \If{$\tau^+=t_k$}
                                \State Add candidate $(\mathrm{up},0)$ to $\mathcal{R}_{k+1}(\vc{r},t+1)$
                                \Comment{Vertical promotion}
                            \ElsIf{$\lambda^x_k(\vc{r},t)=1$}
                                \State $e_{x,k}(\vc{r},t+1)\gets e_{x,k}(\vc{r},t+1)\oplus 1$
                                \State Add candidate $(\mathrm{left},\tau^+)$ to $\mathcal{R}_k(\vc{r}-\uvec{x},t+1)$
                                \Comment{Leftward move}
                            \ElsIf{$\lambda^y_k(\vc{r},t)=1$}
                                \State $e_{y,k}(\vc{r},t+1)\gets e_{y,k}(\vc{r},t+1)\oplus 1$
                                \State Add candidate $(\mathrm{down},\tau^+)$ to $\mathcal{R}_k(\vc{r}-\uvec{y},t+1)$
                                \Comment{Downward move}
                            \Else
                                \State Add candidate $(\mathrm{stay},\tau^+)$ to $\mathcal{R}_k(\vc{r},t+1)$
                                \Comment{Stationary defect}
                            \EndIf
                        \Else
                            \If{$\lambda^x_{K-1}(\vc{r},t)=1$}
                                \State $e_{x,K-1}(\vc{r},t+1)\gets e_{x,K-1}(\vc{r},t+1)\oplus 1$
                                \State Add candidate $(\mathrm{left},\varnothing)$ to $\mathcal{R}_{K-1}(\vc{r}-\uvec{x},t+1)$
                                \Comment{Leftward move in final slice}
                            \ElsIf{$\lambda^y_{K-1}(\vc{r},t)=1$}
                                \State $e_{y,K-1}(\vc{r},t+1)\gets e_{y,K-1}(\vc{r},t+1)\oplus 1$
                                \State Add candidate $(\mathrm{down},\varnothing)$ to $\mathcal{R}_{K-1}(\vc{r}-\uvec{y},t+1)$
                                \Comment{Downward move in final slice}
                            \Else
                                \State Add candidate $(\mathrm{stay},\varnothing)$ to $\mathcal{R}_{K-1}(\vc{r},t+1)$
                                \Comment{Stationary defect in final slice}
                            \EndIf
                        \EndIf
                    \EndFor
                \EndFor
        
                \Statex
                \ForAll{$\vc{r}\in\Lambda$ with $\phi(\vc{r},t + 1)=1$}
                    \State Add candidate $(\mathrm{env},0)$ to $\mathcal{R}_0(\vc{r},t+1)$
                    \Comment{New environmental defect}
                \EndFor
        
                \Statex
                \ForAll{$k=0,\ldots,K-1$}
                    \ForAll{$\vc{r}\in\Lambda$}
                        \If{$|\mathcal{R}_k(\vc{r},t+1)|$ is even}
                            \State $s_k(\vc{r},t+1)\gets 0$
                            \If{$k<K-1$}
                                \State $\tau_k(\vc{r},t+1)\gets 0$
                            \EndIf
                        \Else
                            \State $s_k(\vc{r},t+1)\gets 1$
                            \If{$k<K-1$}
                                \State $\tau_k(\vc{r},t+1)
                                \gets
                                \min\{u:(a,u)\in\mathcal{R}_k(\vc{r},t+1)\}$
                            \EndIf
                        \EndIf
                    \EndFor
                \EndFor
            \end{algorithmic}
        \end{algorithm}

        We next specify the message-sector update. In each timed slice $k<K-1$, the message-sector variables are
        \[
            \xi_k'(\vc{r})
            =
            \bigl(
                \{m_{ij,k}(\vc{r})\}_{(i,j)\in\mathcal{M}_2},
                \,c_k(\vc{r}),
                \,\{\theta_{ij,k}(\vc{r})\}_{(i,j)\in\mathcal{M}_2}
            \bigr),
        \]
        where $m_{ij,k}(\vc{r})$ is the message channel, $c_k(\vc{r})\in\{0,\ldots,q-1\}$ is the clock, and $\theta_{ij,k}(\vc{r})$ is the corresponding message timer. If $m_{ij,k}(\vc{r})=0$, we set $\theta_{ij,k}(\vc{r})=0$; if $m_{ij,k}(\vc{r})=1$, then $\theta_{ij,k}(\vc{r})\in\{0,\ldots,t_k-1\}$. In the final slice, messages are untimed:
        \[
            \xi_{K-1}'(\vc{r})
            =
            \bigl(
                \{m_{ij,K-1}(\vc{r})\}_{(i,j)\in\mathcal{M}_2},
                \,c_{K-1}(\vc{r})
            \bigr).
        \]
        We call $c_k(\vc r, t)$ a message-sector variable even though the defects must examine $c_k(\vc r, t)$ to determine if they are allowed to move or not.

        The message update is performed after the defect-sector update, so the variables $s_k(\vc r,t+1)$, and also $\tau_k(\vc r,t+1)$ for $k<K-1$, have already been determined. We define the update in terms of message candidates. For each channel $(i,j)$, if no message candidate is produced at $(k,\vc{r})$, then the message is erased. If one or more candidates are produced, then a message is written, and in a timed slice its timer is the minimum candidate timer.

        Let
        \[
            N_{ij}(\vc{r})
            \coloneqq
            \{\vc{r}+(-1)^{i+1}\uvec{x},\,
              \vc{r}+(-1)^{j+1}\uvec{y}\}
        \]
        be the two source positions for growth into channel $(i,j)$ at $\vc{r}$, and let
        \[
            \Call{ToomVote}{i,j,k,\vc{r},t}
            \coloneqq
            \Call{Maj}{
                m_{ij,k}(\vc{r},t),\,
                m_{ij,k}(\vc{r}+(-1)^{i+1}\uvec{x},t),\,
                m_{ij,k}(\vc{r}+(-1)^{j+1}\uvec{y},t)
            }.
        \]

        For $k<K-1$, define the valid growth-source set
        \[
            V_{ij,k}(\vc{r},t)
            \coloneqq
            \bigl\{
                \vc{r}'\in N_{ij}(\vc{r})
                :
                m_{ij,k}(\vc{r}',t)=1
                \;\text{and}\;
                \theta_{ij,k}(\vc{r}',t)<t_k-1
            \bigr\},
        \]
        the coupling condition
        \[
            C_k(\vc{r},t)
            \coloneqq
            \bigl[ 
                m_{00,k}(\vc{r},t)=1
                \;\text{and}\;
                \Call{ToomVote}{0,0,k,\vc{r},t}=1
            \bigr],
        \]
        and the persistence condition
        \[
            P_{ij,k}(\vc{r},t)
            \coloneqq
            \bigl[
                m_{ij,k}(\vc{r},t)=1
                \;\text{and}\;
                \theta_{ij,k}(\vc{r},t)<t_k-1
                \;\text{and}\;
                \bigl(
                    \Call{ToomVote}{i,j,k,\vc{r},t}=1
                    \;\text{or}\;
                    C_k(\vc{r},t)=1
                \bigr)
            \bigr].
        \]

        For each timed slice $k<K-1$, each site $\vc{r}$, and each channel $(i,j)\in\mathcal{M}_2$, let $\mathcal{B}_{ij,k}(\vc{r},t+1)$ be the set of message-candidate timers
        \[
        \begin{aligned}
            \mathcal{B}_{ij,k}(\vc{r},t+1)
            \coloneqq\;&
            \{\tau_k(\vc{r},t+1): s_k(\vc{r},t+1)=1\}
            \\
            &\cup
            \{\tau_k(\vc{r},t)+1:
                s_k(\vc{r},t)=1
                \;\text{and}\;
                \tau_k(\vc{r},t)+1<t_k
            \}
            \\
            &\cup
            \bigl\{
                1+\min_{\vc{r}'\in V_{ij,k}(\vc{r},t)}
                \theta_{ij,k}(\vc{r}',t):
                c_k(\vc{r},t)\in\{0,1\}
                \;\text{and}\;
                V_{ij,k}(\vc{r},t)\neq\emptyset
            \bigr\}
            \\
            &\cup
            \{\theta_{ij,k}(\vc{r},t)+1:
                P_{ij,k}(\vc{r},t)=1
            \}.
        \end{aligned}
        \]
        The first two sets are defect-source contributions, from a defect present at $\vc{r}$ at time $t+1$ or a not-yet-promoted defect present at $\vc{r}$ at time $t$; the third is the clock-gated growth contribution; and the fourth is the persistence contribution.

        The timed-slice update is then defined by
        \begin{align*}
            \mathcal{B}_{ij,k}(\vc{r},t+1)=\emptyset
            &\quad\Longrightarrow\quad
            m_{ij,k}(\vc{r},t+1)=0,\qquad
            \theta_{ij,k}(\vc{r},t+1)=0,\\
            \mathcal{B}_{ij,k}(\vc{r},t+1)\neq\emptyset
            &\quad\Longrightarrow\quad
            m_{ij,k}(\vc{r},t+1)=1,\qquad
            \theta_{ij,k}(\vc{r},t+1)
            =
            \min \mathcal{B}_{ij,k}(\vc{r},t+1).
        \end{align*}
        In the final slice $K-1$, the update is the same as above except with all timer fields removed. Finally, in every slice,
        \[
            c_k(\vc{r},t+1)=(c_k(\vc{r},t)+1)\bmod q.
        \]

        \begin{algorithm}[H]
            \caption{Message-sector update for the translation-invariant toric-code streaming decoder at time~$t$.}
            \label{alg:layered-toric-message-update}
            \begin{algorithmic}[1]
                \ForAll{$k=0,\ldots,K-1$}
                    \ForAll{$\vc{r}\in\Lambda$}

                        \If{$k<K-1$}
                            \State $\mathcal{D}\gets\emptyset$
                            \Comment{Defect-source candidate timers}

                            \If{$s_k(\vc{r},t+1)=1$}
                                \State Add $\tau_k(\vc{r},t+1)$ to $\mathcal{D}$
                                \Comment{Surviving defect source}
                            \EndIf

                            \If{$s_k(\vc{r},t)=1$ and $\tau_k(\vc{r},t)+1<t_k$}
                                \State Add $\tau_k(\vc{r},t)+1$ to $\mathcal{D}$
                                \Comment{Old non-promoted defect source}
                            \EndIf

                            \ForAll{$(i,j)\in\mathcal{M}_2$}
                                \State $\mathcal{B}\gets\mathcal{D}$
                                \Comment{Candidate timers for channel $(i,j)$}

                                \State $V\gets
                                \{\vc{r}'\in N_{ij}(\vc{r}):
                                m_{ij,k}(\vc{r}',t)=1
                                \;\text{and}\;
                                \theta_{ij,k}(\vc{r}',t)<t_k-1\}$

                                \If{$c_k(\vc{r},t)\in\{0,1\}$ and $V\neq\emptyset$}
                                    \State Add
                                    $1+\min_{\vc{r}'\in V}\theta_{ij,k}(\vc{r}',t)$
                                    to $\mathcal{B}$
                                    \Comment{Growth candidate}
                                \EndIf

                                \State $C\gets
                                [m_{00,k}(\vc{r},t)=1
                                \;\text{and}\;
                                \Call{ToomVote}{0,0,k,\vc{r},t}=1]$
                                \Comment{Coupling condition}

                                \State $P\gets
                                [m_{ij,k}(\vc{r},t)=1
                                \;\text{and}\;
                                \theta_{ij,k}(\vc{r},t)<t_k-1$
                                \Statex \hspace{2.35cm}
                                $\text{and}\;
                                (\Call{ToomVote}{i,j,k,\vc{r},t}=1
                                \;\text{or}\;
                                C=1)]$
                                \Comment{Persistence condition}

                                \If{$P$}
                                    \State Add $\theta_{ij,k}(\vc{r},t)+1$ to $\mathcal{B}$
                                    \Comment{Persistence candidate}
                                \EndIf

                                \If{$\mathcal{B}=\emptyset$}
                                    \State $m_{ij,k}(\vc{r},t+1)\gets 0$
                                    \State $\theta_{ij,k}(\vc{r},t+1)\gets 0$
                                \Else
                                    \State $m_{ij,k}(\vc{r},t+1)\gets 1$
                                    \State $\theta_{ij,k}(\vc{r},t+1)\gets\min\mathcal{B}$
                                \EndIf
                            \EndFor

                        \Else
                            \State $S\gets [s_k(\vc{r},t+1)=1]\vee[s_k(\vc{r},t)=1]$
                            \Comment{Defect-source condition on the back wall}

                            \ForAll{$(i,j)\in\mathcal{M}_2$}
                                \State $G\gets
                                [c_k(\vc{r},t)\in\{0,1\}
                                \;\text{and}\;
                                \exists\,\vc{r}'\in N_{ij}(\vc{r})
                                \text{ s.t. }
                                m_{ij,k}(\vc{r}',t)=1]$
                                \Comment{Growth condition}

                                \State $C\gets
                                [m_{00,k}(\vc{r},t)=1
                                \;\text{and}\;
                                \Call{ToomVote}{0,0,k,\vc{r},t}=1]$
                                \Comment{Coupling condition}

                                \State $P\gets
                                [m_{ij,k}(\vc{r},t)=1
                                \;\text{and}\;
                                (\Call{ToomVote}{i,j,k,\vc{r},t}=1
                                \;\text{or}\;
                                C=1)]$
                                \Comment{Persistence condition}

                                \State $m_{ij,k}(\vc{r},t+1)\gets S\vee G\vee P$
                            \EndFor
                        \EndIf

                        \State $c_k(\vc{r},t+1)\gets (c_k(\vc{r},t)+1)\bmod q$
                    \EndFor
                \EndFor
            \end{algorithmic}
        \end{algorithm}
    \subsection{Open boundary conditions}

        The above decoders extend to open boundary conditions by applying the open-boundary modifications of Sections~\ref{subsec:rep-code:open-boundary-conditions} and~\ref{subsec:toric-code:open-boundary-conditions} in each slice: the decoders are unchanged except that, at times $t \in q_s\Z$, each slice performs the corresponding splitting step before the usual defect- and message-sector updates.\footnote{For the surface code, the clock period $q$ may need to be taken larger than the clock period used to define the toric-code streaming decoder, since the open-boundary erosion argument of Section~\ref{subsec:toric-code:open-boundary-conditions} requires $q$ to be a sufficiently large constant.}

        Concretely, the splitting step maps the defect, message, and timer configurations of slice $k$ to intermediate configurations; for the repetition code,
        \[
            s_k(x,t),\,\tau_k(x,t),\,m_k(x,t),\,\theta_k(x,t)
            \quad\longmapsto\quad
            s_k'(x,t),\,\tau_k'(x,t),\,m_k'(x,t),\,\theta_k'(x,t)
        \]
        with the correction channels updated accordingly, and for the toric code, the same map is applied with $\vc r$ in place of $x$ and with each message channel $m_{ij,k}$ and its timer $\theta_{ij,k}$ in place of $m_k$ and $\theta_k$. In particular, defects and messages shift toward the condensing boundaries, messages and their timers are copied at the central cut rather than torn apart, and defects, messages, and timers translated out of the system are annihilated. The update from time $t$ to $t+1$ is then computed from the environmental defects $\phi(x,t+1)$ and the split configurations in place of the unsplit ones.

    \subsection{Transversal readout}
    \label{subsec:readout}

        We now describe decoding during transversal readout of a logical $Z$ operator. At the readout time $T$, every data qubit is measured in the computational basis. Without loss of generality, we assume that the readout is noiseless.\footnote{Here, we are using the (standard) fact that readout errors on data qubits have the exact same effect as data-qubit errors occurring on the final time slice of the spacetime noise history, so at most, they renormalize the effective data-qubit error rate upward during the last time step.}
        
        Let $b_e\in\{0,1\}$ be the outcome of measuring qubit $e$. These single-qubit measurement outcomes determine a final reconstructed syndrome configuration of $Z$-type stabilizers:
        \[
            \tilde s_T(a)
            =
            \bigoplus_{e\in\supp(a)} b_e.
        \]
        Comparing this reconstructed syndrome configuration with the previously measured syndrome gives the final syndrome-difference configuration
        \[
            \phi(a, T)
            =
            \tilde s_T(a)
            \oplus
            \tilde s_{T-1}(a).
        \]
        We then insert $\phi(\cdot, T)$ into the bottom slice exactly as in an ordinary syndrome-measurement round.

        After this round, no further measurement data are needed. We set all subsequent environmental inputs to zero and run the decoder for $c_rL$ additional cleanout steps, where $c_r>0$ is a sufficiently large constant whose value will be discussed in the next subsection. If defects remain after these steps, we declare a logical error. Otherwise, the final $X$-type correction is well-defined, and the final value of the logical $Z$ observable is the modulo-two sum of the correction bits and the measurement outcomes $b_e$ along a fixed representative of the logical $Z$ operator. We say a logical error has occurred if the correction configuration differs from the modulo-two sum of errors at each data qubit over all time by a non-trivial logical operator.

        During readout, the computational-basis measurement provides no further $X$-type stabilizer information, so the $X$-type stabilizer sector decoding problem cannot be continued; we therefore discard it and continue decoding only the $Z$-type stabilizer sector. This is harmless because the value of the logical $Z$ observable depends only on the measurement outcomes $b_e$ and the final $X$-type correction, which belongs to the $Z$-type stabilizer sector decoding problem.

    \subsection{Bounds on memory lifetimes}

        To lower-bound the memory lifetimes of our decoders, we apply the sparsity theorem to the spacetime history of data-qubit and stabilizer-measurement-bit errors.

        For the remainder of the paper, let $k_L$ be the smallest integer such that, in a spacetime history consisting solely of a single cluster of initial diameter $w_{k_L}$, the cluster can expand to support a logical operator, or become large enough for periodic-boundary effects to prevent the decoder from terminating. The maximal expansion factor---and hence the precise value of $k_L$---depends on the exact decoder and procedure under consideration; in every case we study, it will be clear a posteriori that this factor is a constant, so $k_L = \Theta(\log L)$ throughout.
        
        We first prove a stretched-exponential lifetime bound in the regime where the number of slices grows logarithmically with the system size.

        \begin{theorem}[stretched-exponential lifetime]
            \label{thm:stretched-exponential-lifetime}
            There exist constants
            \[
                p_*,c,c_r,C',\alpha,t_0>0,
            \]
            sparsity-theorem parameters $(\beta,\gamma,n)$ satisfying inequalities~\eqref{paramineqs}, and integers $q,q_s>0$ such that the following holds.

            Let $q$ be the clock period used by both the repetition-code and toric-code decoders, and let $q_s$ be the splitting period for the open-boundary-condition versions of both decoders. Suppose the data-qubit and stabilizer-measurement-bit noise is $p$-bounded with $p<p_*/2$. Suppose also that the number of slices $K$ satisfies
            \[
                K \geq k_L
            \]
            and that the cleanout step proportionality constant is $c_r$.

            Initialize all data qubits in the all-$0$ product state, so that the value of every logical $Z$ operator is deterministic in the absence of noise. Run any one of the four streaming decoders for $T$ syndrome-extraction rounds, and then perform transversal readout in the computational basis.

            Then, for all four decoders and every $T\geq 1$,
            \[
                \Pr[\textup{logical failure by time }T]
                \leq
                TC'\exp(-cL^\alpha).
            \]
        \end{theorem}

        The same theorem holds for arbitrary toric and surface code encoded states with known initial stabilizer values. We have stated Theorem~\ref{thm:stretched-exponential-lifetime} for only the all-$0$ initial state to keep the statement simple.

        \begin{proof}
            Let $q, q_s > 0$ be integers such that linear cluster erosion holds for all four decoders, and let $(\beta, \gamma, n)$ be sparsity-theorem parameters satisfying inequalities~\eqref{paramineqs}. Next, let $p_* > 0$ be such that if $p < p_*$, then the sparsity theorem holds.

            We begin by assuming that our system size is infinite and that our decoder uses infinitely many slices. Let $w_k = w_0 n^k$ denote the diameter of a $k$-cluster, and let $b_k = b_0 n^k$ denote the corresponding buffer size.
            
            We first consider a noise history consisting of a single $k$-cluster. Since the spacetime diameter of the cluster is $O(w_k)$, there is a constant $c_0$ such that all detector events associated with the cluster have been loaded into the decoder within time $c_0w_k$. Moreover, the total time a defect or message can spend in slices below the $k$th slice is bounded by
            \[
                \sum_{j=0}^{k-1} t_j
                =
                t_0 \sum_{j=0}^{k-1} n^j
                \leq
                2 t_{k-1},
            \]
            where we have used $n \geq 2$. Therefore, by time
            \[
                c_0w_k + 2t_{k-1},
            \]
            all activity sourced by the cluster has either been erased or has reached at least the $k$th slice.
            
            Let us temporarily assume $t_k$ is infinite, so the cluster can never be promoted past the $k$th slice. Since $t_k$ is infinite and the entire cluster has been loaded in, the dynamics now reduce to those of the code-capacity decoder. (Here, we are using the fact that messages are never promoted and are sourced only by defects present in a slice, so the cluster's defects arrive in a region of the $k$th slice containing no defects or non-trivial messages other than those sourced by the cluster's own previously promoted defects.) Since the code-capacity decoder satisfies linear cluster erosion and the $k$th slice contains a single cluster whose defect and message support has diameter $O(w_k)$, it follows that all defects and messages sourced by the cluster are erased within $c_2 w_k$ total time steps, where $c_2 \geq c_1(c_0 + 2t_0/(nw_0))$ for a sufficiently large constant $c_1 \geq 1$. The constant $c_1$ accounts for both the loading time and the fact that the cluster can expand by a constant factor before reaching the $k$th slice.
            
            We now show that we can choose $t_k$ to be finite but sufficiently large so that it does not interfere with the decoding of the $k$-cluster. In particular, we now choose the hierarchy parameters so that the total active decoding time is strictly less than the next timer threshold $t_k$. Since $w_k = w_0 n^k$ and $t_k = t_0 n^k$, this condition is
            \[
                c_1(c_0 + 2 t_0/(n w_0)) w_k \leq c_2 w_k < t_0 n^k,
            \]
            or equivalently
            \[
                c_1(c_0 w_0/t_0 + 2/n) \leq c_2 w_0/t_0 < 1.
            \]
            This can be achieved by taking $n$ and $t_0$ sufficiently large. With these choices, a single $k$-cluster is always erased before any activity sourced by it can be promoted beyond the $k$th slice. Consequently, if the noise history consists of a single $k$-cluster, then this cluster's defects and messages are erased in linear time and its effects never reach beyond the $k$th slice.

            Next, we show that the decoder successfully decodes any noise history consisting only of $k$-clusters. Since the ratio $b_k/w_k = b_0/w_0$ is linear in $n$ (see Eq.~\ref{bkwk}), it follows that for $n$ large enough, the spacetime activity region of each $k$-cluster is disjoint from that of every other $k$-cluster, so each $k$-cluster is decoded exactly as if it were the only cluster present. The required lower bound on $n$ is independent of $k$. Furthermore, since $c_0, c_1, c_2$ can be chosen independently of $n$, this lower bound is compatible with our earlier requirements, and all of these requirements can be simultaneously satisfied by taking $n$ sufficiently large.
            
            We now extend this decoupling property to arbitrary noise histories. We proceed by induction on $k'$, showing that the decoder successfully decodes any history consisting of arbitrary $k$-clusters with $k \leq k'$, where the value of $k$ may vary from cluster to cluster. The base case $k'=0$ follows immediately from the previous paragraph. Assume the inductive hypothesis holds for all $k \leq k'$. Consider a $(k'+1)$-cluster, and suppose that one of its defects is loaded into some slice $k$ with $k \leq k'$. This defect, together with any messages it produces, persists in slice $k$ for time at most $t_k$. If $b_k$ is sufficiently large, equivalently if $b_k/t_k=b_0/t_0$ is sufficiently large, then for all $j \leq k$, these defects and messages remain sufficiently separated in both space and time from every $j$-cluster's defects and messages in slice $k$, so their timers expire before any interaction can occur. This imposes another $k$-independent lower bound on $n$, which is compatible with the conditions on $n$ required by the arguments from prior paragraphs. Thus, by the inductive hypothesis, each $k$-cluster with $k \leq k'$ is decoded exactly as it would be in the absence of the $(k'+1)$-cluster. Furthermore, by the inductive hypothesis, all such $k$-clusters are eroded before reaching slice $k'+1$, and therefore cannot interact with the $(k'+1)$-cluster in that slice, where timers are no longer guaranteed to prevent interaction. In particular, whenever a cluster's activity is promoted into a new slice, it arrives in a region containing no non-trivial messages or defects sourced by other clusters, which is precisely the condition under which the single-cluster analysis above applies. Therefore, the decoder successfully decodes any noise history consisting of arbitrary $k$-clusters with $k \leq k'+1$. The desired result then follows by induction.

            Consequently, for $t_0$ and $n$ sufficiently large, distinct clusters cannot interact before the relevant defects or messages are erased by their timers, so all clusters are decoded independently in the infinite system with infinitely many slices. Informally, for $n$ and $t_0$ sufficiently large, the decoding of each cluster completely decouples from the decoding of all other clusters.

            We also note in passing that local message-erasure rules are not needed for the proofs: in the streaming setting, timers automatically erase messages when they expire, and this is sufficient to ensure linear cluster erosion and linear message erosion. In fact, without local erasure rules, the proofs of linear defect erosion are far simpler; we include local erasure rules only to improve the performance of the decoders.

            For a finite system, it is clear from the arguments above that the maximal expansion factor in the definition of $k_L$ is a constant, and since $w_k = w_0 n^k$, this gives $k_L = \Theta(\log L)$. In particular, there exists $C' > 0$ such that if $K \geq k_L$, then the decoder can fail only if a $k$-cluster with $k \geq k_L$ occurs, and the probability of logical failure satisfies
            \[
                \Pr[\textup{logical failure by time } T] \;\le\; T C' \exp(-c L^{\alpha}),
            \]
            where $T$ is the number of time steps and we have union bounded over all of spacetime. Here, $\alpha$ is the exponent from the sparsity theorem and $c > 0$ is some constant. We also assume that the readout time proportionality constant $c_r > 0$ has been chosen so that it is larger than the relevant erosion constants for all four decoders.
        \end{proof}

        Therefore, with $K=\Theta(\log L)$ slices, the decoder achieves a stretched-exponential memory lifetime. Since
        \[
            t_{K-1}=t_0 n^{K-1}=O(\poly(L)),
        \]
        each timer requires $O(\log L)$ bits, so since there are $O(\log L)$ slices, the total per-site classical overhead is
        \[
            O(\log^2 L) = O(\polylog(L)).
        \]

        We now combine our construction with the back-wall arguments of Ref.~\cite{lake2025localactiveerrorcorrection} to obtain stretched-exponential memory lifetime scaling using only $\polyloglog(L)$ per-site resources. As in that work, the back wall of our construction, namely the final slice $k=K-1$, sees an effectively renormalized noise model because all sufficiently small spacetime clusters are decoded before reaching it.

        \begin{lemma}[filtering before the back wall]
            \label{lem:back-wall-filtering}
                Suppose $K \le k_L$. If $k < K - 1$, then all defects and messages sourced by a $k$-cluster are erased before reaching the back wall, and they never interact with defects or messages from any other cluster.
        \end{lemma}

        The proof is immediate from the proof of the stretched-exponential lifetime theorem above. Because of this lemma and the fact that the code-capacity decoders satisfy linear cluster erosion, the following version of Theorem 3 of Ref.~\cite{lake2025localactiveerrorcorrection} holds.

        \begin{theorem}[polyloglog-overhead stretched-exponential lifetime]
            \label{thm:polyloglog-overhead}
            There exist constants
            \[
                p_*,c,c_r,C',a',\alpha,\zeta,t_0>0,
            \]
            sparsity-theorem parameters $(\beta,\gamma,n)$ satisfying inequalities~\eqref{paramineqs}, and integers $q,q_s>0$ such that the following holds.

            Let $q$ be the clock period used by both the repetition-code and toric-code decoders, and let $q_s$ be the splitting period for the open-boundary-condition versions of both decoders. Suppose the data-qubit and stabilizer-measurement-bit noise is $p$-bounded with $p<p_*/2$. Suppose also that the number of slices $K$ satisfies
            \[
                K \leq k_L
                \qquad\text{and}\qquad
                n^K
                \geq
                a'\left(\frac{\log L}{\log(p_*/p)}\right)^{1/\zeta},
            \]
            and that the cleanout step proportionality constant is $c_r$.

            Initialize all data qubits in the all-$0$ product state, so that the value of every logical $Z$ operator is deterministic in the absence of noise. Run any one of the four streaming decoders for $T$ syndrome-extraction rounds, and then perform destructive readout in the computational basis.

            Then, for all four decoders and every $T\geq 1$,
            \[
                \Pr[\textup{logical failure by time }T]
                \leq
                TC'\exp(-cL^\alpha).
            \]
        \end{theorem}

        The proof is the same as the proof of Theorem 3 of Ref.~\cite{lake2025localactiveerrorcorrection}, so we do not reproduce it here. This theorem implies that $K = O(\log\log L)$ slices, and thus $O(\polyloglog(L))$ per-site resources, are sufficient to achieve a stretched-exponential memory lifetime. 

        The constants in the above theorem are not necessarily the same as those in Theorem~\ref{thm:stretched-exponential-lifetime} (despite some repeated notation). However, one can show that all the constants with overlapping notation in the above theorem and the original theorem can be chosen such that both theorems hold with the same constants.

    \subsection{State preparation}\label{subsec:online-state-preparation}

        We now describe how to locally determine a consistent initial stabilizer frame when the initial stabilizer measurement outcomes are intrinsically random. This situation arises, for example, when preparing a toric-code logical state from the all-$|0\rangle$ state: in the absence of noise, this state has deterministic $Z$-type stabilizer measurement outcomes during the first round of syndrome extraction, but the $X$-type stabilizer measurement outcomes during this round are intrinsically random.

        Our decoder cannot be initialized by treating these random initial $X$-type stabilizer measurement outcomes as an ordinary defect configuration. Such configurations are typically dense, and dense configurations generically overwhelm our decoder and drive it into limit cycles that prevent linear cluster erosion from occurring.\footnote{Dense initial defect configurations generically contain defect pairs separated by distances of order $L$, so the messages sourced by these defects must grow for a time of order $L$ in order for the defects to come into causal contact. Under periodic boundary conditions, messages that grow for times of order $L$ generically flood the entire system and are never erased, which causes all remaining defects to move continuously in one direction, preventing them from being annihilated.
        
        We expect the effect of such pairs to be more benign under open boundary conditions, since boundaries can always erase messages, even when messages have flooded the entire system. However, even if this is the case, rigorously proving that open boundary conditions render such flooding transient (and thus benign) would likely be rather tedious: when flooding occurs, the decoding of different clusters is no longer guaranteed to decouple with high probability, and our proof techniques largely rely on this decoupling.} This is different from the decoder of Ref.~\cite{lake2025localactiveerrorcorrection}, which is capable of performing (a generalized form of) linear cluster erosion even for dense defect configurations, and thus can be trivially initialized by treating these random initial stabilizer measurement outcomes as an ordinary defect configuration.

        To address this, we introduce a local state-preparation procedure that allows our decoder to avoid having to treat the random first-round stabilizer measurement outcomes as ordinary defects. Conceptually, we want to give the decoding problem an open boundary in the time-like direction, so that error chains may terminate on the initial time boundary. For a global decoder, this corresponds to a simple change in boundary conditions, but for our decoder, the situation is more involved because this behavior must be implemented locally. We do this by introducing an absorbing wall in the auxiliary direction and moving this wall upward during state preparation. This rising absorbing wall removes defects associated with the unknown initial stabilizer frame and allows later detector events to be decoded by the usual streaming decoder dynamics. Whenever a defect lands on a site in the absorbing wall, we update the interpretation of the corresponding initial stabilizer measurement at that site.

        This extra structure makes state preparation somewhat more involved than in Ref.~\cite{lake2025localactiveerrorcorrection}. However, the need to manage temporarily unknown stabilizer frames is not unique to state preparation. In particular, the same issue arises during state injection and lattice surgery, where random initial stabilizer measurement outcomes cannot simply be treated as ordinary detector events. The rising-wall construction developed below provides a general local mechanism for assigning a consistent interpretation to such random initial stabilizer measurement outcomes; we will reuse it, with only minor modifications, in both of these settings.

        We first specify how detector events are inserted during state preparation. Apart from the initial round, detector events are handled exactly as in the ordinary streaming decoder. The only change is that the first stabilizer measurement is not treated as a defect configuration, i.e., we set
        \[
            \phi(i,0)=0
        \]
        for every check $i$. For later rounds, we use the usual syndrome differences
        \[
            \phi(i,t)
            =
            \tilde s_t(i)\oplus \tilde s_{t-1}(i),
            \qquad t\geq 1.
        \]
        We also store a variable $\psi(i,t)$, which records the current interpretation of the initial stabilizer measurement. We initialize it as
        \[
            \psi(i,0)=\tilde s_0(i).
        \]

        We next describe the rising absorbing wall. Let $M \geq 1$ be a constant. During preparation, the absorbing wall remains directly above slice $k$ for time $M t_k$. All slices below the wall are active and evolve according to the ordinary streaming decoder. The wall itself is absorbing: any defect candidate promoted into the absorbing region is discarded. Furthermore, whenever a defect is absorbed by site $i$, the corresponding stabilizer interpretation is flipped:
        \[
            \psi(i,t+1)=\psi(i,t)\oplus 1.
        \]

        After the wall has passed the final slice, it hovers above the last slice for an additional time $M t_{K-1}$. During this hovering period, the last slice is temporarily treated like an ordinary timed slice: defects carry timers, and when a timer expires, the defect is promoted into the final absorbing layer and removed. Such an absorption also flips the corresponding stabilizer interpretation. After this final hover period, the absorber is removed and the decoder resumes its standard dynamics.

        Therefore, all slices return to the ordinary streaming decoder dynamics by time
        \[
            T_{\mathrm{prep}}
            =
            M\sum_{k=0}^{K-1}t_k .
        \]
        For $K = k_L$, we have $n^{k_L} = \Theta(L)$, and hence
        \[
            T_{\mathrm{prep}}=\Theta(L).
        \]

        We now specify the modified cellular-automaton rule more explicitly. Set
        \[
            T_k
            =
            M\sum_{j=0}^{k-1}t_j,
            \qquad
            k=0,\ldots,K .
        \]
        Thus, for $k<K$, $T_k$ is the time at which slice $k$ becomes active, and $T_K$ is the time at which the absorber above the final slice is removed. During the interval
        \[
            T_{k}\leq t<T_{k+1},
            \qquad
            k=0,\ldots,K-1,
        \]
        the slices $\ell\leq k$ are active. If $k<K - 1$, the slices $\ell> k$ are absorbing. If $k=K-1$, all slices are active, but the absorbing wall is hovering above slice $K-1$.

        Slice $k = 0$ is active from the beginning and is never absorbing. For $k\geq 1$, while slice $k$ is still absorbing, any defect candidate promoted into slice $k$ is discarded. More precisely, if $t<T_k-1$, then a candidate promoted into slice $k$ during the update from $t$ to $t+1$ is absorbed. If the absorbed candidate is a defect at site $i$, we set
        \[
            \psi(i,t+1)=\psi(i,t)\oplus 1.
        \]
        If no defect is absorbed at $i$, we set
        \[
            \psi(i,t+1)=\psi(i,t).
        \]
        If $t=T_k-1$ and $k < K$, then slice $k$ becomes active at the output time $t+1=T_k$, so the promoted candidate is kept and included in the computation of $s_k(i,t+1)$ and $m_k(i,t+1)$. In this case, the promotion does not flip $\psi$. All active slices below the wall operate according to the ordinary streaming decoder rule.

        Once slice $K-1$ becomes active, it is temporarily treated as a timed slice. Thus, defects and messages in slice $K-1$ carry timers during the interval
        \[
            T_{K-1}\leq t<T_K .
        \]
        If such a defect timer expires before the auxiliary absorber is removed, i.e., during an update with $t<T_K-1$, the corresponding defect candidate is promoted beyond slice $K-1$ and is discarded by the auxiliary absorbing layer. If the promoted candidate is a defect at site $i$, it flips
        \[
            \psi(i,t+1)=\psi(i,t)\oplus 1.
        \]
        On the final update with $t=T_K-1$, the auxiliary absorber is removed. During this update, slice $K-1$ is updated using the ordinary untimed back-wall rule, and from time $T_K$ onward all timers in the final slice are permanently disabled. No promotions from slice $K-1$ into the absorber occur during the update from $t=T_K-1$ to $t+1=T_K$. Consequently,
        \[
            \psi(i,t)=\psi(i,T_K-1)
            \qquad
            \text{for all } t\geq T_K-1 .
        \]

        After time
        \[
            T_{\mathrm{prep}}=T_K,
        \]
        the absorbing wall is completely gone, and the decoder continues as the ordinary streaming decoder. We then commit to $\psi(i,T_K-1)$ as our initial stabilizer frame.

        To make the notion of failure for the state-preparation procedure well-defined, we use the following diagnostic experiment. The experiment may appear somewhat strange at first glance; it is designed in this strange manner only to simplify the formal proofs.
        
        Initialize the data qubits in the all-$|0\rangle$ state, assume that only bit-flip errors on data qubits and $Z$-type stabilizer-measurement outcomes occur, and run the state-preparation decoder in the $Z$-type stabilizer sector. Note that we are using the rising-wall procedure in this sector, even though the initial frame is deterministic in the absence of noise. After $T\geq T_{\mathrm{prep}}$ syndrome-extraction rounds, perform the destructive computational-basis readout procedure.
        
        This experiment induces a spacetime noise history in the $Z$-type stabilizer decoding sector. We will now introduce some definitions in order to precisely define what it means for such a diagnostic experiment to succeed. During the proof of the theorem below, we will show that all of the following definitions are well-defined.

        \begin{definition}[wall-interacting clusters and wall residual]
            \label{def:wall-residual}
            Consider the spacetime noise history from the state-preparation diagnostic experiment, and write its clustering decomposition as
            \[
                N
                =
                \bigsqcup_{k,i} C_k^{(i)}.
            \]
            We say that a cluster occurs before wall removal if it contains a fault location whose time coordinate is less than or equal to $T_{\mathrm{prep}}$.

            Suppose no cluster of level at least $k_L$ occurs before wall removal. Then, a cluster is called wall-interacting if it occurs before wall removal and either (1) some defect sourced by it is absorbed by the rising wall, or (2) it contains a first-round fault on the initial time boundary. For each wall-interacting cluster $C$, let $E_C$ and $F_C$ be $\{0, 1\}$-valued maps on qubits, where $E_C$ is the modulo-two sum of data-qubit errors in $C$ that occurred on that qubit and $F_C$ is the modulo-two sum of correction-bit flips due to $C$ that occurred on that qubit.

            We define the residual associated with a wall-interacting cluster $C$ as
            \[
                R_C
                \coloneqq
                E_C
                \oplus
                F_C,
            \]
            and we define the residual wall error as
            \[
                R_w
                \coloneqq
                \bigoplus_{C\in\mathcal{C}_{w}} R_C,
            \]
            where $\mathcal{C}_{w}$ is the set of wall-interacting clusters. If $\mathcal{C}_{w}=\varnothing$, then $R_{w}=0$.
        \end{definition}

        \begin{definition}[state-preparation failure]
            \label{def:state-preparation-failure}
            Consider the spacetime noise history from a state-preparation diagnostic experiment. Suppose no cluster of level at least $k_L$ occurred before wall removal in the induced spacetime noise history. Let $E$ be the modulo-two sum of all data-qubit errors in this history. We define the wall-adjusted error as
            \[
                E_a
                \coloneqq
                E
                \oplus
                \bigoplus_{C\in\mathcal{C}_{w}} E_C.
            \]
            
            We say a ``logical failure after state preparation'' has occurred if the decoder has not terminated by the end of cleanout, or if the decoder has terminated but 
            \[
                F_a
                \oplus
                E_a
            \]
            is a non-trivial logical operator, where $F$ is the final set of corrections on each qubit produced by the decoder and
            \[
                F_a
                \coloneqq
                F
                \oplus
                \bigoplus_{C\in\mathcal{C}_{w}} F_C
            \]
            is the wall-adjusted correction.
        \end{definition}

        \begin{theorem}[state preparation with a rising absorbing wall]
            \label{thm:state-preparation-rising-wall}
            There exist constants
            \[
                p_*,c,c_r,C',\alpha,t_0>0,
            \]
            sparsity-theorem parameters $(\beta,\gamma,n)$ satisfying inequalities~\eqref{paramineqs}, integers $q,q_s>0$, and constants $M, c_w>0$ such that both Theorem~\ref{thm:stretched-exponential-lifetime} and the following hold with these constants.

            Consider the periodic- and open-boundary versions of the repetition-code and toric-code decoders running the state-preparation procedure with rising-wall proportionality constant $M$. Suppose the data-qubit and stabilizer-measurement-bit noise is $p$-bounded with $p<p_*/2$. Suppose also that
            \[
                K = k_L,
            \]
            and that the cleanout step proportionality constant is $c_r$.

            For a state-preparation diagnostic performed at time $T\geq T_{\mathrm{prep}}$, let $\mathcal{B}_{\mathrm{wall}}$ be the event that the diagnostic spacetime history contains a cluster of level at least $k_L$ that occurs before wall removal. Then,
            \[
                \Pr[\mathcal{B}_{\mathrm{wall}}]
                \leq
                C'\exp(-c_wL^\alpha).
            \]

            On the complementary event $\mathcal{B}_{\mathrm{wall}}^c$, all notions in Definitions~\ref{def:wall-residual} and~\ref{def:state-preparation-failure} are well-defined. Moreover, if $\psi_{\mathrm{true}}$ denotes the ideal initial stabilizer frame with first-round faults removed, then
            \[
                \partial R_w
                =
                \psi_{\mathrm{true}}
                \oplus
                \psi(\cdot,T_{\mathrm{prep}}).
            \]

            Let $\mathsf{Fail}_{\mathrm{sp}}(T)$ denote the event that a logical failure after state preparation occurs in the sense of Definition~\ref{def:state-preparation-failure}. Then, for every $T\geq T_{\mathrm{prep}}$,
            \[
                \Pr[
                    \mathcal{B}_{\mathrm{wall}}^c
                    \cap
                    \mathsf{Fail}_{\mathrm{sp}}(T)
                ]
                \leq
                TC'\exp(-c_wL^\alpha).
            \]
        \end{theorem}

        The diagnostic experiment is only a bookkeeping device to make the failure event in Theorem~\ref{thm:state-preparation-rising-wall} well-defined. In the actual toric- or surface-code state-preparation setting, the same types of arguments apply to the stabilizer sector whose initial frame is unknown. On $\mathcal{B}_{\mathrm{wall}}^c$, the decoder successfully commits to a consistent initial frame, and its discrepancy from the true initial frame is given by $\partial R_w$. After time $T_{\mathrm{prep}}$, the wall has been removed, and the decoder operates according to the ordinary streaming decoder dynamics. Thus, the prepared encoded state is protected as in the usual streaming decoder setting.

        \begin{proof}
            We first record two deterministic estimates. The first is the ordinary single-cluster estimate from the proof of Theorem~\ref{thm:stretched-exponential-lifetime}: there is a constant $C_1\geq 1$ such that, if a $k$-cluster has its first detector event inserted at time $u$, then under the ordinary streaming decoder, all activity sourced by this cluster is erased by time $u+C_1t_k$, and no activity sourced by it is promoted beyond slice $k$.
        
            We will also use the following wall-modified variant of the preceding estimate: there are constants $C_2\geq C_1$ and $M > C_2$ such that, if a $k$-cluster has its first detector event inserted at time $u$, then under the wall-modified dynamics, by time $u+C_2t_k$ all activity sourced by the cluster has either been erased, absorbed by the wall, or removed from slices $0,\ldots,k$ by promotion into slice $k+1$. If $k=K-1$, then ``slice $k+1$'' means the auxiliary absorbing layer above the final slice. For $k = K - 1$, the fact that such constants exist is not clear a priori because in principle the wall could leave the final auxiliary slice before the cluster is completely absorbed; however, we will see such constants exist a posteriori from the arguments below.

            Suppose only a single $k$-cluster $C$ exists in the spacetime history, where $k < k_L$, and that $C$ sources at least one detector event. Let $u$ be the first time at which a detector event sourced by $C$ is inserted.
            
            Suppose further that
            \[
                u+C_2t_k<T_{k+1}.
            \]
            By the wall-modified estimate, by time $u+C_2t_k$ all activity sourced by $C$ has either been erased, absorbed, or promoted into slice $k+1$. But slice $k+1$ is absorbing until time $T_{k+1}$, so any activity promoted into slice $k+1$ before time $u+C_2t_k$ is absorbed. Hence, in this case, all activity sourced by $C$ has either been erased or absorbed before slice $k+1$ becomes active.
        
            Now suppose instead that
            \[
                u+C_2t_k\geq T_{k+1}.
            \]
            Since
            \[
                T_{k+1}=T_k+Mt_k,
            \]
            we have
            \[
                u\geq T_{k+1}-C_2t_k
                =
                T_k+(M-C_2)t_k
                >
                T_k.
            \]
            Thus, slice $k$, and all slices below it, are already active before the first detector event from $C$ is inserted. Moreover, $M$ can be chosen large enough so that $C$ cannot contain any faults on the initial time-like boundary. Therefore, the trajectory of $C$ agrees exactly with its ordinary streaming-decoder trajectory. 
        
            Combining the two cases, every isolated $k$-cluster with $k<k_L$ falls into one of two mutually exclusive groups: either it evolves exactly as it would under the ordinary streaming decoder, or the only deviation from the ordinary trajectory is that some subset of defects sourced by the cluster is absorbed by the rising wall. 
        
            The same induction over cluster level used in the proof of Theorem~\ref{thm:stretched-exponential-lifetime} now applies with ``erased'' replaced by ``erased or absorbed.'' Hence, after choosing the hierarchy parameters with $n$ sufficiently large, distinct clusters of level less than $k_L$ that occur before wall removal do not interact. Consequently, on the event
            \[
                \mathcal{B}_{\mathrm{wall}}^c,
            \]
            every wall-interacting cluster is well-defined. This proves that the wall-interacting cluster set $\mathcal{C}_w$, the maps $E_C$ and $F_C$, and hence $R_C$ and $R_w$, are all well-defined.
        
            Next, the bound on the probability of $\mathcal{B}_{\mathrm{wall}}$ follows immediately from the sparsity theorem.
        
            It remains to prove the identity
            \[
                \partial R_w
                =
                \psi_{\mathrm{true}}
                \oplus
                \psi(\cdot,T_{\mathrm{prep}}).
            \]
            For a wall-interacting cluster $C$, let $I_C$ denote the parity-check pattern due to first-round faults in $C$ on the initial time boundary, and let $A_C$ denote the parity pattern of defect absorptions sourced by $C$ on the rising wall. Thus
            \[
                A_C(i)=1
            \]
            if and only if an odd number of defects sourced by $C$ are absorbed at site $i$. By explicit calculation, one can check that the residual
            \[
                R_C=E_C\oplus F_C
            \]
            has boundary
            \[
                \partial R_C
                =
                I_C\oplus A_C.
            \]
        
            Summing over all wall-interacting clusters gives
            \[
                \partial R_w
                =
                \bigoplus_{C\in\mathcal{C}_w}\partial R_C
                =
                \bigoplus_{C\in\mathcal{C}_w} I_C
                \oplus
                \bigoplus_{C\in\mathcal{C}_w} A_C .
            \]
            Since $\psi(i,0)=\tilde s_0(i)$ and $\psi_{\mathrm{true}}$ is obtained from $\tilde s_0$ by removing the parity-check pattern due to first-round faults,
            \[
                \bigoplus_{C\in\mathcal{C}_w} I_C
                =
                \psi_{\mathrm{true}}
                \oplus
                \psi(\cdot,0).
            \]
            Also, by definition of the wall update rule, $\psi$ changes only when a defect is absorbed, and it changes exactly by the parity of the absorbed defects. Therefore
            \[
                \bigoplus_{C\in\mathcal{C}_w} A_C
                =
                \psi(\cdot,0)
                \oplus
                \psi(\cdot,T_{\mathrm{prep}}).
            \]
            Combining the last three observations yields
            \[
                \partial R_w
                =
                \psi_{\mathrm{true}}
                \oplus
                \psi(\cdot,T_{\mathrm{prep}}),
            \]
            as desired.
        
            Finally, we prove the stated logical-failure bound. On $\mathcal{B}_{\mathrm{wall}}^c$, the rising-wall procedure has the following deterministic effects. First, every wall-interacting cluster contributes the data-qubit error configuration $E_C$ and the correction-bit flip $F_C$, but each such combined contribution is removed from the failure diagnostic by construction. Second, all clusters that are not wall-interacting evolve exactly as they would under the ordinary streaming decoder.
        
            Therefore, conditioned on $\mathcal{B}_{\mathrm{wall}}^c$, a state-preparation failure can occur only if the spacetime history contains a cluster of level at least $k_L$. By the same union bound as in Theorem~\ref{thm:stretched-exponential-lifetime}, we have that for every $T\geq T_{\mathrm{prep}}$,
            \[
                \Pr[
                    \mathcal{B}_{\mathrm{wall}}^c
                    \cap
                    \mathsf{Fail}_{\mathrm{sp}}(T)
                ]
                \leq
                TC'\exp(-c_w L^\alpha),
            \]
            where, if necessary, we shrink $c_w$ so that the same constant applies to both bounds.
        \end{proof}

        With only $K=\Theta(\log\log L)$ slices, the same preparation argument does not imply a stretched-exponential bound on the failure probability. The issue is that the wall leaves the final slice after only $O(\polylog(L))$ time, so the argument above only guarantees that clusters of diameter at most $O(\polylog(L))$ have a benign effect.

        Nevertheless, one can obtain a preparation protocol with similar performance guarantees by performing a modest amount of postselection. Conditioned on postselection succeeding, this protocol obeys the same type of stretched-exponential failure-probability bound as Theorem~\ref{thm:state-preparation-rising-wall}. Let $T_K$ be the time at which the wall would normally leave the auxiliary slice $K$. For $K=\Theta(\log\log L)$, this time is $O(\polylog(L))$. Suppose that instead we keep the auxiliary wall-slice $K$ absorbing until time
        \[
            T_{\mathrm{prep}}=T_K+A L
        \]
        where $A$ is a sufficiently large constant, and only then turn off the wall. Furthermore, suppose that we accept the preparation if and only if no defect candidate is absorbed in the auxiliary slice during the interval
        \[
            [T_K,T_{\mathrm{prep}}].
        \]

        Conditioned on this event and no $k$-clusters with $k \geq k_L$ occurring during the state-preparation procedure, the wall has not partially absorbed any cluster that could influence the subsequent ordinary streaming dynamics. Thus, informally speaking, after the wall is turned off, the decoder behaves essentially identically to the ordinary streaming decoder. By the sparsity theorem and linear cluster erosion, the probability of not accepting the preparation is inverse-superpolynomial but super-inverse-stretched-exponential in $L$. Specifically, for an appropriate choice of $K=\Theta(\log\log L)$, it is at most
        \[
            \exp[-\Omega(\log^\gamma(L))]
        \]
        for some $\gamma > 1$. Thus, the $\polyloglog(L)$-overhead version of our decoder is capable of performing state-preparation and achieving a stretched-exponential memory lifetime with modest postselection. These arguments can be formalized in a straightforward manner, but we will not do so here.\footnote{We note in passing that a variant of this construction may allow state preparation without post-selection using only $\polyloglog(L)$ per-site classical resources. Specifically, one could slowly remove the absorbing region on the final slice row by row, and then introduce additional drifting motion to move defects on the back wall toward this shrinking absorbing region.}
    
    \subsection{State injection}\label{subsec:online-state-injection}

        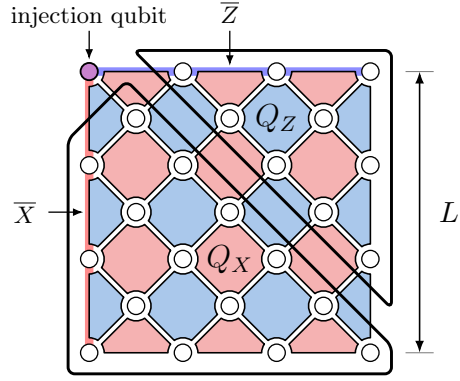
\begin{figure}[t]\centering
   
            \adjustbox{max width=\textwidth}{%
            \tikzsetnextfilename{injection-patch}
            \begin{tikzpicture}[scale=0.62]

            % ---- logical-operator wires (under the stabilisers; peek through the bites) -
            \draw[zstripe] (0,6) -- (6,6);
            \draw[xstripe] (0,6) -- (0,0);

            % ---- stabiliser fills (over the wires; outlines stroked later) ------------
            \fill[xstab] (0.7373,1.1827) -- (0.1827,1.7373) arc[start angle=-55.1821, end angle=-90, radius=0.32] -- (0,0.32) arc[start angle=90, end angle=55.1821, radius=0.32] -- (0.7373,0.8173) arc[start angle=-145.1821, end angle=-214.8179, radius=0.32] -- cycle;
            \fill[xstab] (0.7373,3.1827) -- (0.1827,3.7373) arc[start angle=-55.1821, end angle=-90, radius=0.32] -- (0,2.32) arc[start angle=90, end angle=55.1821, radius=0.32] -- (0.7373,2.8173) arc[start angle=-145.1821, end angle=-214.8179, radius=0.32] -- cycle;
            \fill[xstab] (0.7373,5.1827) -- (0.1827,5.7373) arc[start angle=-55.1821, end angle=-90, radius=0.32] -- (0,4.32) arc[start angle=90, end angle=55.1821, radius=0.32] -- (0.7373,4.8173) arc[start angle=-145.1821, end angle=-214.8179, radius=0.32] -- cycle;
            \fill[zstab] (1.7373,0.1827) -- (1.1827,0.7373) arc[start angle=-55.1821, end angle=-124.8179, radius=0.32] -- (0.2627,0.1827) arc[start angle=34.8179, end angle=-0, radius=0.32] -- (1.68,0) arc[start angle=180, end angle=145.1821, radius=0.32] -- cycle;
            \fill[zstab] (1.7373,2.1827) -- (1.1827,2.7373) arc[start angle=-55.1821, end angle=-124.8179, radius=0.32] -- (0.2627,2.1827) arc[start angle=34.8179, end angle=-34.8179, radius=0.32] -- (0.8173,1.2627) arc[start angle=124.8179, end angle=55.1821, radius=0.32] -- (1.7373,1.8173) arc[start angle=-145.1821, end angle=-214.8179, radius=0.32] -- cycle;
            \fill[zstab] (1.7373,4.1827) -- (1.1827,4.7373) arc[start angle=-55.1821, end angle=-124.8179, radius=0.32] -- (0.2627,4.1827) arc[start angle=34.8179, end angle=-34.8179, radius=0.32] -- (0.8173,3.2627) arc[start angle=124.8179, end angle=55.1821, radius=0.32] -- (1.7373,3.8173) arc[start angle=-145.1821, end angle=-214.8179, radius=0.32] -- cycle;
            \fill[zstab] (1.68,6) -- (0.32,6) arc[start angle=0, end angle=-34.8179, radius=0.32] -- (0.8173,5.2627) arc[start angle=124.8179, end angle=55.1821, radius=0.32] -- (1.7373,5.8173) arc[start angle=-145.1821, end angle=-180, radius=0.32] -- cycle;
            \fill[xstab] (2.7373,1.1827) -- (2.1827,1.7373) arc[start angle=-55.1821, end angle=-124.8179, radius=0.32] -- (1.2627,1.1827) arc[start angle=34.8179, end angle=-34.8179, radius=0.32] -- (1.8173,0.2627) arc[start angle=124.8179, end angle=55.1821, radius=0.32] -- (2.7373,0.8173) arc[start angle=-145.1821, end angle=-214.8179, radius=0.32] -- cycle;
            \fill[xstab] (2.7373,3.1827) -- (2.1827,3.7373) arc[start angle=-55.1821, end angle=-124.8179, radius=0.32] -- (1.2627,3.1827) arc[start angle=34.8179, end angle=-34.8179, radius=0.32] -- (1.8173,2.2627) arc[start angle=124.8179, end angle=55.1821, radius=0.32] -- (2.7373,2.8173) arc[start angle=-145.1821, end angle=-214.8179, radius=0.32] -- cycle;
            \fill[xstab] (2.7373,5.1827) -- (2.1827,5.7373) arc[start angle=-55.1821, end angle=-124.8179, radius=0.32] -- (1.2627,5.1827) arc[start angle=34.8179, end angle=-34.8179, radius=0.32] -- (1.8173,4.2627) arc[start angle=124.8179, end angle=55.1821, radius=0.32] -- (2.7373,4.8173) arc[start angle=-145.1821, end angle=-214.8179, radius=0.32] -- cycle;
            \fill[zstab] (3.7373,0.1827) -- (3.1827,0.7373) arc[start angle=-55.1821, end angle=-124.8179, radius=0.32] -- (2.2627,0.1827) arc[start angle=34.8179, end angle=-0, radius=0.32] -- (3.68,0) arc[start angle=180, end angle=145.1821, radius=0.32] -- cycle;
            \fill[zstab] (3.7373,2.1827) -- (3.1827,2.7373) arc[start angle=-55.1821, end angle=-124.8179, radius=0.32] -- (2.2627,2.1827) arc[start angle=34.8179, end angle=-34.8179, radius=0.32] -- (2.8173,1.2627) arc[start angle=124.8179, end angle=55.1821, radius=0.32] -- (3.7373,1.8173) arc[start angle=-145.1821, end angle=-214.8179, radius=0.32] -- cycle;
            \fill[zstab] (3.7373,4.1827) -- (3.1827,4.7373) arc[start angle=-55.1821, end angle=-124.8179, radius=0.32] -- (2.2627,4.1827) arc[start angle=34.8179, end angle=-34.8179, radius=0.32] -- (2.8173,3.2627) arc[start angle=124.8179, end angle=55.1821, radius=0.32] -- (3.7373,3.8173) arc[start angle=-145.1821, end angle=-214.8179, radius=0.32] -- cycle;
            \fill[zstab] (3.68,6) -- (2.32,6) arc[start angle=0, end angle=-34.8179, radius=0.32] -- (2.8173,5.2627) arc[start angle=124.8179, end angle=55.1821, radius=0.32] -- (3.7373,5.8173) arc[start angle=-145.1821, end angle=-180, radius=0.32] -- cycle;
            \fill[xstab] (4.7373,1.1827) -- (4.1827,1.7373) arc[start angle=-55.1821, end angle=-124.8179, radius=0.32] -- (3.2627,1.1827) arc[start angle=34.8179, end angle=-34.8179, radius=0.32] -- (3.8173,0.2627) arc[start angle=124.8179, end angle=55.1821, radius=0.32] -- (4.7373,0.8173) arc[start angle=-145.1821, end angle=-214.8179, radius=0.32] -- cycle;
            \fill[xstab] (4.7373,3.1827) -- (4.1827,3.7373) arc[start angle=-55.1821, end angle=-124.8179, radius=0.32] -- (3.2627,3.1827) arc[start angle=34.8179, end angle=-34.8179, radius=0.32] -- (3.8173,2.2627) arc[start angle=124.8179, end angle=55.1821, radius=0.32] -- (4.7373,2.8173) arc[start angle=-145.1821, end angle=-214.8179, radius=0.32] -- cycle;
            \fill[xstab] (4.7373,5.1827) -- (4.1827,5.7373) arc[start angle=-55.1821, end angle=-124.8179, radius=0.32] -- (3.2627,5.1827) arc[start angle=34.8179, end angle=-34.8179, radius=0.32] -- (3.8173,4.2627) arc[start angle=124.8179, end angle=55.1821, radius=0.32] -- (4.7373,4.8173) arc[start angle=-145.1821, end angle=-214.8179, radius=0.32] -- cycle;
            \fill[zstab] (5.7373,0.1827) -- (5.1827,0.7373) arc[start angle=-55.1821, end angle=-124.8179, radius=0.32] -- (4.2627,0.1827) arc[start angle=34.8179, end angle=-0, radius=0.32] -- (5.68,0) arc[start angle=180, end angle=145.1821, radius=0.32] -- cycle;
            \fill[zstab] (5.7373,2.1827) -- (5.1827,2.7373) arc[start angle=-55.1821, end angle=-124.8179, radius=0.32] -- (4.2627,2.1827) arc[start angle=34.8179, end angle=-34.8179, radius=0.32] -- (4.8173,1.2627) arc[start angle=124.8179, end angle=55.1821, radius=0.32] -- (5.7373,1.8173) arc[start angle=-145.1821, end angle=-214.8179, radius=0.32] -- cycle;
            \fill[zstab] (5.7373,4.1827) -- (5.1827,4.7373) arc[start angle=-55.1821, end angle=-124.8179, radius=0.32] -- (4.2627,4.1827) arc[start angle=34.8179, end angle=-34.8179, radius=0.32] -- (4.8173,3.2627) arc[start angle=124.8179, end angle=55.1821, radius=0.32] -- (5.7373,3.8173) arc[start angle=-145.1821, end angle=-214.8179, radius=0.32] -- cycle;
            \fill[zstab] (5.68,6) -- (4.32,6) arc[start angle=0, end angle=-34.8179, radius=0.32] -- (4.8173,5.2627) arc[start angle=124.8179, end angle=55.1821, radius=0.32] -- (5.7373,5.8173) arc[start angle=-145.1821, end angle=-180, radius=0.32] -- cycle;
            \fill[xstab] (5.8173,1.7373) -- (5.2627,1.1827) arc[start angle=34.8179, end angle=-34.8179, radius=0.32] -- (5.8173,0.2627) arc[start angle=124.8179, end angle=90, radius=0.32] -- (6,1.68) arc[start angle=-90, end angle=-124.8179, radius=0.32] -- cycle;
            \fill[xstab] (5.8173,3.7373) -- (5.2627,3.1827) arc[start angle=34.8179, end angle=-34.8179, radius=0.32] -- (5.8173,2.2627) arc[start angle=124.8179, end angle=90, radius=0.32] -- (6,3.68) arc[start angle=-90, end angle=-124.8179, radius=0.32] -- cycle;
            \fill[xstab] (5.8173,5.7373) -- (5.2627,5.1827) arc[start angle=34.8179, end angle=-34.8179, radius=0.32] -- (5.8173,4.2627) arc[start angle=124.8179, end angle=90, radius=0.32] -- (6,5.68) arc[start angle=-90, end angle=-124.8179, radius=0.32] -- cycle;

            % ---- stabiliser outlines (black) ------------------------------------------
            \draw[stabline] (0.7373,1.1827) -- (0.1827,1.7373) arc[start angle=-55.1821, end angle=-90, radius=0.32] -- (0,0.32) arc[start angle=90, end angle=55.1821, radius=0.32] -- (0.7373,0.8173) arc[start angle=-145.1821, end angle=-214.8179, radius=0.32] -- cycle;
            \draw[stabline] (0.7373,3.1827) -- (0.1827,3.7373) arc[start angle=-55.1821, end angle=-90, radius=0.32] -- (0,2.32) arc[start angle=90, end angle=55.1821, radius=0.32] -- (0.7373,2.8173) arc[start angle=-145.1821, end angle=-214.8179, radius=0.32] -- cycle;
            \draw[stabline] (0.7373,5.1827) -- (0.1827,5.7373) arc[start angle=-55.1821, end angle=-90, radius=0.32] -- (0,4.32) arc[start angle=90, end angle=55.1821, radius=0.32] -- (0.7373,4.8173) arc[start angle=-145.1821, end angle=-214.8179, radius=0.32] -- cycle;
            \draw[stabline] (1.7373,0.1827) -- (1.1827,0.7373) arc[start angle=-55.1821, end angle=-124.8179, radius=0.32] -- (0.2627,0.1827) arc[start angle=34.8179, end angle=-0, radius=0.32] -- (1.68,0) arc[start angle=180, end angle=145.1821, radius=0.32] -- cycle;
            \draw[stabline] (1.7373,2.1827) -- (1.1827,2.7373) arc[start angle=-55.1821, end angle=-124.8179, radius=0.32] -- (0.2627,2.1827) arc[start angle=34.8179, end angle=-34.8179, radius=0.32] -- (0.8173,1.2627) arc[start angle=124.8179, end angle=55.1821, radius=0.32] -- (1.7373,1.8173) arc[start angle=-145.1821, end angle=-214.8179, radius=0.32] -- cycle;
            \draw[stabline] (1.7373,4.1827) -- (1.1827,4.7373) arc[start angle=-55.1821, end angle=-124.8179, radius=0.32] -- (0.2627,4.1827) arc[start angle=34.8179, end angle=-34.8179, radius=0.32] -- (0.8173,3.2627) arc[start angle=124.8179, end angle=55.1821, radius=0.32] -- (1.7373,3.8173) arc[start angle=-145.1821, end angle=-214.8179, radius=0.32] -- cycle;
            \draw[stabline] (1.68,6) -- (0.32,6) arc[start angle=0, end angle=-34.8179, radius=0.32] -- (0.8173,5.2627) arc[start angle=124.8179, end angle=55.1821, radius=0.32] -- (1.7373,5.8173) arc[start angle=-145.1821, end angle=-180, radius=0.32] -- cycle;
            \draw[stabline] (2.7373,1.1827) -- (2.1827,1.7373) arc[start angle=-55.1821, end angle=-124.8179, radius=0.32] -- (1.2627,1.1827) arc[start angle=34.8179, end angle=-34.8179, radius=0.32] -- (1.8173,0.2627) arc[start angle=124.8179, end angle=55.1821, radius=0.32] -- (2.7373,0.8173) arc[start angle=-145.1821, end angle=-214.8179, radius=0.32] -- cycle;
            \draw[stabline] (2.7373,3.1827) -- (2.1827,3.7373) arc[start angle=-55.1821, end angle=-124.8179, radius=0.32] -- (1.2627,3.1827) arc[start angle=34.8179, end angle=-34.8179, radius=0.32] -- (1.8173,2.2627) arc[start angle=124.8179, end angle=55.1821, radius=0.32] -- (2.7373,2.8173) arc[start angle=-145.1821, end angle=-214.8179, radius=0.32] -- cycle;
            \draw[stabline] (2.7373,5.1827) -- (2.1827,5.7373) arc[start angle=-55.1821, end angle=-124.8179, radius=0.32] -- (1.2627,5.1827) arc[start angle=34.8179, end angle=-34.8179, radius=0.32] -- (1.8173,4.2627) arc[start angle=124.8179, end angle=55.1821, radius=0.32] -- (2.7373,4.8173) arc[start angle=-145.1821, end angle=-214.8179, radius=0.32] -- cycle;
            \draw[stabline] (3.7373,0.1827) -- (3.1827,0.7373) arc[start angle=-55.1821, end angle=-124.8179, radius=0.32] -- (2.2627,0.1827) arc[start angle=34.8179, end angle=-0, radius=0.32] -- (3.68,0) arc[start angle=180, end angle=145.1821, radius=0.32] -- cycle;
            \draw[stabline] (3.7373,2.1827) -- (3.1827,2.7373) arc[start angle=-55.1821, end angle=-124.8179, radius=0.32] -- (2.2627,2.1827) arc[start angle=34.8179, end angle=-34.8179, radius=0.32] -- (2.8173,1.2627) arc[start angle=124.8179, end angle=55.1821, radius=0.32] -- (3.7373,1.8173) arc[start angle=-145.1821, end angle=-214.8179, radius=0.32] -- cycle;
            \draw[stabline] (3.7373,4.1827) -- (3.1827,4.7373) arc[start angle=-55.1821, end angle=-124.8179, radius=0.32] -- (2.2627,4.1827) arc[start angle=34.8179, end angle=-34.8179, radius=0.32] -- (2.8173,3.2627) arc[start angle=124.8179, end angle=55.1821, radius=0.32] -- (3.7373,3.8173) arc[start angle=-145.1821, end angle=-214.8179, radius=0.32] -- cycle;
            \draw[stabline] (3.68,6) -- (2.32,6) arc[start angle=0, end angle=-34.8179, radius=0.32] -- (2.8173,5.2627) arc[start angle=124.8179, end angle=55.1821, radius=0.32] -- (3.7373,5.8173) arc[start angle=-145.1821, end angle=-180, radius=0.32] -- cycle;
            \draw[stabline] (4.7373,1.1827) -- (4.1827,1.7373) arc[start angle=-55.1821, end angle=-124.8179, radius=0.32] -- (3.2627,1.1827) arc[start angle=34.8179, end angle=-34.8179, radius=0.32] -- (3.8173,0.2627) arc[start angle=124.8179, end angle=55.1821, radius=0.32] -- (4.7373,0.8173) arc[start angle=-145.1821, end angle=-214.8179, radius=0.32] -- cycle;
            \draw[stabline] (4.7373,3.1827) -- (4.1827,3.7373) arc[start angle=-55.1821, end angle=-124.8179, radius=0.32] -- (3.2627,3.1827) arc[start angle=34.8179, end angle=-34.8179, radius=0.32] -- (3.8173,2.2627) arc[start angle=124.8179, end angle=55.1821, radius=0.32] -- (4.7373,2.8173) arc[start angle=-145.1821, end angle=-214.8179, radius=0.32] -- cycle;
            \draw[stabline] (4.7373,5.1827) -- (4.1827,5.7373) arc[start angle=-55.1821, end angle=-124.8179, radius=0.32] -- (3.2627,5.1827) arc[start angle=34.8179, end angle=-34.8179, radius=0.32] -- (3.8173,4.2627) arc[start angle=124.8179, end angle=55.1821, radius=0.32] -- (4.7373,4.8173) arc[start angle=-145.1821, end angle=-214.8179, radius=0.32] -- cycle;
            \draw[stabline] (5.7373,0.1827) -- (5.1827,0.7373) arc[start angle=-55.1821, end angle=-124.8179, radius=0.32] -- (4.2627,0.1827) arc[start angle=34.8179, end angle=-0, radius=0.32] -- (5.68,0) arc[start angle=180, end angle=145.1821, radius=0.32] -- cycle;
            \draw[stabline] (5.7373,2.1827) -- (5.1827,2.7373) arc[start angle=-55.1821, end angle=-124.8179, radius=0.32] -- (4.2627,2.1827) arc[start angle=34.8179, end angle=-34.8179, radius=0.32] -- (4.8173,1.2627) arc[start angle=124.8179, end angle=55.1821, radius=0.32] -- (5.7373,1.8173) arc[start angle=-145.1821, end angle=-214.8179, radius=0.32] -- cycle;
            \draw[stabline] (5.7373,4.1827) -- (5.1827,4.7373) arc[start angle=-55.1821, end angle=-124.8179, radius=0.32] -- (4.2627,4.1827) arc[start angle=34.8179, end angle=-34.8179, radius=0.32] -- (4.8173,3.2627) arc[start angle=124.8179, end angle=55.1821, radius=0.32] -- (5.7373,3.8173) arc[start angle=-145.1821, end angle=-214.8179, radius=0.32] -- cycle;
            \draw[stabline] (5.68,6) -- (4.32,6) arc[start angle=0, end angle=-34.8179, radius=0.32] -- (4.8173,5.2627) arc[start angle=124.8179, end angle=55.1821, radius=0.32] -- (5.7373,5.8173) arc[start angle=-145.1821, end angle=-180, radius=0.32] -- cycle;
            \draw[stabline] (5.8173,1.7373) -- (5.2627,1.1827) arc[start angle=34.8179, end angle=-34.8179, radius=0.32] -- (5.8173,0.2627) arc[start angle=124.8179, end angle=90, radius=0.32] -- (6,1.68) arc[start angle=-90, end angle=-124.8179, radius=0.32] -- cycle;
            \draw[stabline] (5.8173,3.7373) -- (5.2627,3.1827) arc[start angle=34.8179, end angle=-34.8179, radius=0.32] -- (5.8173,2.2627) arc[start angle=124.8179, end angle=90, radius=0.32] -- (6,3.68) arc[start angle=-90, end angle=-124.8179, radius=0.32] -- cycle;
            \draw[stabline] (5.8173,5.7373) -- (5.2627,5.1827) arc[start angle=34.8179, end angle=-34.8179, radius=0.32] -- (5.8173,4.2627) arc[start angle=124.8179, end angle=90, radius=0.32] -- (6,5.68) arc[start angle=-90, end angle=-124.8179, radius=0.32] -- cycle;

            % ---- data qubits -----------------------------------------------------------
            \node[dataq] at (0,0) {};
            \node[dataq] at (0,2) {};
            \node[dataq] at (0,4) {};
            \node[dataq] at (1,1) {};
            \node[dataq] at (1,3) {};
            \node[dataq] at (1,5) {};
            \node[dataq] at (2,0) {};
            \node[dataq] at (2,2) {};
            \node[dataq] at (2,4) {};
            \node[dataq] at (2,6) {};
            \node[dataq] at (3,1) {};
            \node[dataq] at (3,3) {};
            \node[dataq] at (3,5) {};
            \node[dataq] at (4,0) {};
            \node[dataq] at (4,2) {};
            \node[dataq] at (4,4) {};
            \node[dataq] at (4,6) {};
            \node[dataq] at (5,1) {};
            \node[dataq] at (5,3) {};
            \node[dataq] at (5,5) {};
            \node[dataq] at (6,0) {};
            \node[dataq] at (6,2) {};
            \node[dataq] at (6,4) {};
            \node[dataq] at (6,6) {};
            \node[magicq] (mq) at (0,6) {};

            % ---- region outlines -------------------------------------------------------
            \draw[rgnoutline] (0.95,6.45) -- (6.45,6.45) -- (6.45,0.95) -- cycle;
            \draw[rgnoutline] (0.8,5.8) -- (6.45,0.15) -- (6.45,-0.45) -- (-0.45,-0.45) -- (-0.45,4.55) -- cycle;

            % ---- region labels (placed on a stabiliser centre, not a qubit) -----------
            \node[arealbl] at (4,5) {$Q_Z$};
            \node[arealbl] at (3,2) {$Q_X$};

            % ---- dimension arrow (L = full lattice height) ----------------------------
            \draw[thinguide] (6.75,0) -- (7.35,0);
            \draw[thinguide] (6.75,6) -- (7.35,6);
            \draw[dimarrow] (7.05,0) -- (7.05,6);
            \node[anchor=west] at (7.25,3) {$L$};

            % ---- text labels -----------------------------------------------------------
            \node[font=\footnotesize, anchor=base] (mslbl) at (0,7.04) {injection qubit};
            \draw[-{Latex[length=4pt]}, line width=0.5pt] (0,6.87) -- (0,6.25);
            \node[font=\footnotesize, anchor=base] (zlbl) at (3,7.04) {$\overline{Z}$};
            \draw[-{Latex[length=4pt]}, line width=0.5pt] (3,6.87) -- (3,6.12);
            \node[font=\footnotesize, anchor=east] (xlbl) at (-0.92,3) {$\overline{X}$};
            \draw[-{Latex[length=4pt]}, line width=0.5pt] (-0.87,3) -- (-0.12,3);

            \end{tikzpicture}%
            }
            \caption{Geometry of the surface-code state-injection procedure. Qubits (white circles) in the region $Q_X$ are prepared in the state $\ket{+}$, while qubits in the region $Q_Z$ are prepared in the state $\ket{0}$. The injection qubit $q_\star$ is the shared endpoint of the two weight-$L$ logical operators $\overline X = X_{q_\star} X_{\gamma_X}$ and $\overline Z = Z_{q_\star} Z_{\gamma_Z}$; it is prepared in the unencoded state $\ket{\psi}$. Blue shapes indicate $Z$-type stabilizers, and red shapes indicate $X$-type stabilizers.}
            \label{fig:injection-geometry}
        \end{figure}

        We now describe how to perform local decoding during state injection. Our injection protocol essentially follows that of Refs.~\cite{Mazurek_2014, Li_2015}. Unlike during state preparation, where ideal first-round stabilizer-measurement outcomes are deterministic throughout one stabilizer sector and intrinsically random throughout the other, during state injection each sector contains both stabilizers with deterministic initial measurement outcomes and stabilizers with random initial measurement outcomes.
        
        To accommodate this, the decoder uses a modified version of the state-preparation procedure. Specifically, it employs one rising absorbing region per sector, and in each sector, this region is restricted to sites with unknown initial stabilizer-measurement outcomes. We call each such region a frame region and refer to stabilizers in frame regions as frame stabilizers.

        We begin by specifying the injection procedure; the geometry of this procedure is illustrated in Fig.~\ref{fig:injection-geometry}. Let $Q$ denote the set of data qubits, let $q_\star$ denote the injection qubit, and let
        \[
            \overline X
            =
            X_{q_\star}X_{\gamma_X},
            \qquad
            \overline Z
            =
            Z_{q_\star}Z_{\gamma_Z}
        \]
        denote the two logical operators shown in Fig.~\ref{fig:injection-geometry}, where $\gamma_X$ and $\gamma_Z$ are the respective supports of these operators excluding $q_\star$. The diagonal line of qubits containing $q_\star$ partitions the remaining data qubits into two regions:
        \[
            Q_X
            \coloneqq
            \{q \in Q\setminus\{q_\star\} : q \text{ lies on or below the diagonal}\},
            \qquad
            Q_Z
            \coloneqq
            Q\setminus(Q_X\cup\{q_\star\}) .
        \]
        Finally, let $\ket{\psi}$ denote the unencoded state to be injected. The injection procedure is as follows:
        \begin{enumerate}
            \item Prepare $q_\star$ in the unencoded state $\ket{\psi}$. For now, we assume this preparation is perfect.
            \item Prepare every qubit in $Q_X$ in the state $\ket{+}$.
            \item Prepare every qubit in $Q_Z$ in the state $\ket{0}$.
            \item Begin measuring the surface code stabilizers at every time step, while applying the decoding procedure described below.
        \end{enumerate}

        Before describing the decoding procedure, we first show that the injection procedure encodes the unencoded state $\ket{\psi}$ as the corresponding logical state~$\ket{\bar \psi}$. Let
        \[
            \ket{\Omega}
            =
            \bigotimes_{q\in Q_X}\ket{+}_q
            \otimes
            \bigotimes_{q\in Q_Z}\ket{0}_q .
        \]
        By construction,
        \[
            X_{\gamma_X}\ket{\Omega}=\ket{\Omega},
            \qquad
            Z_{\gamma_Z}\ket{\Omega}=\ket{\Omega}.
        \]
        Hence, on the initial product state,
        \[
            \overline X
            =
            X_{q_\star}X_{\gamma_X}
            =
            X_{q_\star},
            \qquad
            \overline Z
            =
            Z_{q_\star}Z_{\gamma_Z}
            =
            Z_{q_\star},
        \]
        as desired.

We now describe the decoder for the $Z$-type stabilizer sector; the $X$-type stabilizer sector is decoded analogously. Let $\mathcal Z$ be the set of $Z$-type stabilizers, and define
        \[
            \mathcal Z^{\mathrm{det}} \coloneqq \{i\in\mathcal Z : \supp(i)\subseteq Q_Z\}, \qquad \mathcal Z^{\mathrm{fr}} \coloneqq \mathcal Z\setminus\mathcal Z^{\mathrm{det}}.
        \]
        Stabilizers in $\mathcal Z^{\mathrm{det}}$ have deterministic ideal first-round measurement outcome $0$, while stabilizers in $\mathcal Z^{\mathrm{fr}}$ are frame stabilizers.

        We define the first-round detector events to be
        \[
            \phi(i,0)
            =
            \begin{cases}
                \tilde s_0(i), & i\in\mathcal Z^{\mathrm{det}},\\
                0, & i\in\mathcal Z^{\mathrm{fr}},
            \end{cases}
        \]
        and for $t\geq 1$, we define them to be
        \[
            \phi(i,t)
            =
            \tilde s_t(i)\oplus \tilde s_{t-1}(i), \qquad i\in\mathcal Z.
        \]
        The interpretations of the initial frame-stabilizer measurements (frame interpretations) are stored in a register
        \[
            \psi(i,t)\in\{0,1\},
            \qquad
            i\in\mathcal Z^{\mathrm{fr}},
        \]
        which is initialized as
        \[
            \psi(i,0)
            =
                \tilde s_0(i), \qquad i\in\mathcal Z^{\mathrm{fr}}.
        \]
        The frame interpretation at $i$ is flipped only when an odd number of defects are absorbed at $i$ during a single update step.

        We next define the rising absorbing region. Let $M\geq 1$ be a constant and set
        \[
            T_k
            =
            M\sum_{j=0}^{k-1}t_j,
            \qquad
            k=0,\ldots,K .
        \]
        The injection time is
        \[
            T_{\mathrm{inj}}
            \coloneqq
            T_K
            =
            M\sum_{j=0}^{K-1}t_j.
        \]
        For each slice $k=0,\ldots,K-1$, define the active stabilizer set at time $t$ by
        \[
            \operatorname{Act}_k(t)
            \coloneqq
            \mathcal Z^{\mathrm{det}}
            \cup
            \begin{cases}
                \mathcal Z^{\mathrm{fr}}, & t\geq T_k,\\
                \varnothing, & t<T_k.
            \end{cases}
        \]
        Since $T_0=0$, slice $0$ is active everywhere from the beginning. A site $(k,i)$ with $i\notin\operatorname{Act}_k(t)$ is absorbing at time $t$; absorbing sites store trivial defects, messages, and timers, and are treated as trivial when read by neighboring update rules. During
        \[
            T_j\leq t<T_{j+1},
            \qquad
            j=0,\ldots,K-1,
        \]
        the frame region is active in slices $0,\ldots,j$ and absorbing in slices $j+1,\ldots,K-1$, while the deterministic region is active in every slice.

        We now specify the modified cellular-automaton rule in more detail. During the update from time $t$ to time $t+1$, a defect candidate written to $(k,i)$ is kept if
        \[
            i\in\operatorname{Act}_k(t+1),
        \]
        and is discarded otherwise. If a discarded candidate is a defect at $i\in\mathcal Z^{\mathrm{fr}}$, it is absorbed and flips the frame interpretation. Equivalently, if $P(i,t)$ is the parity of defect candidates absorbed at $i$ during the update from $t$ to $t+1$, then
        \[
            \psi(i,t+1)
            =
            \psi(i,t)\oplus P(i,t),
            \qquad
            i\in\mathcal Z^{\mathrm{fr}}.
        \]

        During injection, the update from $t$ to $t+1$ proceeds in three stages. We begin by fixing a splitting period $q_s$ and a drift period $q_d$.

        \emph{Stage 1: open-boundary splitting.}
        If $t\in q_s\Z$, each slice first performs the ordinary open-boundary splitting step, where sites $i\notin\operatorname{Act}_k(t+1)$ are treated as absorbing. As before, defects and messages shift toward the condensing boundaries, and messages and their timers are copied at the central cut rather than torn apart. Defects, messages, and timers translated out of the system or into an absorbing site are annihilated, with the correction channels updated accordingly. Additionally, for each frame stabilizer $i\in\mathcal Z^{\mathrm{fr}}$, $\psi(i,\cdot)$ is flipped exactly when an odd number of defects are absorbed at $i$ during this step. If $t\notin q_s\Z$, this stage is skipped.

        \emph{Stage 2: uniform drift toward the frame region.}
        If $t\in q_d\Z$, then we apply a uniform drift toward the frame region. During this step, sites $i\notin\operatorname{Act}_k(t + 1)$ are treated as absorbing. Let $\uvc u$ denote a unit vector normal to the condensing boundaries used by the splitting step, and let $\uvc v$ be the unit vector perpendicular to $\uvc u$ oriented from $\mathcal Z^{\mathrm{det}}$ toward $\mathcal Z^{\mathrm{fr}}$. Define
        \[
            D(i)
            =
            \begin{cases}
                i+\uvc v, & i+\uvc v\in\mathcal Z,\\
                i, & i+\uvc v\notin\mathcal Z .
            \end{cases}
        \]
        Every defect at an active site $(k,i)$ is moved to $(k,D(i))$. If $D(i)\neq i$, the correction bits are updated by the fixed local correction corresponding to moving the defect from $i$ to $D(i)$. Messages are moved by the same map $D$. Defect and message timers follow their respective defects and messages during this move. Along the boundary of sites that do not move, messages are combined channelwise by OR, and in each channel the resulting timer is the minimum of the incoming timers and any existing timer at the destination. For each frame stabilizer $i\in\mathcal Z^{\mathrm{fr}}$, $\psi(i,\cdot)$ is flipped exactly when an odd number of defects are absorbed at $i$ during this step. If $t\notin q_d\Z$, this stage is skipped.

        \emph{Stage 3: ordinary update.}
        Finally, we apply the ordinary surface-code streaming decoder update to the preprocessed configuration. The detector-event configuration $\phi(\cdot,t+1)$ is inserted as environmental defect candidates in the bottom slice. For $k<K-1$, a promoted defect from slice $k$ to slice $k+1$ is kept if the destination is active at output time $t+1$, and is otherwise absorbed.

        During injection, the final slice $K-1$ is temporarily treated as a timed slice. If a final-slice defect timer expires during an update with $t<T_{\mathrm{inj}}-1$, the defect is promoted to the auxiliary absorbing layer above the final slice. If this occurs at a frame stabilizer, the defect is absorbed and updates the frame interpretation. If this occurs at a deterministic stabilizer, a rejection flag is raised.

        On the final injection update, from $t=T_{\mathrm{inj}}-1$ to $t=T_{\mathrm{inj}}$, no final-slice promotion is performed. Instead, the final slice is updated using the ordinary untimed back-wall rule.

        If a rejection flag is raised during injection, the injection attempt is discarded and the decoder restarts the entire injection protocol. If no rejection flag is raised, the committed frame interpretation in the $Z$-type sector is
        \[
            \psi_{\mathrm{inj}}(i)
            \coloneqq
            \psi(i,T_{\mathrm{inj}}),
            \qquad
            i\in\mathcal Z^{\mathrm{fr}}.
        \]
        From time $T_{\mathrm{inj}}$ onward, final-slice timers are disabled, the injection-specific absorbing region and uniform drift are removed, and the ordinary surface-code streaming decoder continues.

        To make the notion of failure for the state-injection procedure well-defined, we use the following diagnostic experiment: inject the unencoded state $\ket{0}$, run the injection protocol in the $Z$-type stabilizer decoding sector, continue syndrome extraction until time $T\geq T_{\mathrm{inj}}$, and then perform destructive computational-basis readout of $\overline Z$. This experiment induces a spacetime noise history for the $Z$-type stabilizer decoding sector.
        
        We now introduce the injection analogues of Definitions~\ref{def:wall-residual} and~\ref{def:state-preparation-failure}. During the proof of the theorem below, we will show that all of the following definitions are well-defined.

        \begin{definition}[injection-affected clusters and injection residual]
            \label{def:injection-residual}
            Consider the spacetime noise history induced by the state-injection diagnostic. Write its clustering decomposition as
            \[
                N
                =
                \bigsqcup_{k,i} C_{k}^{(i)} .
            \]
            We say that a cluster occurs during injection if it contains at least one fault location whose time coordinate is less than or equal to $T_{\mathrm{inj}}$.

            Suppose no cluster of level at least $k_L$ occurs during injection. A cluster $C$ is called injection-affected if it occurs during injection and either
            \begin{enumerate}
                \item some defect or message sourced by $C$ is absorbed by a frame stabilizer in $\mathcal Z^{\mathrm{fr}}$, or
                \item $C$ contains a first-round fault that intersects a frame stabilizer in $\mathcal Z^{\mathrm{fr}}$.
            \end{enumerate}

            Let $\mathcal C_{\mathrm{inj}}$ denote the set of injection-affected clusters. For each $C\in\mathcal C_{\mathrm{inj}}$, let $E_C$ and $F_C$ be $\{0,1\}$-valued maps on data qubits. Here, $E_C$ is the modulo-two sum of data-qubit errors in $C$, and $F_C$ is the modulo-two sum of correction-bit flips produced by the decoder due to $C$. Define the residual associated with $C$ by
            \[
                R_C
                \coloneqq
                E_C
                \oplus
                F_C ,
            \]
            and define the injection residual by
            \[
                R_{\mathrm{inj}}
                \coloneqq
                \bigoplus_{C\in\mathcal C_{\mathrm{inj}}} R_C .
            \]
            If $\mathcal C_{\mathrm{inj}}=\varnothing$, then $R_{\mathrm{inj}}=0$.
        \end{definition}

        \begin{definition}[state-injection failure]
            \label{def:state-injection-failure}
            Consider the spacetime fault history from the state-injection diagnostic.

            Suppose no cluster of level at least $k_L$ occurs during injection. Let $E$ be the modulo-two sum of all bit-flip errors on data qubits. We define the injection-adjusted error as
            \[
                E_{\mathrm a}
                \coloneqq
                E
                \oplus
                \bigoplus_{C\in\mathcal C_{\mathrm{inj}}} E_C .
            \]

            We say that a ``logical failure after state injection'' has occurred if the decoder has not terminated by the end of the cleanout, or if the decoder has terminated but
            \[
                F_{\mathrm a}
                \oplus
                E_{\mathrm a}
            \]
            is a non-trivial logical operator, where $F$ is the final correction produced by the decoder in the $Z$-type stabilizer decoding sector and
            \[
                F_{\mathrm a}
                \coloneqq
                F
                \oplus
                \bigoplus_{C\in\mathcal C_{\mathrm{inj}}} F_C
            \]
            is the injection-adjusted correction.
        \end{definition}

        \begin{theorem}[state injection with a rising absorbing region]
            \label{thm:state-injection-rising-absorber}
            There exist constants
            \[
                p_*,c,c_r,C',\alpha,t_0>0,
            \]
            sparsity-theorem parameters $(\beta,\gamma,n)$ satisfying inequalities~\eqref{paramineqs}, integers $q,q_s>0$, an integer drift period $q_d>0$, and constants $M \geq 1$ and $c_{\mathrm{inj}}>0$ such that both Theorem~\ref{thm:stretched-exponential-lifetime} and the following hold with these constants.

            Consider the surface-code streaming decoder running the state-injection protocol in the $Z$-type stabilizer decoding sector. Suppose the bit-flip noise on data qubits and on $Z$-type stabilizer-measurement bits is $p$-bounded with $p<p_* / 2$. Suppose also that
            \[
                K = k_L,
            \]
            and that the cleanout step proportionality constant is $c_r$.

            Let $\mathcal B_{\mathrm{inj}}$ be the event that the diagnostic spacetime history contains a cluster of level at least $k_L$ that occurs during injection. Then,
            \[
                \Pr[\mathcal B_{\mathrm{inj}}]
                \leq
                C'\exp(-c_{\mathrm{inj}}L^\alpha).
            \]

            On the complementary event $\mathcal B_{\mathrm{inj}}^c$, no rejection flag is raised and Definitions~\ref{def:injection-residual} and~\ref{def:state-injection-failure} are well-defined. Moreover,
            \[
                \partial R_{\mathrm{inj}}
                =
                \psi_{\mathrm{true}}
                \oplus
                \psi_{\mathrm{inj}},
            \]
            where $\psi_{\mathrm{true}}$ is the ideal first-round frame on $\mathcal Z^{\mathrm{fr}}$ with first-round faults removed.

            Finally, let $\mathsf{Fail}_{\mathrm{inj}}(T)$ be the event that a logical failure after state injection occurs in the sense of Definition~\ref{def:state-injection-failure}. Then, for every $T\geq T_{\mathrm{inj}}$,
            \[
                \Pr[
                    \mathcal B_{\mathrm{inj}}^c
                    \cap
                    \mathsf{Fail}_{\mathrm{inj}}(T)
                ]
                \leq
                TC'\exp(-c_{\mathrm{inj}}L^\alpha).
            \]
        \end{theorem}

        Again, the diagnostic experiment is only a bookkeeping device to make the failure event in Theorem~\ref{thm:state-injection-rising-absorber} well-defined. For the actual surface-code state-injection protocol, the same type of theorem applies to both stabilizer sectors. In particular, when no $k$-cluster with $k \geq k_L$ occurs in either sector during injection, the decoder successfully commits to a consistent initial frame, and in each sector, its discrepancy from the true initial frame is given by the boundary of the injection residual in that particular sector. Furthermore, a $k$-cluster with $k \geq k_L$ can occur during injection with probability at most inverse-stretched-exponential in $L$. After time $T_{\mathrm{inj}}$, the rising absorbing regions have been completely removed in both sectors, and the decoder operates according to the ordinary streaming decoder dynamics. Thus, conditioned on no large clusters occurring during injection, the injected state is just as protected as in the usual streaming decoder setting.

        \begin{proof}
            Let all the relevant constants be defined as in Theorem~\ref{thm:stretched-exponential-lifetime}.
            
            We begin by proving the following estimate: there are constants $C_0\geq 1$ and $M>C_0$ and a choice of drift period $q_d$ such that the following holds. Suppose the spacetime history consists of a single $k$-cluster $C$ with $k<k_L$, and suppose $C$ sources at least one detector event. Let $u$ be the first time at which a detector event sourced by $C$ is inserted into the decoder. Then, under the injection-modified dynamics, by time $u+C_0t_k$ all activity sourced by $C$ has either been erased; absorbed by a frame stabilizer in slices $0, \ldots, k$; or removed from slices $0, \ldots, k$ by promotion into an absorbing frame-stabilizer site in slice $k+1$. If $k=K-1$, then ``slice $k+1$'' means the auxiliary absorbing layer above the final slice.

            We now prove this estimate. Fix $q_d<q_s$. Suppose first that the cluster does not interact with the rising absorbing region. The only relevant difference from the ordinary dynamics is the drift toward the frame region. This drift can at most translate the cluster and compress it against a non-condensing boundary, and it follows by a slightly modified version of the usual arguments that neither effect prevents linear cluster erosion from holding; this modification may require changing some of the linear erosion constants, but this is harmless since $n$, $t_0$, and $k_L$ can be adjusted to account for these changes. Therefore, in this case, all activity sourced by $C$ is erased by time $u+C_0t_k$ for some constant $C_0\geq 1$.

            Suppose instead that the cluster interacts with the rising absorbing region. By the linear cluster erosion argument from the previous paragraph, the cluster must initially have been within distance $O(w_k)$ of the rising absorbing region. Since $q_d<q_s$, the cluster has an average drift toward the frame region. Therefore, in the worst case, every defect in the cluster that is not pair-annihilated or absorbed by the condensing boundary is either absorbed by a frame stabilizer within time $O(w_k)$ or moved into the frame region within time $O(w_k)$, where it remains until promotion or annihilation. Consequently, $C_0$ can be chosen so that, by time $u+C_0t_k$, every defect sourced by $C$ that has not been erased or absorbed by a frame stabilizer in slices $0, \ldots, k$ has been alive in slice $k$ for at least $t_k$ time steps and has reached that age while located at a frame-stabilizer site.

            Suppose that some defect sourced by $C$ has not been erased or absorbed by a frame stabilizer in slices $0, \ldots, k$ by time $u + C_0 t_k$. By the previous paragraph, every such defect has been alive in slice $k$ for at least $t_k$ time steps and has reached that age while located at a frame-stabilizer site. We now prove that by time $u + C_0 t_k$, every such defect has already been promoted into an absorbing frame-stabilizer site in slice $k+1$.
            
            Suppose first that
            \[
                u+C_0t_k<T_{k+1}.
            \]
            Then, at time $u + C_0 t_k$, slice $k + 1$ is still absorbing, so by this time, all defects that have not been erased or absorbed by a frame stabilizer in slices $0, \ldots, k$ have been promoted into an absorbing frame-stabilizer site.

            Now suppose instead that
            \[
                u+C_0t_k\geq T_{k+1}.
            \]
            Since $T_{k+1}=T_k+Mt_k$, we have
            \[
                u
                \geq
                T_{k+1}-C_0t_k
                =
                T_k+(M-C_0)t_k
                >
                T_k.
            \]
            Thus, slice $k$, and all slices below it, are already active before the first detector event from $C$ is inserted. Moreover, $M$ can be chosen large enough so that $C$ cannot contain any faults on the initial time-like boundary. Therefore, the trajectory of $C$ agrees with its ordinary streaming-decoder trajectory up to a harmless drift toward the frame region, so all activity sourced by $C$ is erased by time $u + C_0 t_k$ without ever being promoted beyond slice $k$.
            
            The arguments above show that every isolated $k$-cluster with $k<k_L$ falls into one of two mutually exclusive groups: up to a harmless drift toward the frame region, either it evolves exactly as it would under the ordinary streaming decoder, or the only deviation from its ordinary trajectory is that some subset of defects sourced by the cluster is absorbed by frame stabilizers.
              
            The same level-by-level induction used in the proof of Theorem~\ref{thm:state-preparation-rising-wall} now applies, with the rising wall replaced by the rising absorbing region. Consequently, on the event $\mathcal{B}_{\mathrm{inj}}^c$, every injection-affected cluster is well-defined. This proves that the injection-affected cluster set $\mathcal{C}_{\mathrm{inj}}$, the maps $E_C$ and $F_C$, and hence $R_C$ and $R_{\mathrm{inj}}$, are all well-defined. Moreover, on this event, no rejection flag is raised.
        
            Next, the bound on the probability of $\mathcal{B}_{\mathrm{inj}}$ follows immediately from the sparsity theorem.
            
            The remainder of the proof then proceeds as in the proof of Theorem~\ref{thm:state-preparation-rising-wall}.
        \end{proof}

        Although the above theorem guarantees that the injected state is protected after injection, it does not bound the probability that a logical error occurs during the injection process itself. We now bound this probability in the theorem below.

        \begin{theorem}[bounded injection error rate]
            \label{thm:state-injection-bounded-error-rate}
            Work in the setting of Theorem~\ref{thm:state-injection-rising-absorber} and suppose all the constants are such that the conclusions of the theorem hold. Let $\mathsf{Wrong}_{\mathrm{inj}}$ be the event that the accepted injected state at time $T_{\mathrm{inj}}$ differs from the ideal encoded state by a non-trivial logical Pauli error. Then, there exist constants $p_*'$, with $0 < p_*' \leq p_* / 2$, and $c_b>0$ such that for all $p\leq p_*'$,
            \[
                \Pr[\mathcal B_{\mathrm{inj}}^c \cap \mathsf{Wrong}_{\mathrm{inj}}]
                \leq
                c_b\, p .
            \]
        \end{theorem}

        \begin{proof}
            In the setting of Theorem~\ref{thm:state-injection-rising-absorber}, the injection qubit $q_\star$ is prepared in the unencoded state $\ket{0}$, and this preparation is treated as perfect. Therefore, since $\overline Z \equiv Z_{q_\star}$ on the initial product state, the value of the logical $Z$ operator is correct at $t=0$. Consequently, on $\mathcal B_{\mathrm{inj}}^c$, the event $\mathsf{Wrong}_{\mathrm{inj}}$ can be caused only by faults during injection. Moreover, only $X$-type errors can flip the value of the logical $Z$ operator, so we may restrict our attention to the $Z$-type stabilizer decoding sector.
        
            By the proof of Theorem~\ref{thm:state-injection-rising-absorber}, on $\mathcal B_{\mathrm{inj}}^c$, the injection-affected clusters are well-defined, distinct clusters do not interact, and the effect of each cluster $C$ on the logical operator is captured by its residual $R_C=E_C\oplus F_C$. Therefore, conditioned on $\mathcal B_{\mathrm{inj}}^c$, a decoder-induced logical Pauli error can occur only if some residual has non-trivial logical action, i.e., it crosses the logical $Z$ operator supported on the condensing boundary in Fig.~\ref{fig:injection-geometry} an odd number of times. Since only injection-affected clusters have non-trivial residual, we may restrict our attention to injection-affected clusters. Moreover, in order for an injection-affected cluster to induce a logical Pauli error, it must interact with the condensing boundary supporting the logical $Z$ operator and at least one site in the rising absorbing region or on the opposite condensing boundary. Interacting with both condensing boundaries would require a cluster that supports a logical operator, i.e., a cluster of level at least $k_L$, which is excluded on $\mathcal B_{\mathrm{inj}}^c$. Therefore, we may restrict our attention to injection-affected clusters that interact with both the rising absorbing region and the condensing boundary supporting the relevant logical $Z$ operator. We call such clusters \emph{bad}. On $\mathcal B_{\mathrm{inj}}^c$, a decoder-induced logical Pauli error requires the occurrence of a bad cluster.
        
            Let $C$ be an injection-affected $k$-cluster with $k<k_L$. It follows from the proof of Theorem~\ref{thm:state-injection-rising-absorber} that $R_C$ is supported within an $O(w_k)$ spatial neighborhood of the spacetime support of $C$. Moreover, a bad $k$-cluster must lie within distance $O(w_k)$ of both the relevant condensing boundary and the frame region, and within an $O(w_k)$ time window of $t=0$. Hence, the number of fault locations at which a bad $k$-cluster can be placed is at most $C_1 w_k^3$ for some constant $C_1>0$. By the sparsity theorem, for any fixed fault location, a $k$-cluster that includes that fault location occurs with probability at most $C_2\,(p/p_*)^{c_2' w_k^\alpha}$ for constants $C_2,c_2'>0$. Union bounding over fault locations and cluster levels then gives
            \[
                \Pr[\mathcal B_{\mathrm{inj}}^c \cap \mathsf{Wrong}_{\mathrm{inj}}]
                \leq
                C_3
                \sum_{k=0}^{k_L-1}
                w_k^3
                \left(\frac{p}{p_*}\right)^{c_2' w_k^\alpha}
                \leq
                C_3
                \sum_{k=0}^\infty
                w_k^3
                \left(\frac{p}{p_*}\right)^{c_2' w_k^\alpha}
            \]
            for some constant $C_3>0$.

            By choosing the sparsity theorem parameters appropriately, we may assume that $c_2' w_0^\alpha\ge 1$. Then, since the series decays supergeometrically, it follows that there exist some $c_b > 0$ and $p_*' \leq p_*/2$ such that if $p \leq p_*'$, then
            \[
                \Pr[\mathcal B_{\mathrm{inj}}^c \cap \mathsf{Wrong}_{\mathrm{inj}}]
                \leq
                c_b\,p,
            \]
            as desired.
        \end{proof}

        Theorem~\ref{thm:state-injection-bounded-error-rate} shows that, apart from errors already present in the unencoded state, the logical error rate accumulated during injection is $O(p)$, up to the inverse-stretched-exponential-in-$L$ probability of $\mathcal B_{\mathrm{inj}}$. Thus, if the error on the unencoded state is also $p$-bounded, then the logical error rate of the injected state becomes arbitrarily small as $p \to 0$ and $L \to \infty$; in particular, the logical error rate can be made smaller than the relevant magic-state or $Y$-state distillation threshold.

        For i.i.d.\ noise, the bound can also be improved by postselection. Suppose, for example, that we accept the injection only if no detector events occur within a radius-$r$ spacetime neighborhood of the injection qubit. The union bound in the proof of Theorem~\ref{thm:state-injection-bounded-error-rate} then loses many of the terms with $w_k\lesssim r$. Of course, increasing $r$ also increases the probability of rejecting the injection attempt.

        We also note in passing that, with only $K=\Theta(\log\log L)$ slices, one can obtain a postselected state-injection protocol analogous to the postselected state-preparation protocol described in the previous subsection. Specifically, one keeps the auxiliary absorbing layer above the final slice present until time $T_{\mathrm{inj}}+AL$, where $A$ is a sufficiently large constant. The injection attempt is accepted only if no defect candidate is absorbed by this auxiliary layer during this additional waiting interval and no rejection flag is raised at a deterministic stabilizer. Conditioned on acceptance, the injected state is just as protected as in the ordinary streaming decoder setting, and, as with state preparation, for an appropriate choice of $K=\Theta(\log\log L)$, the probability of rejecting the injection attempt is at most
        \[
            \exp[-\Omega(\log^\gamma(L))]
        \]
        for some $\gamma > 1$. These arguments can be formalized in a straightforward manner, but we will not do so here.

    \subsection{Numerics}

        \begin{figure}[t]
            \centering
            \includegraphics[width=0.6\linewidth]{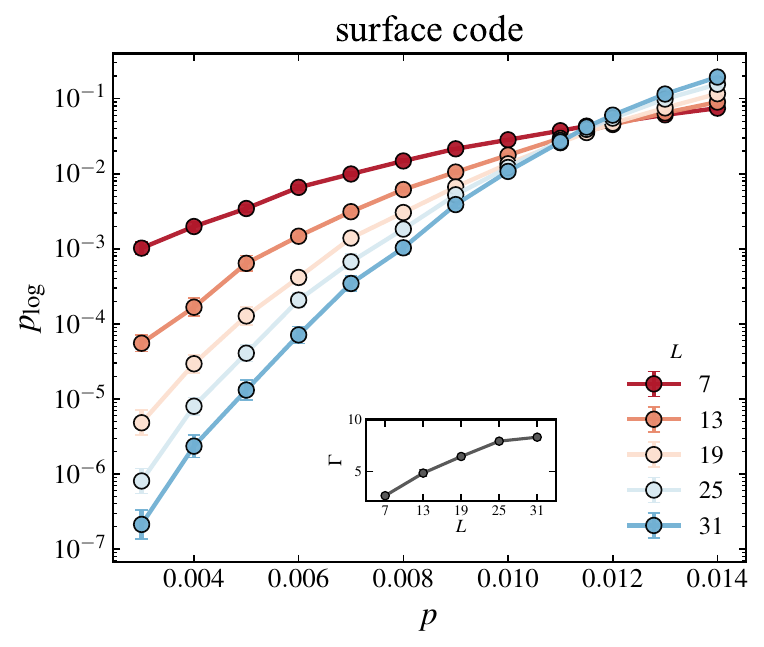}
            \caption{Logical error rate $\plog$ of the surface-code streaming decoder under phenomenological noise, as a function of error rate $p$. Each experiment consists of $L$ rounds of syndrome extraction under i.i.d.\ bit-flip noise with probability $p$ on both data qubits and stabilizer measurement outcomes, followed by noiseless readout of the data qubits. The data are consistent with a threshold of $p_c \approx 1.1\%$. Inset: sub-threshold error suppression exponent $\Gamma_L$ obtained from fits of the form $\plog = C_L (p/p_c)^{\Gamma_L}$ for $p \leq 0.006$. The exponent grows sub-linearly with $L$ over the simulated system sizes.}
            \label{fig:sc_streaming_threshold}
        \end{figure}

        We study the memory-experiment performance of the surface-code streaming decoder using Monte Carlo simulations. Each experiment consists of $L$ rounds of syndrome extraction under i.i.d.\ bit-flip noise with probability $p$ on both data qubits and stabilizer measurement outcomes, followed by noiseless readout of the data qubits. The decoder uses $K = 5$ slices for $L \leq 25$, $K = 6$ slices for $L = 31$, timers $t_k = 2^{k+1}$, and a splitting period $q_s = 10$. We also make two modifications to the decoder in order to improve its performance. First, each time step includes two additional substeps. During each of these substeps, defects may move; existing messages either persist or are erased but do not grow; and timers are not incremented. Second, we set the clock period to be $q = 1$. After the noiseless readout, the decoder runs for $O(L)$ additional time steps, and an experiment is counted as a failure if the residual error is a logical operator or if the decoder fails to terminate by the end of this time period.
        
        The results of these simulations are shown in Fig.~\ref{fig:sc_streaming_threshold}. The data are consistent with a threshold of $p_c \approx 1.1\%$. Below threshold, the logical error rate is suppressed as $\plog \propto (p/p_c)^{\Gamma_L}$ with an exponent $\Gamma_L$ that grows sublinearly in $L$, which is consistent with stretched-exponential suppression of the logical error rate in the code distance. This threshold is lower than the $1.5\%$ threshold of Lake's decoder~\cite{lake2025localactiveerrorcorrection}, which also uses polylogarithmic classical resources per site. However, the mechanism by which that decoder achieves its polylogarithmic overhead is incompatible with stretched-exponential suppression of the failure rate during lattice surgery: a suitably defined analog of our protocols for that decoder would have a failure rate that is inverse-superpolynomial but not inverse-stretched-exponential in $L$, whereas our decoder achieves inverse-stretched-exponential failure rates for lattice surgery.\footnote{Loosely speaking, this is because that decoder attains its polylogarithmic overhead by reducing the time-like distance to $\polylog(L)$. This is benign for a quantum memory, but harmful when measuring lattice-surgery outcomes.}
        
\section{Constant-resource-density streaming decoders} \label{sec:constant-density}

    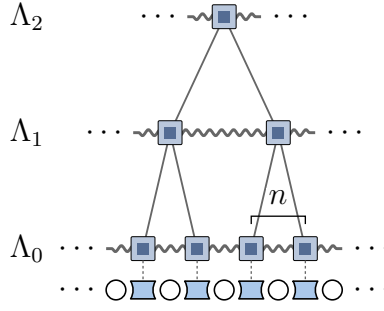
\begin{figure}[t]\centering
        \adjustbox{max width=\textwidth}{%
        \tikzsetnextfilename{coarse-grained-hierarchy}
        \begin{tikzpicture}[
            x=1.3cm, y=1.3cm,
            treewire/.style={black!60, line width=0.7pt}, % parent-child wires (straight, grey)
            samewire/.style={black!60, line width=1.1pt, decorate, % same-level wires (grey squiggle)
                            decoration={snake, amplitude=1.1pt, segment length=4.5pt}},
            cont/.style    ={font=\large},
            levellbl/.style={font=\large}
        ]

        \def\n{2}
        \pgfmathtruncatemacro{\nb}{\n*\n}
        \def\yb{1.18}
        \def\yc{2.36}
        \def\px{0.5506}   % lattice pitch: scrunched so each qubit sits at its neighboring stripes' end-arc centers (2D tile-to-qubit gap)
        \def\ctoff{0.6749}  % ellipsis-ink standoff from the end automata (all levels + lattice row)
        \def\stub{0.3753}   % continuation-stub length: stub tips align with the outer edges of the end qubits
        \def\cfix{0.029}    % \cdots ink sits left of its node centre; nudge nodes right (measured)
        \pgfmathsetmacro{\topx}{(\nb+1)/2}

        % ---- positions ----
        \foreach \i in {1,...,\nb}{ \coordinate (b\i) at ({\i*\px},0); }
        \foreach \j in {1,...,\n}{
            \pgfmathsetmacro{\mx}{(\j-1)*\n + (\n+1)/2}
            \coordinate (m\j) at ({\mx*\px},\yb);
        }
        \coordinate (t1) at ({\topx*\px},\yc);

        % ======== WIRES (squiggles); drawn first so the automata cover their ends ========
        % parent-child (orange)
        \foreach \j in {1,...,\n}{ \draw[treewire] (t1) -- (m\j); }
        \foreach \j in {1,...,\n}{
            \pgfmathtruncatemacro{\lo}{(\j-1)*\n + 1}
            \pgfmathtruncatemacro{\hi}{\j*\n}
            \foreach \k in {\lo,...,\hi}{ \draw[treewire] (m\j) -- (b\k); }
        }
        % same-level (blue)
        \pgfmathtruncatemacro{\nbm}{\nb-1}
        \foreach \i in {1,...,\nbm}{ \pgfmathtruncatemacro{\ii}{\i+1} \draw[samewire] (b\i) -- (b\ii); }
        \pgfmathtruncatemacro{\nm}{\n-1}
        \foreach \j in {1,...,\nm}{ \pgfmathtruncatemacro{\jj}{\j+1} \draw[samewire] (m\j) -- (m\jj); }
        % sideways continuation stubs (blue) + dots
        \draw[samewire] (b1)   -- ($(b1)+(-\stub,0)$);   \node[cont] at ($(b1)+({-\ctoff+\cfix},0)$){$\cdots$};
        \draw[samewire] (b\nb) -- ($(b\nb)+(\stub,0)$);  \node[cont] at ($(b\nb)+({\ctoff+\cfix},0)$){$\cdots$};
        \draw[samewire] (m1)   -- ($(m1)+(-\stub,0)$);   \node[cont] at ($(m1)+({-\ctoff+\cfix},0)$){$\cdots$};
        \draw[samewire] (m\n)  -- ($(m\n)+(\stub,0)$);   \node[cont] at ($(m\n)+({\ctoff+\cfix},0)$){$\cdots$};
        \draw[samewire] (t1)   -- ($(t1)+(-\stub,0)$);   \node[cont] at ($(t1)+({-\ctoff+\cfix},0)$){$\cdots$};
        \draw[samewire] (t1)   -- ($(t1)+(\stub,0)$);    \node[cont] at ($(t1)+({\ctoff+\cfix},0)$){$\cdots$};

        % ======== AUTOMATA (drawn on top of the wire ends) ========
        \foreach \i in {1,...,\nb}{ \aut{\i*\px}{0} }
        \foreach \j in {1,...,\n}{ \pgfmathsetmacro{\mx}{((\j-1)*\n + (\n+1)/2)*\px} \aut{\mx}{\yb} }
        \aut{\topx*\px}{\yc}

        % ======== PHYSICAL QUBIT LATTICE under the finest level (Lambda_0) ========
        % blue stripes = parity checks between adjacent qubits (1D analog of the 2D
        %   stabilizer tiles, with matching arc bites around the qubits);
        % white orbs = qubits
        \pgfmathsetmacro{\autbot}{-0.075*\autscale} % y of an automaton's bottom edge
        \def\ly{-0.421}     % lattice height (closer to the processors)
        \def\lr{0.09}       % check-stripe half-height (drop lines land on the stripe top)
        \def\rr{0.10}       % qubit-orb radius
        \foreach \i in {1,...,\nb}{
            \draw[black!60, line width=0.5pt, dash pattern=on 1pt off 1pt] ({\i*\px},\autbot) -- ({\i*\px},{\ly+\lr});
        }
        % stripe ends reuse the 2D tile-bite arc profile (radius 0.174, corner offset
        % 0.1489 = sqrt(0.174^2-0.09^2)); at pitch \px the end arcs are concentric
        % with the qubits, reproducing the 2D tile-to-qubit gap
        \foreach \i in {1,...,\nb}{
            \draw[fill=pastelbluefig, draw=black, line width=0.6pt, line join=round]
            ({\i*\px-0.1264},{\ly+\lr}) -- ({\i*\px+0.1264},{\ly+\lr})
            arc[start angle=148.85, end angle=211.15, radius=0.174]
            -- ({\i*\px-0.1264},{\ly-\lr})
            arc[start angle=-31.15, end angle=31.15, radius=0.174]
            -- cycle;
        }
        \foreach \i in {0,...,\nb}{
            \filldraw[fill=white, draw=black, line width=0.6pt] ({(\i+0.5)*\px},\ly) circle (\rr);
        }

            % continuation of the physical lattice (aligned under the level dots)
        \node[cont] at ({\px-\ctoff+\cfix},\ly) {$\cdots$};   % same \ctoff standoff as the processor rows; x aligned with the Lambda_0 ellipses
        \node[cont] at ({\nb*\px+\ctoff+\cfix},\ly) {$\cdots$};

        % ======== ANNOTATIONS ========
        \node[levellbl] at (-0.63,0)   {$\Lambda_0$};   % fixed x, clear of the widened ellipses
        \node[levellbl] at (-0.63,\yb) {$\Lambda_1$};
        \node[levellbl] at (-0.63,\yc) {$\Lambda_2$};

        \pgfmathtruncatemacro{\bracea}{\n+1}   % brace over the 2nd family (children of m2)
        \pgfmathtruncatemacro{\braceb}{2*\n}
        \draw[line width=0.5pt]   % rotated square bracket over the block, opening toward the children
            ($(b\bracea)+(0,0.25)$) -- ($(b\bracea)+(0,0.33)$) -- ($(b\braceb)+(0,0.33)$) -- ($(b\braceb)+(0,0.25)$);
        \node[font=\large, fill=white, inner sep=1pt] at ($(b\bracea)!0.5!(b\braceb)+(0,0.52)$) {$n$};

        \end{tikzpicture}%
        }
        \caption{The hierarchy of coarse-grained slices for the repetition-code decoder, drawn for coarse-graining factor $n=2$. Each slice $\Lambda_k$ is a ring of automata, and each slice-$(k+1)$ automaton is the parent of $n$ slice-$k$ automata. Parent--child pairs are joined by straight wires, and within-slice nearest neighbors are joined by wavy wires. The finest slice $\Lambda_0$ sits directly above the system's data qubits (white) and parity checks (blue).}
    \end{figure}

    In this section, we construct a local, two-dimensional, constant-density streaming decoder for the surface code under phenomenological noise. To do so, we replace the $K$ full-resolution slices of its streaming decoder from the previous section with a hierarchy of progressively more coarse-grained slices. The resulting layout is illustrated in Fig.~\ref{fig:toric-coarse-grained-layout}, whose left panel shows the grid-like wiring connecting sites in the same slice and whose right panel shows the tree-like wiring connecting sites in neighboring slices. Our layout is inspired by a known layout for pyramid computers~\cite{doi:10.1137/0216004, dyer1981pyramid}, which is in turn a simple modification of the standard H-tree layout used in very-large-scale integration~\cite{mead1980introduction}.\footnote{A pyramid computer is a parallel computer whose processors form a stack of progressively coarser two-dimensional grids, with nearest-neighbor connections within each grid and parent--child connections between sites in adjacent grids. Very-large-scale integration is the process of creating integrated circuits containing very large numbers of components on a single chip.}

    This coarse-grained hierarchy serves two related purposes. First, it keeps the total number of sites in the hierarchy proportional to the total number of sites in the physical system. Second, it allows all timer variables to be stored using a constant number of bits.\footnote{In Ref.~\cite{breuckmann2016local}, it was suggested that the $O(\log L)$ per-cell memory of Harrington's decoder could in principle be eliminated by distributing the memory and computation of each cell of the hierarchy over the block of physical sites it is responsible for. Our coarse-grained hierarchy allows us to accomplish exactly this.}

    This section is organized as follows. First, we construct a hierarchical repetition-code streaming decoder as a warm-up and give an analogous construction for the toric code. Next, we prove that there is a constant, non-zero threshold below which both decoders achieve stretched-exponential memory lifetimes. We then describe how to adapt the toric-code construction to the surface code and how to modify the hierarchical surface-code decoder to perform local decoding during state preparation and state injection. Finally, we prove that the toric- and surface-code decoders can be implemented in two dimensions using a constant density of bounded-bandwidth, bounded-propagation-speed wires.

    \subsection{Hierarchical repetition-code decoder}
    \label{subsec:cm-online-repetition}

        Consider the repetition code on a ring of length $L$, and let $n>1$ be the parameter from the sparsity theorem. Here, $n$ also serves as a coarse-graining factor for the hierarchy. We assume for simplicity that $n$ is an even integer and that $L$ is divisible by $n^K$, where $K$ is the number of slices used by the decoder. The assumption that $n$ is even is not essential and is only used to simplify the construction.
        
        For $k=0,\ldots,K-1$, define
        \[
            \Lambda_k \coloneqq \mathbb{Z}/(L/n^k)\mathbb{Z}.
        \]
        Each $x\in\Lambda_k$ has a unique representative $r\in\{0,\ldots,L/n^k-1\}$ that can be uniquely written as
        \[
            r = n x' +a,
        \]
        for $a\in\{0,\ldots,n-1\}$ and $x'\in\{0,\ldots,L/n^{k+1}-1\}$. For $k=0,\ldots,K-2$, this allows us to define the parent map
        \[
            \pi_k:\Lambda_k\to\Lambda_{k+1}
        \]
        by
        \[
            \pi_k(x)=x',
        \]
        where $x'$ is determined by the decomposition above. We will also occasionally identify a site $x\in\Lambda_k$ with the site $n^k x + \lfloor n^k/2 \rfloor \in \Lambda_0$.

        The coarse-grained cellular-automaton rule on slice $k$ only fires at times
        \[
            t\in n^k\Z,
        \]
        in a manner that will be made precise below. This timing mechanism can be implemented by bounded-speed signals propagating along wires of length $\Theta(n^k)$.

        For slice $k=0$, the defect-sector variables are
        \[
            \xi_0(x)=\bigl(e_0(x),s_0(x),\tau_0(x)\bigr),
        \]
        and the message-sector variables are
        \[
            \xi'_0(x)=\bigl(m_0(x),c_0(x),\theta_0(x)\bigr).
        \]
        Here, $e_0(x)\in\{0,1\}$ is the correction bit, $s_0(x)\in\{0,1\}$ is the defect bit, $m_0(x)\in\{0,1\}$ is the repetition-code message bit, and $c_0(x)\in\{0,\ldots,q-1\}$ is the clock channel.

        Next, we describe the defect- and message-sector variables for higher slices. For timed slices $0<k<K-1$, the defect-sector variables are
        \[
            \xi_k(x)=\bigl(e_k(x),\mathbf e_k(x),s_k(x),\tau_k(x)\bigr),
        \]
        and the message-sector variables are
        \[
            \xi'_k(x)=\bigl(m_k(x),c_k(x),\theta_k(x)\bigr).
        \]
        The final slice is untimed, so it has no defect or message timer variables:
        \[
            \xi_{K-1}(x)=\bigl(e_{K-1}(x),\mathbf e_{K-1}(x),s_{K-1}(x)\bigr),
            \qquad
            \xi'_{K-1}(x)=\bigl(m_{K-1}(x),c_{K-1}(x)\bigr).
        \]
        Here, $s_k$, $m_k$, and $c_k$ are the defect, message, and clock channels, respectively. On timed slices, defect and message timers have fixed range $\tau_k(x),\theta_k(x)\in \{0,\ldots,t_0-1\}$. If $s_k(x)=0$, we set $\tau_k(x)=0$, and if $m_k(x)=0$, we set $\theta_k(x)=0$.

        The correction bits are defined as follows. The same-slice correction bit $e_k(x)\in\{0,1\}$ is associated with the slice-$k$ correction edge immediately to the left of $x$. The child-to-parent correction bits $\mathbf e_k(x)=\bigl(e_{k,0}(x),\ldots,e_{k,n-1}(x)\bigr)\in\{0,1\}^n$ are indexed by the children of $x$. Specifically, we fix an ordering of the $n$ slice-$(k-1)$ children of each site $x\in\Lambda_k$, denoted $\chi_{k,a}(x)\in\Lambda_{k-1}$ for $a=0,\ldots,n-1$, so that $\pi_{k-1}(\chi_{k,a}(x))=x$. Promoting a defect from the child site $\chi_{k,a}(x)$ to the parent site $x$ toggles $e_{k,a}(x)$. Thus $e_{k,a}(x)$ records, modulo two, how many times the fixed correction string connecting $\chi_{k,a}(x)$ to $x$ has been toggled; this string is chosen to be the shorter of the two strings between these sites on the ring, with a fixed tie-breaking convention if the two strings have equal length.
        
        The full correction on any particular qubit is obtained by expanding both the same-slice correction bits and the child-to-parent correction bits from each slice into their associated fixed correction strings and taking the modulo-two sum of these contributions.
        
        We now describe the update from time $t$ to time $t+1$. The update has three stages. First, defect records arriving during this step are absorbed into local bookkeeping variables. Second, the defect sector is updated on every slice $k$ whose rule fires, i.e., on every slice $k$ for which $t\in n^k\Z$. Third, after the new defect variables have been determined, the message sector is updated on all slices whose rule fires. As usual, these three stages (Algorithms~\ref{alg:cm-rep-defect-record-intake},~\ref{alg:cm-rep-defect-update}, and~\ref{alg:cm-rep-message-update}) are written as global loops over slices and sites for convenience. The update itself can be implemented in a local manner.

        We now describe the first stage in more detail; the resulting defect-record intake rule is given in Algorithm~\ref{alg:cm-rep-defect-record-intake}. Each site has a transient input register
        \[
            \mathcal Q_k(x,t+1)
        \]
        that contains defect records scheduled to arrive at $x\in\Lambda_k$ during the update from time $t$ to time $t+1$. There are two types of defect records that can arrive in this manner:
        \[
            (\mathrm{env},0),
            \qquad
            (\mathrm{up},a,0),
            \qquad
            a\in\{0,\ldots,n-1\}.
        \]
        For slice $0$, a physical detector event $\phi(x,t+1)=1$ inserts the record $(\mathrm{env},0)$ into $\mathcal Q_0(x,t+1)$. For slices $k>0$, a promotion from the $a$th child of $x$ inserts the record $(\mathrm{up},a,0)$ into $\mathcal Q_k(x,t+1)$. The child label $a$ specifies which child-to-parent correction-frame bit $e_{k,a}(x)$ is toggled by the promotion. The input registers are not a long-term memory: records are examined during the intake step and then cleared.

        For each slice $k>0$, each site also carries an arrival-parity bit
        \[
            \rho_k(x)\in\{0,1\}.
        \]
        This bit records the parity of child-promoted defects received from slice $k-1$ since the previous slice-$k$ firing. During intake, an incoming record $(\mathrm{up},a,0)$ toggles both $e_{k,a}(x)$ and $\rho_k(x)$. When slice $k$ fires during the next update, an odd value of $\rho_k(x)$ produces a single defect candidate at $x$, after which $\rho_k(x)$ is reset.

        We initialize all defect-sector, message-sector, and intake variables at time $t=-1$ to zero. The first state at time $t = 0$ is obtained by applying the full update with these initial conditions.

        During a slice-$k$ firing, the defect update is phrased in terms of defect candidates. On timed slices, the possible candidates are
        \[
            (\mathrm{env},0),\qquad
            (\mathrm{up},0),\qquad
            (\mathrm{move},\vartheta),\qquad
            (\mathrm{stay},\vartheta),
            \qquad
            \vartheta\in\{0,\ldots,t_0-1\}.
        \]
        On the final slice, timer fields are omitted. The candidate $(\mathrm{env},0)$ represents a slice-$0$ detector event, while $(\mathrm{up},0)$ represents an odd number of child promotions accumulated since the previous slice-$k$ firing.

        All variables not explicitly changed during an update are copied from time $t$ to time $t+1$. In particular, if slice $k$ does not fire at time $t$, then its same-slice defect and message variables are carried forward, apart from the intake changes to $\mathbf e_k$ and $\rho_k$.
        
        Note that each site has $n=O(1)$ children and $O(1)$ within-slice neighbors, and defect records and timers take values in constant-size alphabets. Therefore, the transient input registers and candidate sets require only $O(1)$ space per site.

        \begin{algorithm}[H]
            \caption{Defect-record intake for the hierarchical repetition-code decoder at time $t$.}
            \label{alg:cm-rep-defect-record-intake}
            \begin{algorithmic}[1]
                \State $\mathcal I_k(x,t+1)\gets\emptyset$ for all $k,x$
                \Comment{Input candidates for the next slice-$k$ firing}

                \Statex
                \ForAll{$x\in\Lambda_0$ with $\phi(x,t+1)=1$}
                    \State Insert $(\mathrm{env},0)$ into $\mathcal Q_0(x,t+1)$
                    \Comment{Physical detector event}
                \EndFor

                \Statex
                \ForAll{$k=0,\ldots,K-1$}
                    \ForAll{$x\in\Lambda_k$}
                        \If{$k>0$}
                            \State $e_{k,a}(x,t+1)\gets e_{k,a}(x,t)$ for all $a=0,\ldots,n-1$
                            \State $\rho_k(x,t+1)\gets \rho_k(x,t)$
                        \EndIf

                        \ForAll{defect records $R\in\mathcal Q_k(x,t+1)$}
                            \If{$R=(\mathrm{env},0)$}
                                \State Add candidate $(\mathrm{env},0)$ to $\mathcal I_0(x,t+1)$
                            \ElsIf{$R=(\mathrm{up},a,0)$}
                                \State $e_{k,a}(x,t+1)\gets e_{k,a}(x,t+1)\oplus 1$
                                \Comment{Child-to-parent frame update}
                                \State $\rho_k(x,t+1)\gets \rho_k(x,t+1)\oplus 1$
                                \Comment{Toggle child-arrival parity}
                            \EndIf
                        \EndFor
                    \EndFor
                \EndFor
            \end{algorithmic}
        \end{algorithm}

        After the intake rule, the defect-sector update described in Algorithm~\ref{alg:cm-rep-defect-update} is applied only to slices whose rule fires. Let $\lambda_k(x,t)\in\{0,1\}$ be the indicator that the base repetition-code decoder moves a defect at $x\in\Lambda_k$ one step to the left, as determined by the current message configuration on slice $k$. All neighbors are understood as neighbors in the coarse-grained lattice $\Lambda_k$.

        If a timed-slice defect has incremented timer $\tau^+=t_0$, then it is promoted to its parent and does not move horizontally during the same slice-$k$ firing. The promoted defect is delivered to slice $k+1$ as a record with its timer reset to $0$. If the timer does not expire, the defect either moves left according to $\lambda_k$ or remains stationary. Defects on the final slice follow the same spatial rule, but with no timer check.

        If an even number of defect candidates meet at a site, they pair-annihilate and no defect remains; if an odd number of candidates meet, a defect survives, and on timed slices its timer is taken to be the minimum timer among the incoming candidates.

        \begin{algorithm}[H]
            \caption{Defect-sector update for the hierarchical repetition-code decoder at time $t$.}
            \label{alg:cm-rep-defect-update}
            \begin{algorithmic}[1]
                \ForAll{$k=0,\ldots,K-1$ with $t\in n^k\Z$}
                    \State $\mathcal R_k(x,t+1)\gets \mathcal I_k(x,t+1)$ for all $x\in\Lambda_k$
                    \Comment{Initialize defect candidates from inputs}
                    \State $e_k(x,t+1)\gets e_k(x,t)$ for all $x\in\Lambda_k$
                    \Statex

                    \If{$k>0$}
                        \ForAll{$x\in\Lambda_k$}
                            \If{$\rho_k(x,t+1)=1$}
                                \If{$k<K-1$}
                                    \State Add candidate $(\mathrm{up},0)$ to $\mathcal R_k(x,t+1)$
                                \Else
                                    \State Add candidate $(\mathrm{up},\varnothing)$ to $\mathcal R_k(x,t+1)$
                                \EndIf
                                \Comment{Odd child-promotion parity}
                            \EndIf
                            \State $\rho_k(x,t+1)\gets 0$
                        \EndFor
                    \EndIf
                    \Statex

                    \ForAll{$x\in\Lambda_k$ with $s_k(x,t)=1$}
                        \If{$k<K-1$}
                            \State $\tau^+\gets \tau_k(x,t)+1$

                            \If{$\tau^+=t_0$}
                                \State $y\gets \pi_k(x)$
                                \State Let $b$ be the child label such that $x=\chi_{k+1,b}(y)$
                                \State Schedule $(\mathrm{up},b,0)$ to arrive in $\mathcal Q_{k+1}(y,t+n^{k+1} + 1)$
                                \Comment{Promotion to parent}

                            \ElsIf{$\lambda_k(x,t)=1$}
                                \State $e_k(x,t+1)\gets e_k(x,t+1)\oplus 1$
                                \State Add candidate $(\mathrm{move},\tau^+)$ to $\mathcal R_k(x-1,t+1)$
                                \Comment{Same-slice move}

                            \Else
                                \State Add candidate $(\mathrm{stay},\tau^+)$ to $\mathcal R_k(x,t+1)$
                                \Comment{Stationary defect}
                            \EndIf

                        \Else
                            \If{$\lambda_{K-1}(x,t)=1$}
                                \State $e_{K-1}(x,t+1)\gets e_{K-1}(x,t+1)\oplus 1$
                                \State Add candidate $(\mathrm{move},\varnothing)$ to $\mathcal R_{K-1}(x-1,t+1)$
                                \Comment{Same-slice move on the final slice}
                            \Else
                                \State Add candidate $(\mathrm{stay},\varnothing)$ to $\mathcal R_{K-1}(x,t+1)$
                                \Comment{Stationary defect on the final slice}
                            \EndIf
                        \EndIf
                    \EndFor
                    \Statex

                    \ForAll{$x\in\Lambda_k$}
                        \If{$|\mathcal R_k(x,t+1)|$ is even}
                            \State $s_k(x,t+1)\gets 0$
                            \If{$k<K-1$}
                                \State $\tau_k(x,t+1)\gets 0$
                            \EndIf
                        \Else
                            \State $s_k(x,t+1)\gets 1$
                            \If{$k<K-1$}
                                \State $\tau_k(x,t+1)\gets
                                \min\{\vartheta:(\cdot,\vartheta)\in\mathcal R_k(x,t+1)\}$
                            \EndIf
                        \EndIf
                    \EndFor
                \EndFor
            \end{algorithmic}
        \end{algorithm}

        The message-sector update described in Algorithm~\ref{alg:cm-rep-message-update} is performed after the defect-sector update on each slice that fires. It is the coarse-grained analogue of the repetition-code message update: the only changes are that the timed slices use the fixed timer range $\{0,\ldots,t_0-1\}$ and that $x-1$ denotes the left neighbor in $\Lambda_k$.

        If no message candidate is produced at a site, then the message is erased. If one or more candidates are produced, then a message is written, and its timer is taken to be the minimum candidate timer. On the final slice, all timer fields are omitted.

        \begin{algorithm}[H]
            \caption{Message-sector update for the hierarchical repetition-code decoder at time $t$.}
            \label{alg:cm-rep-message-update}
            \begin{algorithmic}[1]
                \ForAll{$k=0,\ldots,K-1$ with $t\in n^k\Z$}
                    \ForAll{$x\in\Lambda_k$}

                        \If{$k<K-1$}
                            \State $\mathcal B_k(x,t+1)\gets
                            \{\tau_k(x,t+1): s_k(x,t+1)=1\}$
                            \Statex \hspace{1.95cm}
                            $\cup\;\{\tau_k(x,t)+1: s_k(x,t)=1 \;\text{and}\; \tau_k(x,t)+1<t_0\}$
                            \Statex \hspace{1.95cm}
                            $\cup\;\{\theta_k(x,t)+1: m_k(x,t)=1 \;\text{and}\; m_k(x-1,t)=1 \;\text{and}\; \theta_k(x,t)<t_0-1\}$
                            \Statex \hspace{1.95cm}
                            $\cup\;\{\theta_k(x-1,t)+1: m_k(x-1,t)=1 \;\text{and}\; c_k(x,t)=0 \;\text{and}\; \theta_k(x-1,t)<t_0-1\}$

                            \If{$\mathcal B_k(x,t+1)=\emptyset$}
                                \State $m_k(x,t+1)\gets 0$, \quad $\theta_k(x,t+1)\gets 0$
                            \Else
                                \State $m_k(x,t+1)\gets 1$, \quad $\theta_k(x,t+1)\gets \min\mathcal B_k(x,t+1)$
                            \EndIf

                        \Else
                            \State $m_k(x,t+1)\gets
                            [s_k(x,t+1)=1]
                            \;\vee\;
                            [s_k(x,t)=1]$
                            \Statex \hspace{2.95cm}
                            $\vee\;
                            [m_k(x,t)=1 \;\text{and}\; m_k(x-1,t)=1]$
                            \Statex \hspace{2.95cm}
                            $\vee\;
                            [m_k(x-1,t)=1 \;\text{and}\; c_k(x,t)=0]$
                        \EndIf

                        \State $c_k(x,t+1)\gets (c_k(x,t)+1)\bmod q$
                    \EndFor
                \EndFor
            \end{algorithmic}
        \end{algorithm}

        The notation in Algorithms~\ref{alg:cm-rep-defect-update} and~\ref{alg:cm-rep-message-update} should be interpreted as an effective, non-local description of a local process. For example, adding a same-slice move candidate to $\mathcal R_k(x-1,t+1)$ can be implemented using bounded-speed classical communication between neighboring sites of $\Lambda_k$. Each slice-$k$ site only needs data from an $O(1)$-neighborhood in $\Lambda_k$, and these data can be collected in $O(1)$ communication rounds. Since the relevant sites are separated by physical distance $O(n^k)$ and slice $k$ fires only once every $n^k$ steps, the required communication and local processing can be performed between consecutive slice-$k$ firings.
    
    \subsection{Hierarchical toric-code decoder} \label{subsec:cm-online-toric}
    
        We now give the analogous construction for the toric code. Consider one CSS decoding sector of the toric code on an $L\times L$ torus. For $k=0,\ldots,K-1$, define
        \[
            \Lambda_k
            \coloneqq
            \bigl(\Z/(L/n^k)\Z\bigr)^2 .
        \]
        Each site $\vc r\in\Lambda_k$ has a unique representative
        \[
            (r_x,r_y)
            \in
            \{0,\ldots,L/n^k-1\}^2 ,
        \]
        which can be uniquely written as
        \[
            (n x'+a_x, n y'+a_y),\
        \]
        for
        \[
            a_x,a_y\in\{0,\ldots,n-1\},
            \qquad
            (x',y')\in\{0,\ldots,L/n^{k+1}-1\}^2 .
        \]

        For $k = 0, \ldots, K-2$, define the parent map $\pi_k:\Lambda_k\to\Lambda_{k+1}$ by
        \[
            \pi_k(\vc r)=(x',y').
        \]
        Let~$\mathcal C_n\coloneqq \{0,\ldots,n-1\}^2$ be the set of child labels. For $k>0$, the $\vc a$-child of $\vc r\in\Lambda_k$ is denoted by
        \[
            \chi_{k,\vc a}(\vc r)\in\Lambda_{k-1},
            \qquad
            \vc a = (a_x,a_y) \in\mathcal C_n,
        \]
        so that
        \[
            \pi_{k-1}\bigl(\chi_{k,\vc a}(\vc r)\bigr)=\vc r .
        \]
        We will sometimes identify a slice-$k$ site $\vc r\in\Lambda_k$ with the site $n^k\vc r+\left\lfloor\frac{n^k}{2}\right\rfloor(\uvec x+\uvec y) \in\Lambda_0$.

        The slice-$k$ cellular-automaton rule fires only at times
        \[
            t\in n^k\Z .
        \]
        As in the repetition-code hierarchy, this timing can be implemented by bounded-speed signals propagating along wires of length $\Theta(n^k)$. 

        Throughout this subsection, we use the same toric-code base rule as in Section~\ref{subsec:toric-code:definitions}. In particular, we take $q=3$, so messages grow when the local clock satisfies $c\in\{0,1\}$, while defects move only when $c=0$.

        For slice $k=0$, the defect-sector variables are
        \[
            \xi_0(\vc r)
            =
            \bigl(
                e_{x,0}(\vc r),
                e_{y,0}(\vc r),
                s_0(\vc r),
                \tau_0(\vc r)
            \bigr),
        \]
        and the message-sector variables are
        \[
            \xi'_0(\vc r)
            =
            \bigl(
                \{m_{ij,0}(\vc r)\}_{(i,j)\in\mathcal M_2},
                c_0(\vc r),
                \{\theta_{ij,0}(\vc r)\}_{(i,j)\in\mathcal M_2}
            \bigr).
        \]
        Here, $e_{x,0}(\vc r)$ and $e_{y,0}(\vc r)$ are the correction bits associated with the $x$- and $y$-edges in the negative coordinate directions relative to $\vc r$, $s_0(\vc r)$ is the defect bit, $\tau_0(\vc r)$ is the defect timer, $\{m_{ij,0}(\vc r)\}_{(i,j)\in\mathcal M_2}$ are the three message channels, $c_0(\vc r)\in\{0,\ldots,q-1\}$ is the local clock, and $\{\theta_{ij,0}(\vc r)\}_{(i,j)\in\mathcal M_2}$ are the message timers.

        For higher slices $0<k\leq K-1$, a site $\vc r\in\Lambda_k$ carries two same-slice correction bits
        \[
            e_{x,k}(\vc r),e_{y,k}(\vc r)\in\{0,1\},
        \]
        and a collection of child-to-parent correction bits
        \[
            \mathbf e_k(\vc r)
            =
            \bigl(
                e_{k,\vc a}(\vc r)
            \bigr)_{\vc a\in\mathcal C_n}
            \in\{0,1\}^{n^2}.
        \]
        The same-slice correction bit $e_{x,k}(\vc r)$ records the parity of coarse moves from $\vc r$ to $\vc r-\uvec x$, and $e_{y,k}(\vc r)$ records the parity of coarse moves from $\vc r$ to $\vc r-\uvec y$. The child-to-parent bit $e_{k,\vc a}(\vc r)$ records the parity of promotions from the child $\chi_{k,\vc a}(\vc r)$ to the parent $\vc r$.

        Thus, for $0<k<K-1$, the defect-sector variables are
        \[
            \xi_k(\vc r)
            =
            \bigl(
                e_{x,k}(\vc r),
                e_{y,k}(\vc r),
                \mathbf e_k(\vc r),
                s_k(\vc r),
                \tau_k(\vc r)
            \bigr),
        \]
        while the message-sector variables are
        \[
            \xi'_k(\vc r)
            =
            \bigl(
                \{m_{ij,k}(\vc r)\}_{(i,j)\in\mathcal M_2},
                c_k(\vc r),
                \{\theta_{ij,k}(\vc r)\}_{(i,j)\in\mathcal M_2}
            \bigr).
        \]
        For every timed slice $k<K-1$, the defect and message timers have slice-independent range
        \[
            \tau_k(\vc r),\theta_{ij,k}(\vc r)
            \in
            \{0,\ldots,t_0-1\}.
        \]
        If $s_k(\vc r)=0$, we set $\tau_k(\vc r)=0$, and if $m_{ij,k}(\vc r)=0$, we set $\theta_{ij,k}(\vc r)=0$.

        The final slice $K-1$ is untimed, so we have
        \[
            \xi_{K-1}(\vc r)
            =
            \bigl(
                e_{x,K-1}(\vc r),
                e_{y,K-1}(\vc r),
                \mathbf e_{K-1}(\vc r),
                s_{K-1}(\vc r)
            \bigr),
        \]
        and
        \[
            \xi'_{K-1}(\vc r)
            =
            \bigl(
                \{m_{ij,K-1}(\vc r)\}_{(i,j)\in\mathcal M_2},
                c_{K-1}(\vc r)
            \bigr).
        \]

        As before, a child-to-parent bit $e_{k,\vc a}(\vc r)$ is expanded into a fixed shortest correction string from $\chi_{k,\vc a}(\vc r)$ to $\vc r$. Ties are resolved by a fixed translation-invariant convention. The full microscopic correction is obtained by expanding the coarse correction bits into their associated fixed microscopic correction strings and summing all contributions modulo two. 

        We now describe the update from time $t$ to time $t+1$. As in the repetition-code hierarchy, the update has three stages. First, defect records arriving during the step are absorbed into local bookkeeping variables. Second, the defect sector is updated on every slice whose rule fires. Third, after the new defect variables have been determined, the message sector is updated on every slice whose rule fires.

        Each site has a transient input register
        \[
            \mathcal Q_k(\vc r,t+1),
        \]
        containing defect records that arrive at $\vc r\in\Lambda_k$ during the update from time $t$ to time $t+1$. There are two kinds of records:
        \[
            (\mathrm{env},0),
            \qquad
            (\mathrm{up},\vc a,0),
            \qquad
            \vc a\in\mathcal C_n .
        \]
        For slice $0$, a physical detector event $\phi(\vc r,t+1)=1$ inserts the record $(\mathrm{env},0)$ into $\mathcal Q_0(\vc r,t+1)$. For slices $k>0$, a promotion from the child $\chi_{k,\vc a}(\vc r)$ inserts the record $(\mathrm{up},\vc a,0)$ into $\mathcal Q_k(\vc r,t+1)$.

        For each slice $k>0$, each site also carries an arrival-parity bit
        \[
            \rho_k(\vc r)\in\{0,1\}.
        \]
        This bit records the parity of child-promoted defects received from slice $k-1$ since the previous slice-$k$ firing. During intake, a record $(\mathrm{up},\vc a,0)$ toggles both $e_{k,\vc a}(\vc r)$ and $\rho_k(\vc r)$. When slice $k$ next fires, an odd value of $\rho_k(\vc r)$ produces a single defect candidate at $\vc r$, after which $\rho_k(\vc r)$ is reset.

        All defect-sector, message-sector, and intake-bookkeeping variables are initialized to zero at time $t = -1$. The first state at time $t = 0$ is obtained by applying the full update with these initial conditions.

        During a slice-$k$ firing, the defect update is phrased in terms of defect candidates. On timed slices, the possible candidates are
        \[
            (\mathrm{env},0),
            \qquad
            (\mathrm{up},0),
            \qquad
            (\mathrm{left},\vartheta),
            \qquad
            (\mathrm{down},\vartheta),
            \qquad
            (\mathrm{stay},\vartheta),
        \]
        with $\vartheta\in\{0,\ldots,t_0-1\}$. On the final slice, the timer field is omitted. The candidate $(\mathrm{env},0)$ represents a slice-$0$ detector event, while $(\mathrm{up},0)$ represents an odd number of child promotions accumulated since the previous slice-$k$ firing.

        All variables not explicitly changed during a microscopic update are copied from time $t$ to time $t+1$. In particular, if slice $k$ does not fire at time $t$, then its same-slice defect and message variables are carried forward, apart from the intake changes to $\mathbf e_k$ and $\rho_k$. 

        \begin{algorithm}[H]
            \caption{Defect-record intake for the hierarchical toric-code decoder at time $t$.}
            \label{alg:cm-toric-defect-record-intake}
            \begin{algorithmic}[1]
                \State $\mathcal I_k(\vc r,t+1)\gets\emptyset$ for all $k,\vc r$
                \Comment{Input candidates for the next slice-$k$ firing}

                \Statex
                \ForAll{$\vc r\in\Lambda_0$ with $\phi(\vc r,t+1)=1$}
                    \State Insert $(\mathrm{env},0)$ into $\mathcal Q_0(\vc r,t+1)$
                    \Comment{Physical detector event}
                \EndFor

                \Statex
                \ForAll{$k=0,\ldots,K-1$}
                    \ForAll{$\vc r\in\Lambda_k$}
                        \If{$k>0$}
                            \State $e_{k,\vc a}(\vc r,t+1)\gets e_{k,\vc a}(\vc r,t)$ for all $\vc a\in\mathcal C_n$
                            \State $\rho_k(\vc r,t+1)\gets \rho_k(\vc r,t)$
                        \EndIf

                        \ForAll{defect records $R\in\mathcal Q_k(\vc r,t+1)$}
                            \If{$R=(\mathrm{env},0)$}
                                \State Add candidate $(\mathrm{env},0)$ to $\mathcal I_0(\vc r,t+1)$

                            \ElsIf{$R=(\mathrm{up},\vc a,0)$}
                                \State $e_{k,\vc a}(\vc r,t+1)\gets e_{k,\vc a}(\vc r,t+1)\oplus 1$
                                \Comment{Child-to-parent frame update}
                                \State $\rho_k(\vc r,t+1)\gets \rho_k(\vc r,t+1)\oplus 1$
                                \Comment{Toggle child-arrival parity}
                            \EndIf
                        \EndFor
                        \EndFor
                \EndFor
            \end{algorithmic}
        \end{algorithm}

        After intake, the defect-sector update is applied only on slices whose rule fires. Define the movement indicators on $\Lambda_k$ by
        \[
            \lambda^x_k(\vc r,t)
            \coloneqq
            \bigl[
                c_k(\vc r,t)=0
                \;\text{and}\;
                \exists\,j\in\{0,1\}
                \text{ s.t. }
                m_{0j,k}(\vc r-\uvec x,t)=1
            \bigr],
        \]
        and
        \[
            \lambda^y_k(\vc r,t)
            \coloneqq
            \bigl[
                c_k(\vc r,t)=0
                \;\text{and}\;
                \exists\,i\in\{0,1\}
                \text{ s.t. }
                m_{i0,k}(\vc r-\uvec y,t)=1
            \bigr]
            \;\text{and}\;
            \lnot \lambda^x_k(\vc r,t).
        \]
        Thus a defect moves left if $\lambda^x_k=1$, moves down if $\lambda^y_k=1$, and otherwise remains stationary. All neighbors are understood as neighbors in the coarse-grained lattice $\Lambda_k$.

        If a timed-slice defect has incremented timer $\tau^+=t_0$, then it is promoted to its parent and does not move spatially during the same slice-$k$ firing. The promoted defect is delivered to slice $k+1$ as a record with timer reset to $0$. If the timer does not expire, the defect evolves according to the base toric-code rule on $\Lambda_k$. Defects on the final slice follow the same spatial rule, but without any timer check.

        If an even number of defect candidates meet at a site, they pair-annihilate and no defect remains; if an odd number of candidates meet, a defect survives, and on timed slices its timer is taken to be the minimum timer among the incoming candidates.

                \begin{algorithm}[H]
            \caption{Defect-sector update for the hierarchical toric-code decoder at time $t$.}
            \label{alg:cm-toric-defect-update}
            \begin{algorithmic}[1]
                \ForAll{$k=0,\ldots,K-1$ with $t\in n^k\Z$}
                    \State $\mathcal R_k(\vc r,t+1)\gets\mathcal I_k(\vc r,t+1)$ for all $\vc r\in\Lambda_k$
                    \Comment{Initialize defect candidates from inputs}
                    \State $e_{x,k}(\vc r,t+1)\gets e_{x,k}(\vc r,t)$ and $e_{y,k}(\vc r,t+1)\gets e_{y,k}(\vc r,t)$ for all $\vc r\in\Lambda_k$

                    \If{$k>0$}
                        \ForAll{$\vc r\in\Lambda_k$}
                            \If{$\rho_k(\vc r,t+1)=1$}
                                \If{$k<K-1$}
                                    \State Add candidate $(\mathrm{up},0)$ to $\mathcal R_k(\vc r,t+1)$
                                \Else
                                    \State Add candidate $(\mathrm{up},\varnothing)$ to $\mathcal R_k(\vc r,t+1)$
                                \EndIf
                                \Comment{Odd child-promotion parity}
                            \EndIf
                            \vspace{-0.8em}
                            \State $\rho_k(\vc r,t+1)\gets 0$
                        \EndFor
                    \EndIf
                    \Statex

                    \ForAll{$\vc r\in\Lambda_k$ with $s_k(\vc r,t)=1$}
                        \If{$k<K-1$}
                            \State $\tau^+\gets\tau_k(\vc r,t)+1$

                            \If{$\tau^+=t_0$}
                                \State $\vc y\gets\pi_k(\vc r)$
                                \State Let $\vc b$ be the child label such that $\vc r=\chi_{k+1,\vc b}(\vc y)$
                                \State Schedule $(\mathrm{up},\vc b,0)$ to arrive in $\mathcal Q_{k+1}(\vc y,t+n^{k+1}+1)$
                                \Comment{Promotion to parent}

                            \ElsIf{$\lambda^x_k(\vc r,t)=1$}
                                \State $e_{x,k}(\vc r,t+1)\gets e_{x,k}(\vc r,t+1)\oplus 1$
                                \State Add candidate $(\mathrm{left},\tau^+)$ to $\mathcal R_k(\vc r-\uvec x,t+1)$
                                \Comment{Leftward move}

                            \ElsIf{$\lambda^y_k(\vc r,t)=1$}
                                \State $e_{y,k}(\vc r,t+1)\gets e_{y,k}(\vc r,t+1)\oplus 1$
                                \State Add candidate $(\mathrm{down},\tau^+)$ to $\mathcal R_k(\vc r-\uvec y,t+1)$
                                \Comment{Downward move}

                            \Else
                                \State Add candidate $(\mathrm{stay},\tau^+)$ to $\mathcal R_k(\vc r,t+1)$
                                \Comment{Stationary defect}
                            \EndIf
                        \Else
                            \If{$\lambda^x_{K-1}(\vc r,t)=1$}
                                \State $e_{x,K-1}(\vc r,t+1)\gets e_{x,K-1}(\vc r,t+1)\oplus 1$
                                \State Add candidate $(\mathrm{left},\varnothing)$ to $\mathcal R_{K-1}(\vc r-\uvec x,t+1)$
                                \Comment{Leftward move on the final slice}

                            \ElsIf{$\lambda^y_{K-1}(\vc r,t)=1$}
                                \State $e_{y,K-1}(\vc r,t+1)\gets e_{y,K-1}(\vc r,t+1)\oplus 1$
                                \State Add candidate $(\mathrm{down},\varnothing)$ to $\mathcal R_{K-1}(\vc r-\uvec y,t+1)$
                                \Comment{Downward move on the final slice}

                            \Else
                                \State Add candidate $(\mathrm{stay},\varnothing)$ to $\mathcal R_{K-1}(\vc r,t+1)$
                                \Comment{Stationary defect on the final slice}
                            \EndIf
                        \EndIf
                    \EndFor
                    \Statex

                    \ForAll{$\vc r\in\Lambda_k$}
                        \If{$|\mathcal R_k(\vc r,t+1)|$ is even}
                            \State $s_k(\vc r,t+1)\gets 0$
                            \If{$k<K-1$}
                                \State $\tau_k(\vc r,t+1)\gets 0$
                            \EndIf
                        \Else
                            \State $s_k(\vc r,t+1)\gets 1$
                            \If{$k<K-1$}
                                \State $\tau_k(\vc r,t+1)\gets
                                \min\{\vartheta:(\cdot,\vartheta)\in\mathcal R_k(\vc r,t+1)\}$
                            \EndIf
                        \EndIf
                    \EndFor
                \EndFor
            \end{algorithmic}
        \end{algorithm}

        We next specify the message-sector update. For each channel $(i,j)\in\mathcal M_2$, define the two source positions for growth into $\vc r$ by
        \[
            N_{ij}(\vc r)
            \coloneqq
            \{\vc r+(-1)^{i+1}\uvec x,\,
                \vc r+(-1)^{j+1}\uvec y\},
        \]
        with all positions interpreted in $\Lambda_k$.

        The Toom vote in channel $(i,j)$ is
        \[
            \Call{ToomVote}{i,j,k,\vc r,t}
            \coloneqq
            \Call{Maj}{
                m_{ij,k}(\vc r,t),\,
                m_{ij,k}(\vc r+(-1)^{i+1}\uvec x,t),\,
                m_{ij,k}(\vc r+(-1)^{j+1}\uvec y,t)
            },
        \]
        and the coupling condition is
        \[
            C_k(\vc r,t)
            \coloneqq
            \bigl[
                m_{00,k}(\vc r,t)=1
                \;\text{and}\;
                \Call{ToomVote}{0,0,k,\vc r,t}=1
            \bigr].
        \]

        On timed slices, define the valid growth-source set
        \[
            V_{ij,k}(\vc r,t)
            \coloneqq
            \bigl\{
                \vc r'\in N_{ij}(\vc r):
                m_{ij,k}(\vc r',t)=1
                \;\text{and}\;
                \theta_{ij,k}(\vc r',t)<t_0-1
            \bigr\},
        \]
        and the persistence condition
        \[
            P_{ij,k}(\vc r,t)
            \coloneqq
            \bigl[
                m_{ij,k}(\vc r,t)=1
                \;\text{and}\;
                \theta_{ij,k}(\vc r,t)<t_0-1
                \;\text{and}\;
                \bigl(
                    \Call{ToomVote}{i,j,k,\vc r,t}=1
                    \;\text{or}\;
                    C_k(\vc r,t)=1
                \bigr)
            \bigr].
        \]
        On the final slice, the persistence condition is the same with the timer condition removed:
        \[
            P_{ij,K-1}(\vc r,t)
            \coloneqq
            \bigl[
                m_{ij,K-1}(\vc r,t)=1
                \;\text{and}\;
                \bigl(
                    \Call{ToomVote}{i,j,K-1,\vc r,t}=1
                    \;\text{or}\;
                    C_{K-1}(\vc r,t)=1
                \bigr)
            \bigr].
        \]
        The message update is performed after the defect update on each slice that fires. If no message candidate is produced in a channel, then that message is erased. If one or more candidates are produced, then a message is written, and on timed slices its timer is taken to be the minimum candidate timer. On the final slice, all timer fields are omitted.

        \begin{algorithm}[H]
            \caption{Message-sector update for the hierarchical toric-code decoder at time $t$.}
            \label{alg:cm-toric-message-update}
            \begin{algorithmic}[1]
                \ForAll{$k=0,\ldots,K-1$ with $t\in n^k\Z$}
                    \ForAll{$\vc r\in\Lambda_k$}

                        \If{$k<K-1$}
                            \State $\mathcal D\gets\emptyset$
                            \Comment{Defect-source candidate timers}

                            \If{$s_k(\vc r,t+1)=1$}
                                \State Add $\tau_k(\vc r,t+1)$ to $\mathcal D$
                                \Comment{Surviving defect source}
                            \EndIf

                            \If{$s_k(\vc r,t)=1$ and $\tau_k(\vc r,t)+1<t_0$}
                                \State Add $\tau_k(\vc r,t)+1$ to $\mathcal D$
                                \Comment{Old non-promoted defect source}
                            \EndIf

                            \ForAll{$(i,j)\in\mathcal M_2$}
                                \State $\mathcal B\gets\mathcal D$
                                \Comment{Candidate timers for channel $(i,j)$}

                                \State $V\gets
                                \{\vc r'\in N_{ij}(\vc r):
                                m_{ij,k}(\vc r',t)=1
                                \;\text{and}\;
                                \theta_{ij,k}(\vc r',t)<t_0-1\}$

                                \If{$c_k(\vc r,t)\in\{0,1\}$ and $V\neq\emptyset$}
                                    \State Add
                                    $1+\min_{\vc r'\in V}\theta_{ij,k}(\vc r',t)$
                                    to $\mathcal B$
                                    \Comment{Growth candidate}
                                \EndIf

                                \State $C\gets
                                [m_{00,k}(\vc r,t)=1
                                \;\text{and}\;
                                \Call{ToomVote}{0,0,k,\vc r,t}=1]$
                                \Comment{Coupling condition}

                                \State $P\gets
                                [m_{ij,k}(\vc r,t)=1
                                \;\text{and}\;
                                \theta_{ij,k}(\vc r,t)<t_0-1$
                                \Statex \hspace{2.35cm}
                                $\text{and}\;
                                (\Call{ToomVote}{i,j,k,\vc r,t}=1
                                \;\text{or}\;
                                C=1)]$
                                \Comment{Persistence condition}

                                \If{$P$}
                                    \State Add $\theta_{ij,k}(\vc r,t)+1$ to $\mathcal B$
                                    \Comment{Persistence candidate}
                                \EndIf

                                \If{$\mathcal B=\emptyset$}
                                    \State $m_{ij,k}(\vc r,t+1)\gets 0$
                                    \State $\theta_{ij,k}(\vc r,t+1)\gets 0$
                                \Else
                                    \State $m_{ij,k}(\vc r,t+1)\gets 1$
                                    \State $\theta_{ij,k}(\vc r,t+1)\gets\min\mathcal B$
                                \EndIf
                            \EndFor

                        \Else
                            \State $S\gets [s_k(\vc r,t+1)=1]\vee[s_k(\vc r,t)=1]$
                            \Comment{Defect-source condition on the final slice}

                            \ForAll{$(i,j)\in\mathcal M_2$}
                                \State $G\gets
                                [c_k(\vc r,t)\in\{0,1\}
                                \;\text{and}\;
                                \exists\,\vc r'\in N_{ij}(\vc r)
                                \text{ s.t. }
                                m_{ij,k}(\vc r',t)=1]$
                                \Comment{Growth condition}

                                \State $C\gets
                                [m_{00,k}(\vc r,t)=1
                                \;\text{and}\;
                                \Call{ToomVote}{0,0,k,\vc r,t}=1]$
                                \Comment{Coupling condition}

                                \State $P\gets
                                [m_{ij,k}(\vc r,t)=1
                                \;\text{and}\;
                                (\Call{ToomVote}{i,j,k,\vc r,t}=1
                                \;\text{or}\;
                                C=1)]$
                                \Comment{Persistence condition}

                                \State $m_{ij,k}(\vc r,t+1)\gets S\vee G\vee P$
                            \EndFor
                        \EndIf

                        \State $c_k(\vc r,t+1)\gets (c_k(\vc r,t)+1)\bmod q$
                    \EndFor
                \EndFor
            \end{algorithmic}
        \end{algorithm}

        As before, the notation in Algorithms~\ref{alg:cm-toric-defect-update} and~\ref{alg:cm-toric-message-update} is an effective, non-local description of a local process.
        
    \subsection{Bounds on memory lifetimes} \label{subsec:cm-online-lifetime}

        We now prove memory-lifetime bounds for both hierarchical streaming decoders.

        \begin{theorem}[stretched-exponential lifetime for hierarchical streaming decoders]
            \label{thm:cm-online-threshold}
            There exist constants
            \[
                p_*,c,c_r,C',\alpha,t_0>0,
            \]
            sparsity-theorem parameters $(\beta,\gamma,n)$ satisfying inequalities~\eqref{paramineqs}, and an integer $q>0$ such that the following holds.

            Let $q$ be the clock period used by both hierarchical decoders. Suppose the data-qubit and stabilizer-measurement-bit noise is $p$-bounded with $p<p_*/2$. Suppose also that the number of slices $K$ satisfies
            \[
                K\geq k_L,
            \]
            and that the cleanout step proportionality constant is $c_r$.

            Initialize all data qubits in the all-$0$ product state, so that the value of every logical $Z$ operator is deterministic in the absence of noise. Run either the coarse-grained repetition-code decoder or the coarse-grained toric-code decoder for $T$ syndrome-extraction rounds, and then perform the destructive computational-basis readout procedure of Section~\ref{subsec:readout}.

            Then, for both decoders and every $T\geq 1$,
            \[
                \Pr[\textup{logical failure by time }T]
                \leq
                TC'\exp(-cL^\alpha).
            \]
        \end{theorem}

        The same theorem holds for arbitrary toric-code encoded states with known initial stabilizer values. As in the previous section, we have stated the theorem only for the all-$0$ initial state for simplicity.

        \begin{proof}
            Let $q>0$ be an integer such that linear cluster erosion holds for the code-capacity repetition-code and toric-code decoders, and let $(\beta,\gamma,n)$ be sparsity-theorem parameters satisfying the inequalities~\eqref{paramineqs}. Next, let $p_*>0$ be such that if $p<p_*$, then the sparsity theorem holds.
        
            We begin by temporarily working in an infinite system with infinitely many slices. Suppose the spacetime history contains a single cluster $C$, where $C$ is a $k$-cluster that sources at least one detector event. Let $u$ be the first time at which a detector event sourced by $C$ is inserted into the decoder. Since the spacetime diameter of $C$ is $O(w_k)$, there is a constant $A_0$ such that all detector events due to $C$ are loaded into the decoder by time
            \[
                u+A_0w_k .
            \]
        
            Next, we bound the time needed for activity sourced by $C$ either to be erased below slice $k$ or to reach slice $k$. On a timed slice $j$, a defect or message can survive for at most $t_0$ slice-$j$ firings before it is erased or promoted. Therefore, there are constants $A_1,A_2>0$ such that the time spent at slice $j$ before the activity is either erased or has reached slice $j+1$ is at most
            \[
                A_1t_0n^j+A_2n^{j+1},
            \]
            where the second term accounts for both the transit time to the parent and the possible wait time between slice-$(j+1)$ firings. Consequently, the total time spent below slice $k$ is bounded by
            \[
                \sum_{j=0}^{k-1}
                \left(
                    A_1t_0n^j+A_2n^{j+1}
                \right)
                \leq
                A_3
                \left(
                    \frac{t_0}{n}+1
                \right)
                n^k
            \]
            for some constant $A_3>0$. Hence, by time
            \[
                u+
                A_4
                \left(
                    w_0+
                    \frac{t_0}{n}+1
                \right)
                n^k ,
            \]
            all non-trivial non-correction variables due to $C$ have been erased below slice $k$ and any remaining defects have been loaded into slice $k$, where $A_4>0$ is a constant.

            Since the spatial support of all correction flips and non-trivial non-correction variables generated by $C$ grows by at most a constant amount per unit time (where we consider the correction string flipped at the end of child-to-parent propagation as ``growing'' a constant amount per unit time of travel, even though it technically is flipped instantaneously at the end of the propagation), it follows that by this time the spatial support of all non-trivial non-correction variables due to $C$ is contained in a region of diameter at most
            \[
                A_5
                \left(
                    w_0+\frac{t_0}{n}+1
                \right)n^k
            \]
            for some constant $A_5>0$. Therefore, when viewed on the slice-$k$ lattice, the non-trivial non-correction variables due to $C$ are contained in a region of diameter at most
            \[
                R =
                A_6
                \left(
                    w_0+\frac{t_0}{n}+1
                \right)
            \]
            for some constant $A_6>0$.
            
            Let us temporarily suppress promotion out of slice $k$. Once all slice-$k$ activity sourced by $C$ has arrived, the remaining dynamics reduce to those of the corresponding code-capacity decoder on the slice-$k$ lattice, applied to an input of diameter at most $R$. Since the code-capacity decoder satisfies linear cluster erosion, there is a constant $A_7>0$ such that this remaining activity is erased in at most
            \[
                A_7
                \left(
                    w_0+\frac{t_0}{n}+1
                \right)
            \]
            additional slice-$k$ firings.
        
            Thus, $C$ is erased within at most
            \[
                A_8
                \left(
                    w_0+\frac{t_0}{n}+1
                \right)n^k
            \]
            time steps after the first defect due to $C$ is loaded into the decoder, where $A_8>0$ is a constant. In order to avoid promotion out of slice $k$ before the cluster has been completely decoded, it suffices to have
            \[
                A_8
                \left(
                    w_0+\frac{t_0}{n}+1
                \right)n^k
                <
                t_0 n^k,
            \]
            or equivalently
            \[
                A_8
                \left(
                    \frac{w_0}{t_0}
                    +\frac{1}{n}
                    +\frac{1}{t_0}
                \right)
                <1.
            \]
            This can be accomplished by taking $n$ and $t_0$ sufficiently large. Therefore, in an infinite system with infinitely many slices, $C$ is completely erased in $O(w_k)$ time before any activity sourced by it can be promoted beyond slice $k$.
        
            It remains to pass from the single-cluster estimate to a full noise history. As in the full-resolution hierarchy, we may make $b_k/w_k=b_0/w_0$ and $b_k/t_k = b_0/t_0$ arbitrarily large by taking $n$ sufficiently large, so that any two distinct clusters cannot interact before one of them has been erased. By performing the same induction on cluster levels as in the full-resolution hierarchy, it then follows that all clusters are decoded independently; that is, each cluster evolves exactly as it would if no other clusters were present.
        
            The remainder of the proof then proceeds exactly as in the proof of Theorem~\ref{thm:stretched-exponential-lifetime}.
        \end{proof}

        Thus, with $K=\Theta(\log L)$ slices, the hierarchical streaming decoder achieves a stretched-exponential memory lifetime, just like the non-hierarchical streaming decoder.

        We can also combine the coarse-grained hierarchy with the back-wall argument of Ref.~\cite{lake2025localactiveerrorcorrection}. As in Theorem~\ref{thm:polyloglog-overhead}, the lower slices renormalize the noise experienced by the back wall, so the final slice sees a noise model generated only by larger, rarer clusters of level at least $K-1$.

        Because each slice's decoder satisfies (a suitably defined generalization of) linear cluster erosion, the back-wall analysis of Ref.~\cite{lake2025localactiveerrorcorrection} applies to this renormalized noise model as in Theorem~\ref{thm:polyloglog-overhead}. Thus, for sufficiently small physical error rates, the hierarchical streaming decoder retains a constant, non-zero threshold and achieves the same stretched-exponential memory-lifetime scaling using only
        \[
            K=O(\log\log L)
        \]
        slices. 

        The benefits of this $O(\log \log L)$ scheme are two-fold. First, at finite system sizes, it further reduces the already constant per-site overhead of the decoder. Second, it relaxes the size constraint on the lattice. In particular, the full-depth construction requires $L$ to be divisible by $n^K$ with $K=\Theta(\log L)$, whereas the low-depth version only requires divisibility by $n^K$ for $K=O(\log\log L)$.

    \subsection{State preparation, state injection, and open boundary conditions} \label{subsec:cm-readout-prep-injection-boundaries}
            
        The state-preparation, state-injection, and open-boundary constructions from the translation-invariant decoder have direct coarse-grained analogs. In each case, the full-resolution auxiliary slices are replaced by the coarse-grained slices $\Lambda_k$, and the same type of boundary, rising-wall, rising-absorbing-region, and frame-update mechanisms are implemented on the corresponding coarse-grained lattices.

        We do not give explicit algorithms for these variants, since they require no new ideas and are rather tedious to write down. The only additional bookkeeping comes from promotions, child-to-parent correction-frame bits, and finite signal-propagation delays. All of these effects can be rigorously dealt with by using the exact same techniques introduced in the proof of the memory-lifetime bound for the hierarchical streaming decoders.

        Consequently, the hierarchical streaming decoders support the same local open-boundary decoding, state-preparation, and state-injection procedures as the translation-invariant decoders. With $K=\Theta(\log L)$ slices, the rising-wall preparation and injection protocols inherit the same rigorous stretched-exponential lifetime guarantees. Furthermore, the injection protocol inherits the same type of rigorous injection error probability bounds as the non-hierarchical-decoder injection protocol. With $K=\Theta(\log\log L)$ slices, the postselected preparation and injection variants likewise retain the corresponding rigorous stretched-exponential lifetime bounds and so on.

    \subsection{Constant-density layout and constant-bandwidth control} \label{subsec:cm-online-layout}

        We now prove that the toric- and surface-code decoders can be implemented in two dimensions using a constant density of quantum and classical resources.

        Before proving this result, we give a simple counting argument to show why the existence of such a layout is plausible. The total number of sites in the hierarchy is
        \[
            \sum_{k=0}^{K-1} |\Lambda_k|
            =
            \sum_{k=0}^{K-1}
            \left(\frac{L}{n^k}\right)^2
            =
            O(L^2).
        \]
        Moreover, under a natural coarse-grained placement of sites, neighboring slice-$k$ sites are separated by distance $O(n^k)$, so the length of a wire between two neighboring same-slice sites is $O(n^k)$. Since there are $O((L/n^k)^2)$ slice-$k$ sites, the total same-slice wire length at slice $k$ is
        \[
            O\left(
                \left(\frac{L}{n^k}\right)^2 n^k
            \right)
            =
            O\left(\frac{L^2}{n^k}\right).
        \]
        Promotion wires from slice $k$ to slice $k+1$ have length $O(n^{k+1})$, so their total length is also
        \[
            O\left(
                \left(\frac{L}{n^k}\right)^2 n^{k+1}
            \right)
            =
            O\left(\frac{L^2}{n^k}\right).
        \]
        Summing over $k$ gives total wire length
        \[
            \sum_{k=0}^{K-1}
            O\left(\frac{L^2}{n^k}\right)
            =
            O(L^2).
        \]
        Thus, both the number of sites and the total wire length are extensive in the number of microscopic sites.

        Note that in one spatial dimension, the same argument does not hold: a wire between neighboring slice-$k$ sites has length $\Omega(n^k)$, and since there are $\Theta(L/n^k)$ such wires, every slice contributes $\Omega(L)$ of total wire length. Summing over slices then gives a wire density that grows with the number of slices.

        \begin{lemma}[bounded-density layout]
            \label{lem:bounded-density-layout}
                For every even integer coarse-graining factor $n>1$, there exists a two-dimensional layout for the toric-code hierarchy with the following properties. There are constants $\delta>0$ and $0 < c < \infty$, depending only on $n$, such that:
                \begin{itemize}
                    \item any two distinct sites are separated by distance at least $\delta$;
                    \item slice-$k$ nearest-neighbor sites are separated by distance $O(n^k)$;
                    \item same-slice nearest-neighbor wires for slice $k$ have length $O(n^k)$;
                    \item promotion wires from slice $k$ to slice $k+1$ have length $O(n^{k+1})$;
                    \item at most $c$ wires pass through any point on the torus.
                \end{itemize}
                Here, all distances and lengths are measured with respect to the Euclidean metric.
        \end{lemma}

        In the above lemma, we assume that $n$ is even in order to simplify the proof, but it is not essential. An analogous lemma holds for a suitably defined version of the surface-code hierarchy.

        \begin{proof}
            Consider an $L \times L$ torus. Place each slice-$k$ site $\vc r=(r_x,r_y)\in\Lambda_k$ at
            \[
                P_k(\vc r)
                =
                \left(n^k \left(r_x+\frac{1}{2}\right),\; n^k \left(r_y+\frac{1}{2}\right)\right)
                \pmod L.
            \]
            Define the set of $k$-street coordinates by
            \[
                S_k
                =
                \left\{n^k\left(m + \frac{1}{2}\right) \pmod L : m\in\mathbb Z\right\}
                \subset \mathbb R/(L\mathbb Z),
            \]
            and define the following sets of horizontal and vertical lines:
            \[
                \mathcal H_k
                =
                \mathbb R/(L\mathbb Z)\times S_k,
                \qquad
                \mathcal V_k
                =
                S_k\times \mathbb R/(L\mathbb Z).
            \]
            We call these lines horizontal and vertical $k$-streets, respectively (Fig.~\ref{fig:toric-coarse-grained-layout}).
            
            Slice-$k$ sites lie at intersections of horizontal and vertical $k$-streets. Same-slice nearest neighbors are separated by exactly $n^k$, so same-slice wires have length $O(n^k)$ when routed along the corresponding $k$-streets.

            Consider a promotion wire from a slice-$k$ child to its slice-$(k+1)$ parent. Write
            \[
                \vc r=n\vc R+\vc a,
                \qquad
                \vc a\in\{0,\ldots,n-1\}^2,
            \]
            so that $\pi_k(\vc r)=\vc R$. Route the wire along the following shortest Manhattan-distance path composed of two straight line segments:
            \[
                P_k(\vc r)
                \longrightarrow
                (P_{k+1,x}(\vc R), P_{k,y}(\vc r))
                \longrightarrow
                P_{k+1}(\vc R).
            \]
            Since
            \[
                P_k(\vc r)-P_{k+1}(\vc R)
                =
                \left(
                    n^k \left(a_x + \frac{1 - n}{2}\right),
                    \;
                    n^k \left(a_y + \frac{1 - n}{2}\right)
                \right),
            \]
            each coordinate difference is $O(n^{k+1})$. Thus, every promotion wire has length $O(n^{k+1})$. Its two line segments lie on a horizontal $k$-street and a vertical $(k+1)$-street, respectively.

            It remains to check bounded density. The sets $S_k$ are uniformly separated. Indeed, if $k < j$, then
            \[
                S_k
                =
                n^k\left(\mathbb Z+\frac12\right)
                \pmod L,
            \]
            while
            \[
                S_j
                =
                n^j\left(\mathbb Z+\frac12\right)
                =
                n^k n^{j-k}\left(\mathbb Z+\frac12\right)
                \equiv
                0
                \pmod{n^k}.
            \]
            Thus, points of $S_k$ are congruent to $n^k/2$ modulo $n^k$, whereas points of $S_j$ are congruent to $0$ modulo $n^k$, which implies
            \[
                \operatorname{dist}(S_k,S_j)
                \geq
                \frac{n^k}{2}
                \geq
                \frac12.
            \]
            Thus, distinct streets of the same orientation are separated by a positive constant independent of $k$ and $L$. Furthermore, points within the same set $S_k$ are separated by distance $n^k\geq 1$. Consequently, since sites lie at street intersections, distinct sites are separated by some $\delta>0$ independent of $k$ and $L$.

            Finally, fix a point on the torus. Assume that wires have some finite width that is less than half the separation between streets. Then, this point lies on wires from at most one horizontal street and one vertical street. On a horizontal $m$-street, the only horizontal wires are same-slice $m$-wires and horizontal promotion segments from slice $m$ to slice $m+1$. Similarly, on a vertical $m$-street, the only vertical wires are same-slice $m$-wires and vertical promotion segments from slice $m-1$ to slice $m$. At most $O(n^2)$ wires can pass through the point due to these horizontal and vertical streets. Therefore, there exists some $n$-dependent constant $c > 0$ such that at most $c$ wires pass through any point on the torus.
        \end{proof}

        We have now shown that the hierarchical streaming decoder for the toric code can be implemented using a constant density of sites and wires. All that remains is to show that the wires can be taken to have bounded bandwidth and bounded propagation speed. By the arguments in Section~\ref{subsec:cm-online-toric}, it is clear that we can do this, provided that every child-to-parent record arrives after a fixed delay that depends only on the slice and not on the particular child.

        This arrival schedule can be implemented using wires with a constant number of bounded propagation speeds: by the self-similarity of the layout, the ratio of a wire's length to its required delay depends only on the child label and not on $k$, and there are only $n^2$ child labels. Consequently, the hierarchical streaming decoder for the toric code can be implemented using a constant density of sites and of bounded-bandwidth, bounded-propagation-speed wires. An analogous proof holds for the surface code.

        The toric-code layout admits a natural surface-code analog, obtained by applying the same construction to the same lattice but with open boundary conditions. We now explain how to use this constant-density layout to control the surface-code decoder during readout, state preparation, and state injection.

        First, let us consider readout. By the wire arguments above, we may arrange for a control pulse broadcast from the top of the hierarchy to reach all sites in any given slice simultaneously. In order to implement readout, we begin by sending such a pulse; this pulse carries a constant number of bits, specifying the pulse type and the readout measurement basis. When this pulse reaches slice $0$, all data qubits are measured transversally in the specified basis, and the reconstructed detector events are inserted into the decoder. After $O(L)$ additional time steps, the cleanout is complete (with high probability), and a second global pulse can be sent requesting the logical value.

        Once the second pulse is received, the final logical value is calculated in a distributed manner. Each site computes its local contribution to the logical parity: at slice $0$, the measurement outcomes on qubits that lie on the logical string, corrected by the slice-$0$ correction bits alone; and at higher slices, the correction bits whose expanded correction strings have odd overlap with the logical string. Since the expanded correction strings and the logical string are both predetermined, we can precompute which of a site's outcomes and correction bits should contribute. These contributions can be combined using the child-to-parent wires: each site sends a single bit to its parent, and each parent forwards the parity of its own contribution and the bits received from its children. The aggregate logical value reaches the top of the hierarchy in $O(L)$ time steps, at which point the logical outcome is made available at the IO port at the center of the patch (Fig.~\ref{fig:toric-coarse-grained-layout}).

        Next, we consider stabilizer-state preparation. We again begin by broadcasting a pulse specifying the pulse type and the state to be prepared. Without loss of generality, suppose we prepare a logical $Z$ eigenstate. Let us first consider the $Z$-type stabilizer decoding sector. When the pulse reaches a slice, it resets all decoder variables on that slice. Furthermore, when it reaches slice $0$, the data qubits are initialized in a physical computational basis state that, upon measurement of the code stabilizers, yields the desired logical state. The $Z$-type stabilizer values determined by this product state are used as the known initial $Z$-type stabilizer configuration. Decoding in this sector then proceeds as in the surface-code version of the ordinary hierarchical streaming decoder.

        Now consider the $X$-type stabilizer decoding sector, whose initial stabilizer values are unknown. When the pulse reaches a slice $k>0$, it resets all decoder variables on that slice and marks the slice as absorbing. When it reaches slice $0$, a hierarchical version of the state-preparation procedure of Section~\ref{subsec:online-state-preparation} begins: the first round of post-initialization stabilizer measurements initializes the frame interpretation, and the rising wall then ascends the hierarchy, with each slice's activation time implemented by clock pulses that travel along wires, just as for the slice-$k$ firing schedule. Each absorption at an absorbing slice-$k$ site toggles a locally stored bit, which flips the frame interpretation of the site's associated microscopic check.

        An analogous procedure can be used to perform state injection.

\section{Local fault-tolerant quantum computation} \label{sec:decoding-during-computation}
    
    In this section, we explain how to perform local decoding during fault-tolerant quantum computation.

    We realize universal quantum computation using surface codes and the following standard set of primitives: Clifford gates, $T$ gates, stabilizer-state preparation, and single-qubit Pauli measurements. We generate the Clifford group by implementing logical $\CNOT$ gates using lattice surgery; logical $H$ gates using the surface code's fold-transversal $H$ gate; and logical $S$ gates using $Y$-state distillation. We implement logical $T$ gates using magic state distillation.

    We have already explained how to perform local decoding during single-qubit Pauli measurements, state preparation, and state injection. Recall that (i) $Y$-state distillation can be performed using $Y$-state injection and the Clifford subgroup generated by $H$ and $\CNOT$, and (ii) magic state distillation can be performed using magic state injection and arbitrary Clifford operations. Therefore, all that remains is to explain how to perform local decoding during lattice surgery and logical $H$ gates.

    For simplicity, the local decoding protocols described in this section will be implemented using the translation-invariant surface-code decoder of Section~\ref{sec:online-decoding}. A posteriori, it will be clear that these protocols admit hierarchical analogs with the exact same rigorous performance guarantees as their non-hierarchical counterparts. Since these analogs require no new ideas or proof techniques and are rather tedious to write down, we will not give explicit constructions here.

    The section is organized as follows. First, we describe how to perform local decoding during logical $H$ gates. Then, we describe how to perform local decoding during lattice surgery. Next, we discuss the hierarchical analogs of these protocols and explain how to implement and control them using the constant-density layout of Section~\ref{subsec:cm-online-layout}. Finally, we describe how to construct a local, fault-tolerant quantum computer in two dimensions, and we prove a threshold theorem for this construction.

    The formal arguments in this section will be somewhat abbreviated since they closely follow the proofs of Section~\ref{sec:online-decoding}. For each local decoding procedure introduced below, we describe only the modifications to the arguments of that section needed to obtain rigorous performance guarantees. Each such proof outline can be expanded into a detailed proof in a straightforward but tedious manner. Each such expansion requires no new ideas or proof techniques other than those already used in previous sections.

    \subsection{Hadamard} \label{subsec:computation-hadamard}

        We implement logical Hadamard as follows. First, we apply a physical Hadamard to every data qubit. Then, from the next syndrome-extraction round onward, sites that previously measured $Z$-type stabilizers instead measure $X$-type stabilizers with the same support, and vice versa.

        The decoders for the $X$- and $Z$-type stabilizer sectors are then modified as follows. Let $T-1$ be the last syndrome-extraction round before the physical Hadamards and let $T$ be the first syndrome-extraction round afterward. Suppose that before the Hadamard, the stabilizer at site $a$ was a $Z$-type stabilizer. Then, we modify the detector event $\phi(a,T)$ to be
        \[
            \phi(a,T)
            =
            \tilde s_T^+(a)
            \oplus
            \tilde s_{T-1}^-(a),
        \]
        where $\tilde s_T^+(a)$ is the first post-Hadamard measurement outcome of the now $X$-type stabilizer at site $a$ and $\tilde s_{T - 1}^-(a)$ is the last pre-Hadamard measurement outcome of the formerly $Z$-type stabilizer at site $a$. For $t>T$, the detector event is given by the difference of post-Hadamard $X$-type stabilizer measurements:
        \[
            \phi(a,t)
            =
            \tilde s_t^+(a)
            \oplus
            \tilde s_{t-1}^+(a).
        \]
        Detector events for sites that formerly measured $X$-type stabilizers undergo the same modification with $X$ and $Z$ interchanged.

        We also relabel the $X$-type stabilizer decoding sector as the $Z$-type stabilizer decoding sector and vice versa. Specifically, every stored decoding variable keeps its value and is simply reinterpreted as belonging to the opposite sector. Additionally, the supports of the logical $X$ and $Z$ operators are interchanged.
    
    \subsection{Lattice surgery}

        \begin{figure}[t]
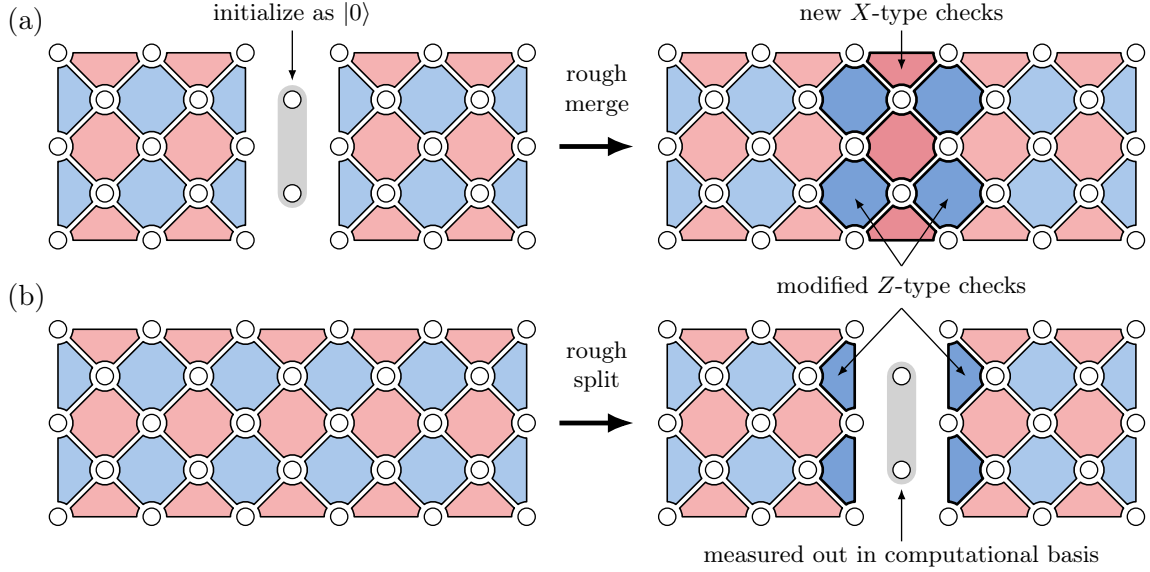
\centering
            \adjustbox{max width=\linewidth}{%
            \tikzsetnextfilename{surgery-merge-split}
%
            }
            \caption{Geometry of lattice-surgery merges and splits. \textbf{(a)}~Rough merge. Seam qubits are initialized in $\ket{0}$, after which modified $Z$-type checks and new $X$-type checks are measured along the seam. This operation merges the two surface-code patches together. \textbf{(b)}~Rough split. Seam qubits are measured out in the computational basis, after which $Z$-type checks along the seam return to being measured in their pre-merge form. This operation splits the merged patch into two separate surface-code patches. Smooth merges and splits are defined analogously, with $X$ and $Z$ interchanged.}
        \end{figure}

        \begin{figure}[t]
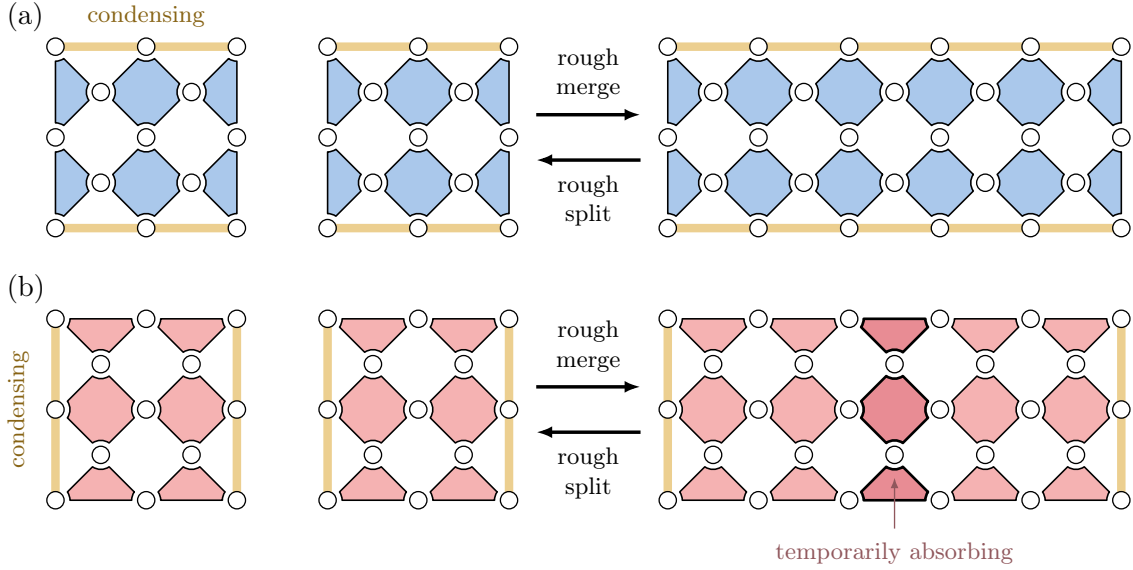

            \centering
            \adjustbox{max width=\linewidth}{%
            \tikzsetnextfilename{surgery-spacetime}
}
            \caption{Spatial decoding graphs for rough merges and splits. \textbf{(a)}~The pre-merge/post-split (left) and post-merge/pre-split (right) spatial geometries for the $Z$-type stabilizer decoding sector. Both operations are called non-condensing for this sector because the condensing boundaries do not participate in the merge or split. \textbf{(b)}~The same spatial geometries for the $X$-type stabilizer decoding sector. Both operations are called condensing for this sector because the boundaries along the seam are themselves condensing. After the merge, the seam is also temporarily treated as absorbing in this sector.}
        \end{figure}

        This section assumes familiarity with lattice surgery for the standard surface code. In particular, we use the merge and split operations described by Horsman et al.\@ in Ref.~\cite{Horsman_2012}, and we refer readers to that work for further details on these operations.

        In what follows, we will apply the sparsity theorem to the full spacetime decoding graph determined by the surgery schedule. Throughout, a $k$-cluster refers to a $k$-cluster in this graph. We will also assume that distinct surgery operations are separated from each other in spacetime by a sufficiently large constant multiple of $L$.
        
        The arguments below assume that $K=\Theta(\log L)$. It may be possible to extend these types of arguments to $K=\Theta(\log\log L)$, e.g., by using a protocol that keeps an absorbing back wall in place until $O(L)$ consecutive time steps pass without any defect being absorbed by the wall; we will not pursue such an extension here.

        \subsubsection{Non-condensing merge}

            For a fixed CSS decoding sector, we say that a lattice-surgery merge operation is \emph{non-condensing} if, before the merge, the seam along which the merge is performed is a non-condensing boundary for that sector's defects, i.e., defects cannot terminate on that boundary.

            We now describe how to perform local decoding during a non-condensing merge. For concreteness, we describe how to perform local decoding for the $Z$-type stabilizer sector during a rough merge; here, we are using the convention that rough boundaries are those whose $Z$-type stabilizers consist primarily of three-body stabilizers, so this is indeed a non-condensing merge for the $Z$-type stabilizer sector. 

            We assume that the new qubits incorporated during the merge are initialized as $\ket{0}$. We refer to these qubits as seam qubits. The first time a non-condensing merge is performed along a given seam, the correction channels for each new seam qubit $q$ are initialized as
            \[
                e_{q,k}(T)=0, \qquad k = 0,\ldots,K-1.
            \]
            In subsequent non-condensing merges and splits along the same seam, these channels are not reset and retain their accumulated values from previous operations.
            
            Let $T$ be the last pre-merge syndrome-extraction round, so $T+1$ is the first post-merge syndrome-extraction round. For each site $a$ whose check has been modified by the merge operation, let $B_a^-$ denote the pre-merge boundary check obtained by deleting the newly introduced qubits from the support of the post-merge check, so
            \[
                B_a^+ = B_a^- Z_{q_a},
            \]
            where $B_a^+$ is the post-merge check and $q_a$ is the new qubit adjacent to $a$.

            We refer to checks modified by the merge as seam checks. For each seam check $a$, we define the detector event $\phi(a, T+1)$ by
            \[
                \phi(a,T+1)
                =
                \tilde s_{T+1}^{+}(a)
                \oplus
                \tilde s_T^{-}(a),
            \]
            where $\tilde s_T^{-}(a)$ is the final pre-merge measurement outcome of $B_a^-$ and $\tilde s_{T+1}^{+}(a)$ is the first post-merge measurement outcome of $B_a^+$. For $t>T+1$, detector events are given by
            \[
                \phi(a,t)
                =
                \tilde s_t^{+}(a)
                \oplus
                \tilde s_{t-1}^{+}(a),
            \]
            where $\tilde s_t^{+}(a)$ and $\tilde s_{t-1}^{+}(a)$ are the measurement outcomes of $B_a^+$ at times $t$ and $t-1$, respectively. Detector events away from the seam are unchanged.

            Next, we specify how the decoder geometry changes. Let $\Gamma^-_Z$ be the pre-merge spatial geometry for the $Z$-type stabilizer decoding sector, and let $\Gamma^+_Z$ be its post-merge spatial geometry. We introduce the post-merge spatial geometry into the decoder slice by slice. Define
            \[
                \Delta_k
                =
                \sum_{j=0}^{k-1} t_j.
            \]
            During the update from time $t$ to time $t+1$, slice $k$ uses the geometry
            \[
                \Gamma_{Z,k}(t)
                =
                \begin{cases}
                    \Gamma^-_Z , & t<T+\Delta_k, \\
                    \Gamma^+_Z , & t\geq T+\Delta_k.
                \end{cases}
            \]
            Apart from the initialization of the new seam-qubit channels; the modified detector events; and the gradual slice-by-slice introduction of the post-merge geometry, the decoder rule is unchanged.

            We now describe how to modify the arguments of Section~\ref{sec:online-decoding} to obtain rigorous performance guarantees for the local decoder during the non-condensing merge. Let us temporarily work in an infinite system with infinitely many slices. Consider a spacetime history consisting of a single cluster $C$, where $C$ is a $k$-cluster that sources at least one detector event. If the cluster is separated in spacetime from the merge operation by more than a constant multiple of $w_k$, then it evolves either entirely in the post-merge spatial geometry $\Gamma^+_Z$ or entirely on one side of the pre-merge spatial geometry $\Gamma^-_Z$, and the ordinary linear cluster-erosion argument for the surface code decoder applies.
    
            It remains to consider a $k$-cluster whose $O(w_k)$ spacetime neighborhood intersects the merge operation. The detector events sourced by such a cluster are inserted over a time interval of length $O(w_k)$. Moreover, the post-merge spatial geometry reaches slice $k$ after the delay
            \[
                \Delta_k
                =
                \sum_{j=0}^{k-1} t_j
                \leq
                2t_{k-1}.
            \]
            Let us temporarily assume that $t_k$ is infinite. Then, after time $O(t_{k-1} + w_k)$, slice $k$ is guaranteed to both contain all activity sourced by the cluster and be using the post-merge spatial geometry. Consequently, since the code-capacity decoder satisfies linear cluster erosion, all defects and messages sourced by the cluster are erased in time $O(w_k)$. As usual, by choosing $t_0$ and $n$ sufficiently large, it follows that when $t_k$ is finite, this erosion finishes before any activity sourced by the cluster can be promoted beyond slice $k$. Therefore, if the noise history consists of a single $k$-cluster, then this cluster's defects and messages are erased in linear time and its effects never reach beyond the $k$th slice. 

            We may then repeat the usual arguments to prove that the decoding of all clusters decouples when the noise history consists only of clusters of diameter at most $c L$ for some constant $c > 0$. Therefore, the same stretched-exponential memory lifetime guarantees for the ordinary surface-code decoder continue to hold during a non-condensing merge.

        \subsubsection{Non-condensing split}

            We next describe the inverse operation: a non-condensing split. For a fixed CSS decoding sector, we say that a lattice-surgery split is \emph{non-condensing} if, after the split, the seam becomes a non-condensing boundary for that sector's defects. For concreteness, we describe how to perform local decoding for the $Z$-type stabilizer sector during a rough split.

            During a rough split, the seam qubits are measured in the $Z$ basis and then removed from the code. Let $T$ be the last pre-split syndrome-extraction round, so $T+1$ is the first post-split syndrome-extraction round. For each check $a$ modified by the split, let $B_a^+$ denote the pre-split check and let $B_a^-$ denote the post-split check obtained by deleting the seam qubit from the support of $B_a^+$. 

            We refer to checks modified by the split as seam checks. For each seam check $a$, we define the detector event $\phi(a, T+1)$ by
            \[
                \phi(a,T+1)
                =
                \tilde s_{T+1}^{-}(a)
                \oplus
                \tilde z_{T+1}(q)
                \oplus
                \tilde s_T^{+}(a),
            \]
            where $\tilde s_T^{+}(a)$ is the final pre-split measurement outcome of $B_a^+$, $\tilde s_{T+1}^{-}(a)$ is the first post-split measurement outcome of $B_a^-$, and $\tilde z_{T+1}(q)$ is the measured $Z$-basis outcome of the seam qubit $q$ in support of $B_a^+$. For $t>T+1$, detector events are given by ordinary post-split syndrome differences:
            \[
                \phi(a,t)
                =
                \tilde s_t^{-}(a)
                \oplus
                \tilde s_{t-1}^{-}(a).
            \]
            Detector events away from the seam are unchanged.

            Next, we specify how the decoder geometry changes. Let $\Gamma^+_Z$ be the pre-split spatial geometry for the $Z$-type stabilizer decoding sector, and let $\Gamma^-_Z$ be its post-split spatial geometry. We introduce the post-split spatial geometry into the decoder slice by slice, with a slightly different schedule as compared to the non-condensing merge case. During the update from time $t$ to time $t+1$, slice $k$ uses the spatial geometry
            \[
                \Gamma_{Z,k}(t)
                =
                \begin{cases}
                    \Gamma^+_Z , & t<T+\Delta_{k+1}, \\
                    \Gamma^-_Z , & t\geq T+\Delta_{k+1}.
                \end{cases}
            \]

            Finally, we describe how to assign consistent interpretations to the seam qubit measurement outcomes. These outcomes are used to assign a consistent interpretation to the post-split seam checks, which are intrinsically non-deterministic. For each seam qubit $q$, we store the measurement outcome $\tilde z_{T+1}(q)$ and fold it into the slice-$0$ correction bit, $e_{q, 0}\to e_{q, 0}\oplus\tilde z_{T+1}(q)$. By construction, the corrections on the seam qubits are finalized after the post-split spatial geometry has been introduced in every slice of the decoder. For each seam check $a$ whose pre-split check contained $q$, the new stabilizer frame is committed as $\psi'(a)=\psi(a)\oplus F(q)$, where $\psi(a)$ is the pre-split stabilizer frame and $F(q)$ is the modulo-two sum over all slices' finalized correction bits on qubit $q$.   

            We now describe how to modify the arguments of Section~\ref{sec:online-decoding} to obtain rigorous performance guarantees for the local decoder during the non-condensing split. Let us temporarily work in an infinite system with infinitely many slices. Consider a spacetime history consisting of a single cluster $C$, where $C$ is a $k$-cluster that sources at least one detector event. If the cluster is separated in spacetime from the split operation by more than a constant multiple of $w_k$, then it evolves either entirely in the pre-split spatial geometry $\Gamma^+_Z$ or entirely on one side of the post-split spatial geometry $\Gamma^-_Z$, and the ordinary linear cluster-erosion argument for the surface code decoder applies.

            It remains to consider a $k$-cluster whose $O(w_k)$ spacetime neighborhood intersects the split operation. The detector events sourced by such a cluster are inserted over a time interval of length $O(w_k)$. Moreover, such a cluster must occur within $O(w_k)$ of the split operation. Therefore, all of the activity sourced by the cluster has either been erased or reached slice $k$ by time $T + O((w_0 + t_0/n)n^k)$.
            
            Let us temporarily assume that the switch from $\Gamma^+_Z$ to $\Gamma^-_Z$ in slice $k$ is postponed indefinitely. Then, by choosing $t_0$ and $n$ sufficiently large, it follows that all defects and messages sourced by $C$ are erased in time $O(w_k)$ before any activity due to $C$ can be promoted beyond slice $k$.

            Now, let us restore the finite switching time. Observe that the hard wall due to the post-split spatial geometry reaches slice $k$ only at time
            \[
                T+\Delta_{k+1}
                =
                T+\Delta_k+t_k.
            \]
            Hence, after all activity due to $C$ has either been erased or promoted to slice $k$, there remains a grace period of length $t_k-O(w_k)$ during which slice $k$ still uses the pre-split spatial geometry $\Gamma^+_Z$. Consequently, by choosing $n$ and $t_0$ sufficiently large, it follows by this observation and the usual types of arguments that the cluster will be completely erased before this grace period has terminated and before any activity sourced by the cluster can be promoted beyond slice $k$.

            Finally, the delayed split is compatible with the usual cluster-decoupling argument. In particular, consider a spacetime history consisting solely of $k$-clusters. Observe that any possible cross-seam interactions in slice $j$ caused by keeping $\Gamma^+_Z$ active in slice $j$ for time $O(t_j)$ after the split remains confined to an $O(t_j)$ spacetime neighborhood of the split operation. Since $b_0$ can be made arbitrarily large relative to $t_0$ (and $w_0$) by increasing $n$, it follows that the buffers between distinct $k$-clusters can be made sufficiently large so that they do not interact. Thus, in an infinite system with infinitely many slices, all $k$-clusters in a spacetime history consisting solely of $k$-clusters are decoded independently. 

            We may then repeat the usual arguments to prove that (i) the decoding of all clusters decouples when the noise history consists only of clusters of diameter at most $c L$ for some constant $c > 0$, and (ii) when no large clusters occur, the measurement outcomes assigned to the qubits on the seam are consistent and allow us to assign a consistent interpretation to the relevant seam stabilizers. Therefore, the same stretched-exponential memory lifetime guarantees for the ordinary surface-code decoder continue to hold during a non-condensing split and the relevant stabilizer frame information is incorrect with probability at most inverse-stretched-exponential-in-$L$.

        \subsubsection{Condensing merge}

            For a fixed CSS decoding sector, we say that a lattice-surgery merge operation is \emph{condensing} if, before the merge, the seam along which the merge is performed is a condensing boundary for that sector's defects, i.e., defects can terminate on that boundary.

            We now describe how to perform local decoding during a condensing merge. For concreteness, we describe how to perform local decoding for the $X$-type stabilizer sector during a rough merge.

             Let $T$ be the last pre-merge syndrome-extraction round, so $T+1$ is the first post-merge round. Let $\Lambda_{\mathrm{seam}}$ be the set of new seam check sites. For each $a\in\Lambda_{\mathrm{seam}}$, let $A_a$ denote the new $X$-type seam stabilizer and write
            \[
                \mathcal Q(a)=\supp(A_a)
            \]
            for its support. For interior seam sites, $A_a$ is one of the usual four-body stars centered on the seam; at the endpoints, it is one of the two usual three-body boundary stars.

            The first seam outcomes are used to initialize the seam frame and are not inserted as detector events. Specifically, for every $a\in\Lambda_{\mathrm{seam}}$, we set
            \[
                \phi(a,T+1)=0 .
            \]
            For $t > T+1$, seam detector events are given by ordinary syndrome differences
            \[
                \phi(a,t)
                =
                \tilde s_t(a)
                \oplus
                \tilde s_{t-1}(a).
            \]
            Detector events away from the seam are unchanged.

            We next specify how the decoder geometry changes. Let $\Gamma^-_X$ be the pre-merge spatial geometry in the $X$-stabilizer sector, and let $\Gamma^+_X$ be its post-merge spatial geometry. The post-merge geometry is introduced slice by slice. Fix a constant $M\geq 1$ and define
            \[
                \Theta_k
                =
                T
                +
                M\sum_{j=0}^{k-1}t_j,
                \qquad
                k=0,\ldots,K .
            \]

            For $k = 0, \ldots, K - 1$, during the update from time $t$ to time $t+1$, slice $k$ treats the seam as absorbing if $t < \Theta_k$ and non-absorbing if $t \geq \Theta_k$. Specifically, if $t < \Theta_k$, all defect-motion indicators and message-sector updates in slice $k$ are computed using the pre-merge spatial geometry, while if $t \geq \Theta_k$, they are computed using the post-merge spatial geometry. Additionally, any candidate promoted into a seam site in slice $k$ is not discarded if and only if $t \geq \Theta_k$. The absorbing seam also hovers above the final slice during 
            \[
                \Theta_{K-1}\leq t<\Theta_K
            \]
            in a manner that will be made precise below. Moreover, even in the post-merge spatial geometry, splitting steps continue to move defects and messages toward the formerly absorbing seam.

            We now describe the modified cellular-automaton rule more explicitly. First, let us describe how we determine the initial seam frame stabilizers. We record the parity of absorbed seam defects in a seam-frame update variable. For each $a\in\Lambda_{\mathrm{seam}}$, let
            \[
                \chi_a(t+1)\in\{0,1\}
            \]
            be the parity of all defect candidates absorbed at seam site $a$ during the update from $t$ to $t+1$; here, we include defects that get merged into the condensing boundary in slices that have not yet transitioned to using the post-merge spatial geometry.

            Let $F_T(q)\in\{0,1\}$ denote the net correction bit on qubit $q$ immediately before the merge, obtained by summing over all slices' correction bits at time $T$ on that qubit modulo-two. We initialize the seam-frame variable by
            \[
                \psi(a,T+1)
                =
                \tilde s_{T+1}(a)
                \oplus
                \left(
                    \bigoplus_{q\in\mathcal Q(a)}F_T(q)
                \right)
                \oplus
                \chi_a(T+1),
                \qquad
                a\in\Lambda_{\mathrm{seam}}.
            \]
            For later updates, the seam frame is updated according to
            \[
                \psi(a,t+1)
                =
                \psi(a,t)
                \oplus
                \chi_a(t+1).
            \]
            
            Next, we describe how the splitting mechanism interacts with the condensing merge protocol. Suppose $t \in q_s \Z$, so the update from $t$ to $t + 1$ includes a splitting step. Then, during this splitting step, seam sites in slice $k$ are treated as absorbing if $t < \Theta_k$ and non-absorbing if $t \geq \Theta_k$. Furthermore, defects absorbed by a seam site $a\in\Lambda_{\mathrm{seam}}$ during this splitting step are counted in $\chi_a(t+1)$.

            Finally, we elaborate on the hovering absorbing seam above the final slice, which is implemented by having the final-slice dynamics be modified by the condensing merge. During the interval
            \[
                T\leq t<\Theta_K-1,
            \]
                the final slice is treated as timed, with all final-slice defect and message timers initialized to zero at the first post-merge update from $T$ to $T+1$. If a final-slice defect timer expires during this interval, the defect is promoted upward. A promoted defect at a seam site $a\in\Lambda_{\mathrm{seam}}$ is absorbed by an auxiliary seam site that hovers above the final slice; this absorption contributes to $\chi_a(t+1)$. A promoted defect at a non-seam site has no absorbing site above it; if such a promotion occurs, the decoder raises a rejection flag and the quantum computation is declared to have failed. On the update from $t=\Theta_K-1$ to $t+1=\Theta_K$, the auxiliary absorbing seam is removed, no final-slice promotions are performed, and the final slice uses the ordinary untimed back-wall rule. From time $\Theta_K$ onward, the usual untimed back-wall rule is used everywhere, and, provided no rejection flag was raised, the seam frame is committed as
            \[
                \psi_{\mathrm{merge}}(a)=\psi(a,\Theta_K),
                \qquad
                a\in\Lambda_{\mathrm{seam}}.
            \]
            The lattice-surgery measurement outcome is obtained from $\psi_{\mathrm{merge}}$ by calculating the parity of the committed stabilizers along the seam.
            
            We now describe how to modify the arguments of Section~\ref{sec:online-decoding} to obtain rigorous performance guarantees for the local decoder during the condensing merge. Work temporarily in an infinite system with infinitely many slices. Consider a spacetime history consisting of a single cluster $C$, where $C$ is a $k$-cluster that sources at least one detector event. If the spacetime support of $C$ is separated in spacetime from the merge operation by more than a constant multiple of $w_k$, then $C$ evolves entirely either in one side of the pre-merge spatial geometry or entirely in the post-merge spatial geometry. In either case, the ordinary linear cluster-erosion argument for the surface code applies.

            It remains to consider a $k$-cluster whose $O(w_k)$ spacetime neighborhood intersects the merge operation. Let $u$ be the first time at which a detector event sourced by $C$ is inserted into the decoder. Observe that there is a constant $C_0$ such that, under the condensing-merge-modified dynamics, by time $u+C_0t_k$ all activity sourced by $C$ has either been erased, absorbed by the seam, or promoted beyond slice $k$. 

            Choose $M>C_0$. Suppose first that
            \[
                u+C_0t_k<\Theta_{k+1}.
            \]
            Then, all activity sourced by $C$ has been erased, absorbed, or promoted into slice $k+1$ before the seam becomes non-absorbing in slice $k+1$. Moreover, one can check that by the standard linear cluster erosion arguments for the surface code, any defect sourced by $C$ promoted into slice $k + 1$ must be located at a seam site, and thus any such promoted defect is absorbed by a seam site.

            Suppose instead that
            \[
                u+C_0t_k\geq \Theta_{k+1}.
            \]
            Since
            \[
                \Theta_{k+1}
                =
                \Theta_k+Mt_k,
            \]
            we have
            \[
                u
                \geq
                \Theta_k+(M-C_0)t_k
                >
                \Theta_k .
            \]
            Thus slice $k$, and all lower slices, already treat the seam as non-absorbing before the first detector event sourced by $C$ is inserted. Moreover, by this time, $C$ cannot contain a first-round fault along the seam, where we have assumed that $M$ has been chosen sufficiently large. Therefore, $C$ does not interact with the seam while it is absorbing and one can check that upon choosing $n$ and $t_0$ sufficiently large, it follows by the standard linear cluster erosion arguments for the surface code that $C$ is completely erased before any activity due to $C$ can be promoted beyond slice $k$.

            We may then repeat the usual arguments to prove that (i) the decoding of all clusters decouples when the noise history consists only of clusters of diameter at most $c L$ for some constant $c > 0$, and (ii) when no large clusters occur, the stabilizer values assigned to the sites on the seam are consistent and their modulo-two sum gives the correct lattice-surgery measurement outcome. Therefore, the same stretched-exponential memory lifetime guarantees for the ordinary surface-code decoder continue to hold during a condensing merge, and the lattice-surgery measurement outcome is incorrect with probability at most inverse-stretched-exponential-in-$L$.

        \subsubsection{Condensing split}

            We next describe the inverse operation: a condensing split. For a fixed CSS decoding sector, we say that a lattice-surgery split is \emph{condensing} if, after the split, the seam becomes a condensing boundary for that sector's defects. 

            We now describe how to perform local decoding during a condensing split. For concreteness, we describe how to perform local decoding for the $X$-type stabilizer sector during a rough split.

            Let $T$ be the last pre-split syndrome-extraction round, so $T+1$ is the first post-split round. During the split, each seam qubit $q$ is measured out in the computational basis, which destabilizes each of the $X$-type seam stabilizers. Consequently, for each seam site $a\in\Lambda_{\mathrm{seam}}$, we set
            \[
                \phi(a,t)=0
            \]
            for all $t \geq T+1$. Detector events away from the seam are unchanged.

            We next specify how the decoder geometry changes. Let $\Gamma^+_X$ be the pre-split spatial geometry and let $\Gamma^-_X$ be the post-split spatial geometry. The post-split spatial geometry is introduced slice by slice, with a slightly different schedule as compared to the condensing merge case. During the update from time $t$ to time $t+1$, slice $k$ uses the spatial geometry
            \[
                \Gamma_{X,k}(t)
                =
                \begin{cases}
                    \Gamma^+_X , & t<T+\Delta_k, \\
                    \Gamma^-_X , & t\geq T+\Delta_k.
                \end{cases}
            \]

            Next, we describe how the splitting mechanism interacts with the condensing split protocol. Suppose $t \in q_s \Z$, so the update from $t$ to $t + 1$ includes a splitting step. Then, during this splitting step, seam sites in slice $k$ are treated as non-absorbing if $t < T +\Delta_k$ and absorbing if $t \geq T +\Delta_k$. 

            We now describe how to modify the arguments of Section~\ref{sec:online-decoding} to obtain rigorous performance guarantees for the local decoder during the condensing split. Work temporarily in an infinite system with infinitely many slices. Consider a spacetime history consisting of a single cluster $C$, where $C$ is a $k$-cluster that sources at least one detector event. If the spacetime support of $C$ is separated in spacetime from the split operation by more than a constant multiple of $w_k$, then $C$ evolves entirely either in one side of the post-split spatial geometry or entirely in the pre-split spatial geometry. In either case, linear cluster erosion holds by arguments we have already applied previously.

            It remains to consider a $k$-cluster whose $O(w_k)$ spacetime neighborhood intersects the split operation. Its detector events are inserted over a time interval of length $O(w_k)$. Moreover, all activity sourced by the cluster has either been erased below slice $k$ or has reached slice $k$ by time
            \[
                T+\Delta_k+O(w_k),
            \]
            and by that time, all remaining defects sourced by the cluster are contained on seam sites.
            
            Let us temporarily assume that $t_k$ is infinite. Since by time $T + \Delta_k$, slice $k$ is using the post-split spatial geometry in which the seam is a condensing boundary, it follows by the argument in the previous paragraph that when $t_k$ is infinite, all defects and messages sourced by the cluster are erased in time $O(w_k)$. As usual, by choosing $t_0$ and $n$ sufficiently large, it follows that when $t_k$ is finite, this erosion finishes before any activity sourced by the cluster can be promoted beyond slice $k$. Therefore, if the noise history consists of a single $k$-cluster, then this cluster's defects and messages are erased in linear time and its effects never reach beyond the $k$th slice. 

            We may then repeat the usual arguments to prove that the decoding of all clusters decouples when the noise history consists only of clusters of diameter at most $c L$ for some constant $c > 0$. Therefore, the same stretched-exponential memory lifetime guarantees for the ordinary surface code decoder continue to hold during a condensing split.

    \subsection{Hierarchical implementation and constant-bandwidth control} \label{subsec:hierarchical-layout}

        The protocols introduced above admit hierarchical, constant-density analogs with the exact same rigorous performance guarantees. For each such analog, we outline the required adaptations and the constant-bandwidth controls sufficient to implement it. Each such outline can be expanded into an explicit construction in a straightforward but tedious manner. We do not outline the proofs of the corresponding performance guarantees because these require no new ideas or techniques other than those already used in previous sections.

        First, let us consider a logical Hadamard. We begin by broadcasting a control pulse from the patch's IO port; when it reaches a slice, the two stabilizer decoding sectors on that slice are relabeled, with every stored decoding variable reinterpreted in place. When it reaches slice $0$, the physical Hadamards are applied and the checks begin to be measured in the interchanged bases.

        Next, let us consider a non-condensing merge. We begin by broadcasting control pulses from the IO ports of the two relevant patches that reach all sites in any given slice simultaneously. When they reach slice $0$, the modified seam checks begin to be measured. The post-merge spatial geometry is then introduced slice by slice; as usual, the associated delays are implemented using clock pulses that travel along constant-bandwidth wires.

        A non-condensing split can be handled similarly. We again begin by broadcasting control pulses through both IO ports of the merged patch; when these reach slice $0$, the seam qubits are measured out and their outcomes are folded into the corresponding slice-$0$ correction bits. The post-split spatial geometry is then introduced slice by slice using the same clock pulse mechanism as before.

        Now we consider a condensing merge. As before, we broadcast control pulses from the IO ports of the two relevant patches, and the post-merge spatial geometry is introduced slice by slice in a delayed manner using the usual mechanisms. On slices whose seam is non-absorbing, the splitting step sweeps defects in the inner halves of the two patches toward the two seam-adjacent columns; a defect promoted out of a site in one of these columns while the seam in the slice above it is still absorbing is absorbed there upon arrival. Only two adaptations require further discussion: (i) how the overall parity of the seam-stabilizer interpretations is tracked, and (ii) how the lattice-surgery measurement outcome is extracted.

        Let us first consider (i). We assign the left patch the role of tracking the seam interpretation; it will be clear in the next subsection that the meaning of ``left'' in our architecture is well-defined. When the pulses reach slice $0$, the new seam checks begin to be measured, and their first-round measurement outcomes immediately absorb the slice-$0$ correction bits along the seam, forming the initial interpretation. This interpretation is then transmitted upward, and each parent site calculates the parity of the interpretations received from its children. The parent site also toggles this combined parity whenever a defect is absorbed at a seam site it is responsible for. When its seam sites become non-absorbing, the parent folds its own seam-relevant correction bits into the interpretation and transmits the result to its parent in the next slice. This process continues slice by slice until the interpretation reaches the top of the hierarchy, after which it can be extracted through the IO port.

        We now consider (ii). The procedure above yields the overall parity of the seam checks, but for a condensing merge, the measurement outcome is defined with respect to the two patches' central logical operators in the relevant basis, so this seam parity must additionally be multiplied by the overall parity of all the relevant stabilizers lying between each central operator and the seam. This overall parity is computed by a second control pulse that aggregates these individual parities upward through each patch's hierarchy, just as in readout; combining the two resulting parities with the seam parity yields the gauge-invariant lattice-surgery outcome. Note that it is at this step that the retained seam correction values along the two boundaries not involved in the merge are finally used: they supply the correct interpretations of the relevant stabilizers along those boundaries.

        Finally, we consider a condensing split. We again broadcast control pulses through both IO ports of the merged patch, and the post-split spatial geometry is introduced slice by slice with the usual delays. 

    \subsection{Architecture and threshold theorem}

        We begin by describing how to construct a local, fault-tolerant, two-dimensional quantum computer. We assume access to a noiseless, local, two-dimensional classical computer that operates in synchronous discrete time. We assume that the classical computer is composed of constant-size classical processors whose classical memory and classical computational capabilities remain bounded as $L \to \infty$. Furthermore, we assume that these classical processors are arranged in a two-dimensional grid and can only communicate with their nearest-neighbors.

        First, we describe the architecture for the quantum computer. Then, we explain how to locally decode this architecture using the protocols introduced in this paper.

        The basic building block of the architecture is a standard surface-code patch that has been folded along one of its diagonals. We arrange these folded patches in alternating orientations along a line, as shown in Fig.~\ref{fig:architecture}. Then, we place this line of folded surface-code patches along the boundary of our two-dimensional classical computer, with each patch's IO port connected to the classical computer by a single constant-bandwidth, constant-propagation-speed wire.
        
        We designate alternating patches as data and auxiliary patches. The auxiliary patches are only used to perform logical $\CNOT$ gates between data patches using lattice surgery. The folded geometry of the patches ensures that fold-transversal logical Hadamards do not interfere with our ability to perform lattice surgery. Therefore, setting aside decoding for the moment, this architecture allows us to realize the universal set of primitives described at the beginning of this section using only geometrically local qubit interactions.
        
        Suppose that our quantum computation has been compiled into a fault-tolerant lattice surgery schedule that is compatible with this architecture. For simplicity, we assume that state preparation, state injection, logical Hadamard, and destructive readout are performed only when a patch is not merged with any neighboring patch.

        The hierarchical implementations and constant-bandwidth controls of the previous subsection, together with those of Section~\ref{subsec:cm-online-layout}, then allow us to locally decode and control the entire quantum computation. In particular, each surface-code patch can be locally decoded, controlled, and read out using only a constant density of constant-size classical processors connected by constant-bandwidth, constant-propagation-speed wires. Moreover, each patch can be controlled and read out using only the single wire connecting it to the classical computer. The classical computer also uses the extracted information to track each patch's logical Pauli frame. This concludes the description of our architecture.
        
        We note in passing that it is straightforward to modify the architecture described in this subsection to allow for a two-dimensional grid-like connectivity pattern between data patches instead of a one-dimensional line-like connectivity pattern.

        Next, we prove that our architecture satisfies a threshold theorem. By the results of this and the previous sections, logical Hadamard, state preparation, state injection, destructive readout, and lattice-surgery-based $\CNOT$ gates can each be performed in $\poly(L)$ time with failure probability at most inverse-stretched-exponential in $L$, provided that the noise is $p$-bounded with $p$ below a constant threshold independent of $L$. Moreover, it is known that high-fidelity $Y$ states can be distilled from injected $Y$ states using only the Clifford subgroup generated by $H$ and $\CNOT$~\cite{Raussendorf_2007}, and that high-fidelity magic states can then be distilled from injected magic states using arbitrary Clifford operations~\cite{Bravyi_2005}. In both cases, the distilled states achieve inverse-stretched-exponential-in-$L$ infidelity in $\poly(L)$ time, with distillation failure probability at most inverse-stretched-exponential in $L$, provided that the residual errors on distinct injected states are independent. This is a strictly stronger requirement than $p$-boundedness and thus requires an additional assumption on the noise model. For the remainder of this argument, we therefore assume that the noise on both data qubits and stabilizer measurement outcomes is independent (but not necessarily identically distributed) Pauli and bit-flip noise, respectively, with error rates at most $p$. Since distinct injection protocols occupy disjoint spacetime regions, their noise histories, and hence their injection errors, are independent. Therefore, each primitive of our universal set can be performed with $\poly(L)$ spacetime overhead and failure probability at most inverse-stretched-exponential in $L$. By a union bound, a quantum computation comprising $N$ such primitives succeeds with probability at least $1 - N e^{-\Omega(L^{\alpha})}$ for some constant $\alpha > 0$; in particular, any computation of size sub-stretched-exponential in $L$ succeeds with high probability, as desired.
    
\section{General translation-invariant stabilizer codes}\label{sec:general}

    In this section, we generalize our constructions to arbitrary translation-invariant topological Pauli stabilizer codes on Euclidean lattices.\footnote{We are grateful to Jeongwan Haah for suggestions about the proofs in this section.} We prove that for every such code, there exists a code-capacity CA decoder with a non-zero threshold and polylogarithmic average decoding time, as well as a streaming decoder that has a non-zero threshold under phenomenological noise. Both decoders use $\poly(\log L)$ classical bits per site; for the code-capacity decoder, this overhead can be reduced to $O(1)$ bits per site at the cost of breaking translation invariance.

    Our starting point is a subroutine originally developed by Bravyi and Haah~\cite{Bravyi_2013} as part of a global renormalization-group decoder. This decoder requires a method of testing whether a given cluster of defects is neutral. In general, this test can be performed with linear algebra in time superlinear (but still polynomial) in the cluster's volume. For the toric code and Haah's code, however, they devised a linear-time alternative called the broom algorithm, which operates by sweeping defects in the cluster toward a designated corner of the cluster. The cluster is neutral if and only if no defects survive the sweep.

    Haah later generalized the broom algorithm to all translation-invariant local stabilizer codes using the polynomial formalism~\cite{Haah_2013}.\footnote{The polynomial formalism has also been used to construct global matching decoders for two-dimensional translation-invariant stabilizer codes; see Refs.~\cite{tan2026generalizedmatchingdecoders2d, sahay2026matchingdecoderbivariatebicycle} for details.} In this language, a cluster is neutral if and only if its syndrome leaves no remainder upon division by a suitable set of polynomials called a Gr\"obner basis. This division process also runs in time linear in the cluster's volume.

    We build on these results by first showing that, for an appropriate choice of term order, this division process can be parallelized to eliminate a neutral cluster in time linear in its diameter, as opposed to its volume. We then show how to combine this parallelized division process with the messages and timers of the previous sections to construct a code-capacity CA decoder that satisfies linear cluster erosion. Finally, we promote the code-capacity decoder to a streaming decoder for phenomenological noise using the slice scheme of Section~\ref{sec:online-decoding}.

    \subsection{Definitions}

        We begin by defining the necessary algebraic objects and introducing the relevant aspects of Haah's polynomial formalism.
        
        Let $R = \F_2[x_1, x_1^{-1}, \ldots, x_D, x_D^{-1}]$ and $S = \F_2[x_1,\ldots,x_D]$ be the rings of Laurent and ordinary polynomials in $D$ variables, respectively, and let $R^j$ and $S^j$ be the free modules of rank $j$ over $R$ and $S$, respectively. We will sometimes interpret multiplication by monomials $x^{\vc u} \in R$ as translations by $\vc u \in \Z^D$ and multiplication by monomials $x^{\vc u} \in S$ as translations by $\vc u \in \Z^D_{\geq 0}$, where $x^{\vc u} \coloneqq x_1^{u_1} \cdots x_D^{u_D}$. We will also sometimes identify elements of $S$ with finite subsets of sites in $\Z_{\geq 0}^D$.

        Fix a translation-invariant stabilizer code on $\Z^D$ with $q$ qubits and $m$ stabilizer generators per unit cell. Let
        \[
            \sigma: R^m \to R^{2q}
        \]
        be the stabilizer map of this code. Here, $R^{2q}$ is the $R$-module of all finitely supported Pauli operators on $\Z^D$, modulo phases, while $R^m$ is the $R$-module of finite subsets of translated stabilizer generators. The map $\sigma$ sends a collection of generators to the Pauli operator given by their product, and $\im \sigma$ corresponds to the stabilizer group. Next, let
        \[
            \epsilon: R^{2q} \to R^m
        \]
        be the excitation map of the code, which sends Pauli errors to their syndromes. We denote the submodule of $S^m$ consisting of all syndromes supported in the non-negative orthant $\Z_{\geq 0}^D$ by
        \[
            \epsilon_+ = \im \epsilon \cap S^m.
        \]

        We now introduce the definitions needed to describe polynomial division in $S^m$. A module monomial of $S^m$ is an element of the form $x^{\vc u} \mathbf{e}_a$, where $\mathbf{e}_a$ is the $a$th standard basis vector. We say that $x^{\vc u} \mathbf{e}_a$ is divisible by $x^{\vc v} \mathbf{e}_b$ if $a = b$ and $v_i \leq u_i$ for every $i$. 
        \begin{definition}[term order]
            \label{def:term-order}
            A \emph{term order} on $S^m$ is a total order $\leq$ on the module monomials satisfying the following properties:
            \begin{enumerate}
                \item if $\mu<\nu$, then
                $x^{\vc w}\mu<x^{\vc w}\nu$
                for every monomial $x^{\vc w}$;
                \item every non-empty collection of module monomials has a least element with respect to $\leq$.
            \end{enumerate}
            The second property is called \emph{well-foundedness}. Given a term order, every non-zero $f \in S^m$ has a unique largest module monomial occurring in it, called its \emph{leading term} $\LT(f)$.
        \end{definition}

        \begin{definition}[Gr\"obner basis]
            \label{def:grobner}
            Fix a term order on $S^m$ and let $N \subseteq S^m$ be a submodule. A finite subset $G = \{g_1, \ldots, g_s\} \subseteq N$ is a \emph{Gr\"obner basis} for $N$ if the leading term of every non-zero element of $N$ is divisible by $\LT(g_j)$ for some $j$.
        \end{definition}

        Given a Gr\"obner basis $G = \{g_1, \ldots, g_s\}$ for a submodule $N \subseteq S^m$, we call a module monomial reducible if it is divisible by $\LT(g_j)$ for some $j$, and irreducible otherwise. In this language, the defining property of a Gr\"obner basis is that the leading term of every non-zero element of $N$ is reducible.

        Gr\"obner bases allow us to perform polynomial division on elements of modules. If $f \in N$ is non-zero, then $\LT(f)$ is reducible, i.e., $\LT(f) = x^{\vc w} \LT(g_j)$ for some $j$. Since multiplication by $x^{\vc w}$ preserves the order, subtracting $x^{\vc w} g_j$ cancels the leading term and replaces $f$ by an element of $N$ whose leading term is strictly smaller. By well-foundedness of the term order, this process terminates after finitely many steps, and since leading terms of non-zero elements of $N$ are always reducible, it can only terminate at zero.

        For every term order on $S^m$ and every submodule $N \subseteq S^m$, there exists a Gr\"obner basis for $N$. Therefore, for every term order on $S^m$, there exists a Gr\"obner basis $G$ of~$\epsilon_+$. We will also use the fact that since $G \subseteq \im \ep$, there exists some set of finitely supported Pauli operators $H = \{h_1, \ldots, h_s\}$ such that $g_j = \ep h_j$ for each $g_j \in G$.

        Finally, we introduce a term order on $S^m$, called a weight order~\cite{schreyer2025introduction}, that will turn out to be useful for our purposes.\footnote{We thank Jeongwan Haah for suggesting the use of a term order based on irrational hyperplanes.} Consider $\R$ as a vector space over $\Q$ and choose positive real numbers $\omega_1,\ldots,\omega_D$ such that
        \[
            E = \{1, \omega_1, \omega_2, \ldots,\omega_D\}
        \]
        is linearly independent over $\Q$. Next, choose distinct positive integers $\eta_1, \ldots, \eta_m$. Define a strictly positive function $\Phi$ on module monomials by
        \[
            \Phi(x^{\vc u}\mathbf e_a)
            =
            \sum_{i=1}^D \omega_i u_i+\eta_a,
            \qquad
            \vc u\in\Z_{\geq0}^D.
        \]
        This induces a relation on $S^m$ given by
        \[
            x^{\vc u} \mathbf e_a < x^{\vc v} \mathbf e_b
            \quad  \iff \quad
            \Phi(x^{\vc u} \mathbf e_a) < \Phi(x^{\vc v} \mathbf e_b),
        \]
        and it is straightforward to check, using the linear independence of $E$ over $\Q$, that this relation is a term order. We will refer to the value of $\Phi$ on a module monomial as the weight of the monomial.

    \subsection{Decoding under code-capacity noise}

        We now use the polynomial formalism to construct local decoders for translation-invariant stabilizer codes under code-capacity noise. First, we state the result of Haah discussed above. Every translation-invariant stabilizer code gives rise to the submodule $\im \ep \subseteq R^m$ of realizable syndromes. A syndrome configuration contained in a box of side length $W$ is neutral---i.e., the syndrome of some finitely supported Pauli error---if and only if its remainder upon division by a Gr\"obner basis of $\im \ep$ is zero.\footnote{Haah works with Laurent-polynomial modules, for which the notions of polynomial division differ from those introduced above. For our purposes, it suffices to consider only ordinary polynomial division.}

        Let $G = \{g_1, \ldots, g_s\}$ be a Gr\"obner basis of $\ep_+$ with respect to the term order induced by $\Phi$, let $H = \{h_1, \ldots, h_s\}$ be a corresponding set of finitely supported Pauli operators with $\ep h_j = g_j$ for each $j$, and write $\LT(g_j) = x^{\vc a_j} \mathbf{e}_{b_j}$.\footnote{Such a Gr\"obner basis can always be computed explicitly for any translation-invariant stabilizer code. The image of $\ep$ is generated over $R$ by the images of the $2q$ standard basis vectors of $R^{2q}$. A finite generating set of $\ep_+ = \ep(R^{2q}) \cap S^m$ can then be computed from these generators using standard elimination-theoretic techniques from commutative algebra, and a Gr\"obner basis with respect to the term order induced by $\Phi$ can then be obtained from this generating set using Buchberger's algorithm~\cite{eisenbud1995commutative}.} The $g_j$ and $h_j$ are finitely supported objects that depend only on the code family and are independent of system size. In particular, applying any translate of an $h_j$ is a geometrically local correction move.

        We now define a parallelized division process.\footnote{We thank Jeongwan Haah for suggesting that his prior work~\cite{haah2013latticequantumcodesexotic} on reinterpreting the broom algorithm~\cite{bravyi2011analytic} as polynomial division could be used to extend our constructions for codes with conventional topological order, the X-cube model, and the cubic code to general translation-invariant stabilizer codes.} First, fix an arbitrary ordering of the elements of $G$, whose only role is to break ties when a monomial is divisible by the leading terms of several basis elements. We call this ordering the priority ordering.

        Suppose $f \in \im \ep$ is contained in a bounding box of side length $W$. Let $\vc w$ be the coordinatewise minimum of the sites in the support of $f$. Since we may always replace $f$ by $x^{-\vc w} f$, we may assume, without loss of generality, that the origin is located at the bounding box's coordinatewise-minimal corner and thus that $f \in \ep_+$.
        
        The division process is defined as follows. In one synchronous round, each monomial $x^{\vc u} \mathbf{e}_a$ of $f$ that is divisible by $\LT(g_j)$ for at least one $j$ triggers the Pauli move
        \[
            x^{\vc u - \vc a_j} h_j
        \]
        for the highest-priority such $g_j$, and all triggered moves are applied simultaneously modulo two. Note that each divisibility test amounts to checking the component index and whether $\vc u \geq \vc a_j$ coordinatewise. Since this check requires inspecting only the individual site's position relative to the origin, the trigger conditions and the corresponding choices of $g_j$ can be evaluated locally and in parallel at every syndrome monomial. Moreover, since the $h_j$ have finite supports, each qubit lies in the support of at most a constant number of triggered moves per round.

        Algebraically, one round updates the syndrome by adding the corresponding translates $x^{\vc u - \vc a_j} g_j$ modulo two. Each targeted monomial cancels against its own move's leading term, and every monomial created in its place has strictly smaller weight.

        Note that all triggers in a round are computed from the pre-update syndrome, each qubit's $X$- and $Z$-corrections are toggled by the parity of the number of triggered moves containing them, and the syndrome is updated by the excitation of this modulo-two sum. Therefore, each round can be implemented using a local update rule with system-size-independent interaction range.

        \begin{theorem}\label{thm:general-parallel-broom}
            Let $f \in \im \ep$ be a syndrome supported in a box of side length at most $W$ whose coordinatewise-minimal corner is the origin. Then, the parallelized division process eliminates $f$ in $O(W)$ rounds, and throughout the process, the syndrome and the accumulated correction are supported in a box of side length $O(W)$.
        \end{theorem}

        \begin{proof}
            Each triggered move adds a term $x^{\vc u - \vc a_j} g_j$ to the syndrome, and since the divisibility condition guarantees $\vc u \geq \vc a_j$, this term is an $S$-multiple of $g_j$. Since $f \in \ep_+$ by assumption and $\ep_+$ is an $S$-submodule of $S^m$, it follows by induction that every intermediate syndrome lies in $\ep_+$.

            For each $g_j$ containing more than one monomial, the leading term has strictly larger weight than every other monomial of $g_j$. Let
            \[
                \delta_j
                =
                \min_{\mu\in\supp(g_j)\setminus\{\LT(g_j)\}}
                \bigl(\Phi(\LT(g_j))-\Phi(\mu)\bigr) > 0
            \]
            be the smallest of these differences, and let $\delta = \min_j \delta_j$ be the minimum of the $\delta_j$ over those $g_j$ with more than one monomial. If every $g_j$ consists of a single monomial, we set $\delta = 1$ instead. Since $G$ is finite and each $\delta_j$ is positive, it follows that $\delta > 0$.

            Call the monomials created by a triggered move the children of the canceled monomial. A basis element consisting of a single monomial creates no children, and since weight differences are translation-invariant, every child lies at least $\delta$ below its parent.

            Let $\Phi_{\max}(t)$ denote the maximum weight over all monomials present after $t$ rounds. We claim that $\Phi_{\max}$ decreases by at least $\delta$ every round, until the syndrome is zero. In one round, every reducible monomial is replaced by its children and every irreducible monomial persists unchanged; some of these monomials may also cancel in pairs. Consequently, every monomial of the new syndrome is either a child or a persisting irreducible monomial.

            Suppose the new syndrome is non-zero and consider its leading term, which attains $\Phi_{\max}(t+1)$. Since the new syndrome is a non-zero element of $\ep_+$, its leading term is reducible, so the leading term must be a child. Therefore, this child lies at least $\delta$ below its parent, which was present in the previous round, so $\Phi_{\max}(t+1) \leq \Phi_{\max}(t) - \delta$. Since the initial maximum weight satisfies
            \[
                \Phi_{\max}(0) \leq \left(\sum_{i=1}^D \omega_i\right) W + \max_a \eta_a
            \]
            and weights are strictly positive, it follows that the syndrome vanishes after at most $\bigl\lceil \Phi_{\max}(0)/\delta \bigr\rceil = O(W)$ rounds.

            Finally, since $\Phi_{\max}$ is monotonically decreasing in $t$, no monomial of weight exceeding $\Phi_{\max}(0)$ ever appears, so any monomial $x^{\vc v} \mathbf{e}_b$ present at any time satisfies $\Phi(x^{\vc v} \mathbf{e}_b) \leq \Phi_{\max}(0)$. Moreover, every term of $\Phi(x^{\vc v} \mathbf{e}_b)$ is non-negative, so $\omega_i v_i \leq \Phi(x^{\vc v} \mathbf{e}_b) \leq \Phi_{\max}(0)$. Additionally, each $v_i$ is itself non-negative since the syndrome lies in $\ep_+$. These facts imply $0 \leq v_i \leq \Phi_{\max}(0)/\omega_i = O(W)$ for each $i$, so the syndrome remains in a box of side length $O(W)$. Since each triggered move applies a translate of some $h_j$ at a monomial within this box and the $h_j$ have finite supports, the accumulated correction is also supported in a box of side length $O(W)$.
        \end{proof}

        We now show how the parallelized division process can be used to construct a local decoder. Specifically, we will use this process to define a CA rule satisfying linear defect erosion and linear message erosion. We work at infinite system size for simplicity.
        
        Suppose the initial syndrome is supported in a box of side length $W$. We will design a CA rule that satisfies linear defect erosion for this simplified case. For now, we will not concern ourselves with linear message erosion, but later on, we will describe how to modify the rule to achieve it.
        
        Each site (unit cell) carries one message bit $m_{\vc s}$ for every orthant $\vc s \in \{0,1\}^D$ such that $\vc s \neq (1, \ldots, 1)$. During the update from time $t$ to time $t+1$, messages grow as follows. The message bit $m_{\vc s}$ becomes non-trivial at time $t+1$ if there exists at least one defect at the site at time $t$, or if $m_{\vc s}$ is non-trivial at the neighbor in the $(-1)^{s_i + 1} \widehat{\vc e}_i$ direction for some $i$ at time $t$. For now, we assume non-trivial messages are never erased, so a defect at $\vc y$ fills the $\vc s$-orthant based at $\vc y$ at unit speed.

        For each coordinate $i$, define the indicator function
        \[
            I_i(\vc r) = \bigvee_{\vc s :\, s_i = 0} m_{\vc s}(\vc r).
        \]
        Note that if $I_i(\vc r) = 1$, then there exists at least one site $\vc r'$ such that $r_i' \leq r_i$ and $\vc r'$ has hosted at least one defect. Let $o_i$ denote the minimum $i$th coordinate over the sites hosting the initial defects. Then, in $O(W)$ time, the messages will have spread across an $O(W)$-neighborhood of the bounding box, and $I_i(\vc r) = 1$ for every site with $r_i \geq o_i$ that lies within this neighborhood.

        For $g \in G$, write $\LT(g) = x^{\vc a} \vc e_b$, and denote the $i$th entry of $\vc a$ by $a_i$. We use the indicator functions to implement the division process without needing to know the location of the coordinatewise-minimal corner of the bounding box. Specifically, we replace the check $\vc u \geq \vc a$ from earlier with the following check: a syndrome monomial $x^{\vc r} \vc e_c$ may trigger a move for $g$ only if $c = b$ and $I_i(\vc r - a_i \uvc e_i) = 1$ for every $i$. By construction, every move permitted by the modified check is also permitted by the original one, and after the $O(W)$ time required for the messages to spread across the aforementioned $O(W)$-neighborhood of the bounding box, the two checks coincide. Therefore, the proof of Theorem~\ref{thm:general-parallel-broom} applies after an additional $O(W)$ delay, so the modified process satisfies linear defect erosion.

        We now modify the rule so that it also satisfies linear message erosion. From now on, we assume that the system size is finite and that each site has $K$ timer registers.

        Let $n \geq 2$ be the integer parameter from the sparsity theorem, let $t_0 \geq 1$ be a constant, and let $t_k = t_0 n^k$. Time is divided into consecutive epochs $k = 0, 1, \ldots, K - 1$, where epoch $k$ lasts $t_k$ steps. Within each epoch, the rule runs unchanged. At the end of each epoch, all message bits are reset to trivial values, while the syndrome and accumulated correction are left untouched. The $k$th timer register counts the steps elapsed during epoch $k$, allowing each site to locally keep track of the current epoch.

        It follows by the arguments of Section~\ref{sec:online-decoding} that for $t_0$ sufficiently large and $K = \Theta(\log L)$, there exists some constant $c > 0$ such that any initial configuration consisting of a single cluster of diameter $r \leq c L$ is eliminated---together with all messages it creates---within time $O(r)$, with all activity confined to an $O(r)$-neighborhood of the cluster. The timed rule therefore satisfies linear cluster erosion for all configurations consisting of single clusters of diameter at most $c L$. Recall that the buffers of the cluster decomposition satisfy $b_k = b_0 n^k$, so by choosing $n$ sufficiently large, the buffer-to-timer ratio $b_k/t_k = b_0 / t_0$ can be made large enough that all clusters of diameter at most $c' L$ are decoded independently, where $c' > 0$ is some constant. It then follows by the arguments of Section~\ref{sec:threshold} that the decoder has a threshold and polylogarithmic average decoding time for every translation-invariant stabilizer code whose distance grows as $\Omega(L^{\alpha})$ for some $\alpha > 0$.

        A priori, the timers require $\poly(\log L)$ bits per site. For $D \geq 2$, however, this overhead can be reduced to $O(1)$ bits per site by implementing the clock pulses using a $D$-dimensional version of the coarse-grained hierarchical layout of Section~\ref{subsec:cm-online-layout}, at the cost of breaking spatial translation invariance. Moreover, restricting to $D \geq 2$ incurs no loss of generality since there are no non-trivial one-dimensional translation-invariant topological stabilizer codes~\cite{Bravyi_2009}. Since this hierarchical implementation requires no ideas beyond those introduced in Section~\ref{subsec:cm-online-layout}, we omit the details.

    \subsection{Decoding under phenomenological noise}

        We now promote the code-capacity decoder of the previous subsection to a streaming decoder for phenomenological noise, following the slice scheme of Section~\ref{sec:online-decoding}. The resulting decoder uses a $\polylog(L)$ density of classical registers and bounded-bandwidth, bounded-propagation-speed wires, and has a rigorous, non-zero threshold against $p$-bounded Pauli and bit-flip noise on data qubits and stabilizer-measurement outcomes, respectively. Since the construction closely follows that of Section~\ref{sec:online-decoding}, we only outline it; the proofs require no ideas beyond those already presented there.

        The decoder consists of $K$ slices indexed by $k = 0, \ldots, K-1$, each running its own copy of a slight modification of the code-capacity decoder. Specifically, each site in slice $k$ carries $m$ binary-valued defect channels, recording which syndrome monomials are present at that site; $2^D - 1$ binary-valued message channels $m_{\vc s}$; $2q$ binary-valued correction registers; and a timer register of range $t_k = t_0 n^k$ for each of its defect and message channels, where $n$ is the parameter from the sparsity theorem and $t_0$ is a constant. The division moves of slice $k$ act on the defect and correction channels of slice $k$ alone. At each time step, every detector event inserts a defect with timer $0$ at the corresponding site and component of slice $0$. A defect that survives in slice $k < K-1$ for $t_k$ steps is promoted: it is removed from slice $k$ and reinserted at the same site and component of slice $k+1$ with its timer reset to $0$. Messages are erased when their timers expire. Defects and messages in the final slice are still timed, but defects are never promoted; the timers there serve only to erase messages---message timers directly, and defect timers by seeding the message-timer updates as in the lower slices. The physical correction at a qubit is the componentwise modulo-two sum of the correction registers at that qubit over all slices.

        As in Section~\ref{sec:online-decoding}, we define the update in terms of defect candidates. During the update of slice $k$, every defect first increments its timer. A defect whose timer expires contributes a candidate at the same site and component of slice $k+1$ with timer $0$ and takes no further part in the update. A defect that triggers a move is consumed and contributes one candidate, carrying its incremented timer, at each of its children; an untriggered defect contributes a stationary candidate carrying its incremented timer. A site and component then hosts a defect if and only if it receives an odd number of candidates, with timer the minimum over those candidates; even numbers of candidates pair-annihilate.

        Messages grow and persist according to the rule of the previous subsection, but their timers are updated using the message-candidate convention of Section~\ref{sec:online-decoding}: each defect present after the defect-sector update sources a candidate carrying its timer; each message whose timer has not expired contributes candidates carrying its incremented timer; and a message is present if and only if at least one candidate is generated, with timer the minimum over the candidates.

        The analysis of Section~\ref{sec:online-decoding} then goes through with the linear cluster erosion argument of the previous subsection in place of the corresponding repetition-, toric-, and surface-code erosion statements. In particular, by choosing $t_0$ and $n$ sufficiently large, every $k$-cluster of the spacetime cluster decomposition is eliminated---together with all defects and messages it creates, in every slice it reaches---before any of its defects can be promoted beyond slice $k$. Consequently, for $K = \Theta(\log L)$, there is a constant $c > 0$ such that all clusters of diameter at most $cL$ are decoded independently. The usual spacetime multiscale clustering argument then implies that, for every translation-invariant stabilizer code whose distance grows as $\Omega(L^{\alpha})$ for some $\alpha > 0$, this decoder has a non-zero threshold under phenomenological noise and achieves a stretched-exponential memory lifetime using only $\poly(\log L)$ classical bits per site.

\section{Acknowledgments}\label{sec:acknowledgments}

    We are grateful to Jeongwan Haah for suggestions on how to extend our constructions for codes with conventional topological order, the X-cube model, and the cubic code to general translation-invariant stabilizer codes using the polynomial formalism. We thank Chris Pattison for conversations about his related, independent work~\cite{pattison_fault_tolerant}. We also thank Ehud Altman, Michael Beverland, Cameron Chang, Margarita Davydova, Gesa D\"unnweber, Hsin-Yuan Huang, Jessica Jiang, Curt von Keyserlingk, Vedika Khemani, Yaodong Li, Xinyu Liu, Sid Parameswaran, Benedikt Placke, Shengqi Sang, Dave Schuster, Jessica Wang, and Victor Wei for valuable discussions and feedback. We used GPT~5.6 Sol to search the literature for existing results on the properties of weight orders. We also used GPT~5.6 Sol and Claude Fable~5 to help us review and revise our paper and prepare figures. GPT~5.6, GPT~6, and Claude Fable~5 were used to assist with numerical simulations.

	A.B. was supported by the Kortschak Scholars Program.
	E.L. was supported by a Miller Research Fellowship. 
	N.S. was supported by the National Science Foundation Graduate Research Fellowship Program under Grant No. 2146752 and the National Defense Science and Engineering Graduate Fellowship Program. This material is based upon work supported by the Air Force Office of Scientific Research under award number FA9550-25-C-B010 in the amount of \$140,900. This work used the DeltaAI system at the National Center for Supercomputing Applications [award OAC 2320345] through allocation PHY250214 from the Advanced Cyberinfrastructure Coordination Ecosystem: Services \& Support (ACCESS) program, which is supported by National Science Foundation grants \#2138259, \#2138286, \#2138307, \#2137603, and \#2138296. Any opinions, findings, and conclusions or recommendations expressed in this material are those of the authors and do not necessarily reflect the views of the National Science Foundation.

\bibliographystyle{unsrtnat}
\bibliography{refs}
\end{document}